\documentclass[11pt]{article}
\usepackage{amsmath, amssymb}
\usepackage{slashed}
\usepackage{verbatim}
\usepackage{latexsym}
\usepackage{euscript}
\usepackage{amsfonts}
\usepackage{verbatim}
\usepackage{cancel}
\usepackage{soul}
\usepackage[normalem]{ulem}
\usepackage[]{array}
\usepackage[]{mathrsfs}
\usepackage[utf8]{inputenc}
\usepackage[T1]{fontenc}
\usepackage{makeidx}
\makeindex
\usepackage{graphicx}
\usepackage{amsmath}
\usepackage{amssymb}
\usepackage{amsxtra}
\usepackage{color}
\def\be{\begin{eqnarray}}
\def\ee{\end{eqnarray}}
\def\0{\nonumber}
\def\d{\partial}
\def\tr{{\rm tr}} 
\def\Tr{{\rm Tr}}

\def\det{{\rm det}}

\def\sfg{{\sf g}}

\def\sfT{{\sf T}}
\usepackage{slashed}
\usepackage{verbatim}
\usepackage{latexsym}\usepackage{color}
\usepackage{euscript}
\usepackage[curve]{xypic}
\newcommand\EW{\EuScript{W}}
\newcommand\EG{\EuScript{G}}
\newcommand\ET{\EuScript{T}}
\newcommand\ER{\EuScript{R}}

\newcommand\ED{\EuScript{D}}
\newcommand\EF{\EuScript{F}}
\newcommand\EB{\EuScript{B}}
\newcommand\EC{\EuScript{C}}
\newcommand\EP{\EuScript{P}}
\newcommand\EV{\EuScript{V}}
\newcommand\EE{\EuScript{E}}
\def\mfs{{\mathfrak s}}

\def\sfP{{\sf P}}

\def\sfG{{\sf G}}
\def\sfM{{\sf M}}

\def\sfT{{\sf T}}
\def\sfD{{\sf D}}

\def\sfd{{\sf d}}\def\sfg{{\sf g}}
\def\sfm{{\sf m}}
\normalfont\large
\begin{document}
\vskip 2cm
\begin{flushright}
{SISSA/20/2024/FISI}
\end{flushright}
\vskip 2cm
\begin{center}

{\LARGE Something Anomalies can tell about SM and Gravity }

\vskip 1cm

{\large  L.~Bonora$^{a}$\footnote{email:bonora@sissa.it}, S.~Giaccari$^{b}$\footnote{email:s.giaccari@inrim.it},\\
\textit{${}^{a}$ International School for Advanced Studies (SISSA),\\Via
Bonomea 265, 34136 Trieste, Italy}\\ 
\textit{${}^{b}$Istituto Nazionale di Ricerca Metrologica,\\ Strada delle Cacce 91,
10135 Torino, Italy. }
}

\end{center}
\vskip2cm
{
{\bf Abstract}. This ia a { review/research paper on} anomalies applied to a bottom-up approach to standard model and gravity. It is divided in two parts. The first consists in a review  proper of anomalies in quantum field theories. Anomalies are analyzed according to three different methods: a perturbative one based on Feynman diagram, a non-perturbative one relying on the Schwinger-DeWitt approach and, third, the one hinging on the Atiyah-Singer family's index theorem. The three methods are applied both to chiral gauge anomalies and trace anomalies. The fundamental distinction that our presentation leads to is between obstructive (O) and non-obstructive (NO) anomalies. The former are tied to the non-existence of fermion propagators, which fatally maim the corresponding theory. In the second part we apply this analysis to the SM and various of its extensions immersed in a gravitational background, and find that they all are plagued by a residual chiral trace anomaly. To completely eliminate all kind of dangerous anomalies in SM-like theories we propose a somewhat unconventional scheme, and exemplify it by means of an explicit model.
The latter is a left-right symmetric model. We embed it in a Weyl geometry to render it conformal-invariant. We then deal with some of its quantum aspects, in particular its even (NO) trace anomalies and the means to preserve its confomal invariance at the quantum level. We briefly review renormalization and unitarity in the framework of similar models discussed in the existing literature. Finally we present a possible (conjectural) application of the model to describe the junction between cosmology and quantum field theory.}
\eject

\tableofcontents
\section{Introduction}

The quantum field theory called the standard model (SM) of particle physics describes the 5\% of the energy/matter of the universe made of baryons and light. Its neutrino sector is still under construction, the dark matter and dark energy sectors are out of its reach, but there is no doubt that the strong, weak and electromagnetic interactions of standard matter find a satisfactory explanation within it. This is well as long as gravity is ignored or, at least, recedes to a remote enough distance that is assumed not to disturb these interactions, and forms a realm of its own described by the so-called standard model of cosmology ($\Lambda$CDM). This situation is clearly unsatisfactory, and the attempts at bridging the gap between the two pictures have abounded since the early times: clearly it is at present the main theoretical physics puzzle, and includes, in particular, the challenge of quantizing gravity. There are two general attitudes in this regard: the top-down one (superstring theories are the prominent examples) and the bottom-up. { In the approach based on a given superstring theory, for example, one starts from a ten dimensional geometric configuration possibly including branes, one considers the field theory (low energy) limit and compactifies to four dimensions; the final field theory includes matter and gravity; the bet is to be able to carefully choose the compactification in such a way as to reproduce the SM spectrum in the final theory; the latter turns out to be automatically UV complete. A bttom-up approach means, instead, starting} from the SM and its classical interaction with gravity to build up step by step a larger effective quantum theory that encompasses all the interactions. Here the key word
is `effective'. We imagine the ultimate theory as a complete quantum field theory that describes the four basic interactions being unitary - that is, it conserves probability - and renormalizable (possibly finite), or more generically, being predictive; but being also able to explain those aspects of the universe which we are so far able to describe phenomenologically, but whose origin and link with ordinary matter and gravity remain mysterious: big bang, inflation, dark matter, dark energy, baryon asymmetry, etc.. No such quantum field theory is yet available. And we have to make do with approximations of it, that is with quantum field theories larger than the SM, comprehending in some form gravity, and able to explain one or more of the above mentioned pieces of cosmology. 

The bottom-up approach in constructing such an effective field theory starting from the SM, parallels the   inverted time evolution of the universe. The SM and classical gravity are probably satisfactory description of the universe below the TeV energy, or after a time of  around $10^{-12}sec$ from the beginning. When the quantum regime of gravity began we do not know, but it must have been much earlier than this. There is a vast stretch in logarithmic time where many phenomena must have taken place, such as baryogenesis, dark matter generation, inflation. All  this part of the history of the universe is still largely conjectural. Trying to reconstruct it is like driving a car in reverse in a dark night while relying only on the information gathered from the headlights. Any tiny bit of information is precious. The purpose of the present paper, { which is in part a review of existing literature and in part an original elaboration}, is to direct the headlights on, and scrutinize, the conditions that make a matter theory like the SM (or an extension thereof) consistent with the (minimal) coupling to gravity. We will find information that has not been sufficiently considered so far, and will inject it { in an attempt to construct} an effective scheme (rather than a specific model) of quantum matter in interaction with quantum gravity. This extra input comes from the analysis of anomalies that arise when matter like that of the SM couples to gravity. { As the title say, we wish to expose how anomalies can help us  in the bottom-up approach.} In quantum field theories there appear two large classes of anomalies, which, short of imagination, we dub type O (O stands for obstructive) and type NO (non-obstructive). The type O is the dangerous kind of anomalies, in that, when present, they compromise renormalization and/or unitarity. In fact, as we shall see, the reason is very down to earth: these anomalies appear in theories with Weyl fermions {(with chiral fermions, in general)}, { they are all odd-parity, and include the so-called consistent chiral gauge and diffeomorphim anomalies and the consistent odd-parity trace anomalies.} They signal the fact that the corresponding fermion propagators do not exist. { Type NO anomalies are not dangerous, they do  not signal any irreparable flaw of the theory and, in general, need not be canceled, in particular the relevant propagtors do exist (possibly after gauge fixing). They include the even-parity trace anomalies and the ABJ anomalies in theories of Dirac fermions.}

{
The first main point stressed by this paper is that the minimal standard model (MSM)  along with its extensions (see below) is free of type O anomalies, as long as we can disregard gravity. But this is not true anymore if the MSM is coupled to gravity in a minimal way. Some type O anomalies, which seem to have been disregarded so far, do not vanish. They are odd-parity (chiral) trace anomalies, which belong to the type O group. The second main purpose is to present a possible type O-anomaly-free scheme, which includes extensions of the SM, and discuss in this setup the role of even-parity trace anomalies, which instead belong to the set of type NO set. Accordingly the paper is divided into two parts, I and II. 

The distinction between type O and type NO anomalies is a basic one particularly in relation to the SM and its extensions. For this reason we have felt the necessity to return to the basics with a concise but  (hopefully) thorough presentation of anomalies and their calculations. This covers part I. We first review the definitions and calculations of the chiral gauge anomalies with perturbative methods based on Feynman diagrams, then with non-perturbative methods, of which we present  in particular the SDW (Schwinger-DeWitt) one. But these two methods, although fundamental for reasons that do not need to be stressed, are unable to unmask the true nature of type O anomalies and why and in what sense they differ from the type NO ones. A third approach is necessary, the one based on the family's index theorem. The latter tells us that type O anomalies are obstructions to the existence of the relevant Weyl fermion propagators, and this reveals their true nature. In particular, that is why they are `dangerous': if the fermion propagators do not exist, the concept  itself of (perturbative) quantization does not make sense. Our subsequent analysis of the SM and its extensions is based primarily on this fact, and on the complementary fact that type NO anomalies do not imply any lack of propagators. This second fact is sanctioned by the family's index theorem, for which Euclidean self-adjoint operators, which are the sources of type NO anomalies, have vanishing family's index, and so do not represent obstructions.

This is the basic distinction to be always kept in mind. It is therefore of paramount importance to recognize type O and type NO anomalies. The well-known consistent gauge anomalies in theories of chiral fermions are type O and their absence in a theory is the first requisite for its consistency. But there are other type O anomalies that correspond to non-vanishing obstructions exposed by the family's index theorem: they are the chiral trace anomalies in theories of Weyl fermions. We devote a sizable portion of part I to derive, by means of different methods, this kind of trace anomalies. Once this is done, we apply the results to the minimal version of the SM and to a set of its extensions, minimally coupled to gravity. It turns out that in all these models there survive some residual chiral trace anomalies, more precisely the chiral trace anomalies generated by the $SU(2)$ gauge fields. This result concludes part I.

In part II we propose a scheme in order to cancel the above anomalies for SM extensions. The cancelation of all type O anomalies requires  a left-right symmetry of the spectrum of fermions, which is however quite different from the L-R extension models present in the literature in that the fermionic L-R symmetry is accompanied by a doubling also of the $SU(3)$ and $U(1)$ gauge fields (but not of the $SU(2)$ ones), by a doubling of the scalars and also by a doubling of the metric (which, incidentally, makes it a bimetric theory).
The resulting theory is denoted $\cal T$. Its formulation becomes remarkably simple if one uses as a bookkeeping device, the AE (axially extended) formalism, which is based on a version of the so-called hypercomplex analysis. This is based on a doubling of variables analogous to the passage from the real to the complex numbers, with the imaginary unit $i$ being replaced by the matrix $\gamma_5$. The action of $\cal T$, without Yukawa couplings, is invariant not only under diffeomorphisms and gauge transformations, like any ordinary extension of the { SM minimally coupled to} gravity, but under doubled AE diffeomorphisms, i.e. diffeomorphisms having two components, the ordinary one and another represented by the coefficient of $\gamma_5$. The chiral symmetry of $\cal T$ splits it into two: ${\cal T}= {\cal T}_R \cup {\cal T}_L$, that are formulated in terms of multiplets of right-handed and left-handed fermions, respectively; but also on sets of `right' and `left' metrics and $SU(3)$ and $U(1)$ gauge fields and scalar fields, respectively, with the only exception of the $SU(2)$ gauge fields, which are in common to both  ${\cal T}_R$ and ${\cal T}_L$. It is important to bear in mind that the pervasive presence of the chiral projectors does not allow the `right' fields to interact directly with the `left' fields , except through the  $SU(2)$ gauge fields.

At this stage of the construction of an anomaly free version of the SM coupled to gravity, invariant under gauge transformations and diffeomorphisms, we face a problem which is common to all analogous attempts.
The SM above the electroweak symmetry breaking energy, is formulated in terms of massless fermion and gauge fields, which as such are naturally coupled to a metric. It is true that the scalars are massive, but at high enough energy these masses can be considered irrelevant. In this situation the absence of dimensionful parameters, the coupling to a metric and the addition of a very simple term, make the action invariant under local Weyl trasformations. The question therefore is whether Weyl invariance has to be subsumed under the local symmetry of the theory, by enlarging it beyond gauge transformations and diffeomorphisms. We believe that this has to be decided in terms of unitarity and renormalizability (predictivity) of the theory. In that sense we do not have a definite answer to the question. After formulating $\cal T$ as $\cal TW$, i.e. as a Weyl conformal field theory, which is a rather simple matter, we observe that Weyl invariance is  a rather peculiar gauge symmetry, in that it admit a minimal realization which does not require the presence of a corresponding vector gauge field; the function of the latter can be carried out by a scalar field which can be identified with the dilaton. We perform the quantization without fixing the conformal gauge; there is no obstruction in that for no propagator is missing without gauge fixing. Then we study the Ward identity of the Weyl symmetry and connect it with the study of trace anomalies. We determine, at least in principle, all the latter and show that they can be canceled by means of Wess-Zumino terms. The obstacles however come from renormalization and unitarity. If we want to oblige the renormalization protocol we must introduce in the action also terms like the square Weyl tensor one, which carries in the particle spectrum a physical ghost, i.e. a particle with wrong sign of the kinetic term, thus affecting unitarity. It would seem that Weyl invariance is an excessive request for our theory $\cal TW$, at least as it is formulated in this paper.

An alternative possibility, is to consider the Weyl symmetry an accidental symmetry that may appear in approximate form at some stage of the history of universe, in any case a non-necessary symmetry in the process of quantization. In this case it seems that the process of renormalization can go on while preserving unitarity. 

But, perhaps, these two alternatives should be thought in sequence, in the sense that they describe two ranges of energies: at lower energy (anyhow higher than the electroweak symmetry breaking scale) the second is a satisfactory description, while at higher energy we have to resort to the first, which requires Weyl invariance. Its lack of unitarity could signal that some degrees of freedom are missing for the theory to be UV complete.

More research is certainly necessary to find a precise answer. But if we do not have an answer in the direction of higher energies, we can perhaps speculate { further} in the opposite direction, that is toward lower energies and the SM in its familiar form. First of all we fix the conformal gauge in the right part of the theory, i.e. in ${\cal TW}_R$, by fixing the corresponding dilaton to a constant value. At this stage the theory has the aspect of the SM minimally coupled to gravity except for a couple of additional (non-minimal) couplings of the Ricci tensor to the scalars. Next we postulate an early breaking of the chiral symmetry (a {\it primitive chiral symmetry breaking}), by means of a Higgs mechanism generated by the non-zero vev {(vacuumm expectation value)} of the real scalar field present in ${\cal TW}_R$. This endows all the `right' fermions and other `right' fields with a large mass, while the ${\cal TW}_L$ remains with all the original symmetries.   Clearly this determines different evolutions for the `left' and `right' theories, which are however not completely unrelated. The point is that they keep interacting, although weakly, via the common $SU(2)$ gauge fields {and Higgs.} The `right' one becomes a theory of heavy quarks and leptons, while the `left'  evolves toward a theory encompassing classical gravity and the SM, together with a real scalar field which can realistically play the role of inflaton. Due to the weak interaction among the two `theories it is natural to think of the `right' as a candidate for the dark matter world. 

As long as all the above is not supported by concrete calculations, it is nothing more than speculations. But fashinating speculations, and not unreasoable.

}
\vskip 0.5cm

\subsection{The standard model}
 
In absence of gravity the world of (visible) matter and light is well described, up to around the TeV energy, by the standard model (SM) of elementary particles. It is a gauge theory with gauge group $SU(3)\!\times\! SU(2)\! \times\! U(1)$. The latter is the symmetry in the unbroken phase, when all the quarks and leptons are massless. The $SU(3)$ gauge bosons mediate the strong interactions. The $SU(2)\!\times\! U(1)$ ones carry the electroweak interactions. In the symmetric phase the spectrum of massless quarks and leptons is specified by three families like the one below,
\be
\begin{matrix} {\sf G}/fields & \quad SU(3)\quad &\quad SU(2)\quad &\quad U(1)\quad\\
\left( \begin{matrix} u \\ d\end{matrix} \right)_{\! L} & 3&2&\frac 16\\
{u_R} &  3 &1&\frac 23\\
{d_R} &  3 &1&-\frac 13\\
\left( \begin{matrix} \nu_e \\ e\end{matrix} \right)_{\! L} & 1&2&-\frac 12\\
 {e_R} &  1 &1&-1
\end{matrix}\label{MSMspectrum}
\ee
The second column specifies the relevant representations of $SU(3)$, the third the ones of $SU(2)$ and the last is the list of $U(1)$ representations, which are denoted by the corresponding hypercharge eigenvalues\footnote{A frequent alternative notation is to use the Lorentz covariant conjugates $ {(u_R)^c}, {(d_R)^c}$ and ${(e_R)^c}$ instead of $u_R,d_R$ and $e_R$, in order to collect all the fields in a unique left-handed multiplet. In this case one has to reverse the signs of their $U(1)$ charges and replace the  representation 3 of $SU(3)$ of $ {u_R}, {d_R}$  with the $\bar 3$ of $ {(u_R)^c}, {(d_R)^c}$.}.  We recall that the hypercharge is defined by
\be
Y=Q-T_3, \quad\quad T_3= \frac 12 \left(\begin{matrix} 1&0\\0&-1 \end{matrix}\right), \label{Y}
\ee
here $Q$ is the electromagnetic charge and $T_3$ the third generator of $SU(2)$ in the doublet representation. { The SM spectrum is completed by the sector of Higgs scalars.} 

The fermion spectrum in \eqref{MSMspectrum} is not the only possible one, it refers to the original formulation of the SM, referred to as the minimal standard model (MSM), which is anyhow the basis of the subsequent (attempted) modifications. We shall discuss several of the latter in the course of the paper, but, for the moment, we focus on \eqref{MSMspectrum}. At suitable low energies the gauge symmetry of the model breaks down to the local $U(1)$ of electromagnetism, and this is achieved thanks to the presence in the model of the Higgs doublet of scalars, which is responsible for the breakdown of symmetry and simultaneous generation of the fermion masses.  For the moment we restrict our consideration to the symmetric phase and to the fermionic spectrum \eqref{MSMspectrum}.

One of the most important criteria followed in the construction of the SM has been the absence of chiral gauge anomalies. This is a well-known fact and, if we review here some well-known material, is because it is necessary for a full appreciation of other less known aspects. Needless to say, the subject of gauge anomalies, after more than fifty years of analyses, papers and attempts still seems to preserve a halo of mystery. But, as anticipated in the introduction, the reason why anomalous theories, i.e. theories infected by type O anomalies, are sick and cannot be taken into serious consideration is not due to a mysterious spell or curse of some malignant jinn, the reason is instead very clear and simple: in such theories, which are theories of Weyl fermions\footnote{{ The same problem exists also for other chiral fermions, for instance for the gravitino, but in this paper we limit ourselves to spin 1/2 chiral fermions.}}: the corresponding fermion propagators do not exist. Anomalies are simply an indirect manifestation of this fact. This is the origin of the conundrum, which explains why the construction of consistent quantum field theories in such cases is impossible. But let us proceed step by step. Needless to say, the reader familiar with this subject may skip this part.

\vskip 0,5cm

\vskip 1cm
\noindent Refs.: \cite{abers1973,bambi2021,baulieu2017,Bertlmann,I,castelo22,Cvetic2022,griffiths1987,
Itzykson,novaes2000,peskin1995,polchinski,Quevedo2024,Weinberg1972,Weinberg2005} .

{ 

\subsection{A summary}

The paper is divided in two parts. The first part is a review of anomalies in field theories. Section 2 is devoted first to the  simplest calculation of chiral gauge anomalies in 4$\sfd$ with perturbative methods (Feynman diagrams). In the next subsection we review the method of Bardeen to arrive at the same result by introducing an auxiliary axial potential. Then we introduce the Wess-Zumino consistency conditions and the BRST cohomology. In the final subsection we show how the previously computed chiral anomalies cancel in the SM. In section 3 we obtain the same Bardeen's anomaly derived before with a non-perturbative method, the Schwinger-DeWitt (SDW) method. Section 4 is devoted to a third method of computing the same chiral anomalies based on the Atiyah-Singer family's index theorem. We  start with a geometrical presentation of anomalies relying on the geometry of principal fiber bundles. We show that their full comprehension requires the introduction of the relevant universal  bundles and classifying spaces. Then we introduce the Atiyah-Singer theorem and the Atiyah-Singer family's index theorem, which is required by the geometry of gauge theories. Next we apply this theorem to the problem of anomalies: the latter turn out to be obstructions to the existence of fermion propagators in theories of chiral fermions, as it clearly appears using the Quillen determinant. This section contains the crucial results for our identification of the meaning of chiral anomalies (both gauge and trace).  Section 5 is an expos\'e of gravitational anomalies.
In section 6 we introduce and compute gauge and gravity-induced odd-parity trace anomalies in theories of Weyl fermions. This first computation is carried out with perturbative methods. At this stage, in  section 7, we evaluate the odd-parity trace anomalies in the standard model minimally coupled to gravity and show that the gauge-induced ones do not cancel completely. In section 8 we recalculate the same trace anomalies by means of the SDW method. To this end we introduce { the MAT (metric-axial-tensor) formalism, wherein we redefine the usual metric tensor $g_{\mu\nu}$ by addint to it another symmetric tensor $f_{\mu\nu}$ multiplied by $\gamma_5$, which makes use of the axial-complex analysis (a form of hypercomplex analysis), see section 8.1 below.} By this method we compute various types of anomalies which may appear in matter theories coupled to gravity. In section 9 we return to the meaning of anomalies and explain the difference between O (obstructive) anomalies and NO (non-obstructive) anomalies. Section 10 is devoted  to our critical assessment of the existing literature on trace anomalies. Part I ends with the application  of the previous results to the SM and its extensions minimally coupled to gravity. In section 11 we compute the (gauge and trace) anomalies  in some of the extensions of the SM and verify that while they are all free of chiral gauge anomalies they all contain a residue of chiral trace anomalies.

Part II begins with a summary of this problem and an enumeration of a few way-outs which are not, however, our preferred choice. In section 13 we propose a scheme, illustrated with a concrete left-right  (or chirally) symmetric model\footnote{We use throughout {\it left-right} (L-R) and {\it chirally symmetric} as synonyms, with some abuse of language because the latter should refer only to fermions, while the former refers to all the fields.}, that completely cancels all the O anomalies. We then construct the full theory, dubbed $\cal T$, which includes the SM extension and gravity. It turns out convenient to use the axial-complex calculus, so that all pieces of the action are in the AE (axially-extended) form, i.e. left-right symmetric form.  Section 14 is devoted to a short review of Weyl geometry and Higgs mechanisms.
In section 15 we embed our model $\cal T$ in Weyl geometry and render it Weyl-invariant. Its conformal-invariant action is called $\cal TW$. In section 16 we start discussing aspects of $\cal TW$ as a quantum theory. Section 17 is devoted to a discussion of even trace anomalies, an introduction of  WZ terms and how to use the latter in order to eliminate all trace anomalies and restore the conformal symmetry at a quantum level.  Section 18 is a short review based on the existing literature on renormalization and unitarity in theories similar to $\cal TW$. Finally in section 19 we argue about the fate of the two halves, ${\cal TW}_L$ and ${\cal TW}_R$, after a spontaneous L-R symmetry breaking and how we may recover at lower energy the familiar minimal SM. 

A few Appendices concern the notations, our rules for the Wick rotation, gauge fixings, an example of trace anomaly due to the scalar fields and a discussion concerning the dilaton.}

\vskip 2cm
\section*{\LARGE \bf Part I}

 \vskip 2cm 
 \section{Review of anomalies (without gravity)}
 
In order to study the chiral anomalies in the massless SM it is not necessary to consider the full Lagrangian. The fermion kinetic terms are enough. Since anomalies are additive it is enough to consider one type of fermion at a time. Let us consider, for instance, the classical kinetic action for a right-handed Weyl fermion coupled to an external gauge field $V_\mu =V_\mu^a T^a$, $T^a$  {being} Hermitean generators, $[T^a, T^b ]= i f^{abc} T^c$ (in the Abelian case $T=1, f^{abc}=0$) in the fundamental representation of a compact group $\sfG$:
\be
S_R[V]= \int d^4x \,i\, \overline {\psi_R} \left( \slashed{\partial}-i \slashed{V}
\right)\psi_R.\label{actionSR}
\ee
{Here $\psi_R$ is a right-handed Weyl spinor, which can be projected out from a Dirac spinor $\psi$  by means of the chiral projector $P_R$, $\psi_R= P_R\psi $ (see Appendix A for the notation)}. This action is invariant under the gauge transformation $\delta \psi_R = - \lambda \psi_R$ and $\delta_\lambda V_\mu= D_\mu \lambda \equiv \partial_\mu \lambda -i [V_\mu,\lambda]$, where $\lambda= \lambda^a(x) T^a$, which  implies the classical conservation of the non-Abelian current  $J^a_{R\mu}=\bar \psi_R \gamma_\mu T^a \psi_R$, i.e.
\be
\left({D\cdot} J_R\right)^a \equiv (\partial^\mu \delta^{ac} +i f^{abc} V^{b\mu}){J_{R\mu}^c}=0.\label{currentconserv}
\ee
This is in general not true after quantization. Our purpose here is to compute such violations. This type of calculations have been done in a number of different ways. First we focus on the perturbative method  in which we treat $V$ as a perturbation and adopt the Feynman diagram technique.

\subsection{The perturbative approach}
\label{ss:perturbative}

The quantum effective action for this theory is given by the generating functional of the connected Green's functions of such currents  in the presence of the source $V^{a\mu}$
\be 
W[V]=W[0]+\sum_{n=1}^\infty \frac {i^{n-1}}{n!} \int \prod_{i=1}^n d{^4}x_i  {V^{a_i\mu_i}(x_i)}\,
\langle 0|{\cal T} J_{R\mu_1}^{a_1}(x_1)\ldots
J_{R\mu_n}^{a_n}(x_n)|0\rangle_c\label{WV}
\ee
and the  full {1}-loop {1}-point function of {$J_{R\mu}^a$} is
\be
\langle\!\langle{J_{R\mu}^a}(x)\rangle\!\rangle = \frac {\delta W[V]}{\delta
V^{a\mu}(x)} = \sum_{n=0}^\infty \frac {i^n}{n!} \int \prod_{i=1}^n d{^4} x_i
{V^{a_i\mu_i}(x_i)} 
\langle 0|{\cal T}J_{R\mu}^a(x) { J_{R\mu_1}^{a_1}(x_1)...J_{R\mu_n}^{a_n}(x_n)} 
|0\rangle_c \label{Jamux}
\ee
 We wish to calculate the odd-parity anomaly of the divergence ${ D\!\cdot\! \langle\!\langle J_R^a\rangle\!\rangle} $, where $D$ represents the covariant derivative defined in \eqref{currentconserv}. The RHS of \eqref{Jamux} contains an infinite series of current amplitudes. The one-point amplitude vanishes  because of translational invariance, the two-point amplitude is non-vanishing, but does not contribute to the odd-parity anomaly. Therefore the first non-trivial contribution comes from the three-point function, which, in terms of Feynman diagrams is represented by the famous triangle diagram, whereby each side represents a fermion propagator and each vertex an interaction vertex. 
 
Here comes the complication, because the Weyl fermion propagator, the inverse of the kinetic operator $i\slashed{\partial}P_R$, simply does not exist. For $P_R$ is a projector and it is not invertible, while $i\slashed{\partial}$ is, but contains both a right-handed and a left-handed part\footnote{It s worth spending a few words to avoid misunderstandings on this issue: this is not the non-invertibility due to the presence of some zero modes of the kinetic operator, a case which is anyhow relevant only to nonperturbative calculations and can be handled by subtracting the zero modes; in the present case we have to do with an operator whose domain and image are different spaces and for which the eigenvalue problem is not even defined.}. 

We can dodge this obstacle by an escamotage: instead of  $i\slashed{\partial}P_R$ we invert  $i\slashed{\partial}$, i.e. we replace the missing propagator with the Dirac propagator. We remark, however, that this corresponds to adding a term $i{\psi_L} \slashed{\partial}\psi_L$ to the integrand of \eqref{actionSR}, which corresponds to adding a {\it free} left-handed Weyl fermion. Whether we have to pay a price or not for this escamotage, it allows us to proceed in the calculation. In momentum space the propagator of a fermion with momentum $p_\mu$ is $\frac i{\slashed {p}+ i\epsilon}$ and the vertices have the form $i\gamma_\mu P_R T^a$. We associate an entering momentum $q$ to the current $j_{R\mu}^a T^a$ to be differentiated and  exiting momenta $k_1$ and $k_2$ to $j_{R\lambda} ^b T ^b$ and $j_{R\rho}^c T^c$; of course, $q=k_1+k_2$. The three-current amplitude factorizes into the $\Tr(T^a T^b T^c)$ factor times the amplitude for three Abelian currents. We forget for the moment the trace factor and proceed with the computation of the Abelian amplitude. In momentum space it is 
\be
q^\mu \widetilde F_{\mu\lambda\rho}^{(R)}(k_1,k_2)\!\! =\!\! \int\!
\frac{d^4p} {(2\pi)^{4}}\mathrm{tr}\left\{\frac{1}{\slashed{p}}     \gamma_{ {\lambda}}
\frac{1+\gamma_5}{2}\frac{1}{\slashed{p}
-\slashed{k}_1} \gamma_{\rho}\frac{1+\gamma_5}{2}
\frac{1}{\slashed{p}-\slashed{q}
}\slashed{q} \frac{1+\gamma_5}{2}\right\}\label{qtriangle}
\ee
The integral over the momentum $p$ running around the loop is UV divergent. It can be dealt with, for instance, by transforming it into a Euclidean integral by means of a Wick rotation, then dimensionally regularizing it and, finally, rotating it back to a Minkowski expression { (see Appendix B for our conventions concerning Wick rotations)}. All of this is well-known. Here we limit ourselves to writing down the result for the odd-parity part. 
 \be
\widetilde F^{(R,odd)}_{\lambda\rho}(k_1,k_2)+ \widetilde F^{(R,odd)}_{\rho\lambda}(k_2,k_1)= \frac 1{12\pi^2}\varepsilon_{\mu\nu\lambda\rho}k_1^\mu k_2^\nu.\label{totaltriangledim0}
\ee
where the second term in the LHS comes from the crossed diagram, which has to be added according to the Feynman rules. 
 
From \eqref{totaltriangledim0}, and ignoring again the group theoretical factor, we can obtain the final result for the purely Abelian case. It is enough to Fourier anti-transform it and insert it in \eqref{Jamux} 
\be
 \partial^\mu\langle\!\langle J_{R\mu}(x)\rangle\!\rangle\!\!&=& \!\!\int\! \frac {d^4q}{(2\pi)^4} e^{-iqx}
\,(- iq^\mu)  \langle\!\langle\tilde  J_{R\mu}(q)\rangle\!\rangle\0\\
&=&-\frac 1 {24\pi^2} \!
 \int \!\frac {d^4q \, d^4k_1\, d^4k_2}{(2\pi)^{12}}
\int\! d^4y d^4 z\, e^{i \left(qx-k_1y-k_2z\right)}\delta(q\!-\!k_1\!-\!k_2)\varepsilon_{\mu\nu\lambda\rho}
k_1^\mu k_2^\nu \, V^\lambda(y) V^\rho(z)\0\\
&=& \frac 1 {24\pi^2} \varepsilon_{\mu\nu\lambda\rho}
 \,\partial^\mu V^\nu(x) \partial^\lambda V^\rho(x).\label{divJRcons}
\ee

\vskip 0.5cm

In the Abelian case the amplitudes \eqref{totaltriangledim0} are all one needs in order to derive the complete result. The analogous anomaly  in the non-Abelian case would require the calculation of  the four-current correlators, but it can be obtained in a simpler way from the Abelian term multiplied by the group-theoretical factor, by means of the Wess-Zumino consistency conditions (see below). In the non-Abelian case, however, it is more expedient to use anti-Hermitean generators, $(T^a)^\dagger = -T^a$ and $[T^a, T^b ]= f^{abc} T^c$, and from now on we will stick to this convention (although the Hermitean convention is more familiar in particle physics) {unless otherwise specified.}
As pointed out before, in the non-Abelian case the three-point correlators are multiplied by
\be
{\rm Tr} (T^a T^b T^c) = \frac 12 {\rm Tr}(T^a [T^b, T^c]) + \frac 12 {\rm Tr}(T^a \{T^b, T^c\})= f^{abc}+ d^{abc}\label{TaTbTc}
\ee
where the normalization used is ${\rm Tr} (T^a T^b)= \delta^{ab}$. Since the three-point function is the sum of two equal pieces with $\lambda \leftrightarrow \rho, k_1 \leftrightarrow k_2$, the first term on the RHS of \eqref{TaTbTc} drops out and only the second remains.
For the right-handed current $J^a_{R\mu}$ we have
\be
\left({D\cdot
\langle\!\langle J_R\rangle\!\rangle}\right)^a = \frac 1{24 \pi^2} \varepsilon_{\mu\nu\lambda\rho}
{\rm Tr}\left[T^a\partial^\mu
\left(V^\nu\partial^\lambda V^\rho+\frac 12 V^\nu V^\lambda
V^\rho\right)\right].\label{consanom}
\ee

{\bf Remarks.} If the fermion is left-handed, instead of right-handed, the anomalies \eqref{divJRcons}and \eqref{consanom} simply change sign. It  follows that for a Dirac fermion, which can be written as the sum of a left-handed and a right-handed Weyl fermion, the consistent gauge anomalies vanish. The same is true for a Majorana fermion, which can also be expressed as the sum of a left-handed and a right-handed Weyl fermion. 

These anomalies affecting Weyl fermions (the consistent chiral gauge anomalies)  are dangerous, in fact deadly, for a theory. A simple example of such a catastrophe is represented by a right-handed fermion coupled to an Abelian gauge field, thus with the anomaly \eqref{divJRcons}.
To see it recall that the Lorentz invariant quantum theory of a gauge vector field inevitably involves a Fock space of states with indefinite norm. In order to select a physical Hilbert subspace of the Fock space, a subsidiary condition is necessary. In the Abelian case, when the fermion current satisfies the continuity equation, the equation of motion leads to  $ \square(\partial\cdot V)=0 $, so that a subspace of states of non-negative norm can be selected through the auxiliary condition
\be
\partial\cdot V^{(-)}(x)\,\vert\,\mathrm{phys}\,\rangle=0\label{dV-=0}
\ee
$V^{(-)}(x)$ being the {annihilation operator, the} positive frequency part of the quantum field.
{On the contrary}, in the above-mentioned chiral model one finds
\be
\square(\partial\cdot V) \sim \varepsilon_{\mu\nu\lambda\rho}
 \,\partial^\mu V^\nu(x) \partial^\lambda V^\rho(x)\neq 0\label{squaredVneq0}
\ee
This is an obstruction to selecting a physical subspace of states with non-negative norm.

\subsection{Bardeen's method}
\label{ss:Bardeen}

Eqs. \eqref{divJRcons} and \eqref{consanom} are the general form of  the consistent Abelian and non-Abelian chiral gauge anomalies in $4\sfd$. But before discussing the Wess-Zumino consistency conditions, let us return to the question of the non-existence of the Weyl propagator, which we have bypassed above by adding to the action of a right-handed Weyl fermion interacting with a gauge potential the action of a left-handed non-interacting fermion. There is another way to overcome this obstacle, Bardeen's way. It consists in a stratagem often resorted to in field theory: when a calculation is too contrived, one had better enlarge the  parameter space of the theory. Bardeen's way, \cite{bardeen1969}, consists of adding new degrees of freedom by considering a theory of Dirac fermions coupled  to a vector potential as well as to an axial one, making the calculation in such a larger setup and taking a specific limit to return to the initial configuration.

Thus let us consider a theory of Dirac fermions coupled to non-Abelian vector and axial potential
\be
S_B[V,A]= \int d^4x \,i\, \overline {\psi} \slashed{\mathscr D}\psi.\label{actionBardeen}
\ee
where $\slashed{\mathscr D}= \slashed{\partial} +\slashed{ V} + \gamma_5 \slashed {A}$. The evident advantage is that now the propagator exists and, perturbatively, corresponds to the usual Dirac propagator. Moreover, taking the limit $V\to V/2, A\to V/2$ in \eqref{actionBardeen} we recover the action of  a right-handed Weyl fermion coupled to $V_\mu$, plus the action of a free left-handed Weyl fermion.

The action\eqref{actionBardeen} is invariant under the gauge transformations
\be
\delta\psi = -(\lambda +\rho\gamma_5)\psi\label{deltapsiAV}
\ee
and
\be
\delta V_\mu = \partial_\mu \lambda +[V_\mu,\lambda]+[A_\mu,\rho],\quad\quad
\delta A_\mu=\partial_\mu \rho +[V_\mu,\rho]+[A_\mu,\lambda]\label{deltaAV}
\ee
where the local functions $\lambda$ and $\rho$ are Lie algebra-valued, $\lambda = \lambda ^a(x) T^a,\,\, \rho = \rho^a(x) T^a$. This implies the classical conservation of the vector and axial current, $j_\mu ^a=i\bar\psi \gamma_\mu T^a\psi,\, j_{5\mu} ^a=i\bar\psi \gamma_\mu \gamma_5 T^a\psi$
\be
(D_V^\mu j_\mu +[A^\mu, j_{5\mu}])^a =0,\quad\quad (D_V^\mu j_{5\mu} +[A^\mu, j_\mu])^a =0 \label{classicalDVA}
\ee
where $D_V$ denotes the covariant derivative with respect to the vector potential.

To compute the anomalies in this case we can proceed as above, via Feynman diagrams, i.e. compute the triangle diagrams $\langle j_5 j j\rangle$ and $\langle j_5 j_5 j_5\rangle$ (the other  distinct three-current amplitudes cannot contribute to the odd-parity anomaly), determine the lowest order contribution to the anomaly and then use the consistency conditions to reconstruct its full expression. It can be done, but it certainly is not the quickest way. Another possibility is to  use a non-perturbative method, see below. Here we anticipate the result. After subtracting trivial terms, if need be, we get vector current conservation
\be
\left(D_V^\mu \langle\!\langle j_\mu(x)\rangle\!\rangle + [A^\mu(x),\langle\!\langle j_{5\mu}(x)\rangle\!\rangle]\right)^a =0\label{vectcons}
\ee
while the axial conservation equation is anomalous:
\be
&&\left(D_V^\mu \langle\!\langle j_{5\mu}(x)\rangle\!\rangle + [A^\mu(x),\langle\!\langle j_\mu(x)\rangle\!\rangle]\right)^a 
  =
\frac 1{4\pi^2} \varepsilon_{\mu\nu\lambda\rho} \tr\left[T^a\left(
\frac 14 F_V^{\mu\nu} F_V^{\lambda\rho} +\frac 1 {12}  F_A^{\mu\nu}
F_A^{\lambda\rho}\right.\right.\0\\
&&\quad\quad\quad-
\frac 16  F_V^{\mu\nu}A^\lambda A^\rho -\left.\left. \frac 16 A^\mu A^\nu F_V^{\lambda\rho} -\frac 23 A^\mu
 F_V^{\nu\lambda} A^\rho-\frac 13 A^\mu A^\nu A^\lambda
A^\rho\right)\right]\label{Bardeenanom}
\ee
where $F_V^{\mu\nu}= \partial^\mu V^\nu -\partial^\nu V^\mu+[V^\mu, V^\nu]$, and
$  F_A^{\mu\nu}= \partial^\mu A^\nu -\partial^\nu A^\mu+[V^\mu, A^\nu]+ [A^\mu,
V^\nu]$. This is Bardeen's anomaly.

From this expression we can derive two results in particular. Setting $A_\mu=0$
we get the so-called covariant  or ABJ anomaly
\be
[D^\mu_V j_{5\mu}]^a =\frac 1 {16\pi^2}  \varepsilon_{\mu\nu\lambda\rho}
 \tr\left(T^a V^{\mu\nu} V^{\lambda\rho}\right)\label{covanom}
\ee
Taking the chiral limit $V\to \frac V2, A\to \frac V2$, and summing
\eqref{vectcons} and \eqref{Bardeenanom}  we get
\be
[D_{V\mu} j^\mu_R]^a= \frac 1{24 \pi^2} \varepsilon_{\mu\nu\lambda\rho}
\tr\left[T^a\partial^\mu
\left(V^\nu\partial^\lambda V^\rho+\frac 12 V^\nu V^\lambda
V^\rho\right)\right]\label{consanom1}
\ee
where $j^a_{R\mu}= i\overline \psi_R \gamma_\mu T^a \psi_R$ with $\psi_R = \frac
{1+\gamma_5}2 \psi$, which is the same consistent non-Abelian gauge anomaly as \eqref{consanom}. 

{\bf Remarks}. { It is worth recalling that both \eqref{Bardeenanom} and \eqref{consanom1} satisfy the appropriate WZ consistency conditions (see below), while \eqref{covanom} does not. The latter is called covariant because the RHS trasforms in the adjoint representation of the gauge Lie algebra. Of course the two anomalies \eqref{consanom1} and  \eqref{covanom} collapse to the same expression in the Abelian case, but the coefficients in front of them are always different.}

Bardeen's anomaly , the covariant anomaly \eqref{covanom} and the consistent anomaly \eqref{consanom1} are all proportional to the third order ad-invariant symmetric tensor $d^{abc}= \frac 12 \tr \left(T^a \{ T^b, T^c\}\right)$. Therefore they all vanish when the latter vanishes.

The covariant anomaly \eqref{covanom} is the anomaly of the chiral current $j_{5\mu}^a$ in a theory of a Dirac fermion coupled to a vector potential, in response to  the gauge variation $\delta\psi= \rho \gamma_5 \psi$. This anomaly cannot show up in the internal lines of Feynman diagrams, therefore it cannot endanger unitarity or renormalizability\footnote{  We reserve the term ABJ anomaly only to this type. It is rather frequent  in the literature (and a source of confusion) to use the ABJ tag for any kind of chiral anomaly, consistent or covariant.}.

\subsection{WZ consistency conditions and BRST cohomology}
\label{ss:WZ-BRST}
At this point we owe the reader a few explanations. We have mentioned already several times consistent and covariant anomalies and their vanishing. It is time to clarify this terminology. 
Gauge invariance of the effective action can be expressed by means of the functional operator $X^a(x)$ defined by
\be
X^a(x) = \partial_\mu \frac{\delta}{\delta V^a_\mu(x)} +f^{abc} V_\mu^b (x)  \frac{\delta}{\delta V^c_\mu(x)},\label{Xax}
\ee
as follows
\be
X^a(x) W[V]=0\label{Xgaugeinvariance}
\ee
This is the full one-loop Ward identity (WI) for gauge symmetry. It can be obtained by expressing the invariance of the effective action as follows
\be
0= \delta_\lambda W[V] = \int d^{\sfd} x \, \delta_\lambda V_\mu^a(x) \frac {\delta}{\delta V_\mu^a(x)} W[V], \label{deltalambdaW}
\ee
integrating by parts and noting that $\lambda^a(x)$ are arbitrary functions. Eq.\eqref{Xgaugeinvariance} is equivalent to the one-loop covariant conservation law
\be 
[D^\mu \langle\!\langle j_\mu(x) \rangle\! \rangle]^a=0\label{Dmujmu}
\ee
In a number of cases the WI \eqref{Xgaugeinvariance} is violated
\be
X^a(x) W[V] = \kappa {\Delta}^a(x)\label{anomalousWI}
\ee
where $\Delta^a (x)$ is a local expressions of the fields and $\kappa$ is a small dimensionless expansion parameter for the one-loop calculation. From now on, for simplicity, we shall set it =1.

Applying $X^b(y)$ to both sides of \eqref{anomalousWI} and then inverting the order of the two operations, we find a remarkable relation of group-theoretical nature
\be
X^a(x) {\Delta}^b(y) - X^b(y) {\Delta}^a(x) + f^{abc} {\Delta}^c(x) \delta(x-y)=0,\label{WZXa}
\ee
that ${\Delta}^a(x)$ must satisfy.  These are the Wess-Zumino (WZ) consistency conditions. 

Now it may happen that
\be
{\Delta}^a(x) = X^a(x) \, {\cal C}\label{triviality}
\ee
where ${\cal C}$ is the spacetime integral of a local expression of $V^a(x)$, in short, a local counterterm. In this case we can redefine the effective action as $W'[V]= W[V] -{\cal C}$, so that
\be
 X^a(x) \, W'[V] =0 \label{newWI}
\ee
and recover invariance. If \eqref{triviality} is not possible for any local counterterm, then $\Delta^a(x)$ is a true anomaly.
Saturating $\Delta^a (x)$ with $\lambda^a(x)$ and integrating over spacetime, we obtain the integrated anomaly
\be
{\cal A}_\lambda =\int d^{\sfd} x \,\lambda^a(x) \Delta^a(x) \label{calA}
\ee
where the sum over $a$ is understood.

The WZ consistency condition \eqref{WZXa} is not very simple to be verified in practice.  It can however be recast into a form which is easier to deal with and at the same time establishes a well-defined mathematical formulation, by means of the BRST formalism. In QFT the BRST formalism is formulated by promoting the gauge variables to anticommuting fields, the FP ghosts. Here we do the same by promoting the gauge parameters $\lambda^a(x)$ to anticommuting fields  $c^a(x)$: $\delta_c V_\mu=  \partial_\mu c + [V_\mu,c]$, and endowing them with the transformation properties
\be
\delta_c c^a=  -\frac 12 f^{abc}c^b c^c, \quad\quad {\rm or} \quad\quad \delta_c c =-\frac 12[c,c]\label{deltall}
\ee
Then it is easy to see that the functional operator
\be
\delta_c= \int d^{\sfd} x \left(\delta_c V_\mu^a(x) \frac {\partial}{\partial  V_\mu^a(x)}+ \delta_c  c^a(x) 
\frac {\partial}{\partial c^a(x)}\right).\label{mfs}
\ee
is nilpotent
\be
\delta_c^2=0\label{snilpotent}
\ee
Using this we can rewrite the WZ consistency condition as follows
\be
\delta_c {\cal A}_c =0\label{mfsA0}
\ee
In order to recover \eqref{WZXa} one can simply differentiate \eqref{mfsA0} with respect to $c^a(x)$ and $c^b(y)$. In this language the anomalous WI \eqref{anomalousWI} can be rewritten
\be
\delta_c W[V] = {\cal A}_c, \label{anomalousWI2}
\ee
from which \eqref{mfsA0} is obtained by simply applying $\delta_c$ to both sides. Similarly we can rewrite the triviality condition \eqref{triviality} as
\be
{\cal A}_c =\delta_c\, {\cal C}\label{triviality2}
\ee
What we have defined in this way is a cohomology problem. $\delta_c$ is the coboundary operator of this problem. The cochain space is the space of local polynomials in $V_\mu^a$ and $c^a$ and their derivatives with the right canonical dimension. A cochain satisfying \eqref{anomalousWI2} is a cocycle, and a cocycle satisfying \eqref{triviality2} is a coboundary. 
In this way we have transformed the anomaly problem into a cohomology problem. Anomalies are non-trivial (non-exact) cocycles  of the operator $\delta_c$. 

This formulation has made it possible to solve the problem of finding all the non-trivial solutions of  eq.\eqref{mfsA0} for all dimensions and all gauge Lie groups. 

To construct such solutions we start from an order $n$ symmetric polynomial
in some representation of the Lie algebra, $P_n (T^{a_1}, . . . , T^{a_n} )$, invariant under the adjoint
transformations:
\be
P_n ([X,T^{a_1}], . . . , T^{a_n} )+\ldots +P_n (T^{a_1}, . . . ,[X, T^{a_n}] )=0\label{PnX}
\ee
for any element $X$ of a Lie algebra ${\mathfrak g}$. In many cases these polynomials are symmetric traces of the generators in the corresponding representation
\be
P_n (T^{a_1}, . . . , T^{a_n} )= Str(T^{a_1} . . . T^{a_n} )= d^{a_1\ldots a_n}\label{Symtrace}
\ee
($Str$ stands for symmetric trace).  For instance 
\be
Tr(T^aT^b)&=& c_2(R) \delta^{ab}\label{trtatb}\\
\frac 12 \Tr (T^a \{T^b,T^c\}) &=& c_3(R) d^{abc} \label{dabc}
\ee
where $c_2(R),c_3(R)$ are representation-dependent numerical coefficients. As these two examples show, the tensors $d^{a_1...a_n}$ are universal, they are characteristic of the Lie algebra; changing the representation only modifies the numerical coefficients in front of them. 

Now, it is convenient to use the differential form notation: $V= V_\mu dx^\mu= V_\mu^a T^a dx^\mu$ and $F= dV+V\wedge V= \frac 12 F_{\mu\nu} dx^\mu\wedge dx^\nu$, where $d$ is the exterior differential operator $d=\frac{\partial}{\partial x^\mu}dx^\mu$ and $F^{\mu\nu}= \partial_\mu V_\nu -\partial_\nu V_\mu +[V_\mu, V_\nu]$. Then one can construct the 2n-form
\be
\Delta_{2n}(V)= P_n(F,F,\ldots F)\label{PnFFF}
\ee
 In this expression the product of
forms is understood to be the exterior product. It is easy to prove, \cite{Chern1969}, that
\be
P_n(F,F,\ldots F)= {d} \left( n \int_0^1 dt\, P_n(V, F_t,\ldots, F_t)\right)
= { d} \Delta_{2n-1}^{(0)}(V)\label{PnAFF}
\ee
where we have introduced the symbols $V_t = t V$ and its curvature $F_t =
{d}V_t +\frac 12  [V_t , V_t ]$, where $0 \leq t \leq 1$. One can prove Eq.\eqref{PnAFF} by noting that $\frac d{dt} F_t= dV+[V_t,V]\equiv d_{V_t}V$ and $d_{V_t}F_t=0$, and exploiting
the symmetry of $P_n$ and the graded commutativity of the exterior product of forms.

Eq.(\ref{PnAFF}) is the first of a sequence of equations that can be shown to hold
\be
&&\Delta_{2n}(V)- {d} \Delta_{2n-1}^{(0)}(V) =0\label{descent1}\\
&&\delta_c \Delta_{2n-1}^{(0)}(V) - { d} \Delta_{2n-2}^{(1)}(V,c)=0\label{descent2}\\
&& \delta_c \Delta_{2n-2}^{(1)}(V,c)- {d} \Delta_{2n-3}^{(2)}(V,c)=0\label{descent3}\\
&&\dots\dots\0\\
&& \delta_c \Delta_{0}^{(2n-1)}(c)=0  \label{descentn}
\ee
These are known as descent equations.
All the expressions $\Delta_k^{(p)} (V, c)$ are polynomials of $ c, dc$, $V, dV$ and their commutators, whose explicit forms are known. The lower index
$k$ is the form degree and the upper one $p$ is the ghost number, i.e. the number
of $c$ factors. The last polynomial $ \Delta_{0}^{(2n-1)}(c)$ is a 0-form and a function of $c$ alone. What matters here is that the
$\Delta_{2n-2}^{(1)}(V,c)$ is the general expression of the consistent gauge anomaly in $\sfd =
2n -2$ dimensions, for, integrating \eqref{descent3} over spacetime, one gets
\be
\delta_c\, {{\cal A}}_c[V]&=&0 \label{boldV}\\
{{\cal A}}_c[V]&=& \int d^{\sfd} x \, \Delta_\sfd^1(c,V), \quad\quad {\rm where}\0\\
\Delta_\sfd^{(1)}(c,V)&=& n(n-1) \int_0^1 dt (1-t) 
P_n( { d}c,V, F_t,\ldots F_t)\label{anom4d}
\ee
${\cal A}_c[V] $ identifies the anomaly up to an overall numerical coefficient. Choosing $n=3$, i.e. $\sfd=4$, the RHS of \eqref{anom4d} is proportional to the RHS of \eqref{consanom} or \eqref{consanom1} multiplied by $c^a$ and summed over $a$. This is a sleek way to show that the chiral anomaly \eqref{consanom1} satisfies the WZ consistency conditions. 

Thus the existence of chiral gauge anomalies relies on the existence of the
adjoint-invariant symmetric  polynomials $P_n$. The latter are completely classified. The space of symmetric ad-invariant polynomials of the simple group $\sfG$ is usually denoted by $\mathbb{I}(\sfG)$. It is a commutative algebra. If $\sfG$ has rank $r$, then $\mathbb{I}(\sfG)$ has $r$ algebraically independent generators of order 
$m_1,...,m_r$. In other words we have $r$ algebraically independent polynomials $P_n$ , or symmetric tensors $d^{a_1...a_r}$, of order $m_1,...,m_r$. The values $m_1,...,m_r$ for the simple Lie algebras are as follows
\be
Lie\quad algebra && m_1,\ldots ,m_r\0\\
A_r&& 2,3,\ldots r+1\0\\
B_r && 2,4,\ldots,2r\0\\
C_r && 2,4,\ldots,2r\0\\
D_r && 2,4,\ldots 2r-2,r\0\\
G_2 && 2,6\label{table}\\
F_4&& 2,6,8,12\0\\
E_6&& 2,5,6,8,9,12\0\\
E_7&& 2,6,8,10,12,14,18\0\\
E_8 && 2,8,12,14,18,20,24,30\0
\ee
For product groups we have simply to consider all possible products of the corresponding symmetric polynomials. When the gauge group contains a $U(1)$ factor, the general form \eqref{PnFFF} can contain as factors any products of the Abelian curvature two-form $F$.

The previous table contains a lot of information about anomalies. If a symmetric tensor is absent in this table, the corresponding cocycle, and thus the corresponding anomaly, does not exist. This is the case, for instance, for ${\sf SU(2)}$ in $4\sfd$. The corresponding Lie algebra is ${A_1}$, which has only the second order symmetric tensor, while in order to construct a consistent chiral anomaly in $4\sfd$ one needs the third order tensor. Analogously, the third order symmetric tensor does not exist for ${\sf D_2}$, which is the Lie algebra of ${\sf SO(4)}$, the compact version of the Lorentz group in $4\sfd$. { But the Lorentz Lie algebra generators are the same as for ${\sf SO(4)}$, except that a few of them are multiplied by $i$.} In fact the local Lorentz symmetry is not anomalous in $4\sfd$.

Other examples of vanishing anomalies may come up even when the adjoint-invariant symmetric tensor in question is non-vanishing, but the fermion representations are real. If a representation is real $T^{a\dagger}=-T^a$. For instance in $4\sfd$, taking the hermitean conjugate of $d^{abc}$ we find $d^{abc}=-d^{abc}$, therefore $d^{abc}=0$.

\subsection{Gauge anomaly cancelation in the SM}
\label{ss:cancelationinSM}

At this point we have enough information to discuss the cancelation of gauge anomalies in the MSM.

As we know, the chiral consistent anomaly is determined by the tensor $d^{abc}= \frac 12 \tr \left(T^a \{ T^b, T^c\}\right)$, where $T^a$ denotes the total 
antihermitean generator of the Lie algebra $\mfs {\mathfrak u}(3)\oplus \mfs {\mathfrak u}(2)\oplus {\mathfrak u}(1)$: dropping labels we can generically write $T= T^{\mfs {\mathfrak u}(3)}\oplus T^{\mfs {\mathfrak u}(2)}\oplus T^{{\mathfrak u}(1)}$. The tensor
$d^{abc}$ decomposes into various independent components, which we list hereafter:

\vskip 1cm
\centerline {Table MSM gauge}
\vskip 0.2cm
\noindent\fbox{
    \parbox{\textwidth}{

\begin{itemize}
\item $ T^{\mfs {\mathfrak u}(3)}\times  T^{\mfs {\mathfrak u}(3)}\times  T^{\mfs {\mathfrak u}(3)}$: there are two left-handed and two right-handed triplet, whose anomalies cancel one another. 

\item $ T^{\mfs {\mathfrak u}(2)}\times  T^{\mfs {\mathfrak u}(2)}\times  T^{\mfs {\mathfrak u}(2)}$,  which vanishes because the tensor $d^{abc}$ vanishes in general for the Lie algebra ${\mfs {\mathfrak u}(2)}$.
\item $ T^{\mfs {\mathfrak u}(2)}\times  T^{\mfs {\mathfrak u}(2)}\times  T^{  {\mathfrak u}(1)}$, in which case we have the trace of two ${\mfs {\mathfrak u}(2)}$ generators in two doublet representations. These traces are non-vanishing because $\tr (T^a T^b) \sim \delta^{ab}$, but they are multiplied by the corresponding ${ {\mathfrak u}(1)}$ charges, whose total value is $6\left(\frac 16\right) -2(\frac 12) =0$.
\item $ T^{\mfs {\mathfrak u}(3)}\times  T^{\mfs {\mathfrak u}(3)}\times  T^{  {\mathfrak u}(1)}$, in which case we have the trace of two ${\mfs {\mathfrak u}(3)}$ left triplet  generators and two right triplet generators. These traces are again non-vanishing, but   they are multiplied by the corresponding ${ {\mathfrak u}(1)}$ charge, whose total value is $3\left( 2\left(\frac 16\right) -\frac 23 +\frac 13\right) =0$.
\item $ T^{ {\mathfrak u}(1)}\times  T^{  {\mathfrak u}(1)}\times  T^{ {\mathfrak u}(1)}$, in this case the tensor is proportional to the overall sum of the charge products:
$6\left(\frac 16\right)^3 - 3 \left(\frac 23\right)^3 -  3 \left(-\frac 13\right)^3+ 2  \left(-\frac 12\right)^3-(-1)^3=0$.
\end{itemize}
}}

\vskip 0.3cm
This completes the analysis of gauge anomalies  in the standard model. There are no local anomalies on a flat background theory.

$${-\!\!\!-\!\!\!-\!\!\!- \,\, o \,\,-\!\!\!-\!\!\!-\!\!\!-}$$
 
The results illustrated so far are generally considered completely satisfactory: the SM is anomaly-free, that is that, and let us move on to other things, notably to its renormalization or to its physics. But there are aspects of the analysis carried out so far that deserve more attention and, in view of the coupling of the SM to gravity, would be superficial to ignore. First, as already promised, we need a non-perturbative method to justify the results shown above beyond the perturbative level. This will turn out to be indispensable for trace anomalies. Second, we need a deeper understanding of why consistent chiral gauge anomalies are so deadly. The obstruction to selecting a physical Hilbert space for a quantum field in the quantization of a Weyl fermion coupled to a gauge potential is a consequence rather the origin of the problem. More insight is necessary.
 
 \vskip 0,3 cm
 
 \noindent Refs.: \cite{adler1969a,alvarezginsparg1984,andrianov1984a, andrianovbonora1984b,andrianovbonorapasti1984,
banerjee1974,bardeen1969,baulieu1984b,bell1969,
 Bertlmann,bonorasoldatizalel2020,breitenlohner1977,BRS1,BRS2,BRS3,duboisviolette1985,
 duboisviolette1986,frampton1983,fujikawa1984a,jackiw1969b,langouchestora1984,manes1985,stora1984,
wesszumino1971,zumino1984}
 
 \section{Non-perturbative approach: the heat-kernel-like methods}
 \label{ss:nonperturbative}
 
 There are several different non-perturbative (a.k.a. functional) methods to calculate anomalies: Schwinger's proper-time method, the Seeley-DeWitt and the zeta-function
regularization method and Fujikawa's method. We comprise them under the term of heat-kernel-like methods. The purpose of all of them is to represent in a mathematically affordable way the path integral corresponding to the quadratic part of a theory, and in particular of a fermionic theory, interacting with external potentials: the central object is therefore the determinant of the full kinetic operator, let us call it $\sf D$, (or the square thereof in the case of Dirac-like operators). It is crucial for these methods to work that the operator in question, after passing from a Minkowski to a Euclidean background metric, be a quadratic elliptic self-adjoint operator. Below we give a synthetic presentation of one of them, which we call SDW (Schwinger-DeWitt). 
 
Our problem is to represent  $\log \det {\sf D}= \Tr \log {\sf D}$. We start from a familiar formula in theoretical physics
 \be
\frac i{{\sf D}+i\epsilon}= \int_0^\infty ds \, e^{is({\sf D}+i\epsilon)}\label{A+iepsilon}
\ee
with infinitesimal $\epsilon > 0$. It certainly holds when $\sf D$ is a real number . We extend it to an operator $\sf D$ by representing it via its eigenvalues. By formally integrating we write
\be
{\sf Tr} \log {\sf D}= - i  \int_0^\infty \frac {d s}{i  s}{\sf Tr}\left(e^{i  {\sf D}  s}\right)+ {\rm const}\label{TrlogA}
\ee
where the $i\epsilon$ prescription is understood, as we will do from now on for the $s$ integrals. 

The Dirac operator  $\slashed{D}= i \gamma^\mu \partial_\mu$, even after the addition of a vector and an axial potential, in a Euclidean background is an example of elliptic operator, but the application of the SDW method to it in the linear form is not known. The best we can do is to apply this method to its square and extract the square root of the determinant. The choice of the square Dirac operator is not always univocal and requires some care. The minimal precaution is that it preserve the same symmetries as the linear operator and become a self-adjoint operator after a Wick rotation, see  below.  Let us call $\EF$ such a square Dirac operator.

We identify the effective action for Dirac fermions with
\be
 W =-\frac i2 {\sf Tr} \left(\ln \,  \EF\right)\label{EAEFg}
\ee
$\sf Tr$ includes  all the traces plus the spacetime integration.
Any variation of \eqref{EAEFg} is given by
\be
\delta W = \frac i2{\sf Tr}\left( \EG\,
\delta\, \EF\right)\label{deltaEA}
\ee
where, in symbols, $\EF \, \EG =-1$. Thus we can formally write
\be
\delta W= -\frac 12 {\sf Tr}\left( \int_0^\infty d s \,e^{i  \EF \, s}\delta  \EF\right). \label{deltaEAW}
\ee
and, like in \eqref{TrlogA},  $W$ can be represented as
\be
 W= -\frac 12 \int_0^\infty \frac {d s}{i  s}{\sf Tr}\left(
e^{i  \EF  s}\right)+ {\rm const}\equiv \int d^{\sfd} x \,   L(  x)
+{\rm const},\label{W+const}
\ee
where
\be
  L(  x) =  -\frac 12 \lim_{x'\to x}\tr \int_0^\infty \frac {d 
s}{i  s}
 K( x,  x',  s),\label{WK}
\ee
and the kernel $  K$ is defined by
\be
  K( x, x ', s)= \langle x|e^{i \EF \,s}|x'\rangle\label{Kxx'}
\ee
Inserted under the symbol ${\sf Tr}$, \eqref{WK} means
integrating over $x$ after taking the limit $x'\to x$. 

{\bf Remark.} We recall that in all the above $s$-integrals the $i\epsilon$ prescription, i.e. the substitution $\EF\to \EF + i\epsilon$, is understood. It is clear that such integrals converge only if $\EF$ is a self-adjoint operator. This is possible in a Euclidean background, i.e after a Wick rotation. In the sequel we make the substitution $\EF \to \widetilde \EF$, where the tilde represents a Wick rotation, and anti-Wick-rotate the final result. However for the sake of simplicity, we understand this substitution and continue with the Minkowski notation. 
 
 \vskip 0,4 cm
 The SDW method proceeds with the definition of  the amplitude
\be
\langle  x, s| x',0\rangle = \langle x| e^{i \EF s}|x'\rangle,\label{hatampA}
\ee
which satisfies the (heat kernel) differential equation
\be
i \frac {\partial }{\partial  s} \langle  x, s| 
x',0\rangle
= - \EF_{ x} \langle
 x, s|  x',0\rangle, \label{hatdiffeqforA}
\ee
$\EF_{ x}$ is a quadratic elliptic differential operator, typically,  $\EF_x = \eta^{\mu\nu}\partial_\mu\partial_\nu+\ldots$, where ellipses denote non-leading terms.

The SDW method now introduces an ansatz for the amplitude \eqref{hatampA}
\be
\langle  x, s|  x',0\rangle = -\lim_{m\to 0}\frac i{(4\pi s)^{\frac  \sfd 2}}
e^{i\left(\frac {( x- x')^2} {4s}-m^2 s\right)}
\Phi( x, x', s),\label{ansatz3}
\ee
$\Phi( x, x', s)$ is a function to be determined. The mass parameter $m$ is introduced to guarantee convergence for the $s$ integration. In the limit $  s\to 0$ the RHS of \eqref{ansatz3} becomes the
definition of a delta function multiplied by $\Phi$. More precisely,  since it must
be $\langle x,0|  x',0\rangle=\delta( x, x')$, and
\be
\lim_{ s\to 0} \frac i{(4\pi s)^{\frac {\sfd} 2}} \, e^{i\left(\frac {( x- x')^2} {4
s}-m^2 s\right)}=
\delta^{(\sfd)}( x, x'),\label{hatlimdelta}
\ee
we must have
\be
\lim_{ s\to 0} \Phi( x, x', s)={\bf
1}.\label{hatlimitPhi}
\ee 
 Looking at \eqref{ansatz3}, in dimension $\sfd$, we can make the identification
\be
K( x, x, s)=\frac i{(4\pi i  s)^{\frac {\sfd}2}}\,\sqrt{ g}\, e^{-im^2 s} [\Phi]( x, x, s).\label{Kxx}
\ee
From now on the symbol $[H](x,\ldots)$ will denote the coincidence limit $\lim_{x'\to x} H(x,x',\ldots)$.

 The heat-kernel equation \eqref{hatdiffeqforA} becomes a quadratic partial differential equation for $\Phi$ in $x_\mu$ and $s$. To find useful solutions one has to expand $\Phi$ is series of $s$:
 \be
\Phi( x, x', s) =\sum_{n=0}^\infty  a_n( x, x') (i s)^n
\label{Phiexp}
\ee
so that \eqref{hatdiffeqforA} becomes a series of recursive differential relations for the coefficients  $a_n$. Starting with the position $[a_0]=1$ they permit us to compute the other coefficients up to the relevant order (which, of course, depends on the dimension $\sfd$).

\subsubsection{Bardeen's anomaly with SDW}
\label{sss:bardeen}

Let us apply the just illustrated method to the calculation of Bardeen's anomaly in  $\sfd=4$. In this case the relevant operator is the square of $\slashed{\mathscr D}= \slashed{\partial} +\slashed{ V} + \gamma_5 \slashed {A}$, which has the properties
\be
\slashed{\ED}^\dagger = \gamma_0 \slashed{\ED} \gamma_0, \quad\quad\left(\slashed{\ED}^2\right)^\dagger = \gamma_0\slashed{\ED}^2\gamma_0,\label{Dsquaredagger}
\ee
but after a Wick rotation (see Appendix B for conventions) we get
\be 
\tilde{\slashed{\mathscr D}}^\dagger =-\tilde{ \slashed{\mathscr D}} , \quad\quad \left(\tilde {\slashed {\mathscr D}}^2\right)^\dagger =  \tilde {\slashed {\mathscr D}}^2\label{DsquareEucl}
\ee

The relevant quadratic Dirac operator is
\be
\slashed{\mathscr D}^2= -\eta^{\mu\nu}\overline {\mathscr D}_\mu {\mathscr D}_\nu
 -\Sigma^{\mu\nu} \bigl(\EB_{\mu\nu} + \bigl(\EC_{\mu\nu}+2( A_\nu \partial_\mu -A_\mu\partial_\nu)\bigr)\gamma_5\bigr)\label{calDsquarebis}
\ee
where $\Sigma_{\mu\nu}=\frac 14[\gamma_\mu,\gamma_\nu]$, $\overline {\mathscr D}_\mu$ is ${\mathscr D}_\mu$ with the change $\gamma_5\to-\gamma_5$,   and
\be
\EB_{\mu\nu}=F_{V\mu\nu} -[A_\mu,A_\nu], \quad\quad \EC_{\mu\nu}= \partial_\mu A_\nu -\partial_\nu A_\mu +\{V_\mu, A_\nu\} - \{A_\mu, V_\nu\}\label{EBEC}
\ee
We use also the notation
\be
\EE=\Sigma^{\mu\nu}\EB_{\mu\nu} +\gamma_5 \Sigma^{\mu\nu}\EC_{\mu\nu}, \label{EE}
\ee
with similar expressions for the Euclidean version.

Under the transformations \eqref{deltaAV} we find
\be
\delta_ \lambda \slashed{\mathscr D}=[ \slashed{\mathscr D},\lambda], \quad\quad \delta_ \rho \slashed{\mathscr D}=[ \slashed{\mathscr D},\rho]+2  \gamma_5 \rho  \slashed{\mathscr D}\label{deltamathscrD}
\ee
The relevant operator is not invariant under the $\rho$ transformation. Therefore an anomaly is to be expected in the divergence of the chiral current. For taking the variation with respect to $\rho$ of the path integral $Z$, we obtain, formally,
\be
\delta_\rho \ln Z = 2 i {\sf Tr} \left( \gamma_5 \rho\right)\equiv {\cal A}^{(unreg)},\label{formalanomaly}
\ee
where ${\sf Tr}$ includes all traces plus spacetime integration. ${\cal A}^{(unreg)}$ is the unregulated anomaly. 
We regularize it with the Schwinger proper time approach. This means using the procedure outlined before with $\EF= -{\slashed{\mathscr D}}^2$ with regard in particular eq.\eqref{deltaEAW}\footnote{The square Dirac operator must respect the properties of the linear one. For instance, in the present case, an alternative quadratic operator would be $\overline {\slashed {\mathscr D}} \slashed {\mathscr D}$, with the transformation property $\delta_{\Upsilon} \left( \overline  {\slashed{\mathscr D} } \slashed{\mathscr D} \right) = [  \overline  {\slashed{\mathscr D} } \slashed{\mathscr D}, \Upsilon] $, with $\Upsilon =\lambda +\gamma_5 \rho$, which is also an elliptic operator when turned Euclidean. This would exclude any anomaly. But this would contradict the linear transformation $\delta_ \Upsilon \slashed{\mathscr D}=[ \slashed{\mathscr D},\Upsilon] +2  \gamma_5 \rho  \slashed{\mathscr D}$.},
 \be
{\cal A}\equiv\delta_\rho W&=& \frac 12 {\sf Tr}\left( \int_0^\infty d s \,e^{-i{\slashed{\mathscr D}}^2\!   s}\delta_\rho  {\slashed{\mathscr D}}^2\right)=2i  {\sf Tr}\left(\rho \gamma_5  \int_0^\infty d s \,e^{-i  {\slashed{\mathscr D}}^2\!  s}\,  {\slashed{\mathscr D}}^2\right)\0\\
&=& - 2 {\sf Tr}\left(\rho \gamma_5  \int_0^\infty d s \,\frac{\partial}{\partial s}e^{-i  {\slashed{\mathscr D}}^2\!  s}\right)= 2  \left.{\sf Tr}\left(\rho \gamma_5  \,e^{-i  {\slashed{\mathscr D}}^2\!  s}\right)\right\vert_{s=0}
\label{deltaEAWa}
\ee
In the last integration convergence at infinity is guaranteed by the factor  $e^{-im^2s}$ (with the  prescription $m^2\to m^2 -i\epsilon$).
A remark is in order here: looking at eqs.\eqref{deltamathscrD} and \eqref{deltaEAWa} these manipulations are possible because the operator $ {\slashed{\mathscr D}}$ in it commute with $e^{i  {\slashed{\mathscr D}}^2\!   s}$. 

Now we write \eqref{deltaEAWa} in coordinate representation 
\be
{\cal A}=-2 \lim_{x'\to x}    {\sf Tr}\left(\rho \gamma_5  \,\langle x|e^{-i  {\slashed{\mathscr D}}^2\!  s}|x' \rangle\right)\Big{\vert}_{s=0}\label{deltaEAW1}
\ee
and insert 
\be
\langle  x, s|  x',0\rangle = \lim_{m\to 0}\frac i{(4\pi s)^{2}}  e^{i\left(\frac {( x- x')^2} {4s}-m^2 s\right)}\Phi( x, x', s),\label{ansatz3bis}
\ee

Next we expand $\Phi(x,x',s)$  in powers of $is$, as in \eqref{Phiexp}, and integrate over $s$. Setting $s=0$ suppresses all higher order terms in the expansion, except the 0-th, first and second order term. The zeroth-order term is annihilated by the trace over $\gamma_5$. The first order term, involving $[a_1]$, diverges for $s\to 0$. $[a_1]$ can be explicitly calculated: its odd-parity part vanishes due to the gamma matrix trace, while the even-parity part survives the tracing, but is trivial and can be subtracted with a counterterm.

Therefore the finite part of \eqref{deltaEAW1} becomes
 \be
{\cal A}=- \frac {2i}{16\pi^2} \int d^4x\Tr \left( \gamma_5 \,\rho\, [a_2](x)\right),\quad\quad [a_2(x)] = \lim_{x'\to x} a_2(x,x')
\label{calAtr2}
\ee
To compute $[a_2]$ we resort to the above-mentioned recursive relation, obtained from the heat-kernel equation:
\be
(n\!+\!1)a_{n+1}+  (x\!-\!x')^\mu\left( \ED_\mu a_{n+1}+ 2\gamma_5\Sigma^{\mu\nu}A_\nu a_{n+1}\right)- \overline{\mathscr D}^\mu {\mathscr D}_\mu a_n-4\gamma_5\Sigma^{\mu\nu}A_\nu \partial_\nu a_n- \EE \,a_n=0 \label{heatkernelflat2}
\ee
Starting from $[a_0]=1$ one can derive $[a_1]$ and $[a_2]$. The calculation is lengthy, but it is not hard to extract the odd-parity part of the trace in \eqref{calAtr2}. It coincides with the RHS of \eqref{Bardeenanom} saturated with $\rho^a$ and integrated over spacetime.\footnote{Alternatively one can arrive directly at the anomaly \eqref{consanom1} by applying the SDW method to  the operator \eqref{DWP++} below, see \cite{I}.}

\vskip 0,3cm 
\noindent Ref.: \cite{bast2009,I,dewitt2003,fujikawa1984a,fujikawa2004,kirsten2022,schwinger1959,seeley1967,vassilevich2003}

\section{Anomalies as obstructions}

Non-perturbative methods, like the one just illustrated, are very useful because they yield the full expressions of consistent anomalies, not just their lowest order approximations. This is extremely useful in cases where the lowest order terms are not enough to unambiguously determine the anomalies. However, returning to  the discussion on fermion propagators after eq.\eqref{Jamux},  they do not differ substantially from the perturbative methods, in that they rely on fermion propagators which are obtained from the original Weyl theory by adding a free Weyl fermion of opposite chirality (or resorting to Dirac fermions, like in Bardeen's approach). The aim of this section is to illustrate a different approach in which there is no need to introduce additional degrees of freedom.  In this new approach anomalies appear as obstructions to the existence of fermion propagators. The method is based on the family's index theorem of Atiyah and Singer. The new approach requires quite a lot of preliminary material. Here we try to squeeze it into a few pages. In this section the framework is entirely Euclidean. We will assume that our conclusions can be transferred to the corresponding Minkowski formulation.

\subsection{A geometrical description of anomalies}
\label{ss:evaluation}

The first element we need is a geometrical description of consistent gauge anomalies. The algebraic formalism of sect.\ref{ss:WZ-BRST}, from this point of view, is unsatisfactory. Looking at eq.\eqref{PnAFF} we see that in the case $n=3$, i.e. $\sfd =4$, both sides of the equation vanish because they are forms of degree 6. One could suspect that the descent equations that follow from this one are built on sand. It is not so, but no doubt an appropriate elaboration is required. It is well-known that the fitting framework for a gauge theory is provided by the geometry of principal and associated fiber bundles. The underlying geometry of a  gauge theory on a spacetime $\sfM$ with gauge group $\sfG$ is grounded on a principal fiber bundle $\sfP(\sfM, \sfG)$, 
$$
\xy\xymatrix{{\sf G}\,\,\ar@{^{(}->}[r]
&{\sf P}\ar[d] ^\pi\\
&{\sf M}}
\endxy
$$
$\pi$ denotes the projection $\pi: {\sf P}\to {\sf M}$. The inverse image of a point in $\sfM$ is a fiber of $\sfP$, and is a copy of $\sfG$. A gauge potential $V_\mu= V_\mu ^a T ^a$, where $T^a$ are the generators of ${\mathfrak g}\equiv Lie(\sfG)$, represents a connection form $V=V_\mu dx^\mu$, with curvature is $F=dV+\frac 12 [V,V]$. A connection $V$ splits the vector fields of $\sfP$ into vertical and horizontal ones\footnote{$V_\mu(x)$ comes from the pull-back of $\sigma_U^\ast V= V_\mu d x^\mu$ by a local section $\sigma_U: U \to \sfP$, where $ \pi\circ\sigma_U (x)= x$ for $\forall x\in U\subset \sfM$.}, and is properly defined on the total space $\sfP$. Thus the six-form polynomial $P_3(F,F,F)$ is also defined on $\sfP$, whose dimension is $\sfd$ times the group dimension. This however does not prevent this form from vanishing, because it is an invariant form that descends on $\sfM$ (a basic form) and thus cannot but vanish. 

The geometry of the total space is not yet enough. We have to enlarge our horizon.

In this setup gauge transformations are represented by vertical automorphisms of $\sfP$. An automorphism is a diffeomorphism of  ${\sf P}$,  $\psi:{\sf P}\to {\sf P}$, such that $\psi(p g) = \psi(p)g$, for any $p\in{\sf P}$ and any $g\in {\sf G}$. Automorphisms form a group which will be denoted $\sf Aut({\sf P})$. For any automorphism $\psi$ the map $\pi \circ \psi$ is a diffeomorphism of the base space, i.e. $\in {\sf Diff(M)}$. Vertical automorphisms are those that do not move the base point. They form a subgroup of  ${\sf Aut(P)}$ denoted by ${\sf Aut_v(P)}$. The corresponding Lie algebras will be denoted by ${\sf aut(P)}$ and ${\sf aut_v(P)}$, respectively; the latter are spaces of vector fields in ${\sf P}$ generated by one-parameters subgroups of  ${\sf Aut(P)}$ and ${\sf Aut_v(P)}$, respectively.
For the time being we  disregard diffeomorphisms of the base space and focus on ${\sf Aut_v(P)}$. The latter is to be identified with the group ${\cal G}$ of gauge transformations, whose Lie algebra $Lie \,{\cal G}$ is, in turn, identified with ${\sf aut_v(P)}$, the Lie algebra of ${\sf Aut_v(P)}$. The reason of this identification is clear from the way a connection transforms under vertical automorphisms. Let $V$ be a connection with curvature $F$ and let $\psi$ be a vertical automorphism: we can associate to it a map $\gamma:\,{\sf P}\to {\sf G}$ defined by $\psi(p) =p\,\gamma(p)$ satisfying $\gamma(pg) = g^{-1} \gamma(u) g$. Then one can show that the pull-back by $\psi$ is
\be
\psi^\ast V = \gamma^{-1} V \gamma + \gamma^{-1} d \gamma,\quad\quad \psi^\ast F= \gamma^{-1} F \gamma\0
\ee
Thus we set ${\sf Aut_v(P)}\equiv {\cal G}$ and introduce the evaluation map
\be
ev:{\sf P}\times {\cal G} \rightarrow {\sf P}, \quad\quad ev(p,\psi)=\psi(p)\label{evmap}
\ee
$\sfP\times {\cal G}$ is a principal fiber bundle over $\sfM\times {\cal G} $, with group $\sfG$. 
This means in particular that, by pulling back a connection $V$ in ${\sf P}$, we obtain a connection ${\bf V} =ev^\ast V$ in  ${\sf P}\times {\cal G}$. This connection splits into a one-form in $\sfP$ plus a one-form in $\cal G$ and contains all the information about the properties of the FP ghosts and BRST transformations. First, it can be shown that  
\be
{\bf V}\equiv ev^\ast V= V+i_{(\cdot)} V\label{evstarA}
\ee
where $i$ is the interior product and $i_{(\cdot)} V$ denotes the map $Z\to i_Z V$, that associates to every $Z\in Lie({\cal G})$ the map $\xi_Z= V(Z): \sfP\to Lie(\sfG)$. $\xi_Z$ is in fact an infinitesimal gauge transformation since the action of $Z$ over the connection $V$ is given by the Lie derivative $L_Z$, which takes the form:
\be
L_ZV = d\xi_Z +[V,\xi_Z] \label{LZA}
\ee

The symbol $i_{(\cdot)} V$ has an anticommuting nature and plays the same role as the FP ghost field $c$ introduced in section \ref{ss:WZ-BRST}. For instance, if $\hat\delta$ represents the exterior differential in ${\cal G}$, one can show that 
\be
\hat\delta V&=& -d i_{(\cdot)}V-\frac 12[V,  i_{(\cdot)}V] -\frac 12  [ i_{(\cdot)}V,V]\label{hatdeltaA}\\
\hat\delta  i_{(\cdot)}V&=&\frac 12 [ i_{(\cdot)}V, i_{(\cdot)}V]\label{hatdeltaiA}
\ee
Therefore if we identify $\hat\delta= (-1)^k \delta_c$, where $k$ is the degree of the form in  ${\sf P}$  it acts on, and $\delta_c$ is the BRST nilpotent operator, we reproduce the BRST transformations introduced in section  \ref{ss:WZ-BRST}.

In other words BRST transformations and coboundary coincide with gauge transformations along the fibers of $\sfP$ and de Rham differential in $\cal G$. Next let us apply these new notions to gauge anomalies. Let $P_n$ be an ad-invariant polynomial with $n$ entries,  $n=\frac {\sfd}2+1$. The expression
\be
TP_n(V) = n\int_0^1 dt P_n(V,F_t,\ldots,F_t) \label{TPnA}
\ee
introduced in section \ref{ss:WZ-BRST} is called {\it transgression formula}, and
\be
d \, TP_n(V) = P_n(F,\ldots ,F) =0\label{dTPnA}
\ee
because of dimensional reasons, for $P_n(F,\ldots,F)$ is a $\sfd+2$ basic form in a $\sfd$ dimensional 
base spacetime ${\sf M}$. Let us pull back \eqref{dTPnA} through the evaluation map to $\sfP\times {\cal G}$. We get
\be
(d+\hat\delta) ev^\ast TP_n(V) =  (d+\hat\delta)  TP_n(ev^\ast V) =0 \label{descent1ev}
\ee
This splits into a set of equations of decreasing degree
\be
&&\hat\delta\,  TP_n(V)+ d\,  i_{(\cdot)} TP_n(V)=0\label{TPdescent2}\\
&&\hat \delta \,  i_{(\cdot)} TP_n(V)+d\, i_{(\cdot)}i_{(\cdot)} TP_n(V) =0\label{TPdescent3}\\
&& \ldots\ldots\0\\
&& \ldots\ldots\0
\ee
These are the descent equations written in geometrical language, and, unlike the algebraic descent equations of section \ref{ss:WZ-BRST}, they are all well defined in $\sfP$. Focusing on $i_Z TP_n(V)$, it can be rewritten
\be 
i_Z TP_n(V)\label{iZTPnA2}=\! -n(n-1)\!\int_0^1\! dt(t-1)\Big(   P_n(di_Z V,V,F_t,\ldots, F_t)-dP_n(V,i_ZV,F_t,\ldots,F_t)\Big)\0
\ee
The first term on the RHS is precisely, up  to global sign, the anomaly \eqref{anom4d} for $\sfd$ generic:
\be
\Delta_\sfd^1(i_{(\cdot)}V,V)= n(n-1)\int_0^1dt(t-1)  P_n(di_{(\cdot)}V,V,F_t,\ldots, F_t).\label{chiralanom'}
\ee
The second piece is a total differential which drops out upon spacetime integration. 

A cocycle $\Delta_{2n-2}^{(1)}(V,c)$ is trivial when there are local cochains $C_{2n-2}^{(0)}(V,c)$ and $C_{2n-3}^{(1)}(V,c)$ such that
\be
\Delta_\sfd^1(V,c)= \delta_c\, C_{2n-2}^{(0)}(V,c) +d C_{2n-3}^{(1)}(V,c)\label{trivial}
\ee
where $C_k^{(p)} (V,c)$ denotes a polynomial $k$-form of ghost number  $p$, built with $V,c$, their commutators, wedge products and exterior differentials. If such local cochains do not exist, the cocycle is said to be non-trivial and defines a non-trivial class of the BRST cohomology, a true anomaly. The latter characterizes local quantum field theories: how does it relate to the familiar de Rham cohomology  { of $\sfM$ and $\sfP$} ?

Let us return to the descent equations \eqref{descent1ev} and, in particular, to the consistency conditions \eqref{TPdescent3}. Since $d\, TP_n(V)=0$, anomalies are to be found among the representatives of the de Rham cohomology group of order $\sfd+1$ in   ${\sf P}$. But do they coincide with them? Now, it is known that the de Rham cohomology group of order ${\sfd+1}$ in  ${\sf P}$, ${\sf H_{de \,Rham}^{\sfd+1}(\sfP)}$, is spanned by forms 
$TP_i(V)\wedge \beta_i$,
where $\beta_i$ are basic and closed forms of order $h_i=2(n\!\!-\!\!i)$, such that $2i\!\!-\!\!1\!\!+\!\!h_i=\sfd\!\!+\!\!1$.  The $\beta_i$ may be forms of the type $P_{n-i}(F,\ldots,F)$ (this is the case of a reducible polynomial $P_n(F)= P_i(F)P_{n-i}(F)$),  or of the type $P'_{n-i}(R,\ldots,R)$, where $R$ is the Riemann (or spin) curvature 2-form on ${\sf M}$, and $P_i'$ are the corresponding ad-invariant symmetric polynomials. But the set of $\beta_i$ may also contain forms that cannot be written in local form like in the previous examples. Such non-local $\beta_i$ do not give rise to anomalies in local field theory. 

On the other hand let us suppose that $\chi= TP_n(V)$ be cohomologically trivial, i.e. $ TP_n(V)= d\eta$. Then, decomposing  \eqref{descent1ev}, we get, in particular,
\be
i_{(\cdot)} TP_n(V) = i_{(\cdot)} d\eta =- d i_{(\cdot)}\eta + \hat \delta \eta\label{trivialcond}
\ee
Since $\hat \delta \eta=-\delta_c \eta$, comparing with \eqref{trivial}, we see that this is a triviality condition for the anomaly $i_{(\cdot)} TP_n(V)$. However, this does not correspond automatically to a trivial field theory anomaly. The condition $ TP_n(V)= d\eta$ means that $TP_n(V)$ is exact in the de Rham cohomology of  ${\sf P}$ . But this does not automatically mean that it is trivial with respect to the local BRST cohomology. For this to be the case it must be that $\eta$ is a local expression of $V$.  If such a local $\eta$ does not exist, we are in presence of a true non-trivial anomaly originated from a cocycle $d\eta$ in ${\sf P}$ which is trivial (a coboundary) in the de Rham cohomology. On the same footing if $\chi$ and $\chi'$ are two closed $\sfd+1$ forms in ${\sf P}$ such that $\chi-\chi'= d\eta$, we can say the corresponding anomalies are the same only if $\eta$ is a local form. 

The previous rather lengthy { and somewhat convoluted} discussion is to stress the fact that the BRST cohomology (i.e. the true anomalies of gauge field theories)  does not coincide with the ordinary cohomology of $\sfP$, and even less with that of $\sfM$. In fact the topology of $\sfM$ does not play a role in determining the local anomalies on it. To geometrically characterize the BRST cohomology of local quantum field theory a further effort is necessary. We need to introduce the { the concept of} classifying space and  universal bundle.

Given a compact Lie group $\sfG$, any principal fiber bundle ${\sf P(M,G)}$ can be obtained as the pull-back of the so-called {\it universal principal bundle} ${\sf EG(BG,G)}$, 
$$
\xy\xymatrix{\sfG\,\,\ar@{^{(}->}[r]
&{\sf EG}\ar[d] ^\pi\\
&{\sf BG}}
\endxy
$$
${\sf BG}$, the basis, is called {\it classifying space}. An important property of $\sf EG$, the total space, is that it is a contractible space, that is, it is continuously deformable to a point. Therefore its cohomology is that of a single point.

For any principal bundle ${\sf P(M,G)}$ there exists a bundle morphisms $(\hat f,f)$ such that the following diagram is commutative
$$
\xy\xymatrix{\sfP\,\,\ar[d]^{\pi} \ar[r]^{\hat f}& {\sf EG} \ar[d]^{\pi}\\
\sfM\ar[r] ^f &{\sf BG}}
\endxy
$$
$f$ is unique up to homotopy\footnote{Two maps $f_1, f_2$ are homotopic if there exist a continuous map $F: \sfM \times I \to {\sf BG}$ such that $F(x,0)=f_1(x)$ and $F(x,1)= f_2(x)$, $\forall x\in \sfm$.} and is called classifying map. Both ${\sf EG}$ and ${\sf BG}$ are generally infinite dimensional, but in many practical applications one can use for them finite-dimensional approximants.

A very important fact is the existence of a universal connection ${\mathfrak v}$ in $\sf EG$ from which any connection on ${\sf P}$ can be derived via pull-back: for any $V$ in ${\sf P}$ there exists a bundle morphisms $(\hat f,f)$ such that $V=\hat f^\ast {\mathfrak v}$. 
For convenience, we shall call, with some abuse of language, {\it universal} the forms $\chi$ in ${\sf P}$ constructed with the connection form $V$  if they can be derived by pulling back an analogous form in ${\sf EG}$ constructed with ${\mathfrak v}$: 
\be
\chi(V) = \chi(\hat f^\ast {\mathfrak v})=\hat f^\ast(\chi({\mathfrak v})),
\ee
for a bundle map $\hat f$.  Therefore forms like $TP_n(V)$ and $P_n(F)$ are universal.
But, of course, in ${\sf P}$ there may be non-universal forms. In particular, the forms $\beta_i$ that appear in the cohomology of $\sfP$, see above, may not be universal. We have already remarked that not all the forms that span the non-trivial cohomology classes of ${\sf H_{de\, Rham} ^{\sfd+1} (P)}$ necessarily correspond to local anomalies. 

Let us now connect the anomalies in quantum field theory to their universal origin.
We pull-back ${\mathfrak v}$ and forms constructed with it to $\sfP\times {\cal G}$ by combining the evaluation map and bundle maps
\be
\sfP\times {\cal G} \stackrel{ev}{\longrightarrow}\sfP\stackrel{\hat f}{\longrightarrow} {\sf EG} \label{evhatf}
\ee 
Now, starting with local expressions of ${\mathfrak v}$ in ${\sf EG}$ and pulling them back with $ev\circ \hat f$ we create universal expressions in $\sfP\times {\cal G}$.  We focus in particular on the universal forms  of order $\sfd+1$, and state the result of the previous discussion: {\it non-trivial  local field theory anomalies are identified by the quotient}
\be
\frac {closed\,\, universal \,\,(\sfd\!+\!1)\!-\!\!forms\,\, in\, \,{\sf P}}{differentials\,\, of\,\, universal\,\,  \sfd\!-\!\!forms\, in\,{\sfP}}\label{closed/exact}
\ee
For instance, consider  $TP_{n_1}(V) P_{n_2}(F)$  and suppose that $P_{n_2}(F) = d\eta$  with $\eta$ a basic form in some manifold ${\sf M}$ of dimension $\sfd=2(n_1\!+\!n_2)\!\!-\!\!2$ . Then  $TP_{n_1}(V) P_{n_2}(F)=d \left(TP_{n_1}(V)\, \eta\right)  $ (because $P_{n_1}\!(F) \,\eta$ is basic and therefore vanishes on $\sfM$ for dimensional reasons). Nevertheless, the anomaly corresponding to $TP_{n_1}(V) P_{n_2}(F)$ is universal, because, even though $P_{n_2}(F) = d\eta$ in ${\sf M}$, the form $\eta$ cannot be universal. To explain the reason for this we need  the concept of Weil homomorphism.

\vskip 1 cm 
{\bf Chern-Weil homomorphism}. For any group $\sfG$ we have introduced earlier the space of symmetric ad-invariant polynomials, usually denoted by $\mathbb{I}(\sfG)$, which is a commutative algebra. Given a principal fiber bundle ${\sf P(M,G)}$ there is a homomorphism $w:{\mathbb I}(\sfG) \rightarrow {{\sf H}^\ast(\sfM)} $ between the algebra of ad-invariant symmetric polynomials and the de Rham cohomology of ${\sf M}$, which associates to any polynomial ${P_n }$ a cohomology class  $ {w(P_n) } =P_n(F,\ldots,F)$.
The homomorphism is constructed by filling in the curvature $F$ of ${\sf P(M,G)}$ in all the entries of $P_n$, so as to build the forms $P_n(F,\ldots,F)$, which are closed in ${\sf M}$. They may or may not be exact according to the topology of ${\sf M}$. But in the case the bundle is the universal one with compact group $\sfG$, and the base manifold is the classifying space, then the Weil homomorphism is an isomorphism, i.e. no form  $P_n(F,\ldots,F)$ is exact.

\vskip 1cm
Then let us explain why the anomaly corresponding to the above-mentioned form $TP_{n_1}(V) P_{n_2}(F)$ is universal. This is so because, even if $P_{n_2}(F) = d\eta$ in ${\sf M}$, the form $\eta$ cannot be universal. 
The reason is that for ${\sf BG}$ the  Weil homomorphism is an isomorphism when the group $\sfG$ is compact. This means that $ P_{n_2}(F)$ is not the differential of any universal form and thus  $TP_{n_1}(V) P_{n_2}(F)$ cannot be written itself as a differential of any universal form. From the field theory point of view this means that  $ P_{n_2}(F)$  cannot be written as the total differential of a local expression in the fields. In other words the anomaly originated from  $TP_{n_1}(V) P_{n_2}(F)$ is a true local field theory anomaly. 

A comment is in order to explain this at first sight surprising correspondence. The origin of it lies in the circumstance that standard perturbative field theory is formulated in a local geometry.  Propagators and vertices are evaluated on plane wave configurations of fields, thus in a unique local patch isomorphic to a Minkowski (or Euclidean) spacetime. In this way the cohomological properties of the base space ${\sf M}$, reflected also in the topology of  ${\sf P}$, are irrelevant. Therefore, as far as local anomalies are concerned, the results of perturbative field theories are independent of the topological features of the base manifold ${\sf M}$. The BRST cohomology of local field theories reflects the de Rham cohomology of the classifying space, which is the same for all spacetimes of given dimension. Once again, the roots of local gauge anomalies in QFT are not in the topology of the spacetime $\sfM$ but in the topology of the classifying space $\sf BG$, \footnote{This is not anymore true if we consider global anomalies, in which the topology of $\sfM$ may play a role. Global anomalies call into play the global  aspects of gauge transformations, i.e. the global aspects of $\cal G$. We will not analyse global anomalies in this paper. Enough is to say that the gauge transformations involved in the various SM's considered here do give rise to global anomalies,  \cite{I}.}.

Now it remains for us one step to close the circle: what is the connection of this geometry with the bugs of the corresponding theories? It is the lack of a fermion propagator, as is made clear by the family's index theorem.

\subsection{The Atiyah-Singer index theorems}

The family's index theorem is the index theorem appropriate for consistent local chiral anomalies. In general an (ordinary) index theorem counts the difference between the graded zero modes of an elliptic operator. For instance, zero modes  of a Dirac operator with a fixed potential and metric\footnote{From now on we introduce in our analysis also the dependence on a non-trivial  metric, which we have so far ignored.} are labeled by + and --, which denote opposite chiralities, and the index counts the difference between the + and the -- zero modes. The family's index theorem does the same, except that the operator, in the case of gauge theories, varies on a Hausdorff space, the moduli space of connections (i.e., the space of orbits under the action of the gauge group). The index is formally represented by the space of zero mode eigenvectors (kernel of the operator) of one chirality {\it minus} the space of zero mode eigenvectors of the opposite chirality (cokernel). The {\it difference} of two vector spaces varying from point to point of the parameter space can be given a mathematical sense in terms of K-theory of vector bundles. What really matters to us in relation to the anomaly problem is not as much the number of zero modes, but the existence or non-existence of the inverse of the Dirac operator. It is of paramount importance that one appreciates the difference between the two situations. The presence of even a finite number of zero modes for a Dirac operator with (twisted by) a fixed potential or metric prevents its inversion, but this problem can be easily fixed by excluding the (finite number of) zero modes from the spectrum. The problem however becomes unwieldy when we face a continuous spectrum of gauge potentials (or, better, their orbit space). Only K-theory can and does provide a sensible treatment. 

We recall that this is the same problem we have already met in the perturbative approach to anomalies, as well as in the non-perturbative approach \`a la Schwinger-DeWitt. In those approaches the inverse of the relevant elliptic operator  (or, which is the same, the existence of an effective action for Weyl fermions) is required, and one is obliged, for instance, to replace the Dirac-Weyl operator
\be
\slashed{\ED}= i (\slashed {\partial} + \slashed {V}) P_+,\quad\quad P_\pm = \frac {1\pm \gamma_5}2\label{DW+}
\ee
by
\be
\slashed{\ED}= i (\slashed {\partial} + \slashed {V} P_+)\label{DWP++}
\ee
in order to guarantee the invertibility of the kinetic operator, i.e. the existence of the fermion propagator. In those approaches anomalies show up as non-conservation of the chiral gauge currents. The family's index theorem, instead, calculates for us the obstructions to the invertibility of the Dirac-Weyl operator.  In other words it tells us what are the topological conditions for the Weyl fermion propagator to exist. 

Although these obstructions manifest themselves in different ways in the two approaches, their origin is the same and can be traced to the geometry of universal bundles and classifying spaces. The added value of the family's index theorem approach is that it provides compact formulas of the obstructions for all dimensions in terms of ad-invariant polynomials, gauge curvatures and/or Riemann curvatures; from which one can derive the consistent gauge anomalies by the transgression formulas and the descent equations - and not only those but also other anomalies, as we shall see. It also provides the numerical coefficients (depending on the fermion field representations) in front of the anomalies, based on which one can study in general the conditions for their cancelation. 

In this limited presentation it is of course impossible to expound all the mathematics underlying the family's index theorem, but we try to provide a summary presentation that is hopefully enough to capture the essential points of the subject.

The family's index theorem applies to elliptic operators. Given two vector bundles $\EE$ and $\EF$  over a Riemannian manifold $X$ and the corresponding spaces of sections $\Gamma(\EE)$ and $\Gamma(\EF)$, respectively,
an elliptic operator is a partial differential operator that maps sections  of $\Gamma(\EE)$  to sections of  $\Gamma(\EF)$ in a way which is fiberwise invertible. In practice we are interested in sections representing matter fields and, in particular, spinors.  For instance, an operator with local representative $A_{ij}\partial_{x_i} \partial_{x_j}$ is elliptic if the matrix $A_{ij}(x)$ is symmetric and positive definite; a Euclidean Dirac operator is elliptic. With reference to field theory a linear or quadratic  elliptic operator locally has an inverse.

An elliptic operator  $P$ on a compact space $X$ is endowed with an index { (\it the analytic index)}
\be
ind\, P = dim (ker\, P)-dim(coker\, P)\label{analyticindex}
\ee
where the kernel of $P$ is the space of its zero eigenvectors and the cokernel is formed by the sections of the target bundle that do not belong to the image of $P$. Remarkable properties of an elliptic operator are: the index of the operator depends on its leading term; if $P$ is self-adjoint (same source and target space) it has real discrete eigenvalues with finite eigenspace consisting of smooth sections.

The Dirac operator is the central object for us now. In physics it is usually (but not exclusively) defined on a Minkowski spacetime (a pseudo-Riemannian manifold), say $X$. After a Wick rotation, i.e. in a Euclidean background, it is an example of linear elliptic operator. Dirac operators act on spinors, which are used in physics to represent fermions. A spinor is a section of a vector bundle with two simultaneous properties: it belongs both to a representation space of the appropriate Clifford algebra and a representation of the appropriate spin group; in addition it may transform also according to the representation of some compact group (the gauge group), in which case the corresponding Dirac operator is called twisted.

To be more precise, let $TX$ be the tangent space of $X$. We call $C\ell(X)$ the Clifford algebra, i.e. the gamma matrix algebra, constructed over $TX$ on the basis of the metric inherited from $X$. And let ${\sf P}_{SO}(X)$ be the {\it orthonormal tangent frame bundle} over $X$, i.e. the principal fiber bundle built with the orthonormal frames of $TX$. We need also the spin bundle ${\sf P}_{Spin}(X)$ which is the double covering of ${\sf P}_{SO}(X)$ (a spin structure):
\be
\xi: {\sf P}_{Spin}(X)\longrightarrow {\sf P}_{SO}(X)\0
\ee
where $\xi(p,g) = \xi(p) \xi_0(g)$, for any $p\in {\sf P}_{Spin}(X)$ and any $g \in Spin_n$, $n=dim(X)$, and $\xi_0: Spin_n \longrightarrow SO(n)$ is the universal covering map (fermions see a doubly ample world with respect to us). A spinor bundle $S$ is a bundle associated to ${\sf P}_{Spin}(X)$

In fact we will deal with complex spinor bundles.  A {\it complex spinor bundle} $S_{\mathbb C}(X)$ is an associated bundle
\be
S_{\mathbb C}(X)=\sfP_{Spin}(X)\times_\mu M_{\mathbb C} \label{spinbundleC}
\ee
where $M_{\mathbb C}$ is a left module of $C\ell(X)\otimes {\mathbb C}$ whose action is represented by $\mu$.

In this context the (Euclidean) Dirac operator, denoted by ${\slashed D}$ (we drop in this section the $\widetilde{\slashed D}$ notation), is
\be
\slashed {D}= i\gamma^\mu (\partial_\mu +\frac 12 \omega_\mu) \label{opDirac}
\ee
where $\gamma^\mu =e_a^\mu\gamma^a$  ($e_a^\mu$ are the inverse vielbein, $\eta^{ab} e_a^\mu e_b ^\nu = g^{\mu\nu}$) and  $\omega_\mu =\omega_\mu^{ab}\Sigma_{ab} $ is the spin connection, with $\Sigma_{ab}= \frac 14[\gamma_a, \gamma_b] $.

If $n$ is even $C\ell(n)(X)$ posseses an element $\omega_{\mathbb C}$ (the chirality operator) such that
\be 
\omega_{\mathbb C} ^2 =1, \quad\quad \omega_{\mathbb C} \, e =-e\, \omega_{\mathbb C}\label{omegaC1}
\ee
for any element $e\in C\ell(n)(X)$. As a consequence any $S_{\mathbb C}$ splits according to
\be
S_{\mathbb C} = S_{\mathbb C}^+\oplus S_{\mathbb C}^-, \quad\quad \quad  S_{\mathbb C}^\pm=\frac 12 (1\pm \omega_{\mathbb C}) S_{\mathbb C}\label{Spmsplit}
\ee
Accordingly $\slashed {D}$ splits as
\be 
 {\slashed {D}}= \left(\begin{matrix} 0 & {\slashed {D}}^- \\  {\slashed {D}}^+ & 0 \end{matrix} \right)\label{Diracsplit}
\ee
Therefore we have $ker\,  {\slashed {D}} = ker \,  {\slashed {D}}^+\oplus ker\,  {\slashed {D}}^-$. Now since $ {\slashed D}$ is self-adjoint, it follows that $( {\slashed {D}}^\pm)^\dagger = {\slashed {D}}^\mp$. And since for any elliptic  operator $T$, $ coker\, T= ker\, T^\dagger$, it follows that
\be
Ind\, {\slashed {D}}^+ = dim(ker\, {\slashed {D}}^+ )- dim (ker\, {\slashed {D}}^-)\label{indexD+}
\ee

In gauge field theory we have to do with spinors that transform according to definite representations 
 of the gauge group $\sfG$. In this case we generalize the previous setup and consider the tensor product of a spinor bundle $S^\pm_{\mathbb C}$ with a vector bundle $E$ corresponding to a representation $\rho$ of the structure group $\sfG$ of ${\sfP}(X,\sfG)$: $S^\pm_{\mathbb C}\otimes E$. The relevant connection will be the spin connection plus a gauge connection $V=V_\mu dx^\mu$ valued in  the representation $\rho$ of the Lie algebra of $\sfG$ with antihermiten generators. The corresponding Dirac operator
\be
\slashed {\ED} = \slashed {D} +i\slashed {V}\label{ED+V}
\ee
acts on the space of sections of $S^\pm_{\mathbb C}(E)\equiv S^\pm_{\mathbb C}\otimes E$ and maps it to itself. In this case
eq.\eqref{Spmsplit} is replaced by
\be
S_{\mathbb C}(E)= S^+_{\mathbb C}(E)\oplus S^-_{\mathbb C}(E)\label{SE+SE-}
\ee
and eq.\eqref{Diracsplit} by
\be 
 {\slashed {\ED}}= \left(\begin{matrix} 0 & {\slashed {\ED}}^- \\  {\slashed {\ED}}^+ & 0 \end{matrix} \right)\label{DiracEDsplit}
\ee
with obvious replacements in the definition of the index.

\subsubsection{The Atiyah-Singer index theorem}

In this context the {\it Atiyah-Singer index theorem} says:
\be
Ind ({\slashed{\ED}}^+) = \int_X \, ch(E) \,\hat A(X)\label{indexED}
\ee
The symbol $ch(E)$ indicates the Chern character of the $E$ bundle, i.e. the rational characteristic class given in terms of the curvature $F$ of $V$ by
\be
ch(E)= r + \frac i{2\pi} \tr\, F + \frac {i^2}{2(2\pi)^2} \tr\, F^2+\ldots+ \frac {i^n}{n! (2\pi)^n} \tr \, F^n +\ldots\label{chE}
\ee
where $r$  is the dimension of the representation $\rho$. The symbol $\widehat A(X)$ denotes the {\it $\widehat A$-genus}, which is the (rational) Pontryagin characteristic class of $X$. It can be expressed in terms of the Riemann curvature $R$ as follows
\be
\widehat A(X)&=& 1 +\frac 1{(4\pi)^2} \frac 1{12}\, \tr \, R^2 + \frac 1{(4\pi)^4}\left[ \frac 1{288}\, \tr\, (R^2)^2 +\frac 1{360}\,\tr\, R^4\right]\0\\
&&\quad+  \frac 1{(4\pi)^6}\left[\frac 1{10368}\, \tr\, (R^2)^3+ \frac 1{4320} \,\tr \, R^2\, \tr\,R^4+\frac 1{5670}\, \tr\, R^6\right]+\ldots \label{hatAX}
\ee

This famous theorem holds for a {\it fixed background}, i.e. for a fixed metric and a fixed gauge potential. But, although it features in several physical applications, it is not what we need in relation to anomalies. As already pointed out we are not interested in the numerical counting of zero modes, but in the invertibility of the Dirac operator over the whole space of metric and potential backgrounds. 

The field theory path integral is a functional of the fields, in particular of the potentials or the metrics or both. Let us focus simply on a gauge theory defined on a principal fiber bundle $\sf P(M,G)$, with a space ${\cal A}$ of connections and a symmetry group ${\cal G}$ of gauge transformations. We denote by ${\cal Z}[V]$ the fermion determinant, i.e. the partial path integral in which we suppose only the quadratic part of the action is considered, and fermions have been integrated out. ${\cal Z}[V]$ is a functional of the gauge potential which can be computed by means of perturbative and non-perturbative methods. But it contains also the gauge degrees of freedom, of which we have to get rid of since the  relevant information for physics is  stored in the orbit space ${\cal A}/{\cal G}$. The full path integral is obtained by integrating over the orbit space
\be
{\cal Z}= \int_{{\cal A}/{\cal G}} d[V] \, {\cal Z}[V]\label{totalZ0}
\ee
Here $[V]$ denotes the orbit  passing through $V$, and ${\cal G}={\sf Aut_v(P)}$.
The issue here is to guarantee that  ${\cal Z}[V]$ is  a well-defined function, integrable over ${\cal A}/{\cal G}$. 

The geometry is represented by the following bundle diagrams:
\be
\xy\xymatrix{{\cal G}\,\,\ar@{^{(}->}[r]
&{\cal A}\ar[d]  \\
& {\cal A}/  {\cal G}}
\endxy\label{graph1}
\ee
which is a principal fiber bundle with structure group ${\cal G}$. Since the actions of $\sfG$ and $\cal G$ commute, another relevant principal fiber bundle is
\be
\xy\xymatrix{{\sf G}\,\,\ar@{^{(}->}[r]
&\frac{{{\sf P}\times \cal A}}{\cal G}\ar[d]  \\
&\frac {\sfM\times\cal A}  {\cal G}}\endxy\label{graph2}
\ee
with structure group $\sfG$. In turn $\frac {\sfM\times\cal A}  {\cal G}$ is a fiber bundle over $\frac {\cal A}{\cal G}$, with fiber $\sf M$. Given a  complex vector space $\sf E$, which is a representation space of $\sfG$,  we have also two associated fiber bundles 
\be
E={\sf P} \times_\sfG {\sf E}, \quad\quad {\rm and}\quad\quad  
 \EE= \frac {{\sf P}\times {\cal A}}{\cal G} \times_{\sfG} {\sf E}\label{twobundle}
\ee

From the above we see that what is required in a gauge theory is an index theorem over the entire orbit space, because we have to deal with an operator that varies from point to point in it. The dimensions of its kernel and cokernel may jump from point to point in the parameter space provided by the orbit space. The relevant object in this case is the difference between the kernel and cokernel, which are two vector spaces, as they vary from point to point, i.e. the difference between two vector bundles over the parameter space.  The appropriate mathematical theory in this situation is $K$-theory. K-theory is precisely the theory of formal differences of vector bundles. Denoting by $[E]$ and $[F]$ the classes of vector bundles isomorphic to $E$ and $F$, respectively, via the Grothendieck construction, K-theory assigns a precise meaning to the difference $[E]-[F]$. Such objects are called virtual bundles.

Let us define first the index for families. The  geometrical environment is the one corresponding to (\ref{graph1},\ref{graph2}). We have a Hausdorff space $\EB$ which is the basis of a principal fiber bundle (like ${\cal A}/{\cal G}$ in \eqref{graph1}). 
A continuous family $\EV$ of smooth vector bundles over $X$, parametrized by $\EB$, is a fiber bundle $\EV \rightarrow \EB$ whose fiber is a smooth vector bundle $V$ over $X$, associated to a principal fiber bundle ${\sf P}(X,\sfG)$, on which the group ${\sf Aut_v(P)}$ acts ($\EV$  is like $\EE$, defined above). Now, let us consider two families $\EV$ and $\EW$ with the respective spaces of sections $\Gamma(\EV)$ and $\Gamma(\EW)$,  and a differential operator $P: \Gamma(\EV)\rightarrow \Gamma(\EW)$.

 $P$ defines a family of elliptic operators which associates at each point of $\EB$ an elliptic operator between $\Gamma(\EE)$ and $\Gamma(\EW)$. The kernel and cokernel of $P$ are vector spaces that vary from point to point in $\EB$. Therefore their difference defines an element of $K(\EB)$:
\be
ind\, P = [ker\, P]-[coker\, P] \in K(\EB)\label{analyticindex}
\ee
 Even though the dimensions of these vector bundles may vary in $\EB$, the definition makes sense. Atiyah and Singer have provided a formula of it in terms of characteristic classes. 

\subsubsection{The family's index and the Quillen determinant}

In the case outlined above in which $P$ is the Dirac operator ${\slashed \ED}^+$ between two families of sections $\Gamma(S^+_{\mathbb C}\otimes \EV)$ and  $\Gamma(S^-_{\mathbb C}\otimes \EV)$ the index theorem can be expressed in the form
\be
ch \left(ind ({\slashed \ED}^+)\right) =\int_{\sfM}  ch( {\EV})\cdot \hat A(T{\cal Q})\label{chindfam}
\ee
In this case $X=\sfM$ is the spacetime manifold, ${\cal Q}\equiv \frac{\cal A}{\cal G}$ is the orbit space of connections, $T{\cal Q}$ is its tangent space and $\EV$ is the gauge bundle. The Chern character of the index measures the extent to which $ker\, {\slashed \ED}^+$ differs from $coker \,{\slashed \ED}^+=ker \,{\slashed \ED}^-$. It is intuitive that, as long as, $ind ({\slashed \ED}^+)$ differs from 0, the inverse of ${\slashed \ED}^+$ cannot exists (see below for a more cogent argument). Therefore, {\it the only way to ensure the existence of this inverse (and, so, the existence of the fermion propagator and, in turn, the existence of the fermion path integral) is that the RHS of \eqref{chindfam} correspond to the 0 class}. This is particularly visible with the first Chern class, i.e. the first non-trivial term of the Chern character. Let us recall that, for a vector bundle $V$
\be
ch(V)=rank(V)+ c_1(V)+\frac 12 \left(c_1(V)^2 -2 c_2(V)\right)+\ldots\label{cherncharacter}
\ee
Therefore 
\be
c_1\left(ind ({\slashed \ED}^+)\right)= \left.\int_M  ch( {\EV})\cdot \hat A(T{\cal Q})\right|_{{\sf d},2}\label{c1chiindfam}
\ee
where ${\sf d}=dim(M)$ and on the RHS we indicate
 the  $(\sfd,2)$ component in $M\times {\cal Q}$. The RHS is a two-form  in ${\cal Q}$, therefore it belongs to a class in $H^2({\cal Q},{\mathbb Z})$. It represents the first Chern class of the so-called Quillen determinant bundle. Given the importance of the latter it is worth spending some time to introduce it in more detail.
 
Let us focus on the case of a twisted Dirac operator, for instance ${\slashed \ED}$ considered above. It is an elliptic operator on a compact space, and so a Fredholm operator, which splits according to eq.\eqref{DiracEDsplit}. If we wish to define the determinant of either ${\slashed \ED}^+$ or  ${\slashed \ED}^-$, as the product of  its eigenvalues, there may be obstacles. As pointed out in section \ref{ss:perturbative}, the first obstacle is the very definition of eigenvalues of a Dirac-Weyl operator. In perturbative and SDW approaches we have bypassed this problem by modifying the Dirac-Weyl operator like in \eqref{DWP++}. Here instead we consider the quadratic operator $({\slashed \ED}^+)^\dagger{\slashed \ED}^+$ \footnote{We use this operator only in order to discuss the eigenvalue splitting, not for a direct calculation for instance by the SDW method, because for that purpose it is inapplicable for an obvious reason: it is not invertible!} . It has the same eigenvalues as  $({\slashed \ED}^-)^\dagger{\slashed \ED}^-$, which are the square of the eigenvalues of the Dirac operator ${\slashed \ED}$, but of course the eigenvectors are different. More precisely, the nonzero eigenvalues of  $({\slashed \ED}^+)^\dagger{\slashed \ED}^+$ are paired with the non-zero eigenvalues of  $({\slashed \ED}^-)^\dagger{\slashed \ED}^-$, in the sense that to each common eigenvalue there correspond two distinct eigenvectors of the two operators with the same multiplicity. But this may not be true anymore for the zero eigenvalues. Of course this is the crucial point in regard to the index theorem.  But it is crucial also for defining the path integral of the two Dirac-Weyl operators. If we wish to define the relevant path integrals as the square root of the product of their eigenvalues, the non-zero eigenvalues do not pose a problem because they are gauge invariant, but for the zero eigenvalues this may not be true. Therefore we have to settle the question of the zero modes in such a way as to produce a well defined integrable ${\cal Z}[V]$.

Let us denote by $\psi_1,\ldots,\psi_r$ a basis for the zero modes of ${\slashed \ED}^+$. In field theory a familiar object,  due to the Pauli principle, is the Slater determinant, i.e. wedge product $\psi_1\wedge\ldots\wedge \psi_r$. It is a complex number (a determinant) that varies as   $ {\slashed \ED}^+$ varies over ${\cal X}\equiv \frac {\sfM\times {\cal A}}{\cal G}$, but, in general, does not form a complex line bundle over ${\cal X}$.  In the same way one can construct an analogous wedge product for $coker \,{{\slashed \ED}}^+=ker\,{{\slashed \ED}^-}$, which also form a complex line, but in general not a line bundle over ${\cal Q}$. However the formal difference of the two, i.e.
\be
\coprod_{[V]\in {\cal X}} \bigwedge^{dim\,{ker {\slashed \ED}_V^+}} ker ( {\slashed \ED}_V^+)^\dagger \otimes  \bigwedge^{dim\,
{coker {\slashed \ED}_V^+}} coker ( {\slashed \ED}_V^+)
\label{Quillen}
\ee
where we have made explicit the dependence on the connection $V$, 
 constitutes an element of $K({\cal X})$, which is called the Quillen determinant line bundle.
This complex line bundle has remarkable topological properties. For us it is a fundamental object because it is trivial, i.e. it has a global non-vanishing section, precisely when  ${\slashed \ED}^+$ is invertible\footnote{{This is what D. Quillen has proved for a Cauchy-Riemann operator over a Riemann surface}. We are not aware of a general proof of this theorem extending to all the cases considered here. We are therefore assuming its validity in the present context, validity which is on the other hand rather plausible, \cite{atiyahsinger1984}, and  heuristically supported by the results in quantum field theory.}.

In the case of the Dirac-Weyl operator ${\slashed \ED}^+$ the Quillen determinant bundle is a bundle over the base space ${\cal X}\equiv \frac {\sfM\times {\cal A}}{\cal G}$. It is a non-trivial bundle as long as $H^2({\cal X},{\mathbb Z})\neq 0$. Upon integrating along the fiber $M$, the relevant cohomology group is $H^2({\cal Q},{\mathbb Z})$ and coincides with $c_1\left(ind ({\slashed \ED}^+)\right)$. If the index vanishes the Quillen bundle is trivial, that is it has a global section, which is a non-vanishing function over  ${\cal A}/{\cal G}$ and can be integrated over. Returning now to the definition \eqref{totalZ0} of the path integral over the orbit space, we conclude that the existence of this section implies that ${\cal Z}[V]$ is well defined.

The formal equation  
\be
\delta \det A = \det A \, \tr \left(A^{-1} {\delta} A\right).\label{deltaderminantofA}
\ee
when applied to $A={\slashed \ED}^+$, implies that, in order for  ${\cal Z}[V]$ to be well defined, the inverse of  ${\slashed \ED}^+$, i.e. the fermion propagator,  must be well defined. The inverse operator is represented by the inverse eigenvalues. For the non-zero eigenvalues invertibility is obvious. For the 0 eigenvalue the inverse is represented by the inverse of the global section of the Quillen determinant bundle. We expect, therefore, that, if  $H^2({\cal Q},{\mathbb Z})$ is trivial, this global section is invertible.
On the contrary a nonzero $c_1\left(ind ({\slashed \ED}^+)\right)$, given by \eqref{c1chiindfam}, is an obstruction to defining the inverse of  ${\slashed \ED}^+$, i.e. the Weyl fermion propagator. The vanishing of  $c_1\left(ind ({\slashed \ED}^+)\right)$ is therefore a necessary condition for the existence of a well-defined partition function.

\subsection{Closing the circle: obstructions and anomalies}

We have just seen that, in a Euclidean $\sfd$-dimensional spacetime $\sfM$, the obstructions to the existence of a Weyl fermion propagator (i.e. the existence of the inverse of ${\slashed \ED}^+$, for instance), is contained in the class
\be
ch( {\EV})\cdot \hat A(T{\cal X})\label{chEhatA}
\ee
where ${\cal X}\equiv \frac {\sf M\times {\cal A}}{\cal G}$.  The question is now: where is the connection, if any, between the just outlined mathematical approach, based  on the obstruction concept, and chiral gauge anomalies? The answer is: the connection is in the topology of the classifying space.  It is now time to establish this relation. The exposition becomes now inevitably more abstract and we will have to assume some results without (even an attempt of a) proof.

For simplicity we consider the pure gauge case. It corresponds to taking \eqref{chEhatA} with $\hat A(T{\cal X})=1$. Then recall that ${\cal G}=\sf Aut_v(P)$ in reference to the principal fiber bundle $\sf P(M,\sfG)$,  that ${\cal A}\rightarrow  {\cal A}/{\cal G}$ is a principal fiber bundle with structure group ${\cal G}$, \eqref{graph1}, that $\frac {{\sf P}\times {\cal A}}{\cal G} \rightarrow \frac {{\sf M}\times {\cal A}}{\cal G}$ is a principal fiber bundle with group $\sf G$, \eqref{graph2}, and, finally, that $ \frac {{ M}\times {\cal A}}{\cal G}$ is a fiber bundle over ${\cal Q}= \frac {\cal A}{\cal G}$ with fiber $\sf M$ (in the case of ${\cal G}=\sf Aut_v(P)$, the bundle  $ \frac {{ M}\times {\cal A}}{\cal G}$ reduces to $M\times \frac {\cal A}{\cal G}$ because the basepoint is not moved by a vertical automorphism). Let us endow $\cal A$ with a connection $\omega$ in the fiber bundle ${\cal A}\rightarrow  {\cal A}/{\cal G}$  (see Appendix 12A of \cite{I} for further details). Then we can define a connection $\eta$ on the $\sfG$ bundle ${\sf P}\times {\cal A}$ over $\sf M\times {\cal A}$ as follows:
\be
\eta_{p,V}(X,Y) = V_p(X)+ V(\omega(Y))_p, \quad\quad X\in T_p{\sf P}, \quad Y \in T_V{\cal A}\label{conneta}
\ee
where $V$ is a connection in $\sfP$ and $\omega$ a connection in ${\cal A}\rightarrow {\cal A}/{\cal G}$. 
To render this construction less abstract let us notice that, looking at the definition \eqref{conneta}, if we restrict $\eta$ to the orbit of $\cal G$ we obtain $\eta= ev^\ast V$, where $ev: \sfP \times {\cal G}\rightarrow \sfP$ is the evaluation map introduced in section \ref{ss:evaluation}.

This connection descends to a connection $\eta'$ on the fiber bundle \eqref{graph2}
\be
\xy\xymatrix{{ \sf G}\,\,\ar@{^{(}->}[r]
&\frac{{{\sf P}\times \cal A}}{\cal G}\ar[d]  \\
&\frac {\sfM\times\cal A}  {\cal G}}
\endxy\label{graph3}
\ee
We can therefore introduce the curvature  $\EF_{\eta'}$ of $\eta'$ and express $ch (\EV)$ in terms of it . Since the Chern character can be expressed as a polynomial in the curvature, the relevant piece for dimension $\sf d$ will take the form $P_n(\EF_{\eta'}, \ldots,\EF_{\eta'})$, where
$n= \frac {\sf d}2+1$ and $P_n$ is an ad-invariant (reducible or irreducible) polynomial for the group $\sf G$. Now let us recall that, in the definition \eqref{conneta} of the connection $\eta$, the connection $V$ in $\sf P$, can be obtained from the universal connection ${\mathfrak v}$ (with curvature $F_{\mathfrak v}$) via a map $\hat f:\sf P\rightarrow \sf EG$, $V=\hat f ^\ast {\mathfrak v}$. In the same way also $\omega$, the connection in ${\cal A}$, can be pulled back from the same universal connection via a map $\hat{\sf f}: {\cal A}\rightarrow \sf EG$: $\omega=\hat{ \sf f}^\ast {\mathfrak v}$. Therefore there exist an overall map $\hat f_\times=(\hat f,\hat{\sf f}):{\sf P}\times{\cal A}\rightarrow {\sf EG}$, such that $\eta=\hat f_\times^\ast {\mathfrak v}$ (for a more detailed discussion of $\eta$ and $\eta'$, see Appendix 12A of \cite{I}). Therefore
\be
P_n(\EF_{\eta}, \ldots,\EF_{\eta})= f_\times^\ast P_n(F_{\mathfrak v},\ldots,F_{\mathfrak v})\label{PnF*Pn}
\ee
The LHS then  descends to $P_n(\EF_{\eta'}, \ldots,\EF_{\eta'})$ on \eqref{graph3}.
Remember that $P_n(F_{\mathfrak v},\ldots,F_{\mathfrak v})$ is a closed form in $\sf EG$ that descends to a closed form in $\sf BG$, which represents a non-trivial cohomology class of the classifying space.

Now, we know that, via the evaluation map and the transgression formula, we can extract from this class a corresponding chiral anomaly, which is given by $i_{(\cdot)}TP_n(V) $. This can be seen directly from the transgression formula ensuing from \eqref{PnF*Pn}
\be
TP_n(\eta)= n\int_0^1 dt\, P_n(\eta, \EF_{\eta},\ldots,\EF_{\eta}),\label{TPneta}
\ee 
which splits in components of order $(2n\!\!-\!\!p\!\!-\!\!1,p)$ in $\sfP$ and ${\cal A}$. The $(2n\!\!-\!\!1,0)$ is precisely $TP_n(V)$.   We also know that any local chiral anomaly takes this form and originates from a class in $\sf BG$. Therefore we see that {\it consistent gauge anomalies and obstructions to the existence of a Weyl fermion propagator coming from the index theorem, are rigidly connected}\footnote{{ The reader should not be misled by the fact that the family's index theorem is formulated for a compact manifold $\sfM$ while  local field theory is formulated in a local patch. As explained above in section \ref{ss:evaluation} the link between the two is natural in the framework of classifying space. The BRST cohomology, that is the cohomology of anomalies, is the same as the deRham cohomology of the classifying space, whose geometry encompasses all compact manifolds.}}
The added bonus of the family's index theorem is that it supplies the precise numerical coefficients of the irreducible polynomials and the product of reducible ones (see \eqref{chE} for the case of spin 1/2 \footnote{For other cases, for instance, for spin 3/2, we need appropriate modifications..}). This allows us to cope with the problem of anomaly cancelation in general.

\subsection{Anomaly cancelation: another look}
\label{ss:anomalycanc}

In the light of the previous results it is worth at this point to return to the issue of anomaly cancelation. 
In general, absence or cancelation of local anomalies takes place:

\begin{itemize}
\item (A) in gauge field theories or sigma models in which the gauge group $\sfG$ has vanishing ad-invariant tensors in the relevant representations. Consistent anomalies are determined by ad-invariant polynomials $P_n$, or by the corresponding ad-invariant tensors $t^{a_1\ldots a_n}$ with a specific coefficient depending on the representation of the matter fields involved. In many simple groups some of these tensors vanish identically. This is the case, as far as the polynomial $P_3$ is concerned,  for all simple groups except $SU(N)$, for $N\geq 3$. This cancelation occurs at the very origin, in the sense that the form $P_3(F(\mathfrak {v}),F(\mathfrak{v}),F(\mathfrak{v}))$, where $\mathfrak{v}$ is the universal connection, identically vanishes in the classifying space. Even in groups where these tensors are non-trivial it may happen they vanish depending on the representation of the fundamental fields. A well-known example is the case of matter in the adjoint representation in 4$\sfd$.

\item (B) in gauge field theories or sigma models with different species of fermion fields, which are separately anomalous, but, when put together in the same theory, the coefficients of the various consistent  anomalies sum up to 0. The standard model of particle physics is an example (see above).

The absence of anomalies in the SM is due to a combination of (A) and (B).

\item (C) there is also a third case, where chiral anomalies due to elementary fermion fields do not cancel completely, but cancelation may take place thanks to other fields in the theory. Such fields are endowed with transformation properties that allow them to cancel the original anomalies of the theory. This is the case of the anomalies canceled by Wess-Zumino terms, or the case of the Green-Schwarz mechanism, vastly used in target space field theories derived from superstrings. 

\end{itemize}

\subsubsection{Other anomalies?}

The family's index theorem of an elliptic operator (like  Dirac-Weyl operator) signals the obstructions to the existence of an inverse (the propagator). So far we have considered the obstructions related to the $({\sf d},2)$ component in $P\times {\cal A}$, see \eqref{c1chiindfam}. We have seen that this obstruction  is linked to the non-triviality of the Quillen determinant. There are other components on the RHS of \eqref{c1chiindfam}. In \eqref{chEhatA} there are forms of any even degree 2,4,6,8,... In $\sfd=4$ it is reasonable to disregard the forms of order 2,8,10,... for dimensional reasons\footnote{But a careful investigation of this point is still lacking.}. We know that chiral gauge anomalies come from the form of degree 6. But what about the degree 4 forms? These forms may in principle couple to the theory. What is their meaning? There is no literature on this subject, but this does not mean that we are allowed to casually shrug off an explanation. 

Then let us focus on the $({\sf d},0)$ component, for ${\sf d}\!=\!4$. 
There are components coming from the $\widehat A$-genus and from $ch(\EV)$.  
The first gives a term $\frac 1{192\pi^2}\tr(R^2)$, the second (in the non-Abelian case) a term $\frac 1{8\pi^2} \tr F^2$.
What can be the meaning of these two terms? The geometrical meaning of the $({\sf d},0)$ component  in $P\times {\cal A}$ could be the `virtual' rank of the index bundle, as one can deduce from the formulas \eqref{chindfam} and \eqref{cherncharacter}. Let us elaborate a bit on this argument. The index, \eqref{analyticindex}, is a local mathematical object that represents an asymmetry between the kernel of $\slashed\ED^+$ and the kernel of  $\slashed\ED^-$. As long as this asymmetry is non-vanishing it represents a threat to the existence of the propagator (and the theory). The above-mentioned densities (of the Chern and Pontryagin classes) are an additional hazard beside the ones contained in the $({\sf d},2)$ component (in $P\times {\cal A}$) of the index, which give rise to the consistent chiral anomalies.\footnote{The integration over $\sfM$ in \eqref{chindfam}  should not mislead the reader: for some $\sfM$ this integral may vanish, but, we should never forget that, in the spirit of locality, such formulas as \eqref{chindfam} hold for any $\sfM$ (see discussion after \eqref{closed/exact}.} This is the mathematical side of the problem. It is to be expected that such a problem will show up also on the field theory side. A physical interpretation is needed.

In specific gauge field theories that couple to a metric, beside the gauge potentials, there is a quantity that naturally couples to the above mentioned odd-parity densities: it is the trace of the energy-momentum tensor of a theory of chiral fermions. 
The corresponding trace anomalies are not obtained via the transgression formula from the universal bundle like the chiral consistent anomalies {(the way we have illustrated above)}, but are pulled back directly to the spacetime $\sfM$ via the classifying map. On the basis of our experience with quantum theories, the correspondence of quantum numbers (form degree, dimension, parity, Lorentz covariance) is perfect and,  barring the presence of  a symmetry that forbids it, a non-vanishing coupling has to be expected.

\vskip .5cm

This will be confirmed in the sequel, but before proceedings along this line of analysis we need an introduction to diffeomorphisms, Weyl transformations and related anomalies.
\vskip 0,3cm

\noindent Ref.:\cite{oalvarezsingerzumino1984,atiyahsinger1984,bonoracottaRS1987,bonoracottaRS1988,
Husemoller1966,Lawson,Quillen}

\section{Gravitational anomalies}
\label{s:gravanom}

When gravity comes into play the main character becomes the metric $g_{\mu\nu}(x)$ and a fundamental symmetry becomes the symmetry under diffeomorphisms. 
The latter correspond to the general coordinate transformations $x^\mu \to x^\mu +\xi^\mu(x)$, where $\xi^\mu(x)$ are generic infinitesimal smooth functions of the coordinates. They act on the metric as follows
\be
\delta_\xi g_{\mu\nu} = D_\mu \xi_\nu +D_\nu \xi_\mu, \quad\quad \xi_\mu= g_{\mu\nu} \xi^\nu\label{deltaxigmunu}
\ee
$D_\mu$ is the covariant derivative: $D_\mu \xi_\nu = \partial_\mu \xi_\nu- \Gamma_{\mu\nu}^\lambda \xi_\lambda$, and $\Gamma_{\mu\nu}^\lambda $ are the Christoffel symbols.

The basic covariant geometrical object in this context is the Riemann tensor
\be
R^\rho{}_{\lambda\mu\nu} &=&\partial_\mu \Gamma_{\nu\lambda}^\rho-\partial_\nu \Gamma_{\mu\lambda}^\rho+ \Gamma_{\mu\sigma}^\rho \Gamma_{\nu\lambda}^\sigma
- \Gamma_{\nu\sigma}^\rho \Gamma_{\mu\lambda}^\sigma\label{bfERcomponents}
\ee 
It is convenient to introduce a compact matrix-form notation for the Christoffel symbols
\be
\Gamma \equiv \{ \Gamma_\nu{}^\lambda\}, \quad\quad \Gamma_\nu{}^\lambda= dx^\mu\,
\Gamma_{\mu\nu}{}^\lambda\label{Christ}
\ee
and for the Riemann curvature tensor
\be
\ER = {d}\Gamma +\Gamma \wedge \Gamma, \quad\quad\ER &=& \{\ER^\rho{}_\lambda\}, \quad\quad \ER^\rho{}_\lambda=\frac 12 dx^\mu\wedge dx^\nu\, R_{\mu\nu}{}^\rho{}_\lambda \label{ER}
\ee
The product between adjacent entries is understood to be the matrix product.

The BRST transformations for the gravitational case are obtained by promoting the parameters $\xi^\mu$ to anticommuting fields. The BRST transformations  of $g_{\mu\nu}$ are the same as above, \eqref{deltaxigmunu}, but now $\xi^\mu$ are anticommuting fields, which transform themselves as
\be
\delta_\xi \xi^\mu= \xi^\lambda\partial_\lambda \xi^\mu\label{deltaximu}
\ee
It is not hard to show that, due to this transformation,  $\delta_\xi$ becomes a nilpotent operation:
\be
\delta_\xi^2=0\label{deltaxisquare}
\ee
This is the coboundary operator for diffeomorphisms. It defines the corresponding cohomology problem, analogous to the one defined above for gauge theories.

In the sequel we will be interested in fermions $\psi$  interacting with gravity. The relevant action is
\be
S[g,\psi]= \int d^{\sfd} x \, \sqrt{g} \, i\overline {\psi} \gamma^\mu(D_\mu +\frac 12 \omega_\mu )\psi \label{actiongpsi}
\ee
where $\gamma^\mu = e^\mu_a \gamma^a$, and $e^\mu _a$ are the vielbein ($\mu,\nu,...$ are world indices, $a,b,...$ are flat indices). $\omega_\mu$ is the spin connection:
\be
\omega_\mu= \omega_\mu^{ab} \Sigma_{ab}\0
\ee
where $\Sigma_{ab} = \frac 14 [\gamma_a,\gamma_b]$ are the Lorentz generators. 
In eq.\eqref{actiongpsi} $\psi$ is a generic fermion (Dirac, Weyl or Majorana). A right-handed Weyl fermion will be more explicitly denoted $\psi_R= \frac {1+\gamma_\ast}2 \psi$, where $\psi$ is a Dirac field and $\gamma_\ast$ denotes the chirality matrix (in 4$\sfd$ it is $\gamma_5$). 

A fermion field transforms, under diffeomorphisms, as
\be
\delta_\xi \psi= \xi^\mu\partial_\mu \psi,\label{deltaxi[si}
\ee
and it is easy to prove that \eqref{actiongpsi} is invariant under diffeomorphisms.

Classically the energy-momentum tensor 
\be
T_{\mu\nu}= \frac 2{\sqrt g} \frac {S[g]}{\delta
g^{\mu\nu}}=\frac i4 \overline {\psi} \gamma_\mu {\stackrel{\leftrightarrow}{\nabla}}_\nu\psi+ \{\mu\leftrightarrow \nu\}-{ \frac i2 \eta_{\mu\nu} \overline {\psi} \gamma^\mu(D_\mu +\frac 12 \omega_\mu )\psi } ,\quad\quad \nabla_\mu =D_\mu +\frac 12 \omega_\mu,
 \label{emt}
\ee
is both conserved on shell and traceless, i.e.
\be
D^\mu T_{\mu\nu}(x) =0, \quad\quad T_\mu^\mu (x) =0\label{conservation}
\ee
The first condition follows from diffeomorphism invariance, the second from invariance of \eqref{actiongpsi} under the Weyl transformation
\be
\delta_\omega g_{\mu\nu}(x)= 2 \omega(x)\, g_{\mu\nu}(x)\label{Weyltransf}
\ee
where $\omega(x)$ is an arbitrary infinitesimal parameter.

The presence of fermions in a theory carries in the game also another local symmetry, the local Lorentz symmetry. { Since it will play a marginal role in this review we limit ourselves to a short mention. Local Lorentz transformations on a spinor $\psi$ are defined by
\be
\delta_\Lambda \psi= -\frac 12  \Lambda \psi, \quad\quad \Lambda= \Sigma_{ab} \Lambda^{ab}\label{delatLambdapsi}
\ee 
where $\Lambda^{ab}(x)$ are arbitrary infinitesimal parameters, antisymmetric in $a,b$. This implies
\be
\delta_\Lambda \omega_\mu=  \partial_\mu\Lambda +\frac 12 [\omega_\mu, \Lambda]\label{deltaLambdaomega}
\ee
On the vielbein $e^a_\mu$ the transformation is $\delta_\Lambda e^a_\mu= \Lambda^a{}_b e^b_\mu$. Like for other symmetries we will promote also $\Lambda_{ab}(x)$ to an anticommuting field, denoted by the same symbol, and endowed with the transformation property $\delta_\Lambda \Lambda_{ab} =- \Lambda_a{}^c \Lambda_{cb}$.
With this property $\delta_\Lambda$ becomes a nilpotent operation: $\delta_\Lambda^2=0$. Consistency requires that $\delta_\xi \Lambda_{ab} = \xi^\mu \partial_\mu \Lambda_{ab}$ and $\delta_\Lambda \xi^\mu=0$, which implies that
{\be 
(\delta_\xi+\delta_\Lambda)^2=0.\label{deltaxideltaLambda}
\ee}}

\vskip 0,5cm

A pause is necessary at this point to make an important recall. In quantum field theory we have normally to do with several symmetries of the action simultaneously, and it does not make sense to to analyse the anomalies of a single symmetry disregarding the others. For the sake of simplicity let us consider two symmetries, which we shall refer to as $R$ and $S$ (for instance vector and axial gauge transformations, or diffeomorphisms and local Lorentz or Weyl transformations). Let $\delta_R$ and $\delta_S$ be the corresponding coboundary operators. They are functional differential  operators
linear in the relevant ghost fields, and satisfy
 \be
 \left(\delta_R+\delta_S\right)^2=0\label{coho1}
 \ee
(see  for instance \eqref{deltaxideltaLambda}), which splits into
\be
\delta_R^2=0\quad\quad \delta_S^2=0,\quad\quad \delta_R\delta_S+\delta_S\delta_R=0\label{coho2}
\ee
A cocycle of $\delta_R+\delta_S$  is in general an overall cocycle, i.e. the sum of two integrated local expressions $\Delta_R$ and  $\Delta_S$ such that
\be
\left(\delta_R+\delta_S\right)W= \Delta_R+\Delta_S\label{coho3}
\ee
Therefore
\be
\left(\delta_R+\delta_S\right)\left(\Delta_R+\Delta_S\right)=0\label{coho4}
\ee
This equation splits into three distinct ones
\be
\delta_R \Delta_R=0,\quad\quad \delta_S\Delta_R+ \delta_R \Delta_S=0,\quad\quad \delta_S\Delta_S=0
\label{coho5}
\ee
In all known cases so far one can find a local expression of the fields (excluding ghosts), ${\cal C}$, such that, for instance,  $\Delta_R=\delta_R {\cal C}$, and we can redefine
$W \rightarrow W'= W-{\cal C}$, so that
\be
\delta_R W'&=& \Delta_R' =0\label{coho6}\\
\delta_S W'&=& \Delta_S -\delta_S {\cal C}\equiv \Delta_S' \label{coho6'}
\ee
That is, in all known cases, the anomaly cocycle can be cast in the form that violates only one symmetry.
\vskip 0.5cm
After the previous technical introduction, it is perhaps useful to give a glance at the anomalies related to diffeomorphisms and Weyl transformations that are in store for us.

Consistent diffeomorphisms and local Lorentz anomalies have the same form as the gauge anomalies introduced in section \ref{ss:WZ-BRST}, in particular in formula \eqref{anom4d}, with $V$ replaced by $\Gamma$ or $\omega$, $F_t$ replaced by $ \ER_t$ or $R_t$,  and $\lambda$ by $\Xi^\mu{}_\nu= \partial_\nu \xi^\mu$ or $\Lambda$, respectively\footnote{$R$ is defined by $R= d\omega +\frac 12 [\omega,\omega],$ it represents the matrix-form $ R=\{ R^{ab}\}$ and it is related to $\ER$ by
$\ER^{\mu}{}_\nu = e_a^\mu\, R^{ab} \, e_{b\nu} $.}.
It is not worth spending more space for these anomalies because in $4\sfd$ they almost all vanish due to the mechanism (A) explained above: the corresponding irreducible ad-invariant symmetric polynomials vanish identically, as  already pointed out after the table \eqref{table}). When the relevant polynomial is reducible, however, we may come across a non-trivial anomaly. This may happen with mixed gauge-gravity anomalies. We shall deal with them in due time.

There is another type of diffeomorphism anomaly which arises as follows. 
The transformation of the Christoffel symbol can be written in the following compact form
\be
\delta_\xi \Gamma^\rho{}_\lambda{} &=&({ i}_\xi{ d}+{ d}\, { i}_\xi) \Gamma^\rho{}_\lambda + { d} \Xi^\rho{}_\lambda +[\Gamma, \Xi]^\rho{}_\lambda \label{deltaGammalr}
\ee
From this and similar formulas it is evident that any diffeomorphism splits into two parts: one looks like an ordinary gauge transformation,  
the other corresponds to the Lie derivative ${\cal L}_\xi = {i}_\xi { d}+{ d} { i}_\xi$. The first transformation gives rise to the just mentioned consistent chiral anomalies, formally similar to the gauge anomalies. The second can also give rise to cocycles of the form
\be
{\cal A}_\sfd[\xi,B] = \int d^{\sfd} x\, \sqrt{g} \, \partial\!\cdot\! \xi \,B\label{diffanomaly2}
\ee
where $B$ is a local expression of the fields that transforms like a scalar: $\delta_\xi B= \xi\!\cdot\partial B$. This cocycle is consistent and, usually, has a Weyl anomaly partner, and form with it a non-trivial cocycle of the coupled cohomology, see below.

Finally we have Weyl anomalies. We have already introduced the Weyl transformation, \eqref{Weyltransf}. For instance, it induces the following transformation for the Christoffel symbols 
\be
\delta_\omega \Gamma_{\mu\nu}^\lambda &=& \partial_\mu\omega \,\delta_\nu^\lambda + 
 \partial_\nu\omega \,\delta_\mu^\lambda -g_{\mu\nu} g^{\lambda\rho}\partial_\rho \omega\, \label{deltaomegaGamma}
 \ee
Promoting $\omega(x)$ to an anticommuting field the Weyl transformation becomes nilpotent, $\delta_\omega^2=0$, and defines a coboundary operator with relative cohomology. We cannot tire to stress again the utmost importance of studying the joint cohomology defined by the coboundary operator $\delta_\xi+\delta_\omega$. 

The possible non-trivial cocycles of $\delta_\omega$  with vanishing diffeomorphism partner in $4\sfd$ are well-known, they take the form
\be
\Delta[g,\omega]=\int d^4 x\, \sqrt{g}\,\omega\, T[g], \quad\quad \delta_\omega \Delta[g,\omega]=0\label{Deltaog}
\ee
where the density $T[g](x)$ can be the quadratic Weyl density
\be
\EW^2=R_{\mu\nu\lambda\rho} R^{\mu\nu\lambda\rho}-2 R_{\mu\nu}R^{\mu\nu} +\frac 13 R^2,\label{quadweyl}
\ee
the Gauss-Bonnet (or Euler) density,
\be
E=R_{\mu\nu\lambda\rho} R^{\mu\nu\lambda\rho}-4 R_{\mu\nu}R^{\mu\nu} +R^2,\label{gausbonnet}
\ee
and the Pontryagin density,
\be
P=\frac 12\left(\varepsilon^{\mu\nu\mu'\nu'}R_{\mu\nu\lambda\rho}R_{\mu'\nu'}{}^{\lambda\rho}\right). \label{pontryagin}
\ee 
Other possible cocycles have densities 
\be
T_e[V]= F_{\mu\nu}F^{\mu\nu}, \label{gaugeaction}
\ee
and
\be
T_o[V]=\varepsilon^{\mu\nu\lambda\rho}F_{\mu\nu}F_{\lambda\rho}.\label{chern}
\ee 
where $F_{\mu\nu}$ is an Abelian field strength, and analogous expressions for the non-Abelian case (a gauge potential is invariant under Weyl transformations in $4\sfd$). Other even trace anomalies will be introduced and discussed in section \ref{s:eventraceanom}.

A Weyl (or conformal) anomaly is the nonvanishing response of the path integral under a Weyl variation: 
\be
 \delta_\omega W[g] =-{\cal A}_\omega,   \quad\quad  {\cal A}_\omega = 
\int d^{\sfd} x\,\sqrt{g}\,\omega(x) T(x)\label{inttraceanom}
\ee
where $T(x)$ is the trace of the energy-momentum tensor
\be
 {T}(x)= g^{\mu\nu}(x) \frac{\delta}{\delta g^{\mu\nu}(x)} W[g].\label{1pttraceanom}
\ee
${\cal A}_\omega$ must take the form of a linear combination of the above cocycles. 

In a theory like \eqref{actiongpsi} all the above trace anomalies may show up. But in the sequel we are principally interested only in those that signal the non-existence of the fermion propagator (the dangerous ones). For instance, the even-parity anomalies show up in many theories (for instance, in a Dirac fermion theory) in which no doubt the fermion propagator exists. These anomalies are interesting for other reasons, but they do not signal any breakdown of the well-definiteness of the theory. For the time being we will focus only on the others, analogous to those that impair gauge theories. On the basis of the above discussion on the family's index theorem we expect them to appear in chirally asymmetric theories (because the family's index for a self-adjoint operator vanishes identically).

\vskip 0,3 cm
\noindent Ref.:\cite{atiyah1989,alvarezwitten1984,bardeen1984,bonoracottaRS1987}

\section{Trace anomalies}
\label{s:traceanomalies}

Historically chiral anomalies were born out of ambiguities. The conservation of a chiral current turned out to be tied to an ambiguous integral. Finally it was decided that it is not wise to freely shift an integration variable in a divergent integral. And the ABJ anomaly was born. For trace anomalies ambiguities are even more abundant and subtle than in the chiral case. It is imperative to have them in mind and try to resolve them before plunging into calculations.

One first ambiguity is in the classical definition of the e.m. tensor. As an example consider the action of a Dirac fermion coupled to a metric and an Abelian vector field
\be
S= \int d^4x \, \sqrt{g} \, i\overline {\psi} \gamma^\mu\left(D_\mu +\frac
12
\omega_\mu- iV_\mu  \right)\psi \label{actionSV}
\ee
with the usual notation. The vector current is $j_\mu= \overline \psi \gamma_\mu \psi$ and { for the stress-energy tensor we may take }
\be
T_{\mu\nu}= \frac i4 \overline {\psi} \gamma_\mu {\stackrel{\leftrightarrow}{\nabla}}_\nu\psi+ \{\mu\leftrightarrow \nu\}, \quad\quad \nabla_\mu =D_\mu +\frac 12 \omega_\mu- iV_\mu
 \label{emt0}
\ee

They are both conserved on shell and $T_{\mu\nu}$ is also traceless on shell. There is in fact another definition of the e.m. tensor, which corresponds to the general formula
\be
T_{\mu\nu}(x)= \frac 2 {\sqrt{g}} \frac {\delta S}{\delta g^{\mu\nu}},\label{Tmunudef}
\ee
and, from the differentiation of the $\sqrt{g}$, in \eqref{actionSV} leads to
\be
\widehat T_{\mu\nu}= \frac i4 \overline {\psi} \gamma_\mu
{\stackrel{\leftrightarrow}{\nabla_\nu}}\psi+(\mu\leftrightarrow\nu)- g_{\mu\nu}\frac i2  \overline \psi \gamma^\lambda {\stackrel{\leftrightarrow}{\nabla_\lambda}}\psi=T_{\mu\nu} - g_{\mu\nu } T_\lambda{}^\lambda,\label{emtmodified}
\ee
which is also conserved and traceless on shell.
This ambiguity in the definition of the e.m. tensor gives rise to an ambiguity in the definition of the trace anomaly (as well as of the diffeomorphism anomaly).  Such an uncertainty is in fact resolved by the definition  \eqref{Duff} below: thanks to it, the second term of $\widehat T_{\mu\nu}$ drops out. For this reason, in the sequel, we will proceed with \eqref{emt0}. But let us comment on  \eqref{Duff}.

The perturbative effective action corresponding to the classical action \eqref{actiongpsi} is expressed in terms of the metric fluctuation $h_{\mu\nu}$, where $g_{\mu\nu}=\eta_{\mu\nu} +h_{\mu\nu}$:
\be
W[g]&=&W[0]  + \sum_{n=1} ^\infty \frac { i^{n-1}}{{2^{n}}n!} \int
\prod_{i=1}^n
d^{\sfd} x_i\sqrt{g(x_i)} h^{\mu_i\nu_i}(x_i)\langle 0| {\cal T}T_{\mu_1\nu_1}(x_1)\ldots  T_{\mu_n\nu_n}(x_n)|0\rangle\label{Wg}
\ee
from which we can derive the full one-loop expression of the e.m. tensor
\be 
\langle\!\langle T_{\mu\nu}(x)\rangle\!\rangle\!\!\! &=&\!\!\! 2\frac {\delta 
W[h]}{\delta
h^{\mu\nu}(x)} = \sum_{n=1}^\infty \frac{i^{n}}{2^n n!}\int
\prod_{i=1}^n
d^{\sfd} x_i\sqrt{g(x_i)} h^{\mu_i\nu_i}(x_i)
 \langle 0|\mathcal{T}T_{\mu\nu}(x)T_{\mu_1\nu_1}(x_1)\ldots
T_{\mu_n\nu_n}(x_n)|0\rangle_c\0\\
\label{1ptem}
\ee

To eliminate the above ambiguity as well as others, our definition of trace anomaly {\it in the perturbative case} will be the following: if $T_{\mu\nu}$ is the stress-energy tensor of  a theory, the trace anomaly is given by the difference, \cite{duff1994,duff2020},
\be
{ g^{\mu\nu}\langle\!\langle T_{\mu\nu}(x) \rangle\!\rangle -\langle\!\langle g^{\mu\nu} T_{\mu\nu}(x) \rangle\!\rangle} ={T}[g](x)\label{Duff}
\ee
Thanks to this formula, the second term of $\widehat T_{\mu\nu}$ drops out. In addition, 
 the field operator $T_\mu^\mu(x)$ vanishes on shell, while in the case a theory contains a conformal soft breaking term (a mass term, for instance) $T_\mu^\mu(x)\neq 0$ even on shell. The second term of \eqref{Duff} is certainly present in such a case and the subtraction in \eqref{Duff} is needed in order to exclude this unwanted term from the anomaly.
 
This is also connected to a notation problem: the difference between the two definitions of the e.m. tensor can be reproduced in the definition of  the corresponding effective action by inserting or not the factor $\sqrt{g(x_i)}$ in the integral $\int d^{\sfd} x_i$ of such formulas as  \eqref{Wg}, \eqref{1ptem} and similar ones. Precisely, when this factor is inserted one should use $\widehat T_{\mu\nu}$, while, if it is not inserted, the definition is valid for $T_{\mu\nu}$. Henceforth we shall be using such simpler version of the effective action.
 
But this is not all concerning the definition \eqref{Duff}.  As a matter of fact, as we shall see, the just mentioned term is non-vanishing in many other instances and in subtler ways. 

The trace anomaly is a violation of the classical tracelessness condition $T_\mu^\mu(x)=0$.  It should be kept in mind that these equations are all valid on shell, while off-shell they do not hold in general. Off-shell  $T_\mu^\mu(x)$ is a non-vanishing quantum operator.  Another important point to be kept in mind is that, in terms of representations of the Lorentz group,  $T_{\mu\nu}(x)$ is a reducible tensor of which the trace $T_\mu^\mu(x)$ is an irreducible component. In the  expression of the effective action the latter is coupled to the field $h(x)=h_\mu^\mu(x)$. Likewise, the amplitude $\langle 0|{\cal T} T_\mu^\mu (x)\Phi(y) \Psi(z)|0\rangle$, where $\Phi$ and $\Psi$ are generic Lorentz covariant fields, is an irreducible component of  $\langle 0|{\cal T} T_{\mu\nu}(x)\Phi(y) \Psi(z)|0\rangle$. Much as we do in field theory when we compute quantum form factors, the former amplitude cannot be computed as a particular case of the latter, but must be computed independently. There are several examples where, when calculated with Feynman diagrams, the amplitudes  $\langle 0|{\cal T} T_\mu^\mu (x)\Phi(y) \Psi(z) |0\rangle$ and $\eta^{\mu\nu}\langle 0|{\cal T}_{\mu\nu}\Phi(y) \Psi(z)|0\rangle$ are different, see \cite{I}.

There are also other reasons that forbid us to merge the two terms of \eqref{Duff}, due to the connection with other amplitudes. For instance, the Feynman amplitude corresponding to the second term is the same as the amplitude of the Kimura-Delbourgo-Salam anomaly, which appears in the current divergence of chiral fermions. Therefore we have to live with the trace anomaly (perturbatively) defined (at the three-point level) by the difference
\be
\eta^{\mu\nu}\langle 0|{\cal T} T_{\mu\nu}(x)\Phi(y) \Psi(z)  |0\rangle-\langle 0|{\cal T} T_\mu^\mu (x) \Phi(y) \Psi(z)  |0\rangle\label{difference}
\ee  
This difference means in particular that the (regularized) effective action has discontinuities: differentiating it with respect to $h_{\mu\nu}$ and then saturating the result with $\eta_{\mu\nu}$ may not be the same as differentiating it with respect to $h(x)=h_\mu^\mu(x)$, which is the conjugate source of $T_\mu^\mu(x)$. 

One may ask what its physical meaning is.
The reason for taking the difference in the LHS of \eqref{Duff} is that two correlators may in general contain extra terms which have nothing to do with the anomaly. These terms are
\begin{itemize}
\item possible soft terms that classically violate conformal invariance;
\item the term $i\eta^{\mu\nu} \bar \psi \slashed {\partial} \psi$ in the modified definition of the e.m. tensor;
\item the semi-local terms in the conformal WI;
\item possible off-shell contributions to the anomaly: the derivative of the effective action with respect to $h_{\mu\nu}$ contracted with $\eta^{\mu\nu}$, or the derivative with respect to $h_\mu^\mu$, do not automatically vanish off-shell. In fact the operator $T_\mu^\mu$ identically vanishes on shell, therefore its contribution can only be off-shell. This means that in formula \eqref{Duff}  the off-shell contributions to the anomaly are subtracted away. In other words the trace anomaly \eqref{Duff} receives only on shell contributions.
\end{itemize}
All these extra terms cancel out in \eqref{Duff}.

The definition \eqref{Duff} of the trace anomaly refers to the perturbative approach. In the non-perturbative approaches of the heat kernel type the trace anomaly is simply the response of the effective action to a Weyl rescaling of the metric\footnote{In the heat kernel method we differentiate the square kinetic operator and ignore the $\sqrt{g}$ factor in the action, thus the e.m. tensor we  are  dealing with is \eqref{emt0}.}. The connection of the two definitions is not simple, in the non-perturbative cases it is hard if not impossible to separately evaluate the two terms of this definition. However, at least intuitively, we know that the anomaly, if any, appears precisely when the kinetic operator is not invertible, that is precisely when the theory is exactly on shell, which corresponds to our previous interpretation. In all cases considered the two approaches lead to the same (final) results. We may nevertheless ask: why is it that in the perturbative calculation we need to compute two separate terms? As we have noted  before the perturbative approach starts from the lowest order of the perturbative cohomology, because higher order calculations are more difficult and often inaccessible. Now the lowest order perturbative cohomology is a much looser mathematical structure than the full BRST cohomology. The former has plenty of non-trivial cocycles, while in the latter non-trivial cocycles are very limited in number. The definition \eqref{Duff} is designed to channel the lowest order perturbative results in the right track so as to coincide with the non-perturbative approaches. In a more formal language one can say that  in the perturbative cohomology each term of \eqref{Duff} is separately unstable, while their difference is stable.

After  this long introduction let us show a few examples of odd-parity trace anomalies.

\subsection{Gauge-induced odd-parity trace anomaly of a Weyl fermion}
\label{ss:gaugeinduced}

We refer to the action \eqref{actionSV} with the fermion $\psi$ being a right-handed Weyl fermion,
$\psi= \psi_R$, coupled to an Abelian vector potential. This time we wish to compute the trace of the quantum e.m. tensor on the basis of the definition \eqref{Duff}. To this end the relevant objects are the classically conserved and traceless (on-shell) energy-momentum tensor 
\be
T^{(R)}_{\mu\nu}= \frac i4 \overline {\psi_R} \gamma_\mu {\stackrel{\leftrightarrow}{\nabla}}_\nu\psi_R+ \{\mu\leftrightarrow \nu\},\quad\quad \nabla_\mu =D_\mu +\frac 12 \omega_\mu +V_\mu
 \label{emtR}
\ee
and the classically conserved current $J_{R\mu}=i\bar \psi_R \gamma_\mu \psi_R$. For a quantum evaluation we must refer to the effective action 
\be
W[h,V]\!\!\!&=&\!\!\!W[0]  + \sum_{n,r=1} ^\infty \frac { i^{n+r-1}}{{2^{n}}n!r!} \int
\prod_{i=1}^n
dx_i\, h^{\mu_i\nu_i}(x_i) \prod_{l=1}^r d y_l\,  e_{a_l}^{\lambda_l}(y_l) V_{\lambda_l}(y_l)\0  \\
&&\quad\quad\quad\cdot \langle 0| {\cal T}T^{(R)}_{\mu_1\nu_1}(x_1)\ldots \widehat T^{(R)}_{\mu_n\nu_n}(x_n)j_R^{a_1}(y_1) \ldots j_R^{a_r}(y_r)|0\rangle\label{WhV}
\ee
The one- and two-point functions do not contribute to the trace anomaly we wish to compute. What we have to find therefore is  the relation between the (odd-parity)  trace $\eta^{\mu\nu}\langle 0|{\cal T}T_{R\mu\nu}(x) j_{R\lambda}(y) j_{R\rho}(z)|0\rangle$ and the (odd-parity) correlator $\langle 0|{\cal T}T_{R\mu}^\mu(x) j_{R\lambda}(y) j_{R\rho}(z)|0\rangle$. The Feynman rule for the vector-fermion-fermion vertex $V_{vff}$ is $i\gamma_\mu P_R$ and the graviton-fermion-fermion vertex $V_{hff}$ is
\be
-\frac i{8} \left[(p-p')_\mu \gamma_\nu + (p-p')_\nu \gamma_\mu\right]P_R \label{2f1g}
\ee
where $p,p'$ are the fermion momenta, one entering the other exiting. The fermion propagator is the usual one\footnote {It is understood that we have added to the action a free left-handed fermion in order to have a well defined propagator.}. We have to compute a triangle diagram in which we associate an incoming momentum $q$ to $T_{R\mu}^\mu(x)$ and outgoing momenta $k_1,k_2$ to the potentials coupled to the two chiral currents, thus $q=k_1+k_2$.

The expression for $\langle 0|{\cal T}T_{R\mu}^\mu(x) j_{R\lambda}(y) j_{R\rho}(z)|0\rangle$  is 
\be
&&\!\!\!\!\!\!\!\!\!\widetilde T^{(R)\mu}_{\mu\lambda\rho}(k_1,k_2) =\!\! \frac 12\int
\frac{d^4p}{(2\pi)^4}\mathrm{tr}\left[\frac{1}{\slashed{p}}     
 \gamma_\lambda P_R\frac{1}{\slashed{p}
-\slashed{k}_1} \gamma_{\rho}P_R
\frac{1}{\slashed{p}-\slashed{k}
_1-\slashed{k}_2}(2 \slashed{p}-\slashed{q}) P_R\right]\label{TRjRjR1}\\
&=&\frac 12 \int
\frac{d^4pd^\delta\ell}{(2\pi)^{4+\delta}}\mathrm{tr}\left\{\frac{\slashed{p} }{{p}^2-{\ell}^2}     
 \gamma_\lambda\frac{\slashed{p} 
-\slashed{k}_1}{(p-k_1)^2 -\ell^2} \gamma_{\rho}
\frac{\slashed{p} -\slashed{q}+\slashed{\ell}}{(p-q)^2-\ell^2}(2 \slashed{p}+2\slashed{\ell}-\slashed{q}) P_R\right\}\0
\ee
The second line is the regularized integral obtained with the adjunction to $p$ of a momentum $\ell$ along $\delta$ extra dimensions. To it the cross term must be added. Evaluating the integral after a Wick rotation yields
\be
\widetilde T^{(R)\mu}_{\mu\lambda\rho}(k_1,k_2)+
\widetilde T^{(R)\mu}_{\mu\rho\lambda}(k_2,k_1)\Big\vert_{odd}=
 \frac 1{24\pi^2}\varepsilon_{\mu\nu\lambda\rho}k_1^\mu k_2^\nu,\label{TmumujRjR}
\ee
This is the result for  $\langle 0|{\cal T}T_{R\mu}^\mu(x) j_{R\lambda}(y) j_{R\rho}(z)|0\rangle$.
In an analogous way we get 
\be
&&\!\!\!\!\!\!\frac 12\int
\frac{d^4pd^\delta\ell}{(2\pi)^{4+\delta}}\mathrm{tr}\left\{\frac{\slashed{p} }{{p}^2-{\ell}^2}     
 \gamma_\lambda\frac{\slashed{p} 
-\slashed{k}_1}{(p-k_1)^2 -\ell^2} \gamma_{\rho}
\frac{\slashed{p} -\slashed{q}}{(p-q)^2-\ell^2}
(2\slashed{p}-\slashed{q} )P_R\right\}\Bigg{\vert}_{odd}+{\it cross}\0\\
&&=  \frac 1{48\pi^2}\varepsilon_{\mu\nu\lambda\rho}k_1^\mu k_2^\nu\label{triangledimtrace2}
\ee
which gives the odd-parity part of $\eta^{\mu\nu}\langle 0|{\cal T}T_{R\mu\nu}(x) j_{R\lambda}(y) j_{R\rho}(z)|0\rangle$. Therefore, after returning to a Lorentzian background, we get 
\be
g^{\mu\nu}\langle\!\langle T_{R\mu\nu}(x) \rangle\!\rangle\Big\vert_{odd} -\langle\!\langle g^{\mu\nu} T_{R\mu\nu}(x) \rangle\!\rangle\Big\vert_{odd} ={-  \frac i{96\pi^2}}\varepsilon_{\mu\nu\lambda\rho} \,\partial^\mu V^\nu(x) \partial^\lambda V^\rho(x)\label{tracedifferenceb}
\ee
The non-Abelian version of this result is
\be
{\cal A}_\omega^{(odd,R)} = -\frac i{384\pi^2} \varepsilon_{\mu\nu\lambda\rho}\tr (F^{\mu\nu} F^{\lambda\rho})\label{AomehanonAbelianR}
\ee
This is the gauge-induced odd-parity trace anomaly for a right-handed Weyl fermion.  For a left-handed Weyl  fermion the only difference is the sign in front of $\gamma_5$ in the initial formula \eqref{TRjRjR1}. Therefore the only change for the odd-parity part is an overall sign:
\be
{\cal A}_\omega^{(odd,L)} = \frac i{384\pi^2} \varepsilon_{\mu\nu\lambda\rho}\tr (F^{\mu\nu} F^{\lambda\rho})\label{AomehanonAbelianL}
\ee
In fact we have not justified the cubic and quartic terms in \eqref{AomehanonAbelianR} and \eqref{AomehanonAbelianL}, which would require the evaluation of higher order amplitudes. They will be derived more confortably with non-perturbative methods.

\subsection{Gravity-induced odd-parity trace anomaly of a Weyl fermion}

Let us consider the case of the action \eqref{actionSV} for a right-handed Weyl fermion $\psi\equiv \psi_R$ coupled only to a metric ($V_\mu=0$).  The effective action is
\be
W^{(R)}[g]&=&W^{(R)}[0]  + \sum_{n=1} ^\infty \frac { i^{n-1}}{{2^{n}}n!} \int
\prod_{i=1}^n
d^{\sfd} x_i h^{\mu_i\nu_i}(x_i)\langle 0| {\cal T}T^{(R)}_{\mu_1\nu_1}(x_1)\ldots  T^{(R)}_{\mu_n\nu_n}(x_n)|0\rangle\0\\&&\label{WRgsimpl}
\ee
Our task is to compute multi-point amplitudes of the e.m. tensor.
To start with, from the action we have to extract the Feynman rules. We write it down more explicitly as follows
\be 
S= \int d^4x \, \sqrt{|g|} \,\left[ \frac i2\overline {\psi_R} \gamma^\mu {\stackrel{\leftrightarrow}{\d}}_\mu  \psi_R -\frac 14\varepsilon^{\mu a b c} \omega_{\mu a b} \overline{\psi_R} \gamma_c \gamma_5\psi_R\right]
\label{actionR1}
\ee
where it is understood that the derivative applies to $\psi_R$ and $\overline {\psi_R}$ only. We have used the relation $\{\gamma^a, \Sigma^{bc}\}=i \, \varepsilon^{abcd}\gamma_d\gamma_5$.

Now we expand
\be
e_\mu^a = \delta_\mu^a +\chi_\mu^a+ ...,\quad\quad e_a^\mu = \delta _a^\mu +\hat \chi_a^\mu +...,\quad\quad {\rm and}\quad
g_{\mu\nu}=\eta_{\mu\nu}+h_{\mu\nu}+...\label{hmunu}
\ee
and make a local Lorentz gauge choice by dropping the antisymmetric part of the vierbein.
Inserting these expansions in the defining relations $e^a_\mu e^\mu_b=\delta_b^a,\quad g_{\mu\nu}= e_\mu^a e_\nu^b \eta_{ab}$,
we find
\be
\hat \chi_\nu^\mu =- \chi_\nu^\mu\quad \quad {\rm and}\quad\quad h_{\mu\nu}=2\,\chi_{\mu\nu}.\label{hatchichi}
\ee
Let us expand accordingly the spin connection:
\be 
\omega_{\mu a b}\, \varepsilon^{\mu a b c}= - \varepsilon^{\mu a b c}\,  \d_\mu \chi_{a\lambda}\,\chi_b^\lambda+...\label{omega}
\ee

Therefore, up to second order the action can be written (by incorporating $\sqrt{|g|}$ in a redefinition of the $\psi_R$ field.)
\be 
S\approx \int d^4x \,  \left[\frac i2 \left(\delta^\mu_a -\chi^\mu_a +\frac 32 (\chi^2)_a^\mu\right) \overline {\psi_L} \gamma^a {\stackrel{\leftrightarrow}{\d}}_\mu  \psi_L +\frac 14\varepsilon^{\mu a b c}\,  \d_\mu \chi_{a\lambda}\,\chi_b^\lambda\,  \overline\psi_L \gamma_c \gamma_5\psi_L\right]\0
\ee
As a consequence  the Feynman rules are as follows (momenta are incoming and  the external gravitational field is assumed to be $h_{\mu\nu}$). The fermion propagator is the usual one as before.
The two-fermion-one-graviton vertex ($V_{ffh}$) is
\be
-\frac i{8} \left[(p-p')_\mu \gamma_\nu + (p-p')_\nu \gamma_\mu\right] \frac {1+\gamma_5}2\label{2f1gch}
\ee
There are two two-fermion-two-graviton vertices: $V_{ffhh}$ 
\be
&& \frac{ 3i}  {64} \Big{[}\left( (p+p')_\mu
\gamma_{\mu'} \eta_{\nu\nu'} +  (p+p')_{\mu} \gamma_{\nu'} \eta_{\nu\mu'}+
\{\mu\leftrightarrow \nu\}\right)\label{2f2g'}\\
&&\quad\quad+ \left(  (p+p')_{\mu'} \gamma_{\mu} \eta_{\nu\nu'}+
(p+p')_{\mu'} \gamma_{\nu} \eta_{\mu\nu'}+ \{\mu'\leftrightarrow
\nu'\}\right)\Big{]} \frac {1+\gamma_5}2 \0
\ee
and $V^\varepsilon_{ffhh}$
\be 
\frac 1{64} t_{\mu\nu\mu'\nu'\kappa\lambda}(k-k')^\lambda\gamma^\kappa\frac {1+\gamma_5}2\label{2f2g}
\ee 
where
\be
t_{\mu\nu\mu'\nu'\kappa\lambda}=\eta_{\mu\mu'} \varepsilon_{\nu\nu'\kappa\lambda} +\eta_{\nu\nu'} \varepsilon_{\mu\mu'\kappa\lambda} +\eta_{\mu\nu'} \varepsilon_{\nu\mu'\kappa\lambda} +\eta_{\nu\mu'} \varepsilon_{\mu\nu'\kappa\lambda}\label{t}
\ee
and the graviton momenta $k,k'$ are incoming.

Let us come next to the explicit calculations. Like in the previous calculations the lowest order contribution to the odd-parity trace anomaly can be proven to come from the three-point e.m. amplitude. In a careless approach this does not seem to make sense, because a well-known result of CFT states that a conformal odd-parity three-point function $\langle 0|{\cal T}T_{\mu\nu}(x) T_{\mu'\nu'}(y) T_{\alpha\beta}(z)|0\rangle^{(odd)}$  vanishes identically for algebraic reasons. Therefore, at the lowest perturbative order, we can write
\be
\eta^{\mu\nu}\langle\! \langle T_{\mu\nu}(x)\rangle\!\rangle^{(odd)} =0, \label{Zhiboedov}
\ee
as can be proven also by a direct calculation. However according to the definition \eqref{Duff} we must compute also the second term with one insertion of a trace of the e.m. tensor.

Here, at variance with the previous examples in this review, the Feynman diagrams to be calculated do not reduce to a single triangle diagram. Due to the existence of the four-legs vertices \eqref{2f2g'} and \eqref{2f2g} there are also two bubble diagrams. They are obtained by joining two vertices, $V_{ffh}$ (on the left) and $V^\varepsilon_{ffhh}$ or $V_{ffhh}$ (on the right) with two fermion propagators. The incoming graviton in $V_{ffh}$ has momentum $q$ and the two outgoing gravitons in $V^\varepsilon_{ffhh}$ or $V_{ffhh}$ are specified by $k_1,\{\mu,\nu\}$ and $k_2,\{\mu',\nu'\}$, respectively, with  $q=k_1+k_2$. The two fermion propagators form a loop. However these diagrams turn out to give a vanishing contribution to the anomaly. Therefore let us focus on the triangle diagram

The triangle diagram with three e.m. insertions is constructed by joining three vertices $V_{ffh}$ with three fermion lines. The external momenta are $q$ (incoming) with labels $\alpha$ and $\beta$, and $k_1,k_2$ (outgoing), with labels $\mu,\nu$ and $\mu',\nu'$, respectively. Of course $q=k_1+k_2$.
The internal momenta are $p $, $p-k_1$ and 
$p-k_1-k_2$, respectively. 

Employing the above Feynman rules of the chiral fermion coupled to an external gravitational field, we can write down the Fourier transform of the three-point e.m. tensor amplitude $\widetilde{\ET}^{(odd)}_{\mu\nu\mu'\nu'\alpha\beta}(k_1,k_2)$. Contracting $\alpha$ and $\beta$ and moving $P_R$ to the rightmost position\footnote{This is, in this case, the simplest prescription, but by no means the only one. In \cite{I} it is shown that the position of $\gamma_5$ is irrelevant for the final result provided we use the same prescription also in computing the diffeomorphism anomaly. In other words, its position is in general irrelevant for the joint Weyl-diffeomorphism cohomology.}, the correlator we are looking for is
\be
\widetilde{\sf T}_{\mu\nu\mu'\nu'}(k_1,k_2) &\equiv&  \widetilde \ET^{(odd)\alpha}_{\mu\nu\mu'\nu'\alpha}(k_1,k_2) = - \frac{1}{256}\int
\frac{d^4p}{(2\pi)^4}\mathrm{Tr}\left\{\left[\frac{\slashed{p}}{p^2}
(2p-k_1)_\mu\gamma_\nu + (\mu
\leftrightarrow
\nu)\right]\frac{(\slashed{p}-\slashed{k}_1)}{(p-k_1)^2}\right.\0\\
&&\left.\times\left[(2p-2k_1-k_2)_{
\mu'}\gamma_{\nu'}(\mu' \leftrightarrow
\nu')\right]\frac{(\slashed{p}-\slashed{k}_1-\slashed{k}_2)}{(p-k_1-k_2)^2}
(2\slashed{p}-\slashed{k}_1-\slashed{k}_2)\left(\frac{
1+\gamma_5}{2}\right)\right\}.\0\\
\label{triangle2}
\end{eqnarray}
which is evidently divergent. We dimensionally regularize it in the usual way. The calculation is lengthy. Finally, after adding the cross term one gets
\begin{equation}
\widehat{ \sfT}^\mathrm{(tot)}_{\mu\nu\mu'\nu'}(k_1,k_2) =-
\frac{i}{3072\pi^2}k^{\alpha}_1 k^{\beta}_2\left(\left(k_1^2+k_2^2+k_1\! \cdot\! k_2\right)
t_{\mu\nu\mu'\nu'\alpha\beta}-t^{(21)}_{\mu\nu\mu'\nu'\alpha\beta}\right)\,.
\label{triangle24}
\end{equation}
where
\be 
t^{(21)}_{\mu\nu\mu'\nu'\kappa\lambda}=k_{2\mu}k_{1\mu'} \varepsilon_{\nu\nu'\kappa\lambda} + k_{2\nu}k_{1\nu'}\varepsilon_{\mu\mu'\kappa\lambda} +k_{2\mu}k_{1\nu'} \varepsilon_{\nu\mu'\kappa\lambda} +k_{2\nu}k_{1\mu'} \varepsilon_{\mu\nu'\kappa\lambda}\label{tij}
\ee

To simplify the derivation we shall set the external lines on shell, $k_1^2=k_2^2=0$. This requires a comment.

Inserted in the formula for the effective action the term $k^{\alpha}_1 k^{\beta}_2\, (k_1^2+k_2^2)t_{\mu\nu\mu'\nu'\alpha\beta}$ gives rise to the cocycle $\Delta_\omega$
\be
\Delta_\omega\sim\int d^4x\, \omega \, \varepsilon^{\mu\nu\lambda\rho} \partial_\mu\square h^\alpha_\nu \partial_\lambda h_{\rho\alpha}\label{fakeDelta}
\ee
This is a consistent Weyl cocycle (at the lowest significant order), but it is not invariant under diffeomorphism transformations, and, of course, there is no covariant extension of it to all orders. In such kind of situation this usually means that there is a partner cocycle of the diffeomorphisms, of the type \eqref{diffanomaly2}, that, together with \eqref{fakeDelta}, form a complete cocycle of the overall Weyl-diffeomorphism cohomology. When this happens usually one can find a local counterterm that cancels one of the two partner cocycles and modifies the other. There are plenty of such examples. But, in this case, if we compute the odd-parity part of the divergence of the e.m. tensor, we find zero, as expected because of the remark around  eq.\eqref{Zhiboedov}.  We can view the problem also from another point of view, we can cancel the cocycle \eqref{fakeDelta} by subtracting from the effective action a counterterm 
\be
\sim \int d^4x \, h \,  \varepsilon^{\mu\nu\lambda\rho} \partial_\mu\square h^\alpha_\nu \partial_\lambda h_{\rho\alpha}\label{fakecounterterm}
\ee
where $h=h_\mu^\mu$. But this counterterm is not invariant under diffeomorphisms, therefore it would generate a diffeomorphism anomaly. This means that at the lowest perturbative order the cross consistency conditions for diffeomorphisms and Weyl transformations are not satisfied. This is not surprising, because one can prove that on algebraic grounds the odd part of the three-point conformal amplitude of the e.m. tensor vanishes identically.  In other words, due to algebraic contraints, the usual cohomology machinery for the two joint symmetries cannot work, at least at the lowest perturbative order (the three-point function).

A way out is to compute higher order diagrams (the four-, five-,... point function may have a nonvanishing odd part)  or resort to a nonperturbative method. The first alternative is rather impervious, the second more accessible and in fact we will exhibit a nonperturbative derivation of this trace anomaly later on. But there is also a way to utilize the result \eqref{triangle24} relying on the fact that should we be able to go beyond the lowest order calculation the rupture of the machinery would disappear. What we have to do is to render the term \eqref{fakeDelta} irrelevant for the derivation of the final result. This is the rationale of putting the external lines on shell.

\subsubsection{On shell conditions}

Putting the external lines on shell means that the corresponding fields have to satisfy the eom of Einstein-Hilbert gravity $R_{\mu\nu}=0$.  In the linearized form this means
\be 
\square \chi_{\mu\nu}- \d_{\mu} \d_{\lambda}\chi^\lambda _\nu + \d_\nu \d_{\lambda}\chi^\lambda _\mu -\d_\mu\d_\nu \chi^\lambda_\lambda=0\label{linrmunu}
\ee
We also choose the De Donder gauge: $\Gamma_{\mu\nu}^\lambda g^{\mu\nu}=0$,
which at the linearized level becomes $2\d_\mu\chi^{\mu}_\lambda -\d_\lambda \chi^\mu_\mu=0$. In this gauge (\ref{linrmunu}) becomes
\be 
\square \chi_{\mu\nu}=0\label{KG}
\ee
In momentum space this implies that $k_1^2=k_2^2=0$. But this is not simply an {\it ad hoc} trick. We are in fact defining a restricted cohomology of the diffeomorphisms and the Weyl transformations: a cohomology defined up to terms $\square  h_{\mu\nu}$ and $\square \xi^\mu$. This is a well defined cohomology, under which we have, in particular,
\be
\delta_\xi \left(2\d_\mu\chi^{\mu}_\lambda -\d_\lambda \chi^\mu_\mu\right) = 2\ \square \xi_\lambda \approx 0\label{Dedonder}
\ee
i.e. in this restricted cohomology the De Donder gauge fixing is irrelevant. Similarly, the term \eqref{fakeDelta} remains null after a restricted diffeomorphism transformation, so it does not play any role in cohomlogy. The restricted cohomology has the same odd class (the Pontryagin one) as the unrestricted one, i.e. it completely determines it (this is not true, for instance, for the even-parity classes).

\subsubsection{Overall contribution}
\label{ssec:overallcontr}

The overall one-loop contribution to the trace anomaly in momentum space, {\it as far as the parity-violating part is concerned}, is given by (\ref{triangle24}). After returning to the Minkowski metric 
and Fourier-anti-transforming it, we can extract the local expression of the trace anomaly, by replacing the results found so far in (\ref{1ptem}). The result, to lowest order, is 
\be 
\langle\!\langle T^{\mu}_{\mu}(x)\rangle\!\rangle^{(odd)} \approx -\frac i{768\pi^2}\varepsilon^{\mu\nu\lambda \rho} \left(\d_\mu\d_\sigma h^\tau_\nu \, \d_\lambda\d_\tau h_{\rho}^\sigma-\d_\mu\d_\sigma h^\tau_\nu \, \d_\lambda\d^\sigma h_{\tau\rho}\right)\label{final1}
\ee
{The factor in front is due to a factor of 8 that comes from the coefficient $\frac 1{2^n n!}$ with $n=2$ in the denominator and a factor of 8 in the numerator
because the Fourier transform of the three point function of the e.m. tensor is 8 times ${\widetilde\ET}_{\mu\nu\mu'\nu'\alpha\beta}(k_1,k_2)$; another factor of 4 in the numerator is due to the symmetry of the tensors $t$ and $t^{(21)}$ in \eqref{triangle24}, which yields four times the same term. Comparing with the expansion 
\be
\varepsilon^{\mu\nu\lambda \rho}R_{\mu\nu}{}^{\sigma\tau}R_{\lambda\rho\sigma\tau}=
 \varepsilon^{\mu\nu\lambda \rho} \left(\d_\mu\d_\sigma \chi^a_\nu \, \d_\lambda\d_a \chi_{\rho}^\sigma-\d_\mu\d_\sigma \chi^a_\nu \, \d_\lambda\d^\sigma \chi_{a\rho}\right)+...\label{epsRR}
\ee
we obtain
\be 
 \langle\! \langle T^{\mu}_{\mu}(x)\rangle\!\rangle^{(odd)} = -\frac i{768\pi^2} \, \frac 12\,\varepsilon^{\mu\nu\lambda \rho}R_{\mu\nu}{}^{\sigma\tau}R_{\lambda\rho\sigma\tau}\label{final2}
\ee
Now applying the definition \eqref{Duff} and recalling \eqref{Zhiboedov},
we obtain the covariant expression of the parity-violating part of the trace anomaly
\be
T[g](x)=  \frac i{768\pi^2} \, \frac 12\,\varepsilon^{\mu\nu\lambda \rho}R_{\mu\nu}{}^{\sigma\tau}R_{\lambda\rho\sigma\tau}.\label{Pontryagintrace}
\ee
It goes without saying that a left-handed Weyl fermion has  the same anomaly with opposite sign in front.

\vskip 0,3cm

{\bf Remark 1.}  On the basis of a not uncommon prejudice in the literature the previous result sounds unexpected. It is sometimes stated that the theory of gravity is chirally blind, meaning that the relevant charge, the mass, is positive, and is thus different from the typical case of a U(1) interaction. However this is based on a misunderstanding. The coupling between gravity and matter is given by the juxtaposition of the metric and the energy-momentum tensor, and the energy-momentum tensors of fermions with opposite chiralities are different. Therefore it is to be expected that at some stage differences  might emerge between fermions with opposite chiralities in their interaction with gravity. An obvious place where such differences may show up are the anomalies, and in particular the trace anomaly, because it involves precisely the coupling between the metric and the energy-momentum tensor.

\vskip 0,3 cm
{\bf Remark 2}. Contrary to what may seem at first sight the anomalies \eqref{chirallimittrace} and \eqref{chirallimittraceleft} violate parity, but, due to the imaginary coefficient in front, do not violate time reversal. Therefore, assuming CPT invariance, they do not violate $CP$ either. The imaginary coefficient  means, however, that the quantum effective Hamiltonian becomes complex, which thus becomes a source of possible violation of unitarity.

\subsection{The KDS anomaly}

It is not irrelevant to notice at this point a fact that has not been given the right importance in the literature. Consider a Dirac fermion $\psi$ coupled to a metric and vector potential $V_\mu$ like in \eqref{actionSV}.
It is invariant under the Abelian local transformation $\psi \to e^{i\gamma_5 \eta}\psi$ and $V_\mu \to  V_\mu + \gamma_5 \partial_\mu \eta$. The corresponding axial current $j_{5\mu}=\overline \psi \gamma_\mu \gamma_5\psi$ is classically conserved. Is it conserved in the quantum theory? Or is it violated due to the gravitational interaction? To answer this question one has to consider the amplitude $\langle \partial\!\cdot\! j_5\, T \,T\rangle$. The amplitude in question is the same as \eqref{triangle2} with the factor $(2\slashed{p}-\slashed {q})$ replaced by $\slashed{q}$ and $P_R$ replaced by $\gamma_5$, and an overall coefficient $\frac 1{64}$ instead of $\frac 1{256}$. If we rewrite $\slashed{q}= 2\slashed{p}-(2\slashed{p}-\slashed{q})$, we see that the second term reproduces the calculation of the previous subsection multiplied by a suitable coefficient, while the term corresponding to $2\slashed{p}$, once regularized, is easily seen to vanish. Therefore we can conclude that
\be
\partial^\mu j_{5\mu}= \frac 1{768\pi^2} \varepsilon^{\mu\nu\lambda \rho}R_{\mu\nu}{}^{\sigma\tau}R_{\lambda\rho\sigma\tau}.\label{Pontryagindiv}
\ee
This is the Kimura-Delbourgo-Salam anomaly, \cite{KDS}. It is an anomaly of a Dirac fermion theory, that is, of the ABJ type\footnote{ ABJ-type anomalies appear in theories involving Dirac fermions. A Dirac fermion can be decomposed into a sum of one left-handed and one right-handed Weyl fermion. If the two Weyl fermions have consistent anomalies, the latter have opposite signs. The consistent anomaly of the  fermion is the sum of these two, thus it vanishes. The ABJ anomaly is instead the difference of the two (the right-handed minus the left-handed one).}  . This derivation of the KDS anomaly is an example of the rigid link that connects chiral ABJ type anomalies to odd-parity trace anomalies. The same holds also for the gauge-induced trace anomaly considered previously, see section \ref{ss:gaugeinduced}, in a Weyl fermion theory, which is rigidly linked in the same way to the standard ABJ gauge anomaly.

\vskip 0,3cm
\noindent Ref.:\cite{bonora2014,bonora2015,bonora2017,bonora2022,DS,KDS}

\section{The standard model in a metric background geometry}
 
We can now analyse the overall panorama of the new anomalies in the SM due to its immersion in a non-trivial metric background. As mentioned before, the  dangerous consistent diffeomorphism  anomalies, which might occur in the divergence of the e.m. tensor, vanish identically for group theoretical reasons. But, as already pointed out, there may appear a mixed gauge-gravitational anomaly.

\subsection{Mixed gauge-gravity anomaly}

The action \eqref{actionSV}, with $\psi\equiv \psi_R$, is invariant under the local $U(1)$ transformation $V_\mu\rightarrow V_\mu+\partial_\mu \lambda$. The relevant current $j_\mu= \bar \psi_R \gamma_\mu \psi_R$ is classically conserved and to find the corresponding anomaly, if any, we have to evaluate the amplitude $\langle \partial\! \cdot \! j_R \,T_R\, T_R\rangle$. The latter can be easily calculated with the usual Feynman diagram. The final result is
\be
\partial^\mu j_{R\mu}= \frac 1{1536\pi^2}\, \varepsilon^{\mu\nu\lambda \rho}R_{\mu\nu}{}^{\sigma\tau}R_{\lambda\rho\sigma\tau}.\label{PontryagindivR}
\ee
Its integrated form is $\sim \Delta_G(\lambda,g)=\int d^4x \lambda\, \varepsilon^{\mu\nu\lambda \rho}R_{\mu\nu}{}^{\sigma\tau}R_{\lambda\rho\sigma\tau}$, which is a diffeomorphism-invariant (trivially) consistent Abelian gauge cocycle. This cocycle can take different forms, for instnace $\Delta_G (\lambda,g)$ is equivalent to 
\be
\Delta_\sfd(\xi,g,V)= \int d^4x \,\sqrt{g}\,\varepsilon^{\mu\nu\lambda\rho}\tr \left(\partial_\mu\Xi \, \Gamma_\nu\right)F_{\lambda\rho}\label{DeltaDxigV}
\ee
where $F_{\mu\nu}=\partial_\mu V_\nu -\partial_\nu V_\mu$, $\Xi$ represents the matrix
$\Xi_\tau{}^\sigma =\partial_\tau \xi^\sigma$ and $\Gamma_\mu$ represents the matrix
$\Gamma_{\mu\sigma}^\tau$. They can be obtained from each other by subtracting a suitable conterterm.

This anomaly can also take the form of a (gauge+diffeomorphism-invariant) Lorentz anomaly. The reason is that all these cocycles descend from the same 6-form $\tr (RR)F$ of the classifying space of the group $SO(4)\times U(1)$.

Let us see the effect of these mixed anomalies in the SM. 
 Adopting the style of section \ref{ss:cancelationinSM} and the notation $\Sigma$ for the generic Lorentz generator, it is easy to see that the only a priori non-vanishing possibility is
 \vskip 0,3cm
 \centerline {Table MSM mixed}
\vskip 0.2cm
 \noindent\fbox{
    \parbox{\textwidth}{
\begin{itemize}
\item $\Sigma\times\Sigma \times  T^{ {\mathfrak u}(1)}$, the trace  $\tr\left(\Sigma^{ab} \Sigma^{cd}\right)$ is non-vanishing, but it is multiplied by the total $U(1)$ charge:
$6\left(\frac 16\right)-3\left(\frac 23\right) - 3\left(-\frac 13\right)+ 2\left(-\frac 12\right)+1=0$.
\end{itemize}
}} 
\vskip 0.3 cm
The addition of sterile neutrinos to (\ref{MSMspectrum}) does not alter this conclusion.

\subsection{Do odd-parity trace anomalies cancel?}

Next let us analyse the situation for trace anomalies in the SM.
First, it is important to clarify that these anomalies are relevant only when gravity is effectively coupled to the model. For, looking at the effective action \eqref{WhV} one can see that any quantum contribution to the trace of the e.m. tensor is multiplied by the metric fluctuation $h=h^\mu_\mu$. If the latter vanishes no quantum contribution from the trace anomaly can affect the effective action. Therefore the previous anomaly analysis is completely satisfactory if gravity can be disregarded. On the other hand, if gravity is assumed to interact with the standard model via the minimal covariant couplings, odd trace anomalies may have significant fallouts, as was pointed out before. Therefore it makes sense to study the conditions under which also these anomalies cancel. Hereafter we broach this subject.

For odd-parity trace anomalies the cancelation takes place in any case if there is a perfect balance between opposite chiralities, between, say, left-handed  and right-handed {  components}. From the above we see that in the multiplet (\ref{MSMspectrum}) there is a balance between the left-handed and right-handed components except for the left-handed  $\nu_e$. 
 \vskip 2cm
\centerline {Table MSM trace-gravity}
\vskip 0.2cm
 \noindent\fbox{
    \parbox{\textwidth}{\begin{itemize}
\item Therefore the multiplet (\ref{MSMspectrum}), when weakly coupled to gravity, will produce an overall non-vanishing (imaginary) coefficient for the Pontryagin density in the trace anomaly.
\end{itemize}
}}
\vskip 0,3cm
 This breakdown is naturally avoided if we add to the above SM multiplet other Weyl fermions (for instance a right-handed sterile neutrino) so as to produce a chirally symmetric model without compromising the cancelation of the chiral gauge and gravity anomalies. 
\vskip 0,3cm
The analysis concerning gauge-induced odd trace anomalies, see (\ref{AomehanonAbelianR}), is more complex. First of all we have three types of such anomalies, constructed with $SU(3), SU(2)$ and $U(1)$ gauge fields, respectively.
\vskip 0,3cm\centerline {Table MSM trace-gauge}
\vskip 0.2cm
 \noindent\fbox{
    \parbox{\textwidth}{\begin{itemize}
    \item   We have six units of the anomaly \eqref{AomehanonAbelianR} with curvature $F\equiv F^{\mathfrak su(3)}$ and six units with opposite sign. Therefore the multiplet (\ref{MSMspectrum}) is free of these anomalies.
    
    \item We have instead 4 units of the same anomaly with gauge field   $F\equiv F^{\mathfrak su(2)}$ and positive sign,  see \eqref{AomehanonAbelianL}, computed in the doublet representation of $\mathfrak su(2)$. 
    \item Finally we have  a $U(1)$ gauge-induced trace anomaly with vanishing total coefficient: $6\left(\frac 16\right)^2-3  \left(\frac 23\right)^2-3 \left(-\frac 13\right)^2 +2 \left(-\frac 12\right)^2-\left(-1\right)^2=0$
\end{itemize}
}}
\vskip 0,3cm
The addition of sterile neutrinos does not change these conclusions. { \it The $SU(2)$ gauge-induced odd trace anomalies do not cancel in the MSM}.
\vskip 0,3 cm

\noindent Ref.: \cite{Weinberg2005,I}

\section{A metric-axial-tensor background}

Before passing to the second part of the paper, we would like to complete our description of the anomalies relevant to the SM, by showing how to compute 
the previous trace anomalies in a non-perturbative way, via the SDW method. To this end we enlarge the background geometry by introducing  the metric-axial-tensor (MAT) gravity. It is a generalization of Bardeen's method for gauge fields to the gravitational environment. It must be said that this extension is not strictly necessary to derive the final results, but, in view of the ambiguities scattered in this type of computations, as pointed out before,  this expedient is { a good bookkeeping device}, certainly less exposed to blunders than the direct method, and, anyhow, it introduces in gravity theories a richer formalism susceptible of new unexplored applications, as we shall see in the second part of the paper.

\subsection{Axial-complex analysis and geometry}
\label{ss:axialcomplex}

Axial-complex numbers are defined by
\be
\hat a = a_1+\gamma_5 a_2\label{hata}
\ee
where $a_1$ and $a_2$ are real numbers ($a_1$ is called axial-real and $a_2$ axial-imaginary). Arithmetic is defined in the obvious way. We have a natural conjugation operation $\overline {\hat a} = a_1-\gamma_5 a_2$.

We shall consider functions $\hat f(\hat x)$ of the axial-complex variable $\widehat x= x_1+ \gamma_5 x_2 $, as well as functions of several axial-complex variables $\widehat x^\mu= x_1^\mu + \gamma_5 x_2^\mu$. We can easily define derivatives:
\be
\frac {\partial}{\partial \hat x^\mu} = \frac 12 \left( \frac
{\partial}{\partial  x_1^\mu}+\gamma_5 \frac {\partial}{\partial
x_2^\mu}\right),
\quad\quad
\frac {\partial}{\partial {\overline{\hat x}} ^\mu}
= \frac 12 \left( \frac {\partial}{\partial  x_1^\mu}-\gamma_5 \frac
{\partial}{\partial  x_2^\mu}\right)
\ee
Notice that for axial-analytic functions (that is, functions defined by their Taylor series expansion)
\be
\frac d{d\hat x} = {\frac \partial{\partial x_1}\equiv\frac {\partial}{\partial
\hat x},\label{dxdx1}}
\ee
whereas $\frac {\partial}{\partial {\overline{\hat x}}}\widehat f(\hat x)=0$.

As for integrals, since we will always deal with fluctuating fields or parameters rapidly decreasing at infinity, we define $\int  d\hat x\,\widehat f(\hat x)$
as the rapidly decreasing primitive $\widehat g(\hat x)$ of $\widehat f(\hat
x)$. Therefore, in particular, the property
\be
\int d\hat x\, \frac {\partial}{\partial\hat x} \hat
f(\hat x)=0\label{stokes}
\ee
follows immediately. 

 In this axial-spacetime we introduce an axial-Riemannian geometry as follows.
We generalize the metric $\widehat g_{\mu\nu} = g_{\mu\nu}+\gamma_5 f_{\mu\nu}$, by adding to the usual metric an axial symmetric tensor. Their background values are $\eta_{\mu\nu}$
and 0, respectively. So, we write as usual $g_{\mu\nu}= \eta_{\mu\nu}+ h_{\mu\nu}$.}
Likewise for the vierbein we write $\widehat e^a_\mu = e^a_\mu+ \gamma_5 c^a_\mu $ and $ \widehat e_a^\mu =\tilde
e_a^\mu+
\gamma_5\tilde c_a^\mu$.
This implies
\be
\eta_{ab}\left(e^a_\mu e^b_\nu + c^a_\mu c^b_\nu\right)= g_{\mu\nu},
\quad\quad \eta_{ab}\left(e^a_\mu c^b_\nu + e^a_\nu c^b_\mu\right)=
f_{\mu\nu}\label{vier2}
\ee
The Christoffel symbols  are defined by
\be
\widehat \Gamma_{\mu\nu}^\lambda &=& \frac 12 \widehat
g^{\lambda\rho}\left(
\frac{\partial}{
\partial{\widehat x^\mu}} \widehat g_{\rho\nu}
+
\frac{\partial}{ 
\partial{\widehat x^\nu}} \widehat g_{\mu\rho}
- 
\frac{\partial}{
\partial{\widehat x^\rho}}
\widehat
g_{\mu\nu}\right)\label{Christ}
\ee
They split as $
\widehat\Gamma_{\nu\lambda}^\mu  = \Gamma_{\nu\lambda}^{(1)\mu}  +\gamma_5
\Gamma_{\nu\lambda}^{(2)\mu}$, 
and are such that the metricity condition is satisfied
\be
\frac{\partial}{\partial \hat x^\mu} \widehat g_{\nu\lambda} =  \widehat
\Gamma_{\mu\nu}^\rho\, \widehat g_{\rho\lambda} +
 \widehat \Gamma_{\mu\lambda}^\rho \,\widehat g_{\nu\rho},\label{metricity1}
\ee
Proceeding the same way one can define the MAT Riemann tensor via $\widehat
R_{\mu\nu\lambda}{}^\rho $:
\be
\widehat R_{\mu\nu\lambda}{}^\rho = -\partial_\mu
\widehat\Gamma_{\nu\lambda}^\rho +
\partial_\nu \widehat\Gamma_{\mu\lambda}^\rho-\widehat \Gamma_{\mu\sigma}^\rho
\widehat\Gamma_{\nu\lambda}^\sigma +
\widehat\Gamma_{\nu\sigma}^\rho\widehat\Gamma_{\mu\lambda}^\sigma \equiv { R}^{(1)}_{\mu\nu\lambda}{}^\rho+\gamma_5 {
R}^{(2)}_{\mu\nu\lambda}{}^\rho\label{hatRiemann}
\ee
The MAT spin connection is introduced in analogy
\be
\widehat \Omega_\mu^{ab} &=& \widehat e_\nu^a\left(\partial_\mu\widehat e^{\nu
b}+\widehat e^{\sigma b}
\widehat \Gamma^\nu_{\sigma \mu}\right) = \Omega_\mu^{(1)ab}
+\gamma_5 \Omega_\mu^{(2)ab}\label{spinconn1}
\ee
In a MAT background one must introduce also axially-extended (AE) diffeomorphisms. They are defined by
\be
\widehat x^\mu\rightarrow \widehat x^\mu+\widehat \xi^\mu(\widehat x^\mu),
\quad\quad\widehat
\xi^\mu=\xi^\mu+\gamma_5
\zeta^\mu\label{axialdiff}
\ee
Operationally these transformations act in the same way as the usual
diffeomorphisms, therefore for the non-covariant part of the transformation 
\be
\delta_{\widehat \xi}^{(n.c.)}\widehat \Gamma_{\mu\nu}^\lambda=
\widehat\partial_\mu\widehat\partial_\nu\widehat \xi^\lambda\label{deltancGamma}
\ee
where the derivatives are understood with respect to $\widehat x^\mu$ and
$\widehat x^\nu$.
We have also
\be
\delta_{\widehat \xi}\widehat g_{\mu\nu} =\widehat D_\mu
\widehat\xi_\nu+\widehat
D_\nu\widehat \xi_\mu\label{deltaxihat}
\ee
where $\widehat\xi_\mu =\widehat g_{\mu\nu} \widehat\xi^\nu$ and ${\widehat
D}_\mu$ is the covariant derivative with respect to $\widehat \Gamma$.

There are two types of Weyl transformations that can be compactly written as 
 \be
\widehat g_{\mu\nu} \longrightarrow e^{2\widehat \omega}  \widehat g_{\mu\nu},
\label{axialWeyl}
\ee
where $\widehat \omega=\omega + \gamma_5 \eta$.

 The SDW method is based on point-splitting along a geodesic. Therefore it is crucial to define geodesics in a MAT background. Their defining equations are
\be
\ddot {\widehat x}^\mu + \widehat \Gamma_{\nu\lambda}^\mu \dot {\widehat
x}^\nu\dot {\widehat
x}^\lambda=0\label{geodesic}
\ee
where a dot denotes a derivative with respect to $\hat t=t_1+\gamma_5 t_2$.
Since  $\widehat g_{\mu\nu} \dot
{\widehat
x}^\mu\dot {\widehat x}^\nu $ is constant for geodesics, we can write for the
arc length parameter $\widehat s$ 
\be
\frac {d\widehat s}{d\hat t} = \sqrt{\widehat g_{\mu\nu} \dot {\widehat
x}^\mu\dot
{\widehat
x}^\nu},\label{dsdt}
\ee
The quantity
\be
\widehat \sigma(\widehat x, \widehat x')= \frac 12 (\widehat s-\widehat s')^2\label{worldfunction}
\ee
is called the {\it world function}. It has remarkable properties, starting with
\be
\widehat \sigma_{;\mu} = \widehat \partial_\mu \widehat \sigma = (\hat t-\hat
t')\widehat
g_{\mu\nu} \dot{\widehat
x}^\nu\equiv -\widehat g_{\mu\nu} \widehat y^\nu\label{s3}
\ee
$\widehat y^\mu$ are the {\it normal coordinates} based at $\widehat x$. 
 One can prove that
\be
\frac 12\widehat \sigma_{;\mu} \widehat \sigma_{;}{}^\mu = \widehat
\sigma\label{smus}
\ee

Covariantly differentiating \eqref{smus} we get
\be
\widehat \sigma_{;\nu} = \widehat \sigma_{;\mu\nu} \widehat
\sigma_{;}{}^\mu\label{diffsigma1}
\ee
In the coincidence ($x'\to x$) limit, $[\widehat \sigma_{;\nu}]=0$. Therefore
\eqref{diffsigma1} is trivial in the coincidence limit. But differentiating further one can show that
\be
[\widehat \sigma_{;\mu\lambda}]=  \widehat g_{\mu\lambda} \label{diffsigma3}
\ee
Now the game consists in repeatedly differentiating these relations and computing the coincidence limit. For instance one can prove
\be
[\widehat\sigma_{;\nu\lambda\rho\tau}] = -\frac 13 \left(\widehat
R_{\nu\tau\lambda\rho}+\widehat R_{\nu\rho\lambda\tau}\right)\label{diffsigma9}
\ee
and so on. 

In the sequel we need the Van Vleck-Morette determinant
\be
\widehat {\cal D}(\widehat x,\widehat x') = \det
(-\widehat\sigma_{{;}\mu\nu'})\label{VVM}
\ee
and the geodetic parallel displacer for spinors,  $\widehat {I}(\widehat x, \widehat {x}')$: the object
$\widehat I (\widehat x,\widehat {x}')\widehat\psi(\widehat x ')$ is the
spinor $\psi(\widehat x)$ obtained by parallel displacing $\widehat\psi(\widehat x ')$ along the
geodesic from $\widehat x '$ to $\widehat x$.  It is a bispinor quantity satisfying
\be
\widehat\sigma_{;}{}^\mu \widehat I_{;\mu}=0 ,\quad\quad [\widehat I]={\bf
1}\label{I}
\ee
Among its remarkable properties we have the following concidence limit
\be
[\widehat I (x,x')_{;[\mu,\nu]}] = [\widehat I (x,x')_{;\mu\nu}] = -\frac 14
\widehat{\cal R}_{\mu\nu}\label{I4}
\ee
where $\widehat {\cal R}_{\mu\nu}= {\widehat R}_{\mu\nu}{}^{ab}
\Sigma_{ab}$.

\subsection{Fermions in a MAT background}

The action of a fermion interacting with a metric and an axial tensor is
\be
\widehat S&=&\int d^4\widehat x \, \left(i\overline {\psi}
\sqrt{\overline{\widehat
g}}\gamma^a\widehat
e_a^\mu
\left(\partial_\mu+\frac 12 \widehat\Omega_\mu \right)\psi\right)(\widehat x)\label{axialaction}
\\
&{=}&  \int d^4\widehat x \,\left(  i\overline {\psi} \sqrt{\overline{\widehat
g}
}\gamma^a(\tilde e_a^\mu+\gamma_5
\tilde c_a^\mu)  \left(\partial_\mu +\frac 12 \left(\Omega^{(1)}_\mu+\gamma_5
\Omega^{(2)}_\mu\right) \right)\psi\right)(\widehat x)   \0\\
\ee
It must be noticed that this action takes on axial-real values. The field $\psi(\widehat x)$
can be understood, classically, as a series of powers of $\widehat x$ 
applied to constant spinors on their right, and the symmetry transformations act
on it 
from the left. The analogous definitions for $\psi^\dagger$ are obtained via
hermitean conjugation.  

Remark the position of the density  $\sqrt{\overline{\widehat g}}$\,: it
must be inserted between $\overline \psi$ and $\psi$, due to the presence in it of the $\gamma_5$
matrix. Of course one has to remember that the kinetic
operator contains a $\gamma$ matrix that anticommutes with $\gamma_5$. 

Let us consider AE (axially extended) diffeomorphisms first, \eqref{axialdiff}.
It is not hard to prove that the action \eqref{axialaction} is invariant under
these transformations, provided $\delta_{\hat \xi} \psi= \hat \xi^\mu \frac {\partial  \psi}{\partial \hat x^\mu}$.  Also for the axial complex Weyl transformation one can
prove that, assuming for the fermion field the transformation rule
\be
\psi \rightarrow e^{-\frac 32 (\omega+\gamma_5 \eta)} \psi,\label{Weylpsi}
\ee
\eqref{axialaction} is invariant.

Now, define the full MAT e.m. tensor by means of
\be
{\bf T}_{\mu\nu} = \frac 2{\sqrt{\widehat g}} \frac {\stackrel{\leftarrow}
{\delta}{\widehat S}}
{\delta \widehat g^{\mu\nu}} \label{fullem0}
\ee
This formula needs a comment, since $\sqrt{\widehat g}$ contains $\gamma_5$. To
give a
meaning to it we understand that
the operator $\frac 2{\sqrt{\widehat g}} \frac {\stackrel{\leftarrow} {\delta} }
{\delta \widehat g^{\mu\nu}}$  on the RHS acts on the operatorial expression,
say
${\cal O}{\sqrt{\widehat g}}$, which is inside the spinor scalar product,  
$\overline \psi
{\cal O}\sqrt{\widehat g} \psi$.
Moreover the functional derivative acts from the right of the action.
 
 The explicit e.m. formula (on shell) is
\be
{\bf T}_{\lambda\rho} = -\frac i2 \overline\psi\widehat \gamma_\lambda
\left(\partial_\rho
+\frac 12 \widehat \Omega_\rho\right)\psi + (\lambda\leftrightarrow \rho)
=- \frac i2 \overline\psi\widehat \gamma_\lambda \widehat\nabla_\rho
\psi+(\lambda\leftrightarrow \rho) \label{Tlambdarho}
\ee
where $\widehat\gamma_\lambda= \gamma_a \widehat e_\lambda^a$. This expression splits into a vector and an axial part, $T_{\mu\nu}$ and $T_{5\mu\nu}$, which in the flat limit are given on shell by
\be
T_{\mu\nu}\approx T^{(flat)}_ {\mu\nu}=-\frac i4 \left(\overline {\psi } \gamma_\mu
{\stackrel{\leftrightarrow}{\partial_\nu}}\psi
 + \mu\leftrightarrow \nu \right),\label{Tmunu0}
\ee
and
\be
T_{5\mu\nu}\approx T^{(flat)}_{5\mu\nu}=\frac i4 \left(\overline {\psi}\gamma_5
\gamma_\mu  {\stackrel{\leftrightarrow}{\partial_\nu}}\psi
 + \mu\leftrightarrow \nu \right),\label{T5munu0}
\ee
Weyl invariance leads to the classical Ward identities:
 \be
{\ET}(x)&\equiv&T_{\mu\nu} g^{\mu\nu} + T_{5\mu\nu} f^{\mu\nu}=0,
\label{WardWeyl1}\\
{\ET}_5(x)&\equiv &T_{\mu\nu} f^{\mu\nu} + T_{5\mu\nu} g^{\mu\nu}=0,
\label{WardWeyl2}
\ee
 
 \subsection{The Dirac operator and its square}
 
 In the action \eqref{axialaction} the Dirac operator is
\be
\widehat F=i \widehat \gamma \! \cdot \! \widehat \nabla= i \widehat
\gamma^\mu \widehat\nabla_\mu = i  \gamma^a \widehat e^\mu_a
\widehat\nabla_\mu\equiv \gamma^a \widehat F_a\label{hatF}
\ee
where  $\widehat \nabla=\widehat D+ \frac 12
\widehat\Omega$ and satisfies $\widehat\nabla_\mu \widehat e^a_\nu=0$. 

Under AE diffeomorphisms $\psi$ transforms as:  $\delta_{\widehat\xi} \psi =
\widehat \xi \!\cdot\! \partial\psi$, while
\be
\delta_{\widehat\xi} \left(i\widehat \gamma\!\cdot\!\widehat \nabla \psi\right) =
\overline{\widehat \xi} \!\cdot\! \partial \left(i\widehat
\gamma\!\cdot\!\widehat \nabla \psi\right) \label{deltahatxiDirac}
\ee
Under AE Weyl transformation $\widehat F$ transform as
\be
\delta_{\hat \omega} \widehat F = -\frac 12 \gamma^a \{ \widehat F_a, \widehat
\omega\}\label{hatFWeyl}
\ee
and it has the following hermiticity property
\be
{\widehat F}^\dagger= \gamma_0 \widehat F \gamma_0\label{Fdagger}
\ee
where  $\gamma_0$ is the non-dynamical (flat) gamma matrix.

In order to be able to apply the SDW method we have to select the square Dirac operator. As pointed out earlier, see section \ref{sss:bardeen}, there are a priori a few possibilities. However, the request of respecting the basic (AE diffeomorphism) symmetry of the theory\footnote{We recall once more that this invariance is needed for the point-splitting technique, which underlies the SDW method, to work properly. We stress that invariance under ordinary diffeomorphisms alone is not enough, because in the chiral limit a violation of the axial diffeomorphisms would precisely affect also the ordinary diffeomorphism conservation.} and the self-adjointness after the Wick rotation, identifies it uniquely. In ordinary gravity, from the diffeomorphism invariance of the fermion action, we can extract the transformation rule
\be
\delta_\xi   \left(i\gamma^\mu \nabla_\mu\psi\right) =\xi\!\cdot\! \partial
\left(i  \gamma\!\cdot\! \nabla\psi\right)   \label{deltaxiDirac}
\ee
while $\delta_\xi \psi =\xi\!\cdot\! \partial\psi$.  Therefore it makes sense to
apply $ \gamma\!\cdot\! \nabla$ to $ \gamma\!\cdot\! \nabla\psi$,
because the latter transforms as $\psi$. This allows us to define the square of
the Dirac operator:
\be
F^2\psi = \left( i \gamma\!\cdot\! \nabla\right)^2 \psi\label{Gpsi}
\ee
The expression $\bar \psi F^2 \psi$ is well defined and invariant and it makes sense to extract the square root of the path integral constructed by exponentiating it.
If we  want to preserve general covariance, it is not possible to repeat the same for MAT because of
\eqref{deltahatxiDirac}, from which we see that $\left(i\widehat
\gamma\!\cdot\!\widehat \nabla \psi\right) $ does not
transform like $\psi$, and an expression like $
\left(i\widehat \gamma\!\cdot\!\widehat \nabla\right)^2 \psi$ would break
general covariance. Noting that
\be
\delta_{\hat\xi} \left(i\overline{\widehat \gamma}\!\cdot\!\overline{\widehat
\nabla} \psi\right) =
{\widehat \xi} \!\cdot\! \partial \left(i\overline{\widehat
\gamma}\!\cdot\!\overline{\widehat \nabla} \psi \right) \label{deltahatxiDirac2}
\ee
when $\delta_{\overline{\widehat \xi}}\psi= \overline{\widehat \xi} \!\cdot\!
\partial\psi$,
we are led to choose, instead, the {\it covariant} quadratic operator
\be
\widehat \EF=\left(i\overline{\widehat \gamma}\!\cdot\!\overline{\widehat
\nabla}\right)\,\left(i {\widehat \gamma}\!\cdot\! {\widehat
\nabla}\right)\label{covDiracsquare}
\ee
This is the first reason that justifies the choice of \eqref{covDiracsquare}.

Let us come next to the Euclidean version of $\widehat \EF$. To deal with it it is easier to remark that
\be
\left(i \widetilde{\widehat\gamma} \widetilde{\widehat \nabla}\right)^\dagger= -
i \widetilde{\widehat{\overline\gamma}} \widetilde{\widehat {\overline\nabla}}\label{eucldaggernabla}
\ee
Therefore
\be
\left(\widetilde {\widehat \EF}\right)^\dagger = \widetilde {\widehat \EF}\label{euclfdaggerEF}
\ee
This is the second fundamental reason for using $\widehat \EF$.

The operator $\widehat \EF$ is the main intermediate
result we need. It is natural to assume that its inverse  $\widehat
\EG$ exists. The differential operator $\widehat \EF$ (after a Wick rotation) is an axial-elliptic operator. In fact its quadratic part can be cast in the form $\partial_i  A_{ij}(x)  
\partial_j$, where $A_{ij}$ is an invertible matrix and its leading term is symmetric and
positive definite. Moreover $\widetilde{\widehat \EF}$ is self-adjoint. Therefore the conditions for the application of the SDW method are satisfied. As we have done before, we will always work with Minkowski quantities, while paying attention, however, that no relation is used that cannot be mapped from the Minkowski to the Euclidean by a  Wick rotation, and back by an inverse Wick rotation.

\subsection{The SDW method for the trace anomaly}

There are significant changes with respect to the flat background application of section \ref{sss:bardeen}.

Again we define  the amplitude
\be
\langle\widehat  x,\widehat s| \widehat x',0\rangle = \langle\widehat  x| e^{i 
\widehat\EF
\widehat s}|\widehat x'\rangle\label{hatampB}
\ee
which satisfies the (heat kernel) differential equation
\be
i \frac {\partial }{\partial \widehat s} \langle \widehat x,\widehat s|\widehat 
x',0\rangle
= - \widehat\EF_{\hat x} \langle
\widehat x,\widehat s|\widehat  x',0\rangle\equiv K(\widehat x, \widehat x',
\widehat s) \label{hatdiffeqforB}
\ee
where $\widehat\EF_{\hat x}$ is the differential operator
\be
\widehat\EF_{\widehat x}=\widehat\nabla_\mu {\widehat g}^{\mu\nu}
\widehat\nabla_\nu- \frac
14 \widehat R\label{hatEHx}
\ee
In formula \eqref{hatampB}, as usual, we understand the $i\epsilon$ prescription. But now, instead of \eqref{ansatz3}, we have the ansatz 
\be
\langle \widehat x,\widehat s| \widehat x',0\rangle =- \lim_{m\to 0}\frac
i{16\pi^2}
\frac {\sqrt{\widehat
D(\widehat x,\widehat x')}} {\widehat s^2}
e^{i\left(\frac {\widehat\sigma(\widehat x,\widehat x')} {2\widehat
s}-m^2\widehat s\right)}\widehat
\Phi(\widehat x,\widehat x',\widehat s)\label{ansatz3ter}
\ee
where $\widehat D(\widehat x,\widehat x')$ is the VVM determinant and
$\widehat\sigma$ is the world function. $\widehat\Phi(\widehat x,\widehat x',\widehat s)$ is to be determined. Since it must
be $\langle\widehat x,0|\widehat  x',0\rangle=\delta(\widehat x,\widehat x')$
and instead of \eqref{hatlimdelta} we have
\be
\lim_{\widehat s\to 0} \frac i{(4\pi)^2} \frac {\sqrt{\widehat D(\widehat
x,\widehat x')}} {\widehat s^2}
\, e^{i\left(\frac {\widehat\sigma(\widehat x,\widehat x')} {2\widehat
s}-m^2\widehat s\right)}
= \sqrt{\widehat g(\widehat x)}\,\,
\delta(\widehat x,\widehat x'),\label{hatlimdeltabis}
\ee
like before it follows that
\be
\lim_{\widehat s\to 0}\widehat \Phi(\widehat x,\widehat x',\widehat s)={\bf
1}\label{hatlimitPhi}
\ee
Eq.\eqref{hatdiffeqforB} becomes an equation for $\widehat \Phi(\widehat
x,\widehat x',\widehat s)$:
\be
i\frac {\partial \widehat\Phi}{\partial\widehat  s} +\frac i{\widehat s}
\widehat \nabla^\mu
\widehat \Phi \widehat \nabla_\mu\widehat \sigma
+\frac 1{\sqrt{\widehat D}} \widehat \nabla^\mu \widehat \nabla_\mu
\left(\sqrt{\widehat D} \widehat \Phi\right)-\left(\frac 14 \widehat
R-m^2\right)\widehat\Phi=0\label{hateqforPhi}
\ee
Again we expand
\be
\widehat\Phi(\widehat x,\widehat x',\widehat s)
= \sum_{n=0}^\infty \widehat a_n(\widehat x,\widehat x') (i\widehat s)^n
\label{hatPhiexp}
\ee
with the boundary condition $[\widehat a_0]=1$. The $\widehat a_n$ must satisfy
the recursive relations:
\be
(n+1)\widehat a_{n+1} +\widehat  \nabla^\mu \widehat a_{n+1} \widehat \nabla_\mu
\widehat \sigma - \frac 1{\sqrt{\widehat D}} \widehat \nabla^\mu \widehat
\nabla_\mu
\left(\sqrt{\widehat D}  \widehat a_n\right) +\left(\frac 14 \widehat R
-m^2\right)\widehat  a_n=0
\label{hatrecursive}
\ee
Using these relations  it is possible to compute each coefficient $a_n$ at the coincidence limit. We are interested in particular in $[\widehat a_2]$, which turns out to be
\be
[\widehat a_2] &=& \frac 12 m^4 -\frac 1{12} m^2 \widehat R +\frac 1 {288} \widehat R^2 -\frac
1{120} \widehat R_{;\mu}{}^\mu
-\frac 1{180} \widehat R_{\mu\nu}\widehat R^{\mu\nu} + \frac 1{180} \widehat
R_{\mu\nu\lambda\rho}\widehat R^{\mu\nu\lambda\rho}\0\\
&&+\frac  1{48}
\widehat{\cal R}_{\mu\nu}\widehat{\cal R}^{\mu\nu}   \label{hata2}
\ee
We recall that $\widehat{\cal R}_{\mu\nu}= {\widehat R}_{\mu\nu}{}^{ab}
\Sigma_{ab}$.

Next we continue as in section \ref{ss:perturbative}. In particular eqs.(\ref{W+const},\ref{WK},\ref{Kxx'}) remain the same and lead to effective action
\be
\frac {\widehat L(x)}{\mu^\sfd}=-\frac i2 (4\pi \mu^2) \,\tr
\int_0^{\infty}d\widehat s\,  (4\pi i
\mu^2 \widehat s)^{-\frac {\sfd}2-1}\sqrt{\widehat g}e^{-im^2\widehat s} [\widehat
\Phi(\widehat x,\widehat x,\widehat s)]\label{Lxmud}
\ee
where $\tr$ denotes the trace over gamma matrices. Our purpose is to analytically continue in $\sfd$. But we can do this only for dimensionless quantities. For this reason $\widehat L$ is multiplied by $\mu^{-\sfd}$, where $\mu$ is a mass parameter. 

Now we make the assumption that $\lim_{s\to\infty} e^{-im^2\widehat s} [\widehat \Phi(\widehat x,\widehat x,\widehat s)]=0$. As a consequence we can integrate by parts
\be
\frac {\widehat L(x)}{\mu^\sfd}
\!\!\!\!&=&\!\!\!\! -\frac{4i}{\sfd(2-\sfd)(4-\sfd)}\frac 1{(4\pi \mu^2)^2} \tr \int_0^{\infty}\!\!\!\!d\widehat
s\,
(2\pi i \mu^2 \widehat s)^{2-\frac {\sfd}2}\sqrt{\widehat g}\frac
{\partial^3}{\partial(i\widehat s)^3}
\left( e^{-im^2\widehat s} [\widehat \Phi(\widehat x,\widehat x,\widehat
s)]\right)\label{Lxmud1}
\ee
Now we use $[\widehat \Phi(\widehat x,\widehat x,\widehat s)]= 1+ {[\widehat a_1]} i\widehat
s+ {[\widehat a_2]}(i\widehat s)^2+\ldots$. 

To expand around $\sfd=4$ we use $\frac 1{\sfd(\sfd-2)(\sfd-4)}\approx \frac 18 \left( \frac 1{\sfd-4}
-\frac 34\right)$. With reference to  \eqref{Lxmud1}, we
differentiate twice $ [\widehat \Phi(x,x,s)]$ and integrate by parts the third
derivative. The result is
\be
\widehat L(\widehat x) &\approx&\frac 1{32\pi^2}\left( \frac 1{\sfd-4} -\frac
34\right)\tr
\left(m^4-2m^2 {[\widehat a_1]}+2 {[\widehat a_2]}\right)\sqrt{\widehat
g}\label{Lxd4}\\
&&+\frac i{64\pi^2}\tr \int_0^{\infty}d\widehat s\,  \ln(4\pi i \mu^2 \widehat
s)
\sqrt{\widehat g}\frac {\partial^3}{\partial(i\widehat s)^3} \left(
e^{-im^2\widehat s} [\widehat
\Phi(\widehat x,\widehat x,\widehat s)]\right)\0
\ee
The last line depends explicitly on the parameter $\mu$ and represents a
part which cannot contribute to the anomaly for dimensional reasons. In the first line of \eqref{Lxd4} one can ignore $m^2$ or $m^4$ terms (they can be subtracted away because they are trivial). Therefore we can limit ourselves to
 \be
\widehat L &=& \frac 1{16\pi^2}\left( \frac 1{\sfd-4} -\frac
34\right)
\int d^\sfd\widehat x\,
\tr
\left({[\widehat a_2]|_{m=0}}\sqrt{\widehat g}\right)\label{LR} 
\ee
We now act with 
$\delta_{\widehat\omega} = \int d^\sfd\widehat x \,
2 \tr \left(\widehat\omega\,\widehat g_{\mu\nu} {\frac{\delta}{\delta {\widehat
g_{\mu\nu}}}}\right)$.
An explicit calculation gives, for example,
\be
\delta_{\widehat\omega} 
\tr\left( 
\sqrt{\widehat g}\,\widehat {\cal R}_{\mu\nu}\widehat{\cal R}^{\mu\nu} \right) 
&=&   {(\sfd-4) \tr\left(\widehat\omega \,
\sqrt{\widehat g}\,\widehat {\cal R}_{\mu\nu}\widehat{\cal R}^{\mu\nu} \right)
+2 \, \tr \left( \sqrt{\widehat g} \, \widehat R \widehat \square \widehat\omega
\right) }
\label{deltaomega7}
\ee
Finally we get
\be
\delta_{\widehat\omega} \tr \left(\sqrt{\widehat g} \,[\widehat
a_2]|_{m=0}\right) 
= (\sfd-4)\tr \left( \sqrt{\widehat g} \, \widehat\omega \, [\widehat a_2]|_{m=0} 
\right) 
-\frac{\sfd-4}{120}\tr \left( \sqrt{\widehat g} \, \widehat R \widehat \square
\widehat\omega \right) 
\label{totald4}
\ee

The second piece can be canceled, e.g. by a counterterm proportional to 
$\tr \left( \sqrt{\widehat g}\widehat R^2 \right)$.

Now the variation of the effective action under the $\widehat \omega$ transformation defines the integrated anomaly. Therefore, defining $\frac{2}{\sqrt{\widehat g}}\frac{\delta}{\delta {\widehat g^{\mu\nu}}}  \widehat L = \widehat\Theta_{\mu\nu}$ and recalling the definition \eqref{inttraceanom}, and using the fact that the second line of \eqref{Lxd4} is Weyl-invariant, we get in the $\sfd\to 4$ limit, 
\be
\int d^4\widehat x \, \tr \left(\widehat\omega {\sqrt{\widehat g}} \, \widehat
g^{\mu\nu} \widehat\Theta_{\mu\nu}\right)
&=& \frac 1{16\pi^2}\int d^4\widehat x\, \tr \left( \sqrt{\widehat g} \,
\widehat\omega\, [\widehat a_2]|_{m=0} \right)  \label{LR2}
\ee
where
the $\sfd-4$ factor in \eqref{totald4} has canceled the pole $\frac 1{\sfd-4}$ in \eqref{LR}.

Clearly, the odd-parity anomaly can come only from the term
$\widehat{\cal R}_{\mu\nu}\widehat{\cal R}^{\mu\nu}$ contained in
$ {[\widehat a_2]}$. For the odd part we have
\be
\int d^4\widehat x  \, \tr {\sqrt{\widehat g}}\,\widehat\omega \, \widehat\ET
= \frac 1{768\pi^2}  \int d^4x \,\tr\sqrt{\widehat g}\, \widehat\omega
\,\widehat{\cal R}_{\mu\nu}\widehat{\cal R}^{\mu\nu}\Big{\vert}_{\rm odd}
\label{intan}
\ee
where we denoted $\widehat \ET =  \widehat g^{\mu\nu}  \widehat \Theta_{\mu\nu}
=  \widehat g^{\mu\nu} \langle\!\langle \widehat T_{\mu\nu}\rangle\!\rangle$.

\subsection{The chiral limit and the chiral trace anomaly}
\label{ss:chirallimit}
{
Let us return to the original problem,
that is the trace anomaly of a Weyl tensor in a chiral fermion theory coupled
to ordinary gravity. To this end we need to take the chiral limit. But before we rewrite 
the anomaly \eqref{intan} by splitting it in the  chiral and anti-chiral parts. Setting $P_\pm= \frac 12(1\pm\gamma_5)$ we can write
\be
\sqrt{\widehat g}
=P_+\sqrt{\det g_+} +P_- \sqrt{\det g_-}, \quad\quad g_{\pm\mu\nu}=g_{\mu\nu}\pm f_{\mu\nu}\label{chiralvolume}
\ee
and 
\be
\widehat\omega =P_+ \omega_+ +P_- \omega_-, \quad\quad \widehat{\cal R}_{\mu\nu} = P_+{\cal R}^{(+)}_{\mu\nu} +   P_-{\cal R}^{(-)}_{\mu\nu} \label{omegaplusminus}
\ee
where
\be
\omega_\pm =\omega \pm \eta, \quad\quad {\cal R}^\pm_{\mu\nu} =  {\cal R}^{(1)}_{\mu\nu} \pm {\cal R}_{\mu\nu}^{(2)}\label{omega+eta}
\ee
Then we can rewrite \eqref{intan} as follows
\be
&&\int d^d\widehat x  \, \tr {\sqrt{\widehat g}}\,\widehat\omega \, \widehat\ET\Big{\vert}_{\rm odd}\label{intan2}\\
&=& \frac 1{1536\pi^2} \int d^4x \, \tr\left(P_+\sqrt{\det g_+}\,\omega_+ \,{\cal R}^{(+)}_{\mu\nu}{\cal R}^{(+)\mu\nu}+P_-\sqrt{\det g_-}\,\omega_- \,{\cal R}^{(-)}_{\mu\nu}{\cal R}^{(-)\mu\nu}\right)\Big{\vert}_{\rm odd}\0
\ee
The chiral limit is defined by making the replacements 
\be \label{colllr}
g_{\mu\nu} \to \eta_{\mu\nu} + \frac
{h_{\mu\nu}}2, \qquad\qquad f_{\mu\nu} \to \frac {h_{\mu\nu}}2
\ee
in the previous formulas. With this choice one has
\be
\hat{g}_{\mu\nu} =P_-\, \eta_{\mu\nu} +P_+\, \sfg_{\mu\nu},
\qquad\qquad \sfg_{\mu\nu} \equiv \eta_{\mu\nu} + h_{\mu\nu}\label{gsfg}
\ee
From this we see that the left-handed part couples to the flat metric, 
while the right-handed part couples to the (generic) metric $\sfg_{\mu\nu}$. As a
consequence we have also
\be
\widehat e^a_m \to \delta^a_m  P_- + e^a_m\,  
P_+,\quad\quad
\widehat e^m_a \to  \delta_a^m  P_- + e^m_a \, P_+\label{vierbechirallimit}
\ee
as well as $\sqrt{\widehat g}\rightarrow P_- +P_+\sqrt{\sfg}.$ Similarly for the Christoffel symbols
\be
 \Gamma_{\mu\nu}^{(1)\lambda} \to \frac 12
\Gamma_{\mu\nu}^\lambda,\quad\quad\Gamma_{\mu\nu}^{(2)\lambda} \to \frac 12
\Gamma_{\mu\nu}^\lambda,\label{Gamma12limit}
\ee
for the spin connections
\be
 \Omega_\mu^{(1)ab}\to \frac 12  \omega_\mu^{ab},\quad\quad 
\Omega_\mu^{(2)ab}\to \frac 12  \omega_\mu^{ab},\label{Omega12limit}
\ee 
and for the curvatures
\be
 R^{(1)}_{\mu\nu\lambda}{}^\rho\to \frac 12 R_{\mu\nu\lambda}{}^\rho,
\quad\quad
R^{(2)}_{\mu\nu\lambda}{}^\rho\to \frac 12
R_{\mu\nu\lambda}{}^\rho,\label{R12limit}
\ee
where all the quantities on the RHS of these limits are built with the metric
$\sfg_{\mu\nu}$.

As a consequence, the action \eqref{axialaction} becomes 
\be
\widehat S& \rightarrow& S'\label{axialactionlimit}\\
S' \!\!\!\! &=& \!\!\!\! \int d^4x \,  i\overline {\psi}\gamma^a
P_-\partial_a \psi+ \int d^4x\, \sqrt{\sfg}\, i\overline
{\psi}\gamma^a e_a^\mu
\left(\partial_\mu +\frac 12 \omega_\mu \right)P_+\psi\0
\ee
where $\gamma^a$ are the flat (non-dynamical) gamma matrices while the vierbein
$e_a^\mu$ 
and the connection 
$\omega_\mu$ are compatible with the metric $\sfg_{\mu\nu}$. The action
$S'$ is the action of a right-handed Weyl fermion coupled to ordinary gravity, except for the term that
represents a decoupled left-handed fermion in the flat spacetime.

Moreover, in the chiral limit we have ${\cal R}_{\mu\nu}^{(-)}\to 0, {\cal R}_{\mu\nu}^{(+)}\to 2 {\cal R}_{\mu\nu}$. Therefore \eqref{intan2} becomes
\be
&&\int d^d\widehat x  \, \tr {\sqrt{\widehat g}}\,\widehat\omega \, \widehat\ET\rightarrow
 -\frac 1{768\pi^2} \int d^4x \, \tr\left( P_+\sqrt{g}\, \omega_+ {\cal R}_{\mu\nu} {\cal R}^{\mu\nu}\right)\0\\
&&= \frac i{2 \times 768\pi^2} \int d^4x \sqrt{g}\, \omega_+\varepsilon^{\mu\nu\lambda\rho} R_{\mu\nu\alpha\beta} R_{\lambda\rho}{}^{\alpha\beta}\label{chirallimit}
\ee
Now notice that an extended Weyl transformation splits as follows
\be
e^{2\widehat\omega} \widehat g = P_+ e^{2\omega_+}g_+ +P_- e^{2\omega_-}g_-\label{Weylsplit}
\ee  
therefore in the chiral limit, for consistency with \eqref{axialactionlimit}, we must have
that $\eta$ coincides with $\omega$,  $\omega\to \omega/2, \eta\to \omega/2$. We conclude that in the chiral limit the e.m. trace is
\be
\ET(x)  =
 \frac{i}{1536\pi^2} \varepsilon^{\mu\nu\lambda
\rho}R_{\mu\nu\alpha\beta}
R_{\lambda\rho}{}^{\alpha\beta}\equiv\ET_R(x)  \label{chirallimittrace}
\ee
which coincides with \eqref{Pontryagintrace}. If, instead of (\ref{colllr}), we take the following chiral limit
\be
g_{\mu\nu} \to \eta_{\mu\nu} + \frac
{h_{\mu\nu}}2, \qquad\qquad f_{\mu\nu} \to -\frac {h_{\mu\nu}}2
\ee
we obtain the
Pontryagin 
Weyl anomaly for left-handed Weyl fermion
\be
\ET_L(x) =- \frac {i}{1536\pi^2} \varepsilon^{\mu\nu\lambda
\rho}R_{\mu\nu\alpha\beta}
R_{\lambda\rho}{}^{\alpha\beta}.\label{chirallimittraceleft}
\ee
\subsubsection{The full MAT trace anomalies}

From eq.\eqref{intan2} it is easy to compute the $\ET$ and $\ET_5$ anomalies in the general MAT background. They are the coefficients of $\omega$ and $\eta$, respectively.:
\be
\ET(x)&=&  \frac i{1536\pi^2}\varepsilon^{\mu\nu\lambda\rho}\left(\sqrt{g_+} R^{(+)}_{\mu\nu\alpha\beta}R^{(+)}_{\lambda\rho}{}^{\alpha\beta} +\sqrt{g_-} R^{(-)}_{\mu\nu\alpha\beta}R^{(-)}_{\lambda\rho}{}^{\alpha\beta}\right)\label{ETx}\\
{\ET}_5(x)&=&  \frac i{1536\pi^2}\varepsilon^{\mu\nu\lambda\rho}\left(\sqrt{g_+} R^{(+)}_{\mu\nu\alpha\beta}R^{(+)}_{\lambda\rho}{}^{\alpha\beta} -\sqrt{g_-} R^{(-)}_{\mu\nu\alpha\beta}R^{(-)}_{\lambda\rho}{}^{\alpha\beta}\right)\label{ET5x}
\ee
We have Wick-rotated back the result: this is the origin of the $i$ in the anomaly coefficient.

\vskip 0.3cm
At this point we can safely { replace $\widehat x^\mu$ with $x^\mu$ everywhere.}

\subsubsection{Other results}

The previous trace anomalies can be obtained with different non-perturbative methods, for instance with the Seeley-DeWitt method, see \cite{I}. On the wake of these derivations one can easily obtain other interesting results. For instance the even-parity trace anomalies for Weyl and Dirac fermions, which we do not report here. 

Another interesting result is the analog of the ABJ anonaly, that is the response of the effective action of a Dirac fermion $\psi$ coupled to ordinary metric $g_{\mu\nu}$ under the transformation
\be
\delta_\eta g_{\mu\nu} =2 \gamma_5 \eta \,g_{\mu\nu},\quad\quad
\delta_\eta \psi =-\frac 32 \gamma_5 \eta\, \psi\label{deltagammafiveeta}
\ee
This response can be extracted from  \eqref{intan2} in the limit $g_{\mu\nu} \to g_{\mu\nu}, f_{\mu\nu}\to 0$. In this limit $ \Gamma_{\mu\nu}^{(1)\lambda} \to \Gamma_{\mu\nu}^\lambda$ and $ \Gamma_{\mu\nu}^{(2)\lambda} \to 0$, $R^{(1)}_{\mu\nu\lambda}{}^\rho\to  R_{\mu\nu\lambda}{}^\rho,
 R^{(2)}_{\mu\nu\lambda}{}^\rho\to 0$. Therefore $g_{\pm\mu\nu} \to g_{\mu\nu}$ and
$R^{\pm}_{\mu\nu\lambda\rho}\to R_{\mu\nu\lambda\rho}$.
As a consequence the odd-parity (ABJ-like) trace anomaly is
\be
\ET_5 (x)= \frac i {768\pi^2}  \varepsilon^{\mu\nu\lambda
\rho}R_{\mu\nu\alpha\beta}
R_{\lambda\rho}{}^{\alpha\beta}.\label{ABJlike2}
\ee
Contrary to the anomalies (\ref{chirallimittrace},\ref{chirallimittraceleft}) the anomaly \eqref{ABJlike2} is not a risk for the consistency of a theory of Dirac fermions coupled to gravity. Rather its absorbitive nature suggests a possible phenomenological application to a decay of a neutral bound state into two gravitons, similar to the one of the ABJ anomaly for a double-photon decay of $\pi^0$.

Finally with a small supplementary effort we can obtain the gauge-induced trace anomaly, for instance for a Weyl fermion coupled to a vector potential $V_\mu$. The kinetic operator in this case is $i\slashed {\mathbb D}$ with
\be
\widehat {\mathbb D}_\mu =\widehat D_\mu +\frac 12 \widehat \Omega_\mu + V_\mu \label{nablamu}
\ee
Proceeding exactly as in the gauge-less case we find
\be
\delta_\omega L= -\frac i{32\pi^2} \int d^4x\, \sqrt{g}\, \omega\,\tr \left( P_+ [a_2](x)\right)\label{intLx}
\ee
The odd part of $\tr \left( P_+ [a_2](x)\right)$ produces precisely the density $-\varepsilon^{\mu\nu\lambda\rho} \partial_{\mu}V_\nu \partial_\lambda V_\rho$.  Therefore \eqref{intLx}, together with the definition \eqref{inttraceanom}, yields for a right-handed fermion
\be
{\cal A}_\omega = -\frac 1{96\pi^2} \int d^4x\,\sqrt{g}\, \omega\, \varepsilon^{\mu\nu\lambda\rho} \partial_{\mu}V_\nu \partial_\lambda V_\rho,\label{Aomega''}
\ee
which coincides with eq.\eqref{tracedifferenceb}.

\section{Recap of anomalies}

The main subdivision of local anomalies is in two types: 
\begin{itemize}
\item type O, comprises all the anomalies that correspond to the non-existence of a (fermion) propagator; these anomalies show up only in theories of chiral fermions; these anomalies are a threat to the consistency of the theory;

\item type NO, includes all the others; they have nothing to do with the lack of any propagator and do not signal {by themselves} any inconsistency of a theory.
\end{itemize}

Type O anomalies are odd-parity, they are present in parity-violating theories of fermions. They may appear as a violation of the classical conservation of the gauge (example. eq. \eqref{consanom1}) or diffeomorphisms symmetries, and of the conformal invariance in the non-vanishing trace of the e.m. tensor (examples: (\ref{pontryagin},\ref{chern}). From a formal point of view they satisfy WZ consistency conditions. In the physical approach and mathematical approach they appear in different, complementary ways. In the former case one faces the problem of the lack of the propagator in a chiral fermion theory and overcomes it by adding a free fermion of opposite chirality. The trick in general works, but it fails if (in an originally classical conformal theory) the gauge current or the e.m. tensor are non-conserved or traceful, in which case inconsistency may manifest itself as a violation of unitarity (through a breakdown of the BRST symmetry in the first case, and a complex effective Hamiltonian in the second). In the mathematical approach, type O anomalies appear as obstructions to the existence of a chiral fermion propagator, as represented by a non-trivial family's index. Between the two approaches there is a perfect correspondence. A non-trivial family's index exactly signals  anomalies that appear in the divergence of the chiral gauge current or of the e.m. tensor, and in the trace of the latter, and vice-versa.

Type NO anomalies are, first, the even-parity anomalies that appear in the trace of the e.m. tensor (see, eqs.(\ref{quadweyl}, \ref{gausbonnet},\ref{gaugeaction} and sec. 17 nd 18 below). They satisfy the appropriate WZ consistency conditions. But type NO include also the odd-parity anomalies of the ABJ type, for instance the anomalies of a chiral current in a Dirac fermion theory (example: eq.\eqref{covanom}), or, anyhow, in theories free of type O anomalies. They are usually called covariant anomalies and appear in the conservation of chiral currents that do not propagate in the internal lines of the Feynman diagrams. Formally they may be viewed as anomalies of chiral currents coupled to auxiliary axial potentials that are subsequently made to vanish. All type NO anomalies are generated from a (Euclidean) self-adjoint kinetic operator and do not signal any absence of propagators (their family's index vanishes, thus no obstructions!). Therefore they are no threat to the consistency of the theory, and, usually, play a role in phenomenological applications. 

{ 
From the calculation point of view the two type of anomalies in the perturbative approach do not differ. As we have explained several times the method based on Feynman diagram is the same, but one may face the difficulties pointed out in ref.\eqref{bonora2022}: the perturbative calculations are rather accessible at one loop of approximation, but may be quite unwieldy at higher loops, especially when gravitons are involved; this is a source of trouble if the one loop calculation has problems like the ones pointed out  in section 6. For this reason it is important to have at hand other methods, in particular non-perturbative ones, like the SDW method. The advantage of these method is that it allows us to compute in one stroke the full expression of the anomaly, non just the lowest perturbative orders.There is a disadvantage with the SDW method: since it is based on the point splitting along a geodesic, it is automatically diffeomorphism-invariant. Therefore it cannot reproduce the consistent diffeomorphism anomalies, but it reproduces all the anomalies that preserve diffeomorphisms. This problem is irrelevant in 4$\sf d$ because such anomalies vanish identically. The SDW method does not distinguish between O and NO anomalies, what makes a difference between the two is the quadratic Dirac (or other) operator. When computing NO anomalies it is of tremendous help to enlarge the space of fields with the adjunction of an axial gauge potential (Bardeen's method) or an axial metric

The calculation of obstructions, corresponding to O anomalies, from the family's index theorem are straightforward. It is enough to consult a textbook such as \cite{Lawson} and copy the formulas of the index there. As far as the chiral gauge anomalies are concerned one further step is needed. One must apply the appropriate transgression formulas \eqref{TPnA} to the polynomials provided by the family's index theorem in order to get the descent equations \eqref{TPdescent2} and companions. Instead, in what concerns the NO anomalies the family's index theorem yields an identically vanishing result, meaning that there are no obstructions to the existence of the relevant propagator. 

A difference between the last and the previous two methods has to be mentioned. In the latter we start from a theory in a Minkowski metric background  and carry out all manipulations with this metric, while we turn to a Euclidean background in order to give a meaning to divergent Feynman integrals or the $s-$integrals in the SDW method. The family's index theorem instead is entirely formulated in the Euclidean: the Dirac (or other) operator is analysed in a Euclidean spacetime.

The O and NO anomalies are studied in quantum field theories for different purposes. The obstructive ones need to be known in order to guarantee that in a well-defined theory they are absent. In this sense the O anomalies do not have any physical application, but they are crucial in determining the physical content of a theory. 
The NO anomalies on the contrary make their appearance (also) in well-defined theories and may have important physical applications. A typical example is the application of the ABJ anomaly of a chiral current, such as \eqref{covanom} in the Abelian case, to the decay $\pi^0$ into two $\gamma$'s. Another application is in the strong $U(1)$ problem, with the same current \eqref{covanom} in the non-Abelian case, but with $T^a=1$.

As for the even trace anomalies, they share the same characteristics as all the NO anomalies. Therefore they do not endanger any theory free of O anomalies, and they might have phenomenological applications similar to ABJ gauge anomalies. One has mentioned after eq.\eqref{ABJlike2}. But this is a ground largely still to be explored. On the other hand even trace anomalies have a special status, because, although they may appear in perfectly well defined theories, they signal the breakdown of Weyl symmetry, therefore they pose the problem of whether that symmetry  have to be restored or not. We will discuss this issue at length in Part II.

\eject

\vskip 1cm
\centerline {Anomaly Recap}
\vskip 0.2cm
\noindent\fbox{
    \parbox{\textwidth}{

\begin{itemize}

\item Site of local anomalies: $\left\{\begin{matrix} {\rm divergence\, of \,current \,or\,of\, e.m.\, tensor} \\ \\{\rm trace \,of \,e.m.\, tensor}\end{matrix}\right.$

\item Local Anomalies are of two types: $\left\{\begin{matrix} {\rm type\,\, O} &\left\{\begin{matrix} {\rm prevent\, existence\, of\, propagators,}\\ {\rm dangerous: must\, be\, canceled} \end{matrix}\right.\\ \quad\\{\rm type\,\, NO} &\left\{ \begin{matrix}{\rm no \,obstruction\, for\, propagators,}\\ {\rm need\, not\, be\, canceled}\end{matrix}\right.\end{matrix}\right.$

\item Where: $\left\{\begin{matrix} {\rm type\,\, O} & {\rm only \, in\, chiral \, theories\, } \\ \\{\rm type\,\, NO} &{\rm in\, any\, theory}\end{matrix}\right.$

\item Examples: $\left\{\begin{matrix} {\rm type\,\, O} &\left\{ \begin{matrix}  {\rm gauge:eq.(\ref{consanom},\ref{consanom1})},\quad {\it cons.}\\ \\{\rm trace: eqs.  (\ref{pontryagin},\ref{chern},\ref{AomehanonAbelianR},\ref{final2},\ref{chirallimittrace})},\quad {\it cons.}\end{matrix}\right.\\ &\\{\rm type\,\, NO} &\left\{ \begin{matrix}{\rm ABJ: eq.\eqref{covanom}},\quad {\it cov.}\\ \\{\rm trace:eqs.(\ref{quadweyl}, \ref{gausbonnet},\ref{gaugeaction})\, and \,secs. 17,\, 18, \, eqs.(\ref{trueAomega1},\ref{scalartraceanomtrue}}),\quad {\it cons.}\\ \\ {\rm ABJ-like\, trace: eq.}\eqref{ABJlike2},\quad {\it cons.}\end{matrix}\right.\end{matrix}\right.$

\item Cancelation:  $\left\{\begin{matrix} {\rm type\,\, O} &\left\{ \begin{matrix} {\rm A:\, group\, theoretical}\,\, ({\footnotesize {\rm unavailable\, for\, trace\, anomalies)}}\\ \\{\rm B:\,  coefficient\, matching}\,\, ({\footnotesize {\rm unlikely\, for\, trace\, anomalies}}) \\ \\{\rm C: \, Wess\!-\!Zumino\, terms \, or \, Green\!-\! Schwarz\, mechanisms}\end{matrix}\right. \\&\\
{\rm type\,\, NO}&
\left\{ \begin{matrix} {\rm in\, general\, not\, required}\\ \\
{\rm for\, even\, trace\, anomalies}\left\{ \begin{matrix}{\rm B:\, coefficient\, matching\, unlikely}\\ \\{\rm C: \,with\,Wess\!-\! Zumino\, terms} \end{matrix}\right.
 \end{matrix}\right.
 \end{matrix}  \right.$
 
\end{itemize}
\vskip 0,5cm
 
Captions: {\it cons.} means they satisfy WZ consistency conditions,\,\,\,  {\it cov.} means covariant

}}}

\vskip 0,3cm
\noindent Ref.: \cite{atiyah1989,bonora2018,bonora2021,I,dewitt2003,Liu2022,Peccei}

\section{Comments}

The results illustrated above concerning trace anomalies are not universally accepted. Some researchs have led to the conclusion that the odd-parity trace anomalies do not show up in field theory models. 
One must recognize that the calculation of these trace anomalies is in general a subtle task, subtler than for other anomalies. 
{ This is not totally surprising because field theory is not an axiomatic theory that allows one to deduce results by means of a unique formalism.  It is rather a theory under construction, which  often requires different approaches. In particular for anomalies it is important to use different methods and compare them in order to get reliable results. Using one single method of calculation may not be safe. One can in fact envisage procedures that yield zero when searching for the odd part of the trace anomaly and even for ordinary chiral gauge anomalies - one example is signalled in the footnote before eq.\eqref{deltaEAWa}. But they are  to be traced back to not infrequent wrong prejudices or misunderstood statements in the literature. Below are some more examples.}

\begin{itemize}
\item type O anomalies may occur only in theories of Weyl fermions, not in theories of Majorana fermions. A quick method to get a vanishing odd-parity trace anomaly is searching for type O anomalies in the latter case. The wrong use of Majorana fermions in the derivation of anomalies arises from a misunderstanding: the misplaced credence that massless Majorana fermions are equivalent to Weyl fermions. A Majorana fermion can be represented as a superposition, satisfying the Majorana condition, of two Weyl fermions of opposite chirality. We have { repeated} many times that Weyl fermions of opposite chirality have opposite type O anomalies, with consequent null overall result for Majorana fermions (no need to do any calculation)\footnote{On the other hand we are not aware of any attempt to formulate the SM in terms of Majorana fermions alone.}.

\item There are several ways to get a vanishing odd-parity trace anomaly via a perturbative calculation, and the calculation in itself may be correct, depending on the regularization procedure. The reason has been explained above. It does not make sense to study confomal anomalies separately from diffeomorphism invariance. The calculation of a conformal anomaly is safe when we can guarantee that diffeomorphisms are conserved. If diffeomorphisms are anomalous their anomaly (in 4$\sfd$) is in general trivial and can be eliminated by a counterterm in the effective action, which in general modifies the trace anomaly. This can be verified in a number of cases, see \cite{I}. But this mechanism does not work in the case of 4$\sfd$ odd trace anomalies at the lowest perturbative order, because the odd part of the conformal 3-point function of the e.m. tensor vanishes identically for algebraic reasons, \cite{zhiboedov}; therefore also its divergence vanishes identically and no (trivial) diffeomorphism anomaly can show up. In other words, this problem is {\it undecidable} at the lowest perturbative order; one must go to higher orders or use non-perturbative methods.

\item In the heat-kernel-like methods applied to, say, a right-handed Weyl fermion one cannot simply use the Dirac kinetic operator $\slashed{D}$ multiplied by the chiral projector $P_R$, because the overall operator  $\slashed{D}P_R$ is not invertible: using formally this operator invalidates all the manipulations necessary to derive anomalies. Its square is zero. Then one may be tempted to use in the SDW method its product with its adjoint, making the (Euclidean) operator $\slashed{D}^2P_R$, and inserting it, for instance, in eq. \eqref{deltaEAW1}. Using mechanically the SDW formulas one gets 0 for the odd-parity trace anomaly.  However the operator $\slashed{D}^2P_R$ is not invertible, therefore, in particular, not elliptic, and cannot be used in the SDW (or in any heat kernel - like) approach.

\item An often confusing issue is Wick rotation. In Appendix B we explain our rules for applying Wick rotation. In perturbative and non-perturbative (SDW) approaches it does not make sense to start from a Euclidean action. Any attempt to Wick-rotate the Minkowski action for Dirac fermions leads to a doubling of degrees of freedom, thus changing the nature of the problem (in the approach with the family's index theorem this difficulty is avoided because the action is not needed, one simply studies the inversion of the linear Dirac-Weyl operator).  The right attitude toward Wick rotation is to consider it a way to make sense of perturbative Feynman diagram integrals or the $s$-integrals in the SDW case.

\item Another confusing issue is that of unitarity. A tenet in physics is that any calculation has to guarantee unitarity. However this is not the correct attitude when { searching for} type O anomalies. The reason is that the latter are not physical quantities.  As we know from above type O anomalies (contrary to type NO ones) not only are not physical but in fact they disrupt unitarity. Type O anomalies signals a clash between correct mathematics and consistent physics (unitarity). As a consequence requiring unitarity in this case is a nonsense: searching for type O anomalies means investigating possible breakdowns of unitarity, and requiring unitarity from the start prevents the possibility to find out if there is any such breach. On the contrary we have to proceed in the most rigorous way to discover if any mathematical obstruction exists.

\item The previous case may be connected with the choice of the square Dirac operator in the SDW approach. It has been suggested that the choice should be $\widehat \EF_s= \left(i\overline{\widehat \gamma}\!\cdot\!\overline{\widehat
\nabla}\right)\,\left(i {\widehat \gamma}\!\cdot\! {\widehat
\nabla}\right)+ \left(i {\widehat \gamma}\!\cdot\! {\widehat
\nabla}\right)\left(i\overline{\widehat \gamma}\!\cdot\!\overline{\widehat
\nabla}\right)$,
instead of our choice \eqref{covDiracsquare}: $
\widehat \EF=\left(i\overline{\widehat \gamma}\!\cdot\!\overline{\widehat
\nabla}\right)\,\left(i {\widehat \gamma}\!\cdot\! {\widehat
\nabla}\right)$.
The choice $\widehat \EF_s$ is symmetric in the exchange $\gamma_5\leftrightarrow -\gamma_5$, thus it excludes any odd-parity result. But it is a forbidden choice because it breaks diffeomorphism invariance, which is the basis of the SDW method and thus disrupts mathematical consistency. The choice of  $\widehat \EF_s$ is a perfect example of what one should not do when computing anomalies. 

\item However, above all, what is missing in all the `vanishing odd trace anomaly' papers is a plausible explanation of why these anomalies should be absent. The densities of Pontryagin and Chern classes have the right quantum numbers (dimensions, form degree, odd-parity, Lorentz covariance) to couple to the trace of the e.m. tensor in a theory of Weyl fermions. Why should this coupling vanish? Usually this requires a protecting symmetry. For instance, a Dirac fermion is free of this anomaly (and of the consistent gauge anomaly as well) because the trace of e.m. tensor (and the divergence of the vector current) is protected by parity conservation. But in the case of a Weyl fermion theory parity is violated. { What other reason is there for such terms not to couple?}

\item { Our final consideration may sound paradoxical, but it is the gist of our anomaly analysis: no doubt eventually an agreement will be reached about odd-parity trace anomalies, but the proof of their existence is only relatively important. The very important issue is the obstructions they represent and are only a symptom of. As explained above the obstructions are those defined by the family's index theorem. Even not believing in the validity of the derivations we have presented, one stumbles anyhow against the existence of the obstructions signalled by the family's index theorem (not a minor result in modern mathematics). Only when these obstructions are removed does the Weyl fermion propagator (and the theory) exist. {The problem represented by such obstructions exists independently of the anomaly problem.}}

\end{itemize}

\vskip 0,3cm
\noindent Ref.: \cite{bastianelli2016,bastianelli2019,bastianelli2020,bastianelli2022,bonorasoldati2019,I,coriano24a,frob2019,frob2021,
larue23,larue24,maalouf2024}

 \vskip 2cm 

\section{Extensions of the MSM}

The vanishing of type O anomalies for the MSM \eqref{MSMspectrum} immersed in a gravitational background have been analyzed in Table MSM gauge, mixed, trace-gravity and trace-gauge. When gravity fluctuating fields become relevant, one gravity-induced trace anomaly and one $SU(2)$ gauge-induced trace anomaly survive cancelation. This may change if we add to the MSM spectrum a right-handed sterile neutrino: it cancels the gravity-induced trace anomaly, but leave the gauge-induced one unaltered.

 We would like to examine now various generalizations of the MSM, see \cite{CPviol}, in relation to the anomaly cancelation when in presence of a non-trivial background metric. For some of them nothing changes because they do not involve modifications of the fermion spectrum. This is the case of the multi-Higgs-doublet models or similar models in which only the spectrum of scalars is modified with respect to the original SM. They have been introduced mostly to describe explicit and spontaneous $CP$ violation. The fermion spectrum is the same as in the original SM. But plenty of extensions of the ordinary SM have been proposed which involve modification of the fermion spectrum. Hereafter we wish to examine some of them.
 
 \subsection{Models with vector-like quarks}
 
 Models with vector-like quarks have been proposed in order to produce non-vanishing but naturally suppressed flavor-changing neutral currents and to simplify the mechanisms of spontaneous $CP$ violation. They also represent an attempt to solve the strong $CP$ problem. The quark spectrum together with an ordinary lepton spectrum are summarized in the following table
 
 \be
\begin{matrix} \quad&{\sf G}/fields & \quad SU(3)\quad &\quad SU(2)\quad &\quad U(1)\quad\\
n_g&\left( \begin{matrix} u \\ d\end{matrix} \right)_{\! L} & 3&2&\frac 16\\
n_g &{u_R} & \bar 3 &1&\frac 23\\
n_g &{d_R} & \bar 3 &1&-\frac 13\\
n_u &{U_L} &  3 &1&\frac 23\\
n_d &{D_L} & 3 &1&-\frac 13\\
n_g+n_u &{U_R} & \bar 3 &1&\frac 23\\
n_g+n_d &{D_R} & \bar 3 &1&-\frac 13\\
n_g&\left( \begin{matrix} \nu_e \\ e\end{matrix} \right)_{\! L} & 1&2&-\frac 12\\
n_g& {e_R} &  1 &1&-1
\end{matrix}\label{SMvectorlikespectrum}
\ee
where $n_g$ is the number of generations (that is, for instance, $u_L$ represents a set of field numbered from 1 to $n_g$: $u_{Li}$, $i=1,\ldots, n_g$, and so on). Analogously $n_u$ and $n_d$ are the numbers of left-handed singlets $U_{L\alpha}$, $\alpha=1,\ldots,n_u$, etc.
These models are the old SM to which a bunch of vector-like $SU(2)$ singlet quarks have been added.  By vector-like quarks it is meant that they are quarks whose left- and right-handed components appear in a symmetric form. In other words, the kinetic term of the left and right components of each of them conjure up the kinetic term of a Dirac fermion.

On the basis of our previous conclusions, the addition of chiral symmetric fermions does not add any type O anomaly, neither gauge nor trace. Therefore the balance of anomalies is the same as in the original MSM (with or without sterile neutrinos).

\vskip 0,2cm

\noindent Refs.: \cite{aguilar13,castelo22}

\subsection{Left-right (parity) symmetric extensions of the SM model}
\label{ss:LR}

The MSM is $CP$ invariant, but $P$ non-invariant. Some generalizations have been considered with the aim to realize parity-invariant models. The models are built in such a way that parity is spontaneously broken in the vacuum, so that in the broken phase the original MSM turns up. A parity-invariant SM can be realized 
by introducing a second $SU(2)$ group, realizing in this way a left-right symmetry. The gauge  group is
$SU(3) \times SU(2)_L \times SU(2)_R \times U(1)$ and the spectrum of fields is

\be
\begin{matrix} {\sf G}/fields & \quad SU(3)\quad &\quad SU(2)_L\quad&\quad SU(2)_R\quad &\quad U(1)\quad\\
\left( \begin{matrix} u \\ d\end{matrix} \right)_{\! L} & 3&2&1&\frac 13\\
\left( \begin{matrix}u \\ d\end{matrix} \right)_{\! R} & 3&1&2&\frac 13\\
\left( \begin{matrix} \nu_e \\ e\end{matrix} \right)_{\! L} &1& 2&1&-1\\
\left( \begin{matrix} \nu_e \\ e\end{matrix} \right)_{\! R}&1 & 1&2&-1 
\end{matrix}\label{SMLRspectrum}
\ee
\vskip 1cm
The relation \eqref{Y} in this case is
\be
\widetilde Y=Q-T_{3L}-T_{3R},  \label{YLR}
\ee
and $\widetilde Y$ is interpreted as $\frac {B-L}2$ {($B$ is the baryon, $L$ thelepton number).}

Warning: in this model left and right fermions couple to different (left and right) gauge fields, say $W_{R\mu}$ and $W_{L\mu}$. To make this point more clear we notice that this situation is the same as considered in connection with Bardeen's method in section \ref{ss:Bardeen}. For the action term
\be
\overline{\psi_R} i\gamma^\mu \left(\partial_\mu + W_{R\mu}\right) \psi_R +\overline{\psi_L} i\gamma^\mu \left(\partial_\mu + W_{L\mu}\right) \psi_L\label{psiL+psiR}
\ee
can be easily rewritten in the form
\be
\overline\psi   i\gamma^\mu \left(\partial_\mu +V_\mu +\gamma_5 A_\mu\right)\psi \label{V+gamma5A}
\ee
where $ W_{R/L\mu}=V_\mu\pm A_\mu$.

The analysis of gauge and diffeomorphism anomalies in this model leads to:

\vskip 2cm
\centerline {Table LR gauge and mixed}
\vskip 0.2cm
\noindent\fbox{
    \parbox{\textwidth}{
    \begin{itemize}
\item $ T^{\mfs {\mathfrak u}(3)}\times  T^{\mfs {\mathfrak u}(3)}\times  T^{\mfs {\mathfrak u}(3)}$:  there are two left-handed and two right-handed triplet, whose anomalies cancel one another. 

\item $ T_L^{\mfs {\mathfrak u}(2)}\times  T_L^{\mfs {\mathfrak u}(2)}\times  T_L^{\mfs {\mathfrak u}(2)}$  and $ T_R^{\mfs {\mathfrak u}(2)}\times  T_R^{\mfs {\mathfrak u}(2)}\times  T_R^{\mfs {\mathfrak u}(2)}$ both vanish because the tensor $d^{abc}$ vanishes in general for the Lie algebra ${\mfs {\mathfrak u}(2)}$.
\item $ T^{\mfs {\mathfrak u}(3)}\times  T^{\mfs {\mathfrak u}(3)}\times  T^{  {\mathfrak u}(1)}$, in which case we have the trace of two ${\mfs {\mathfrak u}(3)}$ left triplet  generators and two right triplet generators. These traces are again non-vanishing, but   they are multiplied by the corresponding ${ {\mathfrak u}(1)}$ charge, whose total value is $3\left( 2\left(\frac 13\right) -2 \left(\frac 13\right)\right) =0$.
\item $ T_L^{\mfs {\mathfrak u}(2)}\times  T_L^{\mfs {\mathfrak u}(2)}\times  T^{  {\mathfrak u}(1)}$, in which case we have the trace of two ${\mfs {\mathfrak u}(2)}$ generators in two doublet representations. These traces are non-vanishing because $\tr (T^a T^b) \sim \delta^{ab}$, but they are multiplied by the corresponding ${ {\mathfrak u}(1)}$ charges, whose total value is $6\left(\frac 13\right) +2(-1) =0$.

 \item Analogously $ T_R^{\mfs {\mathfrak u}(2)}\times  T_R^{\mfs {\mathfrak u}(2)}\times  T^{  {\mathfrak u}(1)}$, leads to: $-6\left(\frac 13\right) -2 (-1)=0$

\item $ T^{ {\mathfrak u}(1)}\times  T^{  {\mathfrak u}(1)}\times  T^{ {\mathfrak u}(1)}$: in this case the tensor is proportional to the overall sum of the charge products:
$6\left(\frac 13\right)^3 +2 \left(-1\right)^3 -  6 \left(\frac 13\right)^3- 2  \left(-1\right)^3=0$.

\item $\Sigma\times\Sigma \times  T^{ {\mathfrak u}(1)}$: the trace  $\tr\left(\Sigma^{ab} \Sigma^{cd}\right)$ is non-vanishing, but it is multiplied by the total $U(1)$ charge:
$6\left(\frac 13\right) +2 \left(-1\right) -  6 \left(\frac 13\right)- 2  \left(-1\right)=0$.
\end{itemize}
}}
 \vskip 2cm
The analysis of gravity and gauge-induced trace anomalies yields
\vskip 0.5cm
  
  \centerline {Table LR trace-gravity and trace-gauge}
\vskip 0.2cm
 \noindent\fbox{
    \parbox{\textwidth}{\begin{itemize}
\item The multiplet (\ref{SMLRspectrum}), when weakly coupled to gravity, will produce (6+2) units of  trace anomaly with  Pontryagin density with + sign and (6+2) units with - sign, with vanishing total result.
\item   We have six units of the anomaly \eqref{AomehanonAbelianR} with curvature $F\equiv F^{\mathfrak su(3)}$ and six units with opposite sign. Therefore the multiplet (\ref{MSMspectrum}) is free of these anomalies.
    
    \item We have instead 4 units of the same anomaly with gauge field   $F\equiv F_L^{\mathfrak su(2)}$ and positive sign,  see \eqref{AomehanonAbelianL}, computed in the doublet representation of $\mathfrak su(2)$.
      \item We have also 4 units of the same anomaly with gauge field   $F\equiv F_R^{\mathfrak su(2)}$ and negative sign,  see \eqref{AomehanonAbelianL},   computed in the doublet representation of $\mathfrak su(2)$.
    \item Finally we have  a $U(1)$ gauge-induced trace anomaly with vanishing total coefficient: $6\left(\frac 13\right)^2 +2 \left(-1\right)^2 -  6 \left(\frac 13\right)^2- 2  \left(-1\right)^2=0$
\end{itemize}
}}
  \vskip 0,2cm
  \noindent Refs.: \cite{grimus93,mohapatra75,mohapatra80,pati74}.

 \subsection{Other hypothetical extensions of the MSM}
 
 All the extensions of the SM considered so far have a residual trace anomaly. Let us try to realize a combination of elementary fields which are completely anomaly-free. To this end we start from the generation \eqref{MSMspectrum} and add doublets or singlets of leptons.

 \be
\begin{matrix} \quad&{\sf G}/fields & \quad SU(3)\quad &\quad SU(2)\quad &\quad U(1)\quad\\
1 &\left( \begin{matrix} u \\ d\end{matrix} \right)_{\! L} & 3&2&\frac 16\\
1 &{u_R} &  3 &1&\frac 23\\
1 &{d_R} & 3 &1&-\frac 13\\
1 &\left( \begin{matrix} \nu_e \\ e\end{matrix} \right)_{\! L} & 1&2&-\frac 12\\
1& {e_R} &  1 &1&-1\\
n_L& \left( \begin{matrix} \nu \\ e\end{matrix} \right)_{\! L} & 1&2&y_L\\
n_R&\left( \begin{matrix} \nu \\ e\end{matrix} \right)_{\! R} & 1&2&y_R\\
m_L& {\nu_L}&1&1&0\\
m_R& {\nu_R}&1&1&0
\end{matrix}\label{SMhypospectrum}
\ee
 where $n_L,n_R,m_L,m_R$ are integers and $y_L,y_R$ are rationals to be determined.
 We notice that the modifications involves only the leptonic sector, therefore all the data involving $SU(3)$ 
 are unchanged. Below we report only the differences $\Delta_i$'s with respect to the entries in Tables MSM and the equations that have to be satisfied in order to cancel all anomalies  
 \vskip 0.3cm
\centerline {Table of differences with MSM}
\vskip 0.2cm
\noindent\fbox{
    \parbox{\textwidth}{
\begin{itemize}
\item $ T^{\mfs{\mathfrak u}(2)}\times  T^{\mfs{\mathfrak u}(2)}\times  T^{  {\mathfrak u}(1)}$: $\Delta_1=2\left(n_L\, y_L -n_R\, y_R\right)$, \quad\quad \quad\quad\quad\quad  \quad\quad $\Delta_1 =0$
\item $ T^{ {\mathfrak u}(1)}\times  T^{  {\mathfrak u}(1)}\times  T^{ {\mathfrak u}(1)}$: $\Delta_2= n_L \left(y_L\right)^3- n_R \left(y_R\right)^3$,\quad\quad \,\quad\quad\quad\quad \quad\quad  $ \Delta_2 =0$
\item $\Sigma\times\Sigma \times  T^{ {\mathfrak u}(1)}$: $\Delta_3=n_L\, y_L -n_R\, y_R$, \quad\quad\quad\quad\quad\quad\quad\quad\quad\quad \quad\quad\,\,  \quad $\Delta_3 =0$
\item $\Sigma\times\Sigma \times {\rm number\,\, of\,\, units}$: $\Delta_4 = n_L-n_R+m_L-m_R$, \quad\quad \quad\quad\quad\quad$\Delta_4 +1=0$
   \item $T^{\mfs {\mathfrak u}(2)}\times  T^{\mfs {\mathfrak u}(2)}\times {\rm number\,\, of\,\, units}$:
   $\Delta_5= n_L-n_R $,\quad\quad  \quad\quad \quad\quad\quad\quad\,\,$ \Delta_5+4 =0$
  \item $ T^{ {\mathfrak u}(1)}\times  T^{  {\mathfrak u}(1)}\times  {\rm number\,\, of\,\, units}$: $\Delta_6 =n_L \left(y_L\right)^2- n_R \left(y_R\right)^2$, \quad\quad \quad\quad $\Delta_6 =0$
\end{itemize}
}}
\vskip 0,3cm
The simplest solution is $n_L=0$, $n_R=4$, $y_R=0$ and $m_R=0, m_L=3$. I.e. we should add 4 right-handed $U(1)$-sterile doublets and 3 left-handed sterile neutrinos. $y_R=0$ means that the two components of the doublet have charge $1/2$ and $- 1/2$, very exotic animals.

\subsection{GUT models}

Among the extensions of the SM a particular role is reserved to the grand-unified theories. They have the major merit of unifying strong, weak and electromagnetic interactions. The most prominent among them is probably the $SO(10)$ GUT. In what concerns the problem of the anomaly analysis it turns out to be particularly simple because all fermions of the MSM of each family
plus a right-handed neutrino are collected in one single multiplet, the ${\bf 16}$ irreducible representation of $SO(10)$. Using the same notation as for the MSM the multiplet is:
\be
(u_L,d_L,(u_R)^c ,(d_R)^c, \nu_L,e_L, (\nu_R)^c,(e_R)^c)\label{so10}
\ee
where $X^c$ represents the conjugate of $X$ (and $u$ and $d$ carry three components each).
This left-handed multiplet is potentially anomalous for a consistent gauge anomaly, but such a gauge anomaly is proportional to $Str \left(T^{\mathfrak so(10)} T^{\mathfrak so(10)}T^{\mathfrak so(10)}\right)$, the symmetric trace of three generators $T^{\mathfrak so(10)}$ of the Lie algebra of $SO(10)$.
As is clear from the table \eqref{table} from the entry $D_5$, this trace vanishes identically. When we couple this model to gravity the consistent diffeomorphism anomalies are absent for an analogous and usual reason. It is free of any mixed gauge-gravity anomaly because the latter is proportional either to  
 $\tr \left(T^{\mathfrak so(10)} T^{\mathfrak so(10)}\right) \tr \Sigma$ or to $\tr \left(T^{\mathfrak so(10)}\right) \tr \left(\Sigma\Sigma\right)$, where $\Sigma$ denotes a generator of the Lorentz group.
 
The situation for the odd trace anomalies is different. They do not cancel completely. We may have two types of trace anomalies, one proportional to the Pontryagin density $ \epsilon^{\mu\nu\lambda\rho} R_{\mu\nu}{}^{\sigma\tau} R_{\lambda\rho}{}_{\sigma\tau}$ and the other to the Chern class density $\epsilon^{\mu\nu\lambda\rho}\,\tr_{\bf 16}\left(F_{\mu\nu}^{\mfs{\mathfrak u}(10)}F_{\lambda\rho}^{\mfs{\mathfrak u}(10)}\right)$.
There are 16 units of the former and one of the latter (corresponding to the $\bf 16$ representation of $SO(10)$). The former cancel out, let us recall why: the symbol such as $(\psi_R)^c$ (for instance $u_R$) in a generic metric can be rewritten as
\be
 (\psi_R)^c= \gamma^0 C \psi_R^*= \gamma^0 C P_R^* \psi^*=P_L \gamma^0 C\psi^*=P_L \psi^c = (\psi^c)_L.\label{psiLpsiR}
\ee
Inserted into the kinetic term, it gives
\be
\sqrt{g}\, \overline {(\psi^c)_L} \,\gamma^\mu(\partial_\mu +\frac 12 \omega_\mu )(\psi^c)_L=
\sqrt{g} \, \overline{(\psi_R)^c}\, \gamma^\mu(\partial_\mu +\frac 12 \omega_\mu )(\psi_R)^c=\sqrt{g}\, \overline {\psi_R}\, \gamma^\mu(\partial_\mu +\frac 12 \omega_\mu )\psi_R\label{LcR}
\ee
where $\omega_\mu= \omega_\mu^{ab}\Sigma_{ab}$ and $\Sigma_{ab}$ are anti-hermitean. The last passage requires an overall transposition and a partial integration. Therefore the kinetic term of the multiplet \eqref{so10}, coupled only to the metric, splits into 16 independent Weyl fermion kinetic terms, 8 left-handed and 8 right-handed, with opposite contribution to the trace anomaly. 

As for the other anomaly, the Chern class one, this `sleight of hand' is not possible, because the multiplet \eqref{so10} couples to a unique potential valued in the Lie algebra of $SO(10)$ and it is not possible to split the kinetic term in individual components like in the metric case.

\vskip 2cm
\noindent{\LARGE \bf Part II}

\section{What should we do then?}

The conclusion from the previous analysis is that it is possible to cancel all the anomalies in the extended SMS only at the price of introducing new exotic particles, which make up a rather unlikely spectrum. Barring these possibilities, which are all comprised in the schemes (A) and (B) of section \ref{ss:anomalycanc}, there remains the scheme (C), i.e. introducing in the theory fields with suitable transformation properties. The troubling anomalies are those related to the densities of the Chern and Pontryagin classes:
\be
{\cal A}^{\mfs{\mathfrak u}(2)}_\omega =  i\int d^4x \sqrt{g} \, \omega\, \tr\left(F^{\mfs{\mathfrak u}(2)}*F^{\mfs{\mathfrak u}(2)}\right)\equiv  \int d^4x \sqrt{g} \,\omega\, \epsilon^{\mu\nu\lambda\rho}\,\tr\left(F_{\mu\nu}^{\mfs{\mathfrak u}(2)}F_{\lambda\rho}^{\mfs{\mathfrak u}(2)}\right)\label{su2traceanom}
\ee
and
\be
{\cal A}^{grav}_\omega= i\int d^4x \sqrt{g}\, \omega\, \tr\left(\ER *\ER\right)\equiv  \int d^4x \sqrt{g} \,\omega\, \epsilon^{\mu\nu\lambda\rho} R_{\mu\nu}{}^{\sigma\tau} R_{\lambda\rho}{}_{\sigma\tau}\label{gravtraceano}
\ee
The latter can be canceled, for instance, by adding to the MSM a right handed sterile neutrino. The former anomalies are more `resistant'. In any case both can be canceled by a WZ term\footnote{For a discussion of WZ terms see section \ref{ss:WZ}.}, that is a term proportional to
\be
{\cal C}^{\mfs{\mathfrak u}(2)}_{WZ} =i \int d^4x \sqrt{g}\, \sigma\,\tr\left(F^{\mfs{\mathfrak u}(2)}*F^{\mfs{\mathfrak u}(2)}\right)\label{WZsu2}
\ee
and
\be
{\cal C}^{grav}_{WZ} =i \int d^4x \sqrt{g}\, \sigma\,\tr\left(\ER *\ER\right)\label{WZgrav}
\ee
respectively.  $\sigma$ is a scalar field which transforms as
\be
\delta_\omega \sigma = \omega\label{transfsigma}
\ee
under a Weyl transformation. This scheme looks very much like the Peccei-Quinn solution for the strong CP problem. But the two schemes refer to different energy scales. The Peccei-Quinn solution applies at the electroweak scale ($\sim 200 \,GeV$), corresponding to $10^{-11} sec.$ in the history of the universe, when gravity was already classical. The cancelation of trace anomalies by \eqref{WZsu2} and \eqref{WZgrav} should take place at much higher energies and earlier times. But the trouble here is with the $i$ which renders both counterterms imaginary. It is very likely that this term may backreact and endanger unitarity. 

This imaginary unit is important and it is worth spending a comment on it. It comes out explicitly from the calculation. But there is also an indirect argument that supports its presence. As was said above the $i$ guarantees the invariance under time reversal. If the $i$ were not there time reversal symmetry would be violated, together with parity. But while parity violation is expected since the classical theory is not parity-invariant, $T$ symmetry violation is totally unexpected because the classical theory is $T$ symmetric and there is no time reversal violation in the process of deriving the anomaly, for instance by dimensional regularization. 

This imaginary unit marks also the difference with the even-parity trace anomalies, where it is absent. As we have explained above, even trace anomalies do not affect unitarity, and a WZ term, like above, can be easily implemented in order to restore conformal invariance {(but see below for a more accurate discussion).}

It is perhaps not useless to recall once again the difference between type O and type NO anomalies. The just mentioned odd-parity anomalies are type O, they are a spy of the lack of a Weyl fermion propagator, which hinders the idea itself of quantization. They are unveiled by obstructions that manifest themselves in the family's index theorem. However - this is the important  point - the corresponding anomalies are the symptoms of a disease, not the disease itself. The topological obstructions are.  We can suppress the symptoms, i.e. these anomalies, with clever WZ terms, but they do not heal the disease.

Resuming the previous discussion, one possible attitude is to assume that gravity couples to matter in a non-minimal way, which does not give rise to anomalies. This implies a reformulation of the SM and its coupling to gravity, which is not covered by this article.

Finally, a more radical possibility is that gravity is entirely classical. As explained above the coupling of the anomalous e.m. trace in the effective action is via the fluctuating field $h=h_\mu^\mu$. This field is ineffective if gravity is classical. As a consequence the anomalies in question can be disregarded.

Summarizing we have the following alternatives for odd trace anomaly cancelation:

\begin{itemize}
\item minimal coupling matter-gravity: cancelation by exotic SM spectrum;
\item minimal coupling matter-gravity: cancelation by WZ terms, unitarity in danger;

\item non-minimal coupling gravity-matter: to be investigated;

\item classical gravity: trace anomalies irrelevant;
\end{itemize}

Altogether none of these possibilities is appealing. The present paper is clearly inspired by a bottom-up attitude. It is an analysis of the consistency conditions that arise as soon as we try to immerse the SM, in one of its versions, into a background of dynamical gravity. In such context we meet new anomalies which are not present when gravity is irrelevant. These new anomalies  are both of type O and of type NO. As we have explained the former may endanger unitarity. Therefore one has to make sure that they are absent. The possibilities we have presented above are unconvincing. Even freely playing with the still unknown SM sector of neutrinos, it does not seem possible to meet the requirement of full freedom from type O anomalies by simply enlarging the spectrum of the MSM. If our anomaly analysis is correct it is clear that we have to change something in the previous pictures of the SM. In the rest of this paper we would like to present an (unconventional) proposal in this direction. It certainly does not solve all the problems, but it is a hint of where the anomaly analysis, if taken seriously, may lead to.

\section{A chirally symmetric model}

As pointed out above the L-R model is the one that comes closest to the cancelation of all the anomalies. That would be the case if instead of two $SU(2)$ gauge fields, one left and one right, we had a unique $SU(2)$ gauge symmetry and a unique gauge field $V_\mu =V_\mu^a T^a$. However this would lead to  a theory of Dirac fermions, and at that point we would not have a mechanism to break parity and recover the MSM in the broken phase.

We would like to construct a  L-R symmetric (or chirally symmetric) model which is certainly anomaly-free. It is based on the same multiplet as the MSM with the addition of a right-handed sterile neutrino. In the usual SM notation it is 
\be
\begin{matrix} {\sf G}/fields & \quad SU(3)\quad &\quad SU(2)\quad &\quad U(1)\quad\\
\left( \begin{matrix} u \\ d\end{matrix} \right)_{\! L} & 3&2&\frac 16\\
{(u_R)^c} &  \bar 3 &1&-\frac 23\\
{(d_R)^c} & \bar 3 &1&\frac 13\\
\left( \begin{matrix} \nu_e \\ e\end{matrix} \right)_{\! L} & 1&2&-\frac 12\\
{(e_R)^c} &  1 &1&1\\
{(\nu_R)^c} &  1 &1&0
\end{matrix}\label{Lspectrum}
\ee
where $ X^c$ represents the Lorentz conjugate spinor of $X$, i.e. $X^c=\gamma_0 C X^\ast$. This multiplet couples to a left gravitational metric and connection, and to the $SU(3)_L\times SU(2)\times U(1)_L$ gauge fields. We have seen that all the anomalies cancel out except for  4 units of the trace anomaly due to the gauge field   $F\equiv F^{\mathfrak su(2)}$,  \eqref{AomehanonAbelianL}, computed in the doublet representation of $\mathfrak su(2)$. 

The multiplet \eqref{Lspectrum} describes left-handed particles and right-handed antiparticles.

There is also a right-handed multiplet
\be
\begin{matrix} {\sf G}/fields & \quad SU(3)\quad &\quad SU(2)\quad &\quad U(1)\quad\\
\left( \begin{matrix} u' \\ d'\end{matrix} \right)_{\! R} & 3&2&\frac 16\\
{(u'_L)^c} &  \bar 3 &1&-\frac 23\\
{(d'_L)^c} & \bar 3 &1&\frac 13\\
\left( \begin{matrix} \nu'_e \\ e'\end{matrix} \right)_{\! R} & 1&2&-\frac 12\\
{(e'_L)^c} &  1 &1&1\\
{(\nu'_L)^c} &  1 &1&0
\end{matrix}\label{Rspectrum}
\ee
coupled to a right gravitational metric and connection. This multiplet couples to the $SU(3)_R\times SU(2)\times U(1)_R$ gauge fields. The anomaly analysis of this right-handed multiplet is the same as for the left-handed one except for the sign of the trace anomaly due to the gauge field   $F\equiv F^{\mathfrak su(2)}$,  see \eqref{AomehanonAbelianL}, which is opposite. Therefore the overall sum of the anomalies of the system vanishes.

The multiplet \eqref{Rspectrum} describes right-handed particles and left-handed antiparticles.

We shall call these two intertwined theories, with field content \eqref{Lspectrum} and \eqref{Rspectrum}, ${\cal T}_L$ and ${\cal T}_R$, respectively. The overall theory is free of type O anomalies.
We denote it simply by ${\cal T}={\cal T}_L \cup {\cal T}_R$.

{\bf Important.} Both multiplets couple to the same $SU(2)$  gauge fields. Only in this case do all anomalies cancel! This is the reason why we use the symbol $\cup$ intead of +.

{
Let us see explicitly in the sequel the various possible pieces of the relevant actions. The MAT (hypercomplex) formalism turns out to be a very effective bookkeeping device to write them down and we will use it as broadly as possible.}

\subsection{The quadratic fermion action}

The quadratic fermionic action is modeled on the one already introduced above, \eqref{axialaction},
\be
\widehat S_f&=&\int d^4\widehat x \, \left(i\overline {\psi}
\sqrt{\overline{\widehat
g}}\gamma^a\widehat
e_a^\mu
\left(\ED_\mu+\frac 12 \widehat\Omega_\mu \right)\psi\right)(\widehat x)\label{axialactionSM}
\\
&{=}&  \int d^4\widehat x \,\left(  i\overline {\psi} \sqrt{\overline{\widehat
g}
}\gamma^a(\tilde e_a^\mu+\gamma_5
\tilde c_a^\mu)  \left(\ED_\mu +\frac 12 \left(\Omega^{(1)}_\mu+\gamma_5
\Omega^{(2)}_\mu\right) \right)\psi\right)(\widehat x)   \0
\ee
where $\psi$ is a 16-component spinor field, which encompasses the two left- and right-handed multiplets above, as will be explained shortly. In \eqref{axialactionSM} we have introduced the covariant derivative $\widehat{\ED}_\mu=\widehat{\partial}_\mu + \EV_\mu $, where $\widehat{\partial}_\mu$ here and below has to be understood as $\frac {\partial}{\partial{\widehat x^\mu}}$, and $\EV_\mu$ is valued in the Lie algebra $SU(3)_L\times SU(3)_R \times SU(2)\times U(1)_L \times U(1)_R$ in the relevant fundamental representations of $SU(3)$ and $SU(2)$ and in the representations of $U(1)$ specified by their hypercharges. 
{
Now let us recall that
\be
\widehat g = \{\widehat g_{\mu\nu}\}=g+\gamma_5 f, \quad \quad\widehat g_{\mu\nu} = g_{\mu\nu} +\gamma_5 f_{\mu\nu} = g_{+\mu\nu} P_+ + g_{-\mu\nu}P_-, \quad\quad g_{\pm\mu\nu} = g_{\mu\nu}\pm f_{\mu\nu}\label{g+-}
\ee
and
\be
\widehat g^{-1} =\{ \widehat g^{\mu\nu}\},\quad\quad       \widehat g^{-1} = {\rm g} +\gamma_5{\rm  f},\quad\quad \widehat g^{-1} \widehat
g=1,\quad\quad
\widehat g^{\mu\lambda}\widehat g_{\lambda\nu}= \delta^\mu_\nu\label{GG-1}
\ee
so that\footnote{Careful! ${\rm g}^{\mu\nu}, {\rm f}^{\mu\nu}$ are {\it not} the inverse of $g_{\mu\nu}, f_{\mu\nu}$, respectively, while $g_\pm^{\mu\nu}$ is the inverse of $g_{\pm\mu\nu}$.}
\be
\widehat g^{\mu\nu}= {\rm g}^{\mu\nu}+\gamma_5{\rm f}^{\mu\nu} =g_+^{\mu\nu} P_+ + g_-^{\mu\nu} P_-, \quad\quad  g_\pm^{\mu\nu}={\rm g}^{\mu\nu}\pm {\rm f}^{\mu\nu}, \quad\quad  g_\pm^{\mu\lambda} g_{\pm \lambda \nu}= \delta^\mu_\nu\label{g-1+-}
\ee
Moreover
\be
\sqrt{\widehat g}
= \sqrt{\det(g_+)}P_+ +\sqrt{\det(g_-)} P_-, 
\label{volume+-}
\ee
\be
&&\widehat e^a_\mu = e_\mu^{(+)a}P_+ + e_\mu^{(-)a}P_-  ,\quad\quad e_\mu^{(\pm) a} = e_\mu ^a \pm c_\mu ^a,\quad\quad  e_\mu^{(\pm) a} e_a^{(\pm) \nu}=\delta_\mu^\nu \0\\
&&\widehat e_a^\mu=\tilde e_a^{(+)\mu}P_++
{ e}_a^{(-)\mu}P_-, \quad\quad e_a^{(\pm) \mu} = {\rm e}_a ^\mu \pm {\rm c}_a ^\mu,\quad\quad e_a^{(\pm) \mu}  e_\mu^{(\pm) b}=\delta_a^b\label{vier+-}
\ee
\be
\widehat \Omega_\mu^{ab}  = \Omega_\mu^{(+)ab}P_+ +\Omega_\mu^{(-)ab}P_-, \quad\quad \Omega_\mu^{(\pm)ab}=  \Omega_\mu^{(1)ab} \pm  \Omega_\mu^{(2)ab}\label{spinconn+-}
\ee
Using these we can split $\widehat S_f$ into $\widehat S_f = S_f^{(+)} + S_f^{(-)}$ where
\be
S_f^{(\pm)}&=&\int d^4\widehat x \, \left(\sqrt{g_\pm}\, i\overline {\psi}P_\mp 
\gamma^a
 e_a^{(\pm)\mu}P_\pm
\left(\ED_\mu^{(\pm)}+\frac 12 \Omega_\mu ^{(\pm)}P_\pm\right)\psi\right)(\widehat x)\label{axialaction+-}
\ee
More in detail,
\be
S_f^{(+)}\equiv S_{fR}
&=&\int d^4\widehat x \, \left(\sqrt{g_+}\, i\overline {{\psi'}_{R}}  
\gamma^a
 e_a^{(+)\mu}
\left(\ED_\mu^{(+)}+\frac 12 \Omega_\mu ^{(+)}\right)\psi'_{R}\right)(\widehat x)\label{fermionaction+}
\ee}
where $\psi'_R$ represents the right-handed multiplet \eqref{Rspectrum}, and 
\be
\ED^{(+)}_\mu=\partial_\mu +{\sfg}_X^+ X^{(+)}_\mu +{\sfg}_WW_\mu +{\sfg}_B^+B_\mu^{(+)} \label{XWQ+}
\ee
while
\be
S_f^{(-)}\equiv S_{fL}
&=&\int d^4\widehat x \, \left(\sqrt{g_-}\, i\overline {{\psi}_L}  
\gamma^a
 e_a^{(-)\mu}
\left(\ED^{(-)}_\mu+\frac 12 \Omega_\mu ^{(-)}\right)\psi_{L}\right)(\widehat x)\label{fermionaction-}
\ee
where $\psi_L$ represents the left-handed multiplet \eqref{Lspectrum}, and
\be
\ED^{(-)}_\mu=\partial_\mu + X^{(-)}_\mu +W_\mu +B_\mu^{(-)}
\ee
The symbols $X^{(\pm)}_\mu,W_\mu,B^{(\pm)}_\mu$ refer to the $SU(3)_{R/L}, SU(2)$ and $U(1)_{R/L}$ potentials, respectively. { Of course each potential has its own distinct coupling to the fermions, which can be made explicit through a rededinition.}

\vskip 0.2cm

{\bf Remark.} Here and below, once the hypercomplex variable $\widehat x$ has done its job as a bookkeeping device, we can safely replace it everywhere with $x$.

\vskip 0.2cm

 $S_f^{(-)}$ and $S_f^{(+)}$ are separately $CP$-invariant.

\subsection{The gauge field action}

The action for the gauge fields is
\be
\widehat S_g =- \frac 1{8} \int d^4\widehat x \,\Tr \left(\frac 1{\widehat {\sf g}^2}\sqrt{\widehat g}\, \widehat g^{\mu\mu'} \widehat g^{\nu\nu'} F_{\mu\nu } F_{\mu'\nu'} \right),\quad\quad F_{\mu\nu} =F_{\mu\nu}^a T^a\label{actiongf}
\ee
where $\Tr$ denotes the trace over all matrices, including $\gamma_5$, and $T^a$ are the generators of the overall gauge Lie algebra.{ The coupling constant can be split into left and right parts. This can be made explicit in the action  as follows: define
\be
\widehat \sfg=\sfg+\gamma_5 {\sf  h}= \sfg_+P_++\sf g_-P_-, \quad\quad \sfg_\pm=\sfg\pm {\sf h}\label{gaugecoupling}
\ee
then the factor ${\widehat {\sf g}}^{-2}$ inside the round brackets of \eqref{gaugecoupling} becomes
\be
\frac 1{{\widehat \sfg}^{2}}= \frac 1{\sfg_+^2}P_+ + \frac 1{\sfg_-^2}P_-\label{gaugecoupling1}
\ee
Therefore \eqref{actiongf} splits as follows
\be
S_g^{(\pm)} =- \frac 1{4{\sf g}^2_\pm}  \int d^4\widehat x \,\sqrt{g_{\pm}} \, \tr \left( g_\pm^{\mu\mu'} g_\pm ^{\nu\nu'}  F_{\mu\nu } F_{\mu'\nu'} \right)\label{actiongf+-}
\ee

The action $\widehat S_g$ is appropriate for the $SU(2)$ gauge fields, which are the same both for ${\cal T}_L$ and ${\cal T}_R$. $S_g^{(+)}$ and $S_g^{(-)} $ are separately $CP$-invariant.

\subsection{AE gauge field action}

We can generalize the previous action by extending it to a field $\widehat V_\mu=V_\mu+ \gamma_5 A_\mu$
\be
\widehat S_{aeg} =- \frac 1{8} \int d^4\widehat x \,\Tr \left(\frac 1{\widehat {\sf g}^2}\sqrt{\widehat g}\, \widehat g^{\mu\mu'} \widehat g^{\nu\nu'} \widehat F_{\mu\nu } \widehat F_{\mu'\nu'} \right),\quad\quad \widehat F_{\mu\nu} =\widehat F_{\mu\nu}^a T^a\label{actiongfae}
\ee
where $\widehat F= \widehat d \widehat V +\frac 12[\widehat V,\widehat V]$, and $\Tr$  and $T^a$ are as above. The action \eqref{actiongfae} splits as
\be
S_{aeg}^{(\pm)} =- \frac 1{4{\sf g}^2_\pm}  \int d^4\widehat x \,\sqrt{g_{\pm}} \, \tr \left( g_\pm^{\mu\mu'} g_\pm ^{\nu\nu'}  F^{(\pm)}_{\mu\nu } F^{(\pm)}_{\mu'\nu'} \right)\label{actiongfae+-}
\ee
where $ F^{(\pm)}_{\mu\nu }= \hat d V^{(\pm)} +\frac 12[ V^{(\pm)}, V^{(\pm)}]$ and $ V_\mu^{(\pm)}= V_\mu \pm A_\mu$, and $\widehat{\sf g}= \sfg+ \gamma_5 {\sf h}= \sfg_+P_++ \sfg_-P_-$. As an example, in the case of $SU(2)_R\times SU(2)_L$ considered in section \ref{ss:LR} above, $ V_\mu^{(\pm)}$ has been denoted $W_{R/L\mu}$.

In the case of ${\cal T}={\cal T}_L +{\cal T}_R$, the action $S_{aeg}$ is appropriate for the groups { $SU(3)_L\times SU(3)_R$ and $ U(1)_L\times U(1)_R$, each with its own coupling $\widehat \sfg$,} since we have distinct left and right potentials.
As before, the action splits in the sum $S_{aeg}^{(+)}+ S_{aeg}^{(-)}$ with
\be
S_{aeg}^{(\pm)} =- \frac 1{4 \sfg_\pm^2} \int d^4\widehat x \,\sqrt{g_{\pm}} \, \tr \left( g_\pm^{\mu\mu'} g_\pm ^{\nu\nu'}  F^{(\pm)}_{\mu\nu } F^{(\pm)}_{\mu'\nu'} \right)\label{actiongfae+-2}
\ee
where $F^{\pm}$ denotes the curvatures of the $SU(3)_R$ and $ U(1)_R$,  and $SU(3)_L$ and $U(1)_L$ potentials, respectively.
Of course the factors $\sfg_\pm^{-2}$ can be absorbed, as usual, in a redefinition of the gauge potentials.}

{
 $S_{aeg}^{(+)}$ and $S_{aeg}^{(-)}$ are separately $CP$-invariant. If $\sfg_+=\sfg_-$,  $\widehat S_g$ and $\widehat S_{aeg}$ are parity-invariant.
}

{

\subsection{AE real scalar field}

We will need also the action for an axially-extended (AE) real scalar field $\widehat\Phi= \phi +\gamma_5 \pi$, where $\phi$ is an ordinary real and $\pi$ a pseudoreal scalar field. We write the action as
\be
\widehat S_{aes}=\frac 12\int d^4\widehat x \,\tr \left[\sqrt{\widehat g}\,\left( \, \widehat g^{\mu\nu} \partial_\mu\widehat \Phi \partial_\nu \widehat\Phi - \widehat  m^2 \widehat\Phi^2-\frac {\widehat\lambda}4 \widehat\Phi^4\right) \right]\label{hataescalar}
\ee
We can decompose $\widehat\Phi$ as $\widehat\Phi= \Phi_+ P_++ \Phi_- P_-$ where $\Phi_\pm = \phi \pm \pi$, as well as $\widehat m^2= m^2_+P_+ + m^2_- P_-$ and $\widehat \lambda = \lambda _+ P_+ +\lambda_- P_-$. Then we have the splitting $\widehat S_{aes}=S_{aes}^{(+)}+ S_{aes}^{(-)}$, with
\be
S_{aes}^{(\pm)}=\frac 12 \int d^4 \widehat x \,\sqrt{ g_\pm}\, \left(  g_\pm^{\mu\nu} \partial_\mu \Phi_\pm \partial_\nu \Phi_\pm - m_\pm^2 \Phi^2_\pm -\frac {\lambda_\pm}4 \Phi_\pm^4\right)\label{hataescalar+-}
\ee

\subsection{AE scalar doublet action}

We will employ later also an AE complex scalar field $\widehat H= {\sf h}\pm \gamma_5 {\sf k}$ that couples to a metric $\widehat g_{\mu\nu}$ and is a doublet under $SU(2)$. The action is 
\be
\widehat S_{aed}=\int d^4\widehat x \,\tr \left[\sqrt{\widehat g}\, \left( \widehat g^{\mu\nu} \ED_\mu \widehat H^\dagger \ED_\nu\widehat H -  \widehat {M}^2 \widehat H^\dagger \widehat H -\frac {\widehat\lambda}4 \left(\widehat H^\dagger \widehat H\right)^2\right)\right]\label{Hscalar}
\ee
where $\ED_\mu = \partial_\mu -i\sfg W_\mu$, and $W_\mu$ is an $SU(2)$ gauge field (notice that $W_\mu$ is not AE). We can decompose $\widehat H$ as $\widehat H= H_+ P_++ H_- P_-$ where $H_\pm ={\sf  h} \pm {\sf k}$, as well as $\widehat M^2= M^2_+P_+ + M^2_- P_-$ and $\widehat \lambda = \lambda _+ P_+ +\lambda_- P_-$. Then 
$\widehat S_{aed}$ splits as $\widehat S_{aed}=S^{(+)}_{aed}+S^{(-)}_{aed}$ with
\be
S_{aed}^{(\pm)}= \int d^4\widehat x \,\sqrt{g_\pm} \left[ g_\pm^{\mu\nu} \ED_\mu H_\pm^\dagger \ED_\nu H_\pm -  M_\pm^2 H_\pm^\dagger H_\pm -\frac {\lambda_\pm}4 \left(H_\pm^\dagger H_\pm\right)^2\right]\label{Hscalar+-}
\ee
}
{
$ S_{aes}^{(+)}$ and $S_{aes}^{(-)}$ as well as  $S_{aed}^{(+)}$ and $S_{aed}^{(-)}$ are separately $CP$-invariant.}

\subsection{The EH-like action}

The Einstein-Hilbert-like action in this context takes the form
\be
\widehat S_{EH} =\frac 14\int d^4\widehat x \,\Tr  \left( \frac 1 {\widehat\kappa} \sqrt{\widehat g}\, \widehat R\right),  \label{EHlike}
\ee
where $\widehat R$ is the Ricci scalar, obtained from \eqref{hatRiemann} by contracting $\nu$ with $\rho$ to obtain the Ricci tensor 
\be
\widehat R_{\mu\lambda}= R^{(+)}_{\mu\lambda} P_+ + R^{(-)}_{\mu\lambda} P_-, \quad\quad
R^{(\pm)}_{\mu\lambda}=R^{(1)}_{\mu\lambda}\pm R^{(2)}_{\mu\lambda}\label{Rmunu+-}
\ee
and then
\be
\widehat R = \widehat g^{\mu\lambda} \widehat R_{\mu\lambda}= \widehat g^{\mu\lambda}\left( \widehat R^{(1)}_{\mu\lambda}+ \gamma_5 R^{(2)}_{\mu\lambda}\right)= R^{(+)}P_+ + R^{(-)} P_-
\label{Ricci+-}
\ee
where $R^{(\pm)}= g_\pm^{\mu\nu}  R^{(\pm)}_{\mu\lambda} $.  Moreover $\frac 1{\widehat \kappa} = \frac 1{\kappa_+} P_+ + \frac 1 {\kappa_-}P_-$.

The action \eqref{EHlike} splits as $\widehat S_{EH}= S^{(+)}_{EH}+S^{(-)}_{EH}$ with
\be
S^{(\pm)}_{EH}= \frac 1{2\kappa_\pm}   \int d^4\widehat x \,\sqrt{g_{\pm}} R^{(\pm)}\label{EHlike+-}
\ee

\vskip 0.3cm

{
{\bf{Remark}}. The actions  $S_{EH}^{(+)}$ and $ S_{EH}^{(-)}$ are separately $CP$-invariant.  The previous formalism works even if there is only one metric, that is $g_+=g_-$. Here and in the sequel we wish to keep track of the more general possibility of a bimetric theory.
}

\subsubsection{AE diffeomorphisms}

Axially-extended diffeomorphisms, see section \ref{ss:axialcomplex}, are defined by
\be
\widehat x^\mu\rightarrow \widehat x^\mu+\widehat \xi^\mu(\widehat x^\mu),
\quad\quad\widehat
\xi^\mu=\xi^\mu+\gamma_5
\zeta^\mu= \xi^\mu_+ P_+ + \xi^\mu_- P_- , \quad\quad \xi^\mu_\pm = \xi^\mu \pm \zeta^\mu\label{axialdiff1}
\ee
 The way the various fields transform under these transformations is formally the same as for the corresponding ordinary fields, for instance, 
 \be
 \delta_{\widehat \xi} \psi = \widehat \xi^\mu \partial_{ \mu}\psi, \quad\quad
  \delta_{\widehat \xi}\widehat \Phi = \widehat \xi^\mu \partial_{\mu}\widehat\Phi,\quad\quad \delta_{\widehat \xi}\widehat g_{\mu\nu} =\widehat D_\mu
\widehat\xi_\nu+\widehat
D_\nu\widehat \xi_\mu,\quad\quad{\rm etc.}
 \ee
In particular $\sqrt{\widehat g}$ has the basic property that, under AE diffeomorphisms:
\be
\delta_{\hat\xi} \sqrt{\widehat g}= \widehat\xi^\lambda \widehat\partial_\lambda
\sqrt{\widehat g} +
\sqrt{\widehat g}\, \widehat\partial_\lambda \widehat\xi^\lambda\label{volume4}
\ee
Since the form of the actions (\ref{axialactionSM},\ref{actiongf},\ref{hataescalar},\ref{EHlike}) above are formally the same as that of the corresponding ordinary actions, it follows that  the former are invariant under the extended diffeomorphisms.

In turn we have 
\be
\delta_{\xi_\pm} \psi = \xi_\pm^\mu \partial_\mu \psi, \quad\quad \delta_{\xi_\pm} \Phi_\pm = \xi_\pm^\mu  \partial_\mu \Phi_\pm, \quad\quad {\rm etc.}\label{deltaxipm}
\ee
It follows that the actions $S_f^{(\pm)},S_{g}^{(\pm)}, S_{aeg}^{(\pm)},S_{aes}^{(\pm)}$ and $S_{EH}^{(\pm)}$ are invariant under the (ordinary) diffeomorphisms spanned by the parameters $\xi_\pm^\mu$ respectively.

A system invariant under AE diffeomorphisms is automatically {\it chirally symmetric}, but need not be parity-invariant because parity invariance requires also the equality of the left and right couplings. The systems defined by the actions (\ref{axialactionSM},\ref{actiongf},\ref{Hscalar},\ref{hataescalar},\ref{EHlike}) are chirally symmetric.

The actions \eqref{actiongf} and \eqref{Hscalar} deserve a special comment. Since they involve an $SU(2)$ gauge field, say $W_\mu$, valued in the Lie algebra $\mathfrak {su(2)}$ without any axial counterpart, it would seem inappropriate to consider the transformations
\be
\delta_{\widehat \xi} W_\mu = \widehat \xi^\lambda  \widehat\partial_\lambda  W_\mu+ \widehat\partial_\mu \widehat \xi^\lambda W_\lambda\label{deltaxiWmu}
\ee
because $\widehat\xi^\lambda$ splits according to eq.\eqref{axialdiff1}. But, first, one should recall that  $W_\mu$ are functions of $\widehat x$, and, second, the operation \eqref{deltaxiWmu} is akin to a similar very familiar operation in quantum field theory when, in an ordinary theory of Dirac fermions coupled to an Abelian vector potential $V_\mu$, we consider a chiral transformation $\delta V_\mu = i \gamma_5 \partial_\mu \lambda$. This transformation formally does not make any sense, because $V_\mu$ is not a $4\times 4$ matrix, but it is accepted with the understanding that we can slyly introduce a phantom axial companion of $V_\mu$ and eventually drop it, as we have explained several times.

\subsubsection{AE Weyl transformations}

The axially extended Weyl transformations are defined by
 \be
\widehat g_{\mu\nu} \longrightarrow e^{2\widehat \omega}  \widehat g_{\mu\nu},\quad\quad\widehat \omega=\omega + \gamma_5 \eta=\omega_+ P_++\omega_- P_-
\label{axialWeyl}
\ee
where  $\omega_\pm=\omega \pm\eta$. Under these transformations we have
\be
\delta_{\widehat\omega} \sqrt{\widehat g}&=& \sfd \,\widehat\omega\, \sqrt{\widehat
g}\label{deltaomega1}\\
\delta_{\widehat\omega} \widehat R&=& -2\widehat \omega \,\widehat R -2 (\sfd-1)
\widehat \square \widehat \omega\label{deltaomega2}\\
\delta_{\widehat\omega} \widehat R_{\mu\nu\lambda}{}^\rho &=&-\delta_\nu^\rho
\widehat D_\mu \widehat D_\lambda \widehat \omega +  \delta_\mu^\rho \widehat
D_\nu \widehat D_\lambda \widehat \omega+ \widehat D_\mu \widehat D_\sigma
\widehat \omega \,\widehat g^{\rho\sigma} \widehat g_{\nu\lambda}-\widehat D_\nu
\widehat D_\sigma \widehat \omega\, \widehat g^{\rho\sigma} \widehat
g_{\mu\lambda}\label{deltaomega3}
\ee
The fermions transform as
\be
\psi \rightarrow e^{-\frac 32 (\omega+\gamma_5 \eta)} \psi,\label{Weylpsi}
\ee
the scalars as
\be
\widehat \Phi \rightarrow e^{-\widehat \omega} \widehat \Phi, \label{WeylPhi}
\ee
while the gauge fields are invariant
\be
\widehat V_\mu \rightarrow \widehat V_\mu.\label{WeylVmu}
\ee
Under AE Weyl transformations the actions (\ref{axialactionSM},\ref{actiongf},\ref{actiongfae}), are invariant, while ({\ref{Hscalar}, \ref{hataescalar}, \ref{EHlike}) are not. 

 $S_f^{(\pm)},S_{g}^{(\pm)}, S_{aeg}^{(\pm)}$ are invariant under the Weyl transformations with parameter $\omega_\pm$,respectively.

\subsection{Yukawa couplings}
\label{s:yukawacoupling}

The Yukawa coupling between a fermion $\psi$, an AE complex scalar $\widehat\Phi$ and another fermion $\chi$ is defined as follows. Set
\be
\widehat S_Y= S_Y + \overline S_Y\label{yukawa}
\ee
Then
\be
S_{Y} = \int d^4\widehat x \,\left( \overline\psi{\widehat y}\,\sqrt{\overline{\widehat g}} \,\overline{\widehat \Phi}\, \chi^c \right)\label{yukawa0}
\ee
where $\widehat y$ is the {Yukawa} coupling constant. 
Recall that $\overline{\widehat \Phi} =\phi -\gamma_5 \pi$ denotes the axial conjugate , therefore
\be
\delta_{\widehat \xi}\overline{\widehat \Phi} = \overline{\widehat\xi}^\mu \widehat\partial_\mu \overline{\widehat \Phi} \label{deltaoverlinePhi}
\ee
Moreover
\be
\delta_{\widehat \xi}\chi^c =(\xi^\mu -\gamma_5 \zeta^\mu) \gamma_0 C  \widehat\partial_\mu \chi^\ast=
\overline{\widehat\xi}^\mu  \widehat\partial_\mu \chi^c \label{deltachic}
\ee
Therefore 
\be
\delta_{\widehat \xi}\left( \overline\psi\sqrt{\overline{\widehat g}} \,\overline{\widehat \Phi} \chi^c\,\right)=   \widehat\partial_\mu \left( \overline\psi\sqrt{\overline{\widehat g}} \, \overline{\widehat\xi}^\mu\,\overline{\widehat \Phi} \chi^c\right)\label{integrand1}
\ee
Thus \eqref{yukawa} is invariant under AE diffeomorphisms. Its Hermitean conjugate is
\be
\overline S'_{Y} = \frac 12 \int d^4\widehat x \,\left( \overline{\chi^c}\sqrt{\widehat g} \,{\widehat \Phi}^\dagger \psi\right)\label{yukawahc}
\ee
Proceeding as above one can prove that
\be
\delta_{\widehat \xi}\left( \overline{\chi^c}\sqrt{\widehat g} \,{\widehat \Phi}^\dagger \psi\right)=  \widehat\partial_\mu
 \left( \overline{\chi^c}\widehat\xi^\mu\sqrt{\widehat g} \,{\widehat \Phi}^\dagger \psi\right)\label{delatintegrand2}
 \ee
which proves that \eqref{yukawahc} is invariant under AE diffeomorphisms.

Similarly, applying the transformation rules \eqref{Weylpsi} and  \eqref{WeylPhi}, one can prove that both  \eqref{yukawa} and  \eqref{yukawahc} are invariant under AE Weyl transformations.

\vskip 0.3cm
{\bf Remark.} Notice that such terms as $\int d^4\widehat x \,\left( \overline\psi\sqrt{\overline{\widehat g}} \,\overline{\widehat \Phi} \chi\right)$ and the conjugate, are not AE diffeomorphism-invariant.

\vskip 0.3cm

Now we specify the previous results for our case. If $\psi_L$ be the (reducible) left multiplet of ${\cal T}_L$. Then we define
\be
S_{YL}= \frac 12 \int d^4\widehat x \,\left( \overline{\psi_L} {\widehat y}\sqrt{\overline{\widehat g}} \,\overline{\widehat \Phi} (\psi_L)^c\right)= \frac {y_-}2 \int d^4\widehat x \sqrt{g_-}  \overline{\psi_L}\Phi_-  (\psi_L)^c\label{yukawaL}
\ee
Its hermitean conjugate is
 \be
\overline S_{YL}=\frac {y_-^*}2 \int d^4\widehat x \sqrt{g_-}  \overline{(\psi_L)^c}\Phi^\dagger_- \psi_L\label{yukawaLhc}
\ee

Much in the same way we can introduce the Yukawa couplings for the model ${\cal T}_R$, i.e for the multiplet \eqref{Rspectrum}:
\be
S_{YR}=  \frac {y_+}2 \int d^4\widehat x \sqrt{g_+}  \overline{\psi'_R}\Phi_+ (\psi'_R)^c,\quad\quad \overline S_{YR}= \frac {y_+^*}2 \int d^4\widehat x \sqrt{g_+}  \overline{(\psi'_R)^c}\Phi^\dagger_+ (\psi'_R)\label{yukawaR}
\ee
{ In the sequel we shall need these formulas in particular in the case where $\Phi_\pm$ are real fields. We shall denote them with the symbol
\be
S_Y= S_{YL}+S_{YR}+ h.c.\label{yukawaconj}
\ee}

\subsubsection{Symmetry properties of Yukawa couplings}

\subsubsection*{Local Lorentz invariance of Yukawa terms}

An infinitesimal local Lorentz transformation of $\psi$ is defined by
\be
\delta_L \psi = \Lambda_{ab}\Sigma^{ab} \psi, \quad\quad \Sigma^{ab} =\frac 14 [\gamma^a, \gamma^b] \label{deltaLpsi}
\ee 
It follows that
\be
\delta_L \psi^c = \Lambda_{ab}\Sigma^{ab} \psi^c, \quad\quad \delta_L \overline\psi^c =- \overline\psi \Lambda_{ab}\Sigma^{ab}, \label{deltaLpsic}
\ee
therefore \eqref{yukawa} is unchanged.

\subsubsection*{Gauge invariance of Yukawa terms}

Suppose $\psi$ is a (reducible) multiplet that transforms under a local $U(1)$ group with Hermitean generator $Y$:
\be 
\psi \rightarrow e^{i\widehat \alpha Y } \psi,\quad\quad \widehat \alpha = \alpha +\gamma_5 \beta \label{U1transfpsi}
\ee 
Then 
\be
\psi^c \rightarrow  e^{i\overline{\widehat \alpha}Y} \psi^c, \quad\quad \overline\psi\rightarrow  \overline\psi\, e^{-i\overline{\widehat \alpha}Y}.\label{U1transfpsic}
\ee
In order for \eqref{yukawa} to be invariant under these transformation, $\widehat \Phi$ must be invariant too
\be
\widehat\Phi \rightarrow \widehat \Phi \label{U1transPhi}
\ee 
{ which holds in the case of the AE scalar field case we shall need below. The same holds in the case of $SU(2)$ or $SU(3)$ gauge transformations. For let the Hermitean generators be $T^a$. A local gauge transformation on $\psi$ is
\be
\psi \rightarrow e^{i\alpha_a T^a} \psi\label{SU2transfpsi}
\ee
which implies
\be
\psi^c \rightarrow e^{i \overline{\widehat\alpha}_a T^a} \psi^c ,\quad\quad \overline \psi \rightarrow
\overline \psi\,  e^{-i \overline{\widehat\alpha}_a T^a} \label{SU2transfpsic}
\ee
Thus, if $\widehat \Phi$ is invariant, \eqref{yukawa} does not change.

\subsubsection*{Discrete symmetry properties of Yukawa terms}
\label{ss:discretesymmetry}

The above Yukawa couplings  are symmetric under AE diffeomorphisms, local Lorentz transformations and also under AE Weyl transformations. We have not examined yet their properties under  $C, P$, and $CP$ transformations. Using the formulas of Appendix A, we see that they are not invariant under $C$ and $P$, but this is in line with all the terms quadratic in the spinors, which have the same lack of $C$ and $P$ invariance, see \eqref{axialactionSM}, but are $CP$ invariant. On the contrary the above Yukawa's may not be $CP$ invariant. If we disregard the phases of the $P$ and $C$ transformations  the sum $S'_{YL}+ \overline S'_{YL}$ is neither parity nor charge conjugation-invariant, and changes sign under $CP$. Analogously the sum $S'_{YR}+ \overline S'_{YR}$ is neither parity nor charge conjugation-invariant, and reverses its sign under  $CP$. {Of course we can arrange the transformation phases $\widehat \beta_p, \widehat \beta_c$ of $\psi$  in such a way as to absorb this minus sign and restore $CP$ invariance, for instance setting $\beta_p+\beta_c=\pi$. This choice does not affect the fermion kinetic terms or the Yukawa couplings of the next subsection. With this choice the total action \eqref{yukawaconj} is parity-invariant. However the above Yukawa couplings do not reproduce the SM Yukawa terms. For this reason we resort to other Yukawa couplings, which are $CP$ invariant but oblige us to sacrifice in part the overall symmetry of the theory, see next subsection.

\subsubsection*{Mass terms}

Let us finally add a comment that will be useful later on. Yukawa couplings of scalar and fermions are in particular designed to produce fermion masses via the Higgs mechanism. Here we consider the one illustrated in subsection \ref{ss:singlethiggsmechanism} below. Let us consider as an example the mass terms produced by the vev $v$ of the field $\Phi_+$ in $S'_{YR}+h.c.$, \eqref{yukawaR}. We take $y_+$ real and we limit ourselves to the doublet $\left(\begin{matrix} d_R\\u_R\end{matrix}\right)$ and the singlets $(d_L)^c, (u_L)^c$. Plugging them  in $S'_{YR}+h.c.$ the integrand turns out to be proportional to (disregarding $\sqrt{g_+}$)
\be
\frac {vy_+}2\left( \overline {d_L} (d_L)^c + \overline {(d_L)^c} d_L+  \overline {d_R} (d_R)^c + \overline {(d_R)^c} d_R\right)\label{majod}
\ee
and the same expression for $u$. It is well-known that the kinetic terms in $S_f^{(+)}$, eq. \eqref{fermionaction+}, constitute the kinetic terms of a massless Dirac spinor $\psi'_d=d_L+d_R$ and $\psi'_u =u_L+u_R$. The terms \eqref{majod} form the Majorana mass term of that spinor, so the complete eom is the Majorana equation, which in the free case takes the form
\be
\slashed {\partial} \psi'_{d/u} + m {\psi'}_{d/u}^c =0\label{freeMajoranaeq}
\ee}

\subsection{SM Yukawa couplings}

To reproduce the SM couplings (at least in the left part of our theory) we must use
\be
S_{YdL} = \frac {y^-_{H_d}}2 \int d^4\widehat x \,\sqrt{g_-}\left( \overline{\psi_{dL}}\,{H}_{d-} \chi_{sR}\right)+ h.c.\label{yukawadL+conj}
\ee
where $\psi_{dL}$ is a left-handed $SU(2)$ doublet, $\widehat  H_{d-}$ is also an $SU(2)$ doublet, conjugate to the $\psi_{dL}$ one in the inner product of the $SU(2)$ doublet representation space, while $\chi_{sR}$ is a right-handed singlet, all of them belonging to ${\cal T}_L$. Similarly, for ${\cal T}_R$,
\be
S_{YdR} = \frac {y^+_{H_d}}2 \int d^4\widehat x \,\sqrt{g}\left( \overline{\chi'_{dR}}\,{H}_{d+} \psi'_{sL}\right)+ h.c.\label{yukawadR+conj}
\ee
Eqs.\eqref{yukawadR+conj} and \eqref{yukawadL+conj} are (separately) invariant under diffeomorphisms parametrized by $\xi_\pm^\mu$ and conformal transformations parametrized by $\omega_\pm$, respectively. They are invariant under $SU(2)$ gauge transformations and the other SM gauge transformations, and are $CP$ invariant. Again it is not possible to write them in compact form in terms of $\widehat H$ and $\widehat g_{\mu\nu}$ as in the previous actions, that is by means of the full-fledged hypercomplex formalism. We stress that this does not spoil the L-R symmetry.

We shall denote the  set of Yukawa couplings introduced in this subsection by
\be
S_{Yd} = S_{YdL} +S_{YdR} \label{Yukawad}
\ee  
and its hermitean conjugate by $\overline S_{Yd}$.} An example of Yukawa coupling of $S_{YdL}$ is
\be
 \frac {y^-_{H_d}}2 \int d^4\widehat x \,\sqrt{g_-}\left( \overline{Q_{L}}\,{H}_{d-} \chi_{sR}\right)\label{yukawahiggsdoublet}
 \ee
 where $Q_L= \left( \begin{matrix} u_L\\
  d_L\end{matrix} \right)$, $H_d =  \left( \begin{matrix} \phi_1\\
  \phi_2\end{matrix} \right)$ and $\chi_{sR} = u_R$ or $d_R$. {Due to the breakdown of the electroweak symmetry the term \eqref{yukawahiggsdoublet} gives rise to a mass term proportional to the vev of $H_d$ and to $\frac {y^-_{H_d}}2$, see subsection \ref{ss:doublethiggsmechanism}. Together with the kinetic terms in $S_f^{(+)}$, eq. \eqref{fermionaction+}, it forms a Dirac fermion action, for instance for $\psi_d=d_L+d_R$, whose free eom is the ordinary massive Dirac equation:
\be
\slashed {\partial} \psi_{d} +m \psi_{d} =0\label{freeDiraceq}
\ee}

\subsection{The overall ${\cal T}$ model}

The action of the chirally symmetric model is {
\be
\widehat S_{ch-sym}=\widehat S_f  + \widehat S_g + \widehat S_{aeg} +  (S_Y+ S_{Yd}+ h.c.)+\widehat S_{aes}+\widehat S_{aed}+ \widehat S_{EH}\label{actionchir-symm}
\ee}
where $\widehat S_g$ has gauge group $SU(2)$, while $ \widehat S_{aeg}$ has gauge group $SU(3)_L\times SU(3)_R \times U(1)_L \times U(1)_R$. { Concerning the discrete symmetries, with the choice made in subsection \ref{ss:discretesymmetry}, ${\cal T}$ is $CP$ and $T$ invariant.} This theory is chirally symmetric and non-anomalous.  It is in particular a {\it bimetric theory}\footnote{ For a review see\cite{bimetric2016}. In general a bimetric theory is introduced in order to realize a ghost-free massive spin 2 particle, see \cite{hassan12}, a motivation different from the present one.}.  Since the fermion spectrum is L-R symmetric by construction, it is completely free of type O anomalies.  {Excluding the term $S_{Yd}+h.c.$, the action \eqref{actionchir-symm} is invariant under AE diffeomorphisms. $S_{YdL}+h.c.$ and $S_{YdR}+ h.c.$ are separately invariant under $\xi_-^\mu$ and  $\xi_+^\mu$.  Concerning the Weyl transfomations: $\widehat S_f  + \widehat S_g + \widehat S_{aeg}+(S_Y+h.c) $ are invariant under AE transformations.  $S_{YdL}+h.c.$ and $S_{YdR}+ h.c.$ are invariant under Weyl transformations with papameters $\omega_-$ and $\omega_+$, respectively.}
 
 But the scalar plus gravity action, $\widehat S_{aes}+\widehat S_{aed}+ \widehat S_{EH} $ is not invariant under AE Weyl transformations, only the remaining part is. We should add that, in addition, $\widehat S_{EH}$ is not renormalizable. We need an enhancement of symmetry in this part of the action, a formulation that preserves AE Weyl symmetry and, hopefully, achieves renormalizability. This is indeed possible by embedding the model in the so-called Weyl geometry, \cite{Weyl1}. All this constitutes a very well-known subject. We are moving in a considerably explored ground and we do not have anything new to say about it. We limit ourselves to a short summary of a rather vast literature, see \cite{ghilencea21,ghilencea24,mohammedi2024,rachwal2022,roumelioti24,scholz2018}, and to apply it to our context. The resulting model will be called ${\cal TW}$.
\vskip 0,2cm

\noindent Ref.: \cite{bimetric2016,rachwal2022,scholz2018,Weyl1,Weyl2}

\section{The Weyl geometry and gravity embedding}
\label{s:weylgeometry}

In an ordinary gravitational background geometry the Weyl transformation is given by
\be
g_{\mu\nu} \to e^{2\omega} g_{\mu\nu}\label{WT}
\ee
The Christoffel symbols transform as
\be
\Gamma_{\mu\nu}^\lambda \to \Gamma_{\mu\nu}^\lambda + \delta_\mu^\lambda\, \partial_\nu \omega +  \delta_\nu^\lambda\, \partial_\mu \omega- g_{\mu\nu} g^{\lambda\rho} \partial_\rho\omega\label{Christtransf}
\ee
We can construct Weyl-invariant Christoffel symbols as follows
\be
\widetilde \Gamma_{\mu\nu}^{(1)\lambda} = \Gamma_{\mu\nu}^\lambda - \left(\delta_\mu^\lambda\, \partial_\nu \varphi +  \delta_\nu^\lambda\, \partial_\mu \varphi - g_{\mu\nu} g^{\lambda\rho} \partial_\rho\varphi\right)\label{Christ1}
\ee
where the field $\varphi$ (a dilaton) under Weyl transforms as
\be
\varphi \to \varphi + \omega \label{deltavarphi}
\ee
The field $\varphi$ is dimensionless.

Alternatively, other Weyl-invariant Christoffel symbols are
\be
\widetilde \Gamma_{\mu\nu}^{(2)\lambda} = \Gamma_{\mu\nu}^\lambda - q\left( \delta_\mu^\lambda\, C_\nu +  \delta_\nu^\lambda\, C_\mu  - g_{\mu\nu} g^{\lambda\rho} C_\rho\right)\label{Christ2}
\ee
where the vector field $C_\mu$ transforms as
\be
C_\mu \to C_\mu +\frac 1{q} \partial_\mu \omega,\label{delatCmu}
\ee
and $q$ is a dimensionless parameter.
Still another set of equivalent Christoffel symbols is
\be
\widetilde \Gamma_{\mu\nu}^{\lambda}&=& \Gamma_{\mu\nu}^\lambda - \,\alpha \left(\delta_\mu^\lambda\, \partial_\nu \varphi +  \delta_\nu^\lambda\, \partial_\mu \varphi - g_{\mu\nu} g^{\lambda\rho} \partial_\rho\varphi\right)- q(1-\alpha)\left( \delta_\mu^\lambda\, C_\nu +  \delta_\nu^\lambda\, C_\mu  - g_{\mu\nu} g^{\lambda\rho} C_\rho\right)\0\\
&=& \Gamma_{\mu\nu}^\lambda - \,\alpha \left(\delta_\mu^\lambda\, \sfD_\nu \varphi +  \delta_\nu^\lambda\, \sfD_\mu \varphi - g_{\mu\nu} g^{\lambda\rho} \sfD_\rho\varphi\right)- q\left( \delta_\mu^\lambda\, C_\nu +  \delta_\nu^\lambda\, C_\mu  - g_{\mu\nu} g^{\lambda\rho} C_\rho\right)
\label{Christ3}
\ee
where $\alpha$ is a dimensionless parameter and  we have introduced the Weyl-invariant derivative
\be
{\sf D}_\mu \varphi= \partial_\mu \varphi - q C_\mu\label{EDmu}
\ee
 We can use these Christoffel symbols to build the Riemann and Ricci tensors. For instance,
\be
\widetilde R_{\mu\nu} = R_{\mu\nu} +3D_\nu S_\mu - D_\mu S_\nu +g_{\mu\nu} D\cdot S +2 S_\mu S_\nu -2 g_{\mu\nu} S \cdot S\label{tildeRiccimunu}
\ee
and
\be
\widetilde R = R + 6\left( D\cdot S - S\cdot S\right), \label{Rscalar}
\ee
where 
\be
 S_\mu =\alpha \, \partial_\mu \varphi+ (1-\alpha) \,q\, C_\mu \label{defSmu}
\ee
The conformal Weyl formalism allows also another decomposition of $\widetilde R_{\mu\nu}$. Summing $\widetilde R$ with $\alpha=1$ multiplied by $\epsilon$ to $\widetilde R$ with $\alpha=0$ multiplied by $1-\epsilon$ one can write
\be
\widetilde R = R + 6\epsilon\, \left( D\cdot \partial\varphi -  \partial \varphi\cdot \partial\varphi\right)+6 (1-\epsilon)\left(q D\cdot C -{q}^2C\cdot C\right), \label{Rscalarepsilon}
\ee
where $\epsilon $ is a new dimensionless parameter (not to be confused with $\alpha$).

Notice that, from \eqref{tildeRiccimunu},
\be
\widetilde R_{[\mu,\nu]}=\frac 12 \left(\widetilde R_{\mu\nu}- \widetilde R_{\nu\mu}\right)= 2( D_\nu S_\mu - D_\mu S_\nu) = 2(\partial_\nu S_\mu - \partial_\mu S_\nu)= 2q(\partial_\nu C_\mu - \partial_\mu C_\nu)\label{RmunuRnumu}
\ee

By construction, $\widetilde R_{\mu\nu}$ is conformal-invariant, but
\be
\widetilde R = g^{\mu\nu} \widetilde R_{\mu\nu}\rightarrow e^{-2 \omega} \widetilde R\label{widetildeRW}
\ee
Thus we can construct a Weyl-invariant gravitational action as follows. The first piece is the conformally enhanced EH action
\be
S^{(c)}_{EH} =\frac 1{2\kappa} \int d^4x \sqrt{g}\, e^{-2\varphi} \left(\widetilde R + {\mathfrak c}\, e^{-2\,\varphi}\right)\label{ScEH}
\ee
To it we can add
\be
S^{(c)}_C =  -\frac 14 \int d^4x \sqrt{g}\,C_{\mu\nu}C^{\mu\nu}, \quad\quad C_{\mu\nu}=\partial_\mu C_\nu - \partial_\nu C_\mu \label{ScS}
\ee
Finally there is also
\be
S^{(c)}_W =  -\frac 1\eta \int d^4x \sqrt{g}\, C_{\mu\nu\lambda\rho} C^{\mu\nu\lambda \rho} \label{ScW}
\ee 
where $ C_{\mu\nu\lambda\rho}$ is the Weyl tensor, which also is Weyl-invariant, and $\eta $ is a dimensionless constant. If we can disregard total derivatives in the action, \eqref{ScW} can be replaced by
\be
S^{(c')}_W =  -\frac 2\eta \int d^4x \sqrt{g} \left( - R_{\mu\nu} R^{\mu\nu}+ \frac 13 R^2\right)  \label{Sc'W}
\ee 
These have to be understood as interacting terms, while the kinetic terms are contained in \eqref{ScEH} and \eqref{ScS}.

We can easily embed also matter in Weyl geometry. For instance consider a real scalar field $\Phi$: the action
\be
 S^{(c)}_{s}= \frac 12 \int d^4x \,\sqrt{ g}\, \left[g^{\mu\nu} {\sf D}_\mu \Phi {\sf D}_\nu \Phi -m^2 e^{-2\varphi}\Phi^2-\frac {\lambda}4 \Phi^4 \right]\label{hataescalarc}
\ee
is Weyl-invariant  provided $\Phi$ transform as
\be
\Phi \rightarrow e^{-\omega}\Phi  \label{HWtr}
\ee

In general ${\sf D}_\mu$ is the Weyl-covariant derivative of the scalar $\Phi$
\be
{\sf D}_\mu \Phi= \left(\partial_\mu +\alpha\, \partial_\mu \varphi +{q}(1-\alpha)\, C_\mu\right)\Phi \label{EDmugen}
\ee

Putting together \eqref{ScEH} and \eqref{hataescalarc} we can write down the conformal-invariant scalar-gravity action
\be
S^{(c)}_{EH+s+C} &=&\frac 1{2\kappa} \int d^4x \sqrt{g}\, \left(e^{-2\varphi}+\zeta \Phi^2\right) \left(\widetilde R + {\mathfrak c}\, e^{-2\varphi}\right) \0\\
&& +\frac 12 \int d^4x \,\sqrt{ g}\, \left[ g^{\mu\nu} {\sf D}_\mu \Phi {\sf D}_\nu \Phi  -m^2 e^{-2\varphi}\Phi^2-\frac {\lambda}4 \Phi^4\right]\0\\
&& -\frac 14 \int d^4x \sqrt{g}\,C_{\mu\nu}C^{\mu\nu} -\frac 1\eta \int d^4x \sqrt{g}\, C_{\mu\nu\lambda\rho} C^{\mu\nu\lambda \rho},  \label{ScEHs}
\ee 
where $\zeta$ is a constant that measures the non-minimal coupling of the scalars to gravity, and $\widetilde R$ is given by \eqref{Rscalarepsilon}, that is
\be
\widetilde R = R + 6\epsilon\,  g^{\mu\nu}\left(D_\mu \partial_\nu\varphi -\partial_\mu \varphi \partial_\nu\varphi\right)+6(1\!-\!\epsilon)g^{\mu\nu}\left( q D_\mu C_\nu - {q}^{2} C_\mu C_\nu\right), \label{Rscalarepsilonext}
\ee
 with $\epsilon$ a constant to be determined.

It should be clear that $\zeta, \kappa^{-1}$ and $\mathfrak c$ all have the dimension of  a mass square, while $\lambda$ and $\eta$ are dimensionless. 

\subsection{Weyl geometry: other field theory formulations}
\label{ss:weylgeometry2}

Another formulation of Weyl geometry in field theory is possible in terms of a scalar field $\chi$ of physical dimension 1. Instead of \eqref{ScEH} the basic action is
\be
S'{}^{c}_{\!\!\!EH}=  \frac 1{2\xi} \int d^4x \sqrt{g}\,\chi^2 \left(\widetilde R + {\mathfrak c}\,\chi^2\right)\label{ScEH'}
\ee 
where $\xi$ is a dimensionless constant, while $\mathfrak c$ is as before\footnote{Of course $\chi$ cannot be identified with $e^{q\varphi}$ because they have different physical dimension, 1 and 0, respectively.}. $\chi$ transforms as
\be
\chi \to e^{\omega} \chi\label{chitransform}
\ee
The relation between $\chi$ and $\varphi$ can be written
\be
\varphi= \log (\ell \chi) \label{chivarphi}
\ee
where $\ell$ has the dimension of a length.
Using this one can translate all the above formulas in terms of $\chi$ ($C_\mu$ does not undergo any change). In the sequel we will use the formulation in terms of $\varphi$. It should be noticed that we need a dimensional scale to relate $\chi$ to the Weyl geometry. 
\vskip 1cm

Returning to the main subject of the embedding in a Weyl geometry, the embedding of fermions is automatic because the corresponding kinetic term is Weyl-invariant. The corresponding action term does not need the introduction of either $\varphi$ or $C_\mu$. As a consequence the anomalies remain the same as before the Weyl geometry embedding. No  new anomalies are generated due to fermions.

The embedding of the gauge kinetic terms and of the Yukawa couplings does not require any modifications, because they are Weyl-invariant.

\vskip 0.5cm
{\bf Remark} Concerning the (local) Weyl symmetry there are two possible attitudes. The first is to consider it as an ordinary gauge symmetry, characterized not only by a local parameter $\omega(x)$ but also by an associated (independent) gauge field $C_\mu$ with relevant curvature and covariant derivatives, as we have done above (in this case we understand $\alpha=\epsilon=0$ in all the previous formulas). Another attitude is to drop $C_\mu$ and consider the resulting theory as an unusual gauge theory symmetric under a local parameter but without corresponding gauge field. This means we set $C_\mu=0, \alpha=\epsilon=1$\footnote{It is actually not necessary to set $C_\mu=0$, it suffices to require $C_\mu\sim \partial_\mu \varphi$, which defines an integrable Weyl structure.}. Both attitudes are legitimate. But it is worth remarking that, contrary to what occurs in ordinary gauge theories, it is not necessary to introduce a vector gauge potential $C_\mu$ in order to guarantee the gauge invariance of the action, $\partial_\mu\varphi$ can do the same job. { But, in such a case, we do not introduce new degrees of fredom (say $C_\mu$), the field $\varphi$ has the same number of degrees of freedom as the gauge parameter.

The most frequent approach in the literature consists in introducing $C_\mu$ and treating it in the usual way as in gauge theories with a Higgs mechanism, which makes it a massive vector field by absorbing a scalar.  We deem worth recalling the latter mechanism, although it is extremely well-known, because it  plays a prominent role in the sequel. We consider the two cases that will appear below: a real  singlet and a doublet Higgs mechanism with complex scalars.
}
\vskip 0,5 cm
\noindent Refs. \cite{Brans,cecchini24,Dirac,Jordan,omote71,utiyama75}, for early applications of Weyl geometry to the SM, see \cite{bars,cheng88,drechsler,englert75,englert76,ohanian,nishino,nishino11,quiros,smolin}

{
\subsection{The real singlet Higgs mechanism}
\label{ss:singlethiggsmechanism}

In a theory like \eqref{ScEHs} the Higgs mechanism is realized as follows. We single out the action 
\be
S_{s}= \frac 12 \int d^4 \widehat x \,\sqrt{ g}\, \left( \, g^{\mu\nu} {\sf D}_\mu \Phi {\sf D}_\nu \Phi + m^2 \Phi^2-\frac {\lambda}4 \Phi^4\right)\label{scalar}
\ee
where
\be
{\sf D}_\mu \Phi = \left(\partial_\mu+q C_\mu\right)\Phi\label{covderH}
\ee
and  the term $m^2$ is produced, for instance, by the gauge choice $\varphi=0$, i.e. $m^2= \frac {\zeta\mathfrak c}\kappa$. 

Setting $\Phi={v+\rho}$, where $ \rho$ is a real field and $v= \sqrt{\frac {2m^2}\lambda}$, a real number, denotes the minimum of the potential, we have
\be
\sfD_\mu \Phi \sfD_\mu \Phi&=& \left(\partial_\mu   +q C_\mu \right)(v+\rho)\left(\partial^\mu +q C^\mu\right)(v+\rho)\label{DmuHDmuH}
\ee
The quadratic terms in $\frac 12\sfD_\mu \Phi \sfD_\mu \Phi- \sf V$ ($\sf V$ is the potential) are
\be 
\frac 12 \partial_\mu \rho \partial^\mu\rho+\frac 12 q^2 C_\mu C^\mu v^2-\frac 12 \lambda v^2 \rho^2\label{quadraticaction}
\ee
The result is a massive scalar $\rho$ as well as a massive vector field $C_\mu$.

Including this case among the Higgs mechanisms is somewhat of an abuse of language, because  there is no gauge field eating a scalar, but we can consider it as a sort of limiting case. A massive $C_\mu$ produced by a Higgs mechanism, is often considered in the literature, see \cite{ghilencea21,ghilencea24,mohammedi2024}.

Finally, the Yukawa coupling \eqref{yukawaR}, for $\Phi$ real,  when $\langle \Phi\rangle =v$, produce mass terms proportional to $v$ such as those in eq. \eqref{majod}.

 \vskip 0.3cm
{\bf Remark}. If $\psi$ denotes a singlet or a multiplet, that transforms under a gauge transformation
\be
 \psi \rightarrow e^{i \alpha_a T^a} \psi , \quad\quad  (T^a)^\dagger = T^a,\label{finitegaugetransf1}
 \ee
 then
 \be
 \psi^c \rightarrow  e^{i \alpha_a T^a} \psi^c, \quad\quad \overline \psi \rightarrow \overline \psi  e^{-i \alpha_a T^a}\label{finitegaugetransf2}
 \ee
Therefore a mass term like \eqref{majod} remains gauge invariant after the breaking.

\subsection{The doublet Higgs mechanism}
\label{ss:doublethiggsmechanism}

{
Suppose $\Phi$ is a complex doublet field, transforming as 
\be
\Phi \rightarrow e^{i\alpha_a T^a} \Phi,\quad\quad \Phi^\dagger \rightarrow \Phi^\dagger e^{-i\alpha_a T^a} \label{HSU2trans}
\ee
under a local $SU(2)$ transformation, $a=1,2,3$. The relevant terms of the action are 
\be
S^{(c)}_{d} &=&\int d^4x \,\sqrt{ g}\, \left[ g^{\mu\nu} {\sf D}_\mu \Phi^\dagger {\sf D}_\nu \Phi+ M^2 \Phi^\dagger \Phi-\frac {\lambda}4 \left(\Phi^\dagger \Phi\right)^2\right] \label{doubletaction}
\ee 
where
\be
\sfD_\mu \Phi = \left(\partial_\mu -i\sfg V_\mu +q C_\mu\right)\Phi\label{covderH}
\ee}
The symmetry breaking is determined by the decomposition
\be
\Phi=\left(\begin{matrix}\phi\\v+ \frac {h+i\chi}{\sqrt{2}}\end{matrix}\right) \quad\quad\langle\Phi\rangle=\left(\begin{matrix}0\\v\end{matrix}\right)\label{Phidoublet}
\ee
where $\phi$ is a complex scalar field, $h$ and $\chi$ are real scalar fields. $\phi$ and $\chi$ are Goldstone bosons destined to become components of the three $SU(2)$ gauge fields, while $h$ is bound to become the massive Higgs field.  The three Goldstone degrees of freedom represented by $\phi$ and $\chi$ can be absorbed by an $SU(2)$ gauge transformation, so that, with a suitable choice of the parameters $\alpha_a$, $\Phi$ can be represented as
\be
\Phi =  e^{i\alpha_a T^a} \Phi_h, \quad\quad \Phi_h=\left(\begin{matrix}0\\v+ \frac {h}{\sqrt{2}}\end{matrix}\right) \label{Phidoubletred}
\ee
We recall that if $M^2$ is the mass of $\Phi$ and $\lambda$ the strength of the quartic coupling in the action, $v$ is given by
\be
v^2= \frac {2M^2}{\lambda}\label{v2M2lambda}
\ee
Under this $SU(2)$ gauge transformation the covariant derivative of $\Phi$,
\be
\sfD_\mu \Phi = \left(\partial_\mu -i\sfg W_\mu +q C_\mu\right)\Phi, \quad\quad W_\mu= W_\mu^a T^a\label{covderH}
\ee
becomes
\be
{\sf D}_\mu \Phi \rightarrow  e^{i\alpha_a T^a}{\sf D}'_\mu \Phi_h\label{DmuPhitransf}
\ee
where, in the covariant derivative, $i\sfg W_\mu$ is replaced by
\be
i\sfg X_\mu = e^{-i\alpha_a T^a}\left(-\partial_\mu +i\sfg W_\mu\right) e^{i\alpha_a T^a}\label{W'mu}
\ee
In this way $\chi$ and the two components of $\phi$ are absorbed into  $X_\mu$.} Disregarding for simplicity $C_\mu$, we have
\be
\sfD_\mu \Phi^\dagger \sfD^\mu \Phi \rightarrow \frac 12 \partial_\mu h \partial^\mu h +\frac {{\sfg}^2}2 X_\mu^a X^{a\mu} \left(v +\frac h{\sqrt 2}\right)^2\label{DmuPhiD,muPhi}
\ee
summed over $a$. Since the kinetic term in the action is gauge invariant, this assigns a mass square $v^2 \sfg^2$ to the three gauge bosons $X^a_\mu$.

On the other hand
\be
\Phi^\dagger \Phi \rightarrow \Phi_h^\dagger \Phi_h = \left(v+\frac h{\sqrt 2}\right)^2\label{PhiPhi}
\ee
The scalar potential $\sf V$ becomes
\be
{\sf V} = -\frac 12 M^2 v^2 + M^2 h^2 + {\cal O}(h^3)\label{potentialh}
\ee

This gives the mass of the Higgs field $h$. Via the Yukawa term $S_{Yd}$ also the fermions coupled to $\Phi$ acquire a mass proportional to $v$.

\vskip 1cm

In the sequel we shall indeed consider two types of scalar field actions. In one, $ \Phi$ is a real scalar field in $\cal T$ denoted $\widehat \Phi$. In the second case,  $\Phi$ is a complex $SU(2)$ multiplet, which in the AE case will be denoted $\widehat H$. The first action will be denoted $\widehat S_{aes}$ and the second $\widehat S_{aed}$.

\vskip 1cm
 The next step consists in embedding $\cal T$ in the AE Weyl geometry, see \cite{ghilencea21,mohammedi2024,rachwal2022, scholz2018}.

\section{Embedding in AE Weyl geometry}
\label{embeddingweyl}

The embedding in axially extended  (AE) Weyl geometry is straightforward. The embedding of the fermion kinetic terms and of Yukawa couplings does not require modifications. We can write
\be
\widehat S^{(c)}_f=\widehat S_f, \quad\quad \widehat S^{(c)}_{g} = \widehat S_{g},\quad\quad \widehat S^{(c)}_{aeg} = \widehat S_{aeg},\quad\quad S_Y{}^{(c)}= S_Y,\quad\quad  S^{(c)}_{Yd}= S_{Yd}\label{Scf=Sf}
\ee
For the gravitational part we can write
\be
\widehat S^{(c)}_{EH} =\frac 1{4} \int d^4\widehat{x}\,\tr\left[\frac 1{\widehat\kappa} \sqrt{\widehat g}\, e^{-2\widehat\varphi} \left(\widetilde{\widehat R} + \widehat{\mathfrak c}\, e^{-2\widehat\varphi}\right)\right] \label{ScEHhat}
\ee 
where $\widehat \varphi= \varphi +\gamma_5 \varpi$, $ \widehat \kappa = \kappa_+ P_++\kappa_- P_-$, and
\be
\widetilde{\widehat R} =\widehat R + 6\left( \widehat D\cdot \widehat S -\widehat S\cdot \widehat S\right)\label{Rscalarhat}
\ee
where\footnote{Here and below we treat the two $\widehat S_\mu$ as separate cases. One could just as well consider a combination of the two with coefficients $\alpha$ and $1-\alpha$, respectively, as was done before.}
\be
\widehat S_\mu = \partial_\mu \widehat\varphi,\quad\quad {\rm or} \quad\quad  \widehat S_\mu = q\,\widehat C_\mu, \quad \quad \widehat C_\mu= C_\mu +\gamma_5 E_\mu\label{defSmuhat}
\ee

We can of course easily write down the splitting in $+$ and - part of the actions. To start with
\be
\widehat S_{EH}^{(c)} = S_{EH}^{(c+)}+  S_{EH}^{(c-)}\label{SEHcsplit}
\ee
where
\be
 S_{EH}^{(c\pm)}= \frac 1{2\kappa_\pm} \int d^4x \sqrt{g_\pm} e^{-2\varphi_\pm} \left( \widetilde R^{(\pm)}+\widehat{\mathfrak c} e^{-2\varphi_\pm}\right) \label{SEHc+-}
 \ee
which is invariant both under diffeomorphisms with parameters $\xi_\pm$ and under Weyl transformations with parameters $\omega_\pm$. Here
\be
 \widetilde R^{(\pm)}= R^{(\pm)}+ 6\left( D^{(\pm)}\cdot S^{(\pm)} -  S^{(\pm)}\cdot  S^{(\pm)}\right)
 \label{Rpmtilde}
 \ee
 where
 \be
 S_\mu^{(\pm)}=q C_\mu^{(\pm)} \quad\quad {\rm or} \quad\quad S_\mu^{(\pm)}=\partial_\mu\varphi^{(\pm)}\label{Smupm}
 \ee
and
\be
\varphi_\pm = \varphi \pm \varpi, \quad\quad C_\mu^{(\pm)}= C_\mu\pm E_\mu\label{varphiSmu}
\ee

Similarly
\be
\widehat S^{(c)}_C=  -\frac 18 \int d^4\widehat x \, \tr\left[\sqrt{\widehat g}\,\widehat g^{\mu\mu'} \widehat g^{\nu\nu'}\widehat  C_{\mu\nu}\widehat C_{\mu'\nu'}\right]= S^{(c+)}_C+ S^{(c-)}_C  \label{Scpmhat}
\ee
where $\widehat C_{\mu\nu}=\widehat \partial_\mu \widehat C_\nu -\widehat \partial_\nu\widehat C_\mu,$ and
\be
S^{(\pm)}_C=  -\frac 14 \int d^4x \sqrt{g_\pm}\,g^{\mu\mu'}_\pm g^{\nu\nu'}_\pm C^{(\pm)}_{\mu\nu}C^{(\pm)}_{\mu'\nu'},\label{SpmC}
\ee

{

For a real AE  scalar $\widehat \Phi$:
\be
 \widehat S^{(c)}_{aes}=\frac 12 \int d^4\widehat x \,\tr \left\{\sqrt{\widehat g}\, \left[ \widehat g^{\mu\nu} \widehat {\sf D}_\mu \widehat \Phi  \widehat{\sf D}_\nu\widehat \Phi +\widehat m^2 e^{-2\widehat \varphi}\widehat \Phi^2 -\frac {\widehat\lambda}4  \widehat \Phi^4\right]\right\}= \widehat S^{(c+)}_{aes}+ \widehat S^{(c-)}_{aes} \label{hataescalarcs}
\ee
where $ \widehat\sfD_\mu= \partial_\mu +q\widehat C_\mu$ or  $ \widehat\sfD_\mu= \partial_\mu +\partial_\mu\widehat \varphi$, $\widehat \Phi= \phi+\gamma_5 \pi$ and $\widehat m^2 =m^2_+ P_++ m^2_-P_-$ . Finally
\be
 S^{(c\pm)}_{aes}= \frac 12  \int d^4x\, \sqrt{g_\pm} \left[ g_\pm^{\mu\nu}\left({\sf D}^\pm_\mu \Phi_{\pm}\right)\left({\sf D}^\pm_\nu \Phi_{\pm}\right)+ m^2_\pm e^{2\varphi_\pm}  \Phi^2_\pm-\frac {\lambda_{\pm}} 4\Phi_{\pm} ^4 \right]\label{Scpmhat}
 \ee
 with $\sfD_\mu^\pm= \partial_\mu +qC_\mu^{(\pm)}$  or  $\sfD_\mu^\pm= \partial_\mu +\partial_\mu\varphi^{(\pm)}$ and $\Phi_{\pm}  = \phi\pm \pi$.

Finally, for an $SU(2)$ doublet $\widehat H$:

\be
 \widehat S^{(c)}_{aed}= \int d^4\widehat x \,\tr \left\{\sqrt{\widehat g}\, \left[ \widehat g^{\mu\nu} \left(\widehat {\sf D} _\mu \widehat H\right)^\dagger \left(\widehat {\sf D} _\nu \widehat H\right)+\widehat M^2 \widehat H ^\dagger \widehat H -\frac {\widehat\lambda_h}4 \left(\widehat H^\dagger  \widehat H\right)^2\right]\right\}= \widehat S^{(c+)}_{aed}+ \widehat S^{(c-)}_{aed} \label{hataescalarcs}
\ee
where $\widehat \sfD_\mu = \partial_\mu +q\widehat C_\mu- i\sfg W_\mu$ or $\widehat\sfD_\mu= \partial_\mu + \partial_\mu \widehat\varphi- i\sfg W_\mu$ , $W_\mu$ being a gauge field valued in the $SU(2)$ Lie algebra representation  to which $\widehat H$ belongs (beware, $W_\mu$ is not split).
\be
 S^{(c\pm)}_{sm}=   \int d^4x\, \sqrt{g_\pm} \left[ g_\pm^{\mu\nu}\left({\sf D}_\mu H_\pm\right)^\dagger \left({\sf D}_\nu H_\pm\right)+M^2_\pm H^\dagger_\pm H_\pm-\frac {\lambda_{h\pm}} 4 \left( H_\pm^\dagger  H_\pm \right) \right]\label{Scpmhat}
 \ee
 with $\sfD_\mu^\pm= \partial_\mu + q C_\mu^{(\pm)}-i\sfg W_\mu$ or $\sfD_\mu^\pm= \partial_\mu + \partial_\mu\varphi^{(\pm)}-i\sfg W_\mu$.

Putting everything together we can write down the most general AE scalar plus gravity action
\be
\widehat S^{(c)}_{EH+aes+aed+C} &=&\frac 1{2} \int d^4\widehat x\, \tr\left[\frac 1{\widehat \kappa}\sqrt{\widehat g}\, \left(e^{-2\widehat\varphi}+\widehat\zeta_h    \widehat H^\dagger \widehat H+ \widehat\zeta  \widehat \Phi^2\right) \left(\widetilde{\widehat R} +\widehat {\mathfrak c}\, e^{-2\widehat\varphi}\right)\right] \0\\
&& +\int d^4\widehat x \,\tr\left\{\sqrt{\widehat g}\, \left[\widehat g^{\mu\nu} \left(\widehat{\sf D}_\mu  \widehat H\right)^\dagger \left(\widehat{\sf D}_\nu \widehat H\right) +\widehat M^2 e^{-2\widehat \varphi} \widehat H ^\dagger \widehat H-\frac {\widehat\lambda_h}4 \left(\widehat  H^\dagger \widehat H\right)^2\right]\right\}\0\\
&& +\frac 12 \int d^4\widehat x \,\tr\left\{\sqrt{\widehat g}\, \left[\widehat g^{\mu\nu}  \widehat{\sf D}_\mu \widehat \Phi  \widehat{\sf D}_\nu\widehat \Phi +\widehat m^2 e^{-2\widehat \varphi}\widehat \Phi^2 -\frac {\widehat\lambda}4 \widehat \Phi^4\right]\right\}\0\\
&& -\frac 14 \int d^4\widehat x \,\tr\left[\sqrt{\widehat g}\,\widehat C_{\mu\nu}\widehat C^{\mu\nu}\right] - \int d^4x\,\tr\left[\frac 1{\widehat \eta} \sqrt{\widehat g}\, \widehat C_{\mu\nu\lambda\rho}\widehat C^{\mu\nu\lambda \rho}\right],  \label{ScEHsC}
\ee 
where
\be
\widetilde {\widehat R} =\widehat  R + 6\,\epsilon\, \widehat g^{\mu\nu}\left(\widehat D_\mu \widehat \partial_\nu\widehat \varphi - \widehat \partial_\mu \widehat \varphi \widehat \partial_\nu\widehat \varphi\right)+6\, (1\!-\!\epsilon)\,\widehat g^{\mu\nu}\left(q\widehat D_\mu\widehat  C_\nu -{q}^2\widehat C_\mu\widehat C_\nu\right), \label{Rscalarepsilonext}
\ee
and
\be
\widehat C_{\mu\nu}\widehat C^{\mu\nu}= \widehat g^{\mu\mu'} \widehat g^{\nu\nu'} \widehat C_{\mu\nu} \widehat C_{\mu'\nu'}, \quad\quad {\rm etc.}\0
\ee
We have also added the non-minimal coupling strenths $\widehat \zeta_h$ and $\widehat \zeta$, with $\widehat \zeta_h = \zeta_{h+} P_+ + \zeta_{h-} P_-$, $\widehat \zeta = \zeta_+ P_+ + \zeta_- P_-$, etc.

It is also possible to replace $\widehat\Phi$ with a complex field. The action has the same terms as the field  $\widehat H$, by replacing it with $\widehat \Phi$ and $W_\mu $ with a $U(1)$ gauge field $V_\mu$.}

\vskip 1cm 
{
In conclusion, the overall conformal-invariant model $\cal TW$ is defined by the action
\be
\widehat S^{(c)}_{ch-sym}=\widehat S_f  + \widehat S_g + \widehat S_{aeg} +    (S_Y+ S_{Yd}+ h.c.)+ \widehat S^{(c)}_{EH+aes+aed+C} \label{actionchir-symmconf}
\ee
It splits into 
\be
S^{(c-)}_{ch-sym}= S^{(-)}_f  + S^{(-)}_g +  S^{(-)}_{aeg} +  ( S_{YL}+S_{YdL}+h.c.)+   S^{(c-)}_{EH+aes+aed+C} \label{actionchir-symmconf-}
\ee
and
\be
S^{(c+)}_{ch-sym}= S^{(+)}_f  +  S^{(+)}_g +   S^{(+)}_{aeg} +  ( S_{YR}+S_{YdR}+ h.c.)+  S^{(c+)}_{EH+aes+aed+C} \label{actionchir-symmconf-}
\ee
The label (-/+) stands for left/right.}

\vskip 0.5cm

To summarize, $\cal TW$ is a theory of massless fields, which is not only classically symmetric under the already mentioned symmetries, i.e.  under diffeomorphisms, local Lorentz and Weyl transformations as well as the $SU(3)_L\times SU(3)_R \times SU(2)\times U(1)_L \times U(1)_R$ gauge transformations, but has also the right interaction terms to be classified as renormalizable. However we have no guarantee about its unitarity. This is due, in the perturbative approach, to the possible appearance in the gravity propagators of ghosts (wrong sign of the quadratic kinetic operator). 

Therefore let us proceed to illustrate some consequences of the quantization of $\cal TW$.

\section{${\cal TW}$ as a quantum theory}

The $\cal TW$ theory has several local symmetries: diffeomorphism symmetries, local Lorentz invariance,  Weyl invariance, beside the symmetries under $SU(3)_L\times SU(3)_R\times SU(2)\times U(1)_L\times U(1)_R$ gauge transformations. Quantization requires gauge fixing. The reason is that the gauge degrees of freedom are unphysical and have to be eliminated. But there is also another reason: if the gauge is not fixed the propagators of the gauge fields and of the metric are not defined (the corresponding kinetic operators are not invertible) and, therefore, quantization is impossible. The two things are connected: gauge invariance means that, in the space of field configurations, beside the physical degrees of freedom we have  a set of unphysical configurations of infinite volume (in the path integral sense). This infinity is the origin of the non-invertibility of the relevant kinetic operator, being related to the (infinite volume) space of its zero modes. An educated guess for gauge fixing, accompanied by the introduction of FP ghosts { in order to fix the mismatch between the unphysical degrees of freedom and the dofs eliminated by the gauge fixing}, can lead to the solution of both problems. If we now look at the Weyl invariance we see a difference: {  in section \ref{s:weylgeometry} we have already noticed that the introduction of a vector gauge field $C_\mu$ is not required in order to implement a Weyl symmetry, the scalar field $\varphi$ alone can do the job. In $\cal TW$ we would like to exploit this possibility and explore the formulation where the vector gauge potential $\widehat C_\mu$ is absent. In such a case} the propagator of $\widehat\varphi$ exists even when the gauge has not been fixed { and there is a perfect matching between the unphysical dofs and those eliminated for instance by the gauge choice $\widehat \varphi=0$ (thus no FP ghosts are necessary)}. This seems to suggest that the nature of Weyl symmetry is different from that of ordinary gauge symmetries. In the sequel, therefore, we would like to explore an unusual possibility: we will not fix this gauge, but keep it unfixed (in particular, therefore, there will be neither a ghost nor a BRST symmetry corresponding to it), but much like we do  for rigid symmetries we wish to preserve this local symmetry throughout the process of quantization. { In the literature the attitude is generally different and  the Weyl symmetry is treated as an ordinary gauge symmetry and quantized with gauge fixing, ghost and corresponding BRST symmetry, see notably \cite{oda22a,oda22b,oda23}. In the last reference the authors signal the difficulty of producing compatible BRST Weyl, diffeomorphism and St${\rm \ddot u}$ckelberg gauge symmetry fixings, see sec.  \ref{s:unitarity} for a further comment. In our approach this problem is avoided, and it should be recalled that we are not violating orthodoxy: solving a gauge theory without fixing the gauge is not in itself so heretical, it is simply extremely complicated and perhaps even impossible when gauge invariance implies the lack of a propagator; but, we repeat, in our case the relevant $\varphi$ propagator exists  and, so, it makes sense to preserve Weyl gauge invariance through (the perturbative) quantization. } 

Our choice to ignore the field $\widehat C_\mu$ is tantamount to setting $\widehat C_\mu =\partial_\mu \widehat \varphi$, which, in the literature, defines an {\it integrable Weyl structure}. This structure, among other things, has the virtue to avoid the SCE (the second clock effect), which  { revives} an objection raised by Einstein against non-metricity, \cite{Weyl2}: the parallel transport of a unit vector along curves with the same initial and final point would lead to different results, involving both space and time intervals. In particular the time difference (SCE) would be given by the integral of $C_{\mu\nu}$ over the surface comprised between the two curves, which, of course, in the integrable Weyl structure vanishes. For more details and references, see \cite{ghilencea23,ghilencea24,hobson20,pala22}.

Summarizing, the quantum $\cal TW$ theory is dealt with by fixing the gauge for diffeomorphism and gauge invariance, introducing the corresponding FP ghosts and recovering the appropriate BRST symmetries, { while trying to preserve Weyl symmetry. But quantization gives rise to a new problem involving precisely the preservation of Weyl symmetry due to even trace anomalies.}

\section{The problem of even trace anomalies}
\label{s:eventraceanom}

The subject of this section is the problem of even trace anomalies in $\cal TW$ in its symmetric phase (i.e. before the {\it primitive chiral symmetry breaking}, see below). 

Trace anomalies have been discussed above at various stages, and especially in section \ref{s:traceanomalies}. The analysis was limited there to  theories of free fermions coupled to background potentials (metric or gauge fields), and mostly to odd-parity anomalies. Now we have to do with a fully interacting theory, not only for matter fields but also for the gauge potentials, the metrics and the dilatons. 

In general, trace anomalies for matter fields interacting with background potentials arise as a response of the effective action to a Weyl variation of the metric: $g_{\mu\nu} \rightarrow e^{2\omega} g_{\mu\nu}$ for a local parameter $\omega$. The classical definition of the e.m. tensor for matter fields interacting with a background metric is
\be
 T_{\mu\nu}= \frac 2{\sqrt g}\frac {\delta S}{\delta g^{\mu\nu}}\label{Tmunucl}
\ee
$S$ being the classical action. If the latter is Weyl-invariant the e.m. tensor is traceless. This follows from the classical WI
\be
{\delta_\omega S}= \int d^\sfd x\, \left(\frac {\delta S}{\delta g^{\mu\nu}} \delta_\omega g^{\mu\nu} +\sum_i \frac  {\delta S}{\delta f_i} \delta_\omega f_i\right)=0\label{completeWIcl}
\ee
where $f_i$ denotes generic matter fields, and, for infinitesimal $\omega$, $\delta_\omega g^{\mu\nu}= -2\omega g^{\mu\nu}, \delta_\omega f_i= -2 y_i \omega f_i$ (where $y_i$ is 0 for gauge fields, $\frac {\sfd -2}2$ for scalars and $\frac {\sfd -1}2$ for fermions, etc.). If the matter fields are on shell, i.e.  $\frac  {\delta S}{\delta f_i} =0$, it follows that $T_{\mu\nu} g^{\mu\nu}=0$ due to the arbitrariness of $\omega$.

When the metric is dynamical the interpretation of \eqref{Tmunucl} changes  for $\frac {\delta S}{\delta g^{\mu\nu}}$ does not represent the e.m. tensor of the metric field, but the LHS of its equation of motion. It is well-known that in the EH gravity this leads to the Einstein equation
\be
R_{\mu\nu}-\frac 12 g_{\mu\nu} R +\Lambda g_{\mu\nu} = -\kappa \, T_{\mu\nu}\label{Einsteineq}
\ee 
where $\Lambda$ is the cosmological constant and $T_{\mu\nu}$ is the e.m. tensor of the matter fields.

Therefore in an interacting theory like $\cal TW$\footnote{In this section we refer to a general abstract theory with the same general features as $\cal TW$. For this reason we use a generic notation for the fields and couplings, dropping in particular the $_\pm$ labels, but the following discussion applies as well to $\cal TW$, and to ${\cal TW}_L$ or ${\cal TW}_R$ separately or in conjunction, in the symmetric phase.} the WI for Weyl transformations has the same form as \eqref{completeWIcl}, but the interpretation of e.m. tensor of eq.\eqref{Tmunucl}, while it still holds for matter fields, is not valid for the metric itself. If we refer, for instance, to \eqref{ScEH}  we recall that, since we have dropped $C_\mu$, the tilded Ricci scalar is 
\be
\widetilde R = R+ 6\, \square \varphi -6 \partial\varphi\! \cdot\! \partial \varphi\label{tildeRiccimunuvarphi}
\ee
Therefore the equation of motion for the metric is
\be
 R_{\mu\nu} - \frac 12 g_{\mu\nu}\left( R+{\mathfrak c}\,e^{-2\varphi}\right)=-2\kappa\,e^{2  \varphi}T_{\mu\nu}^{(m)}\label{Einsteineqmod2}
\ee 
where $T_{\mu\nu}^{(m)}$ denotes the e.m. tensor of all the matter fields (including the dilaton) coupled to the metric. Let us recall that this equation is Weyl-covariant. 

Taking the trace of \eqref{Einsteineqmod2} we
get that, when all the other fields are on shell, the equation
\be
 R + 2\,{\mathfrak c}\,e^{-2\varphi}= 2\kappa\, e^{2\varphi} \,T^{(m)}, \quad\quad\quad T^{(m)}= g^{\mu\nu} T^{(m)}_{\mu\nu}\label{Riccieqmod}
\ee
should hold as a consequence of \eqref{completeWIcl} (since $\omega$ is arbitrary). 

It must be remarked that, as required by \eqref{completeWIcl}, we obtain a dynamical equation for $g_{\mu\nu}$ that does not imply automatically the vanishing of the trace of $T_{\mu\nu}^{(m)}$. That is, in order to preserve Weyl invariance it is, in general, not necessary for the matter e.m. tensor to be traceless. { There may exist situations of this type, but, in view of the discussion that follows, they fall in the realm of non-perturbative solutions. For simplicity we exclude them from our analysis.}

Now, the matter e.m. tensor $T_{\mu\nu}^{(m)}$ contains the contributions from the e.m. tensors of all the fields, including $\varphi$, except the metric. However, in general, $T_{\mu\nu}^{(m)}$ is not simply the direct sum of the e.m. tensors of the various fields coupled only to the metrics (like, for instance, in  \eqref{emtmodified}), because there are also interaction terms that mix different fields beside the metrics. This means that the problem we face with the trace of the RHS of \eqref{Einsteineqmod2} is not simply the sum of the distinct problems of computing the trace of each field species separately\footnote{ For the same reason we cannot factor the overall path integral of the theory as a product of separate path integrals, each of them involving the quadratic action of each field species interacting with other fields treated as external sources.}.  When dealing with a free field interacting with external potentials we have the possibility to use either a perturbative or a non-perturbative approach. The latter consists in resorting to a
one-loop quantum energy-momentum tensor $\langle\!\langle T_{\mu\nu}\rangle\!\rangle$, which, for  matter fields, is defined by
\be
\langle\!\langle T_{\mu\nu}\rangle\!\rangle = \frac 2{\sqrt g}\frac {\delta W}{\delta g^{\mu\nu}}\label{<<Tmunu>>}
\ee
where $W$ is the effective action, and then using the WI 
\be
{\delta_\omega W}=- \int d^\sfd x \sqrt{g} \, \omega \langle\!\langle T_{\mu\nu}\rangle\!\rangle g^{\mu\nu}\label{<<Tmumu>>}
\ee
in order to identify a possible trace anomaly.

But this approach is not available when dealing with a fully interacting theory like $\cal TW$. We must turn to the perturbative method, which consists in general in fixing the gauges { and introducing the corresponding ghosts, in determining the Feynman rules and, in particular, writing down  propagators and vertices, and, finally, after introducing an appropriate external source for each elementary field, writing the appropriate Slavnov-Taylor identity for each BRST symmetry (this will be commented further on in section \ref{s:unitarity}). However rewriting \eqref{completeWIcl} as a Slavnov-Taylor identity would obscure the role of the e.m. tensors, which are composite operators (not elementary fields). In order to highlight this role we will proceed in another way. Concerning Weyl anomalies, the available quantum results} at present are (perturbative and/or non-perturbative) local expressions of trace anomalies obtained by integrating out in each separate path integral various matter fields, for instance fermions, scalars, etc., which feature quadratically in expressions where they interact with other fields considered as background. The question is whether these results are of any use when we consider a fully interacting theory such as $\cal TW$. In the sequel we would like to discuss how they can in fact be brought to bear in a perturbative approach.
 
The $\cal TW$ action contains terms coming from the quadratic kinetic terms of the various dynamical fields coupled to the metric, the dilaton and the gauge potentials, but also several interaction terms: non-Abelian gauge fields interacting among themselves, scalar fields interacting with fermions, gauge fields  and the dilaton, gravitons interacting with the other fields and among themselves.
The perturbative approach singles out kinetic terms, each being quadratic in one species of fields, from which the propagators are extracted, and deals with all the other terms as interaction terms from which the vertices are derived. It is important to recall that in the perturbative approach the fields we consider are fluctuations around a background; for most fields (fermions, vectors and also scalars in the symmetric phase) the fluctuating fields are { usually} identified with the fields themselves, while for the metric the fluctuating field is $h_{\mu\nu}$ with $g_{\mu\nu}=\eta_{\mu\nu} +h_{\mu\nu}$\footnote{The dilaton could be expanded around a non-vanishing value $ \varphi_0$, but in the present discussion we set $\varphi_0=0$.}. In the following discussion we set, for the moment,  $\mathfrak c=0$. 

Next we proceed as follows. We expand both sides of \eqref{Riccieqmod} in series of $h_{\mu\nu}$ and forget the interaction terms of all the other fields. The zeroth-order term on the LHS vanishes\footnote{Because its lowest order is linear in $h_{\mu\nu}$, for, at the lowest order, $R = \partial_\mu\partial_\nu h^{\mu\nu}-\square h_\mu^\mu$.}. Therefore at the lowest order the RHS, which is the sum of the traces of the e.m. tensors of the various species of fields, must vanish too. But since the e.m. tensors of the various species are non connected by interaction terms, this implies that they   must  be separately traceless. This is all classical.

To quantize \eqref{Einsteineqmod2} and \eqref{Riccieqmod} the perturbative approach allows us to start from the classical zeroth-order expression of the matter e.m. tensor or its trace, which is the sum of the free e.m. tensors for each distinct field species and treat all the other terms as interactions. To extract explicit formulas we expand the full action in terms of the fluctuating fields and forget their interactions, the action becomes a free field action (and the equations of motion are the free ones). From it we can extract the lowest order (on shell conserved and traceless) e.m. tensor for each species via the canonical formulas. Alternatively we can use \eqref{Tmunucl} by coupling every single species with the metric $g_{\mu\nu}$ in such a way as to realize Weyl invariance, using \eqref{Tmunucl} and eventually setting the metric equal to the flat one. { Said another way, we expand the full action in series of $h_{\mu\nu}$: the zeroth-order term is made of the free actions while the coefficient of  $h^{\mu\nu}$ in the first-order term identifies the on-shell conserved and traceless e.m. tensors}.

The zero-th order e.m. tensors obtained in this way are
\be
 T^{(f)}_{\mu\nu}= \frac i4 \overline {\psi} \gamma_\mu
{\stackrel{\leftrightarrow}{\partial_\nu}}\psi+(\mu\leftrightarrow\nu)- \eta_{\mu\nu}\frac i2  \overline \psi \gamma^\lambda {\stackrel{\leftrightarrow}{\partial_\lambda}}\psi, \label{Tmunufermion}
\ee
for fermions (which, in particular, may be chiral), and
\be
T^{(g)}_{\mu\nu} =-\frac 1{g^2}\left(\partial_\mu A_\lambda \partial_\nu A^\lambda + \partial_\lambda A_\mu \partial^\lambda A_\nu- \frac 12 \eta_{\mu\nu} \partial_\lambda A_\rho \partial^\lambda A^\rho\right)\label{Tmunugauge}
\ee
for Abelian gauge fields. This is obtained after fixing the Lorenz gauge. The non-Abelian formula  differs from it only for the trace symbol in front of the RHS, the trace being applied to the gauge Lie algebra generators.

The above e.m. tensors are (on the shell of the free equations of motion) conserved and traceless. At variance with these examples the derivation of the e.m. tensors for scalar fields via formula \eqref{Tmunucl} applied to the scalar fields in \eqref{ScEHsC} does not automatically lead to conservation and tracelessness. However if we now return to formula \eqref{Riccieqmod} in the spirit of perturbation theory, we have seen that to the lowest order the LHS vanishes. Therefore, as already  pointed out, the matter e.m. tensor must indeed be traceless at this order. It follows that also each e.m. tensor for scalars must be separately traceless. 

Therefore for the complex scalar field $\Phi$ we expect
\be 
T^{(s)}_{\mu\nu} =\partial_\mu \Phi^\dagger \,\partial_\nu \Phi+\partial_\nu \Phi^\dagger \,\partial_\mu \Phi- \eta_{\mu\nu}\, \partial_\lambda \Phi^\dagger \,\partial^\lambda \Phi+ \frac 13 \left(\eta_{\mu\nu} \square -\partial_\mu\partial_\nu\right) \Phi^\dagger \Phi\label{TmunuscalarPhi0th}
\ee
and an analogous formula for $H$. In analogy we expect a similar expression for the dilaton
\be
T_{\mu\nu}^{(\varphi)} = \frac {6}{\kappa}\left(  \partial_\mu \varphi \partial_\nu \varphi - \frac 12\, \eta_{\mu\nu}\partial\varphi\! \cdot\! \partial \varphi+ \frac 16 \left(\eta_{\mu\nu} \square -\partial_\mu\partial_\nu\right) \varphi^2\right)\label{Tmunuvarphi0th}
\ee
These expressions are known as improved e.m. tensors. But what do they have to do with the action 
of $\cal TW$ and in particular with \eqref{ScEHsC}? 

Consider a free scalar field $\phi$ in dimension $\sfd$. Its improved (on-shell conserved and traceless) e.m. tensor is
\be 
T^{(s)}_{\mu\nu} =\partial_\mu \phi \,\partial_\nu \phi - \frac 12 \eta_{\mu\nu}\, \partial_\lambda \phi \,\partial^\lambda \phi+ \frac {\sfd -2}{4(\sfd -1)}  \left(\eta_{\mu\nu} \square -\partial_\mu\partial_\nu\right) \phi^2,\label{Tmunuscalarphi}
\ee
assuming the free equation of motion is $\square \phi=0$.
It can be derived with the improved canonical formula. But it can also be derived from the action
\be
S_\phi = \frac 12 \int d^\sfd x \,\sqrt{g} \left(\partial_\mu \phi \partial^\mu \phi +\frac {\sfd -2}{4(\sfd -1)} R\, \phi ^2\right) \label{scalaraction}
\ee
where $R$ is the Ricci scalar. This action is conformal-invariant. Applying \eqref{Tmunucl} to it, integrating by parts and setting $g_{\mu\nu} =\eta_{\mu\nu}$ one obtains \eqref{Tmunuscalarphi}. Now what we do is the following: to the free action of $\phi$ we add and subtract a term $\int d^4 x \,\sqrt{g}  R\, \phi ^2$ (with the appropriate coefficient). The expansion in $h_{\mu\nu}$ gives
\be
2  \int d^4 x \, h^{\mu\nu}  \left(\eta_{\mu\nu} \square -\partial_\mu\partial_\nu\right) \phi^2+ {\cal O}(h^2)\label{1stapproximant}
\ee
from which one can derive \eqref{Tmunuscalarphi}. Adding and subtracting the same term of course does not change anything, but while the lowest order approximant \eqref{1stapproximant} allows us to write \eqref{Tmunuscalarphi}, the term with opposite sign will remain and contribute an interaction term that was not present in the original action, and must be taken into account when computing the trace anomaly. All the terms of higher order in \eqref{1stapproximant} cancel one another. Said another way, we expand the metric on the RHS of \eqref{scalaraction}: the zeroth-order term is the free action of $\phi$ and is left unchanged by the addition of  $\int d^4 x \,\sqrt{g}  R\, \phi ^2$, the coefficient of  $h^{\mu\nu}$ in the first order term identifies the on-shell conserved and traceless e.m. tensor \eqref{Tmunuscalarphi}; the first order term of  $-\int d^4 x \,\sqrt{g}  R\, \phi ^2$ is accounted for as an interaction term. All the higher order terms of   $\int d^4 x \,\sqrt{g}  R\, \phi ^2$ disappear.
On the other hand if in the original action there are mass terms like in (\ref{scalar},\ref{doubletaction}) and (\ref{Scpmhat},\ref{hataescalarcs}), these terms are to be accounted for among the interaction terms.

This procedure can be adopted for the scalar fields $\varphi,\Phi$ and $H$. 

Finally we have to consider also the contribution to the total trace anomaly due to the Faddeev-Popov ghosts. There are two types of these ghosts, those coming from the gauge fixing of  non-Abelian gauge symmetries, denoted $c^a$ and $\overline c^a$ and those derived from fixing the diffeomorphism symmetry, denoted by $\xi^\mu$ and $\overline \xi^\mu$. 

A gauge fixed and FP ghost action for the former can be found in Appendix C, below. The ghost action to the 0-th and first order in $h_{\mu\nu}$, excluding all the other interaction terms, can be extracted from 
\be
S^{c,\bar c}= \int d^4 x \, \sqrt{g} \, g^{\mu\nu}\partial_\mu \overline c^a \partial_\nu c^a \label{cbarcgf}
\ee
summed over $a$. But the e.m. we obtain  in this way 
\be
T_{0\mu\nu}^{(c,\bar c) } =\partial_\mu \overline c^a\, \partial_\nu c^a +  \partial_\nu \overline c^a\, \partial_\mu c^a-  \eta_{\mu\nu} \partial_\lambda \overline c^a\partial^\lambda c^a\label{Tmunughg1}
\ee
is neither conserved nor traceless, in other words we face the same problem as with the scalar field when trying to extract a conserved and traceless e.m. tensor (\eqref{cbarcgf} is not Weyl-invariant). Therefore we proceed as in the scalar field case and add to the Lagrangian term in \eqref{cbarcgf} the term $\frac 13 \sqrt{g} R\, \overline c^a c^a$. In this way the corresponding lowest order e.m. tensor is
\be
T_{\mu\nu}^{(c,\bar c)} =\partial_\mu \overline c^a\, \partial_\nu c^a +  \partial_\nu \overline c^a\, \partial_\mu c^a-  \eta_{\mu\nu} \partial_\lambda \overline c^a\partial^\lambda c^a+ \frac 13 \left(\eta_{\mu\nu} \square -\partial_\mu\partial_\nu\right) \overline c^a c^a\label{Tmunughg}
\ee
which is conserved and traceless.

For diffeomorphisms we use the DeDonder gauge, whose lowest order form is: $2\partial_\mu h^\mu_\lambda -\partial_\lambda h_\mu^\mu=0$. Proceeding as in the gauge case we find a ghost action (see Appendix C, below),
\be
S^{(\xi,\bar \xi)}=- \int d^4 x\, \sqrt{g}\, g^{\mu\nu}  \overline \xi_\lambda \,\partial_\mu\partial_\nu \xi^\lambda\label{xixibaraction}
\ee
Eq.\eqref{xixibaraction} is not Weyl-invariant. Adding, as in the scalar field case, a term $ \int d^4 x\, \sqrt{g}\,R \, \overline\xi_\lambda \xi^\lambda$ to this action, at the lowest order we find the traceless e.m. tensor
 \be
T_{\mu\nu}^{(\xi,\overline \xi)} =\partial_\mu \overline \xi_\lambda \partial_\nu \xi^\lambda +  \partial_\nu \overline \xi_\lambda \partial_\mu \xi^\lambda -\eta_{\mu\nu} \partial_\lambda \overline \xi _\rho \partial^\lambda \xi^\rho + \frac 13 \left(\eta_{\mu\nu} \square -\partial_\mu\partial_\nu\right) \overline \xi_\lambda \xi^\lambda\label{Tmunughd}
\ee

\vskip 0.5cm
{\bf Comment. Other backgrounds}
\vskip 0.2cm
{
In the perturbative approach described so far we have started from the background $g_{\mu\nu}=\eta_{\mu\nu}, {\mathfrak c}=0$ and $\varphi =0$. This is the simplest background. Other more general backgrounds are possible, $g_{\mu\nu}= g_{0\mu\nu}\neq \eta_{\mu\nu},  {\mathfrak c}\neq 0$ and $\varphi= \varphi_0\neq 0$ and even  $V_\mu=V_{0\mu}\neq 0, W_\mu=W_{0\mu}\neq 0$.  They are supposed to satisfy the background equation \eqref{Einsteineqmod2}
\be
 \widetilde R_{0\mu\nu} - \frac 12 g_{0\mu\nu}\left( \widetilde R_0+{\mathfrak c}\,e^{-2\varphi_0}\right)=-2\kappa\,e^{2  \varphi_0}T_{0\mu\nu}^{(m)}\label{Einsteineqmod3}
\ee 
whose trace is
\be
 \widetilde R_0 +2 {\mathfrak c} \, e^{-2\varphi_0} =2\kappa\, e^{-2\varphi_0} T^{(m)}_0 \label{tracebackground}
\ee
The new fluctuating fields are ${\sf h}_{\mu\nu}=g_{\mu\nu}-g_{0\mu\nu}$ and $\varphi-\varphi_0,V_\mu-V_{0\mu} $ and $ W_\mu-W_{0\mu}$. Now we expand  \eqref{Einsteineqmod2} in terms of  the fluctuating fields around the classical solution of \eqref{Einsteineqmod3}, in the same way we have done before around the trivial flat solution. As before the zeroth-order term of the expansion of the full action in ${\sf h}_{\mu\nu}$ provides the free field equations of motion and the first order one the free e.m. tensors.  The quantization is carried out as before, except that e.m. tensors and vertices will contain insertions of the background fields. We will not pursue this here. We only notice that since the calculation of the trace anomalies must be carried out according to the formula \eqref{Duff}, the two sides of eq.\eqref{tracebackground}, will drop out of the final result.
\vskip 0,3cm

\noindent Refs.: \cite{birrell1984,parker2009}.
}
 
\subsection{Even trace anomalies: further examples}

The previous argument allows us to conclude that the first perturbative level of the RHS of eq.\eqref{Riccieqmod} can be computed for each species separately. To this end we have to single out the propagators and vertices. The propagators are well-known, some have been already written down. We summarize them for the reader's convenience, they are: $\frac i {\slashed p}$ for fermions, $- \frac i{p^2} \eta_{\mu\nu} \delta^{ab}$ for gluons (in the Lorenz gauge), $\frac i{p^2}$ for real or complex scalars, $\frac i{p^2}\delta^{ab}$ for FP gauge ghost. { The propagator between $h_{\mu\nu}$ and $h_{\alpha\beta}$ has a complicated form depending on the classical action we start from and on the gauge fixing conditions. For the present purposes enough is to know that the generic form of this propagator in the UV limit is $\frac i{2\kappa p^2} t_{\mu\nu \alpha\beta}$, where the constant tensor $t$ is constructed out of the flat Lorentz metric $\eta_{\mu\nu}$. This simplified form is sufficient to recognize at a glance the order of divergence of a Feynman diagram.} The  propagator for the dilaton $\varphi$ is $ \frac {i\kappa}{6 }\frac 1 {p^2}$. 

The relevant vertices for fermions have been introduced earlier, for instance the vertex $V_{ffh}$, of two fermions and one $h_{\mu\nu}$ gravity fluctuation field in eq.\eqref{2f1g}. The fermion--fermion--gluon vertex is $V_{ffg}: ig\gamma_\mu T^a$.  A fermion--fermion--scalar vertex generated by a Yukawa term with coupling $y$ takes the simple form
 \be
 V_{ffs}&:& iy\label{Vffs}
 \ee
 Similarly the scalar--scalar--$h_{\mu\nu}$-graviton vertex  is given by
\be
V_{ssh} &:&-\frac i2\left(\eta_{\mu\nu} p\!\cdot\! p'-  \left(p_\mu p'_\nu + p_\nu p'_\mu\right)\right)\label{Vssh}
\ee
 where $p'$ is an entering scalar particle momentum and $p$ an exiting one. The subtraction of the term \eqref{1stapproximant} entails the addition of another scalar--scalar--graviton vertex 
 \be
V'_{ssh} &:&  \frac i6\left (\eta_{\mu\nu} p\!\cdot\! p'- \left(p_\mu p'_\nu + p_\nu p'_\mu\right)\right)\label{V'ssh}
\ee
 
 The complex scalar--complex scalar--gluon vertex $V_{ssg}$ is
\be
 V_{ssg}&:&  i T^a \sfg\left(p_\mu-p_\mu' \right)\label{Vssg}
 \ee
 where the gluon leg is represented by $V_\mu^a T ^a$ and $\sf g$ is the relevant gauge coupling. The gluon--gluon--graviton vertex is
 \be
 V_{ggh}&:& \frac i2 \delta^{ab}\left[ (p_\alpha p'_\beta + p_\beta p'_\alpha)\, \eta_{\mu\nu} -p\!\cdot\! p'\left(\eta_{\alpha\mu} \eta _{\beta \nu}  +\eta_{\alpha\nu} \eta _{\beta \mu}+ \frac 12 \eta_{\alpha\beta}  \eta_{\mu\nu}\right)\right] \label{Vggh}
 \ee
  referring to the legs $h_{\alpha\beta}, V_\mu^a T^a, V_\nu^b T^b$ with $p$ entering, $p'$ exiting for the last two. Here we have assumed the normalization $\Tr(T^a T^b)=\delta^{ab}$. The triple-gluon coupling, $V_{ggg}$, can be found in any gauge field theory textbook.

 The ghost--ghost--$V_\mu$-gluon vertex is
 \be
 V_{c\bar c g}&:&  -\sfg f^{abc} p_\mu \label{Vcbarcg}
 \ee 
 The ghost--ghost--$h_{\mu\nu}$-graviton vertex is
  \be
 V_{c\bar c h}&:&  -\frac i2( p_\mu p'_\nu + p_\nu p'_\mu - \eta_{\mu\nu}\, p\!\cdot\! p')\label{Vcbarch}
 \ee 
 The vertex $V_{\xi \bar \xi h}$ is
 \be
 V_{\xi \overline \xi h} &:&  -\frac i2( p_\mu p'_\nu + p_\nu p'_\mu - \eta_{\mu\nu}\, p\!\cdot\! p')\eta_{\lambda \rho}\label{Vxixibarh}
 \ee
where the legs are $h_{\mu\nu}, {\overline\xi}^\lambda, \xi^\rho$ with momenta $q,p,p'$ respectively $(q+p+p'=0)$.

 As for the dilaton $\varphi$, $V_{\varphi\varphi h}$ is the same as $V_{ssh}$ multiplied by $ \frac {3}\kappa$. Then 
\be
 V_{H^\dagger H\varphi}&:& i\, p\!\cdot\! p'\label{VHHvarphi}
\ee
where $p$ is the momentum of $H^\dagger$ and $p'$ is the momentum of $\varphi$. The vertex $V_{\Phi^\dagger \Phi \varphi}$ is the same. {The mass terms of scalar fields such as those in  (\ref{scalar},\ref{Scpmhat},\ref{hataescalarcs}) originate interaction terms proportional to the square mass. }

There are also vertices due to terms like \eqref{1stapproximant} and higher order vertices. But the ones  shown above are sufficient to illustrate the main point.

Thus, at least at the lowest loop order, we can compute the overall anomaly of $\cal TW$  using the results and methods presented in the first part of this article. As for the odd-parity (i.e. type O) trace anomalies in $\cal TW$ this confirms what has already been said above: they cancel out by construction. But in $\cal TW$ there are even (type NO) trace anomalies. It is worth recalling the difference between the two types of anomalies we are talking about here. 

Odd trace anomalies are produced in models of chiral fermions coupled to a background metric or gauge field. They are a consequence of the lack of the corresponding fermion propagators. With this kind of anomalies a theory simply does not make sense.  Even trace anomalies, on the contrary, do not correspond to a lack of propagator and the corresponding theory is well-defined. It is true that, for instance, in the calculation of the even trace anomaly in a theory of a Maxwell field coupled to a metric the propagator is ill-defined, because the kinetic operator is not invertible due to gauge invariance, but this fact is remedied by suitably fixing a gauge (and taking into account the contribution of the appropriate FP ghosts). Once the relevant propagator is defined the calculation is not unlike the odd-parity case we have seen above. The results are anomalies taking the form of eq.\eqref{Deltaog}. The most well-known have density $T[g]$ given by  (\ref{quadweyl},\ref{gausbonnet}). In perturbation theory their lowest order terms are found from the triangle diagram contribution of the one-loop e.m. tensor $\langle\!\langle T^{(f)}_{\mu\nu}\rangle\!\rangle$, according to the prescription \eqref{Duff}. But there are others, perhaps less well-known or plainly ignored, anomalies.
They are characterized by the fact that their density are covariant pieces of actions. An example is represented by eq.\eqref{gaugeaction}. Its explicit computation can be found in \cite{I}. Here we summarize the main facts. It shows up, for instance, in a theory of Dirac fermions coupled both to an external metric and to an external Maxwell potential. At the lowest level it is again born out of $\langle\!\langle T^{(f)}_{\mu\nu}\rangle\!\rangle$, but from a triangle diagram amplitude formed by joining a lowest level chiral e.m. tensor to two $V_{ffg}$ vertices and  joining each other by means of three fermion propagators, with the addition of the crossed amplitude. These amplitudes are divergent and extracting the finite part one finds 
\be
\!\!\!{\cal A}^{(g)}_\omega =-\frac 1{12\pi^2}  \int d^4x\, \omega \, \left(-\partial_\nu V_\lambda \partial^\nu V^\lambda + \, \partial_\nu V_\lambda \partial^\lambda V^\nu\right)=\frac 1{24\pi^2}  \int d^4x\, \omega \, F_{\nu\lambda }F^{\nu\lambda}\label{trueAomega1}
\ee
This is the lowest order result. The all-order expression invariant under both diffeomorphisms and gauge transformations is
\be
{\cal A}^{(g)}_\omega =\frac 1{24\pi^2}  \int d^4x\,\sqrt{g}\, \omega \, F_{\nu\lambda }F^{\nu\lambda},\label{trueAomega2}
\ee
Its non-Abelian extension is  
\be
{\cal A}^{(g)}_\omega =\frac 1{24\pi^2}  \int d^4x\,\sqrt{g}\, \omega \, \tr \left(F_{\nu\lambda }F^{\nu\lambda}\right).\label{trueAomeganonAbelian}
\ee

From this example it is clear what has to be done in order to identify the trace anomalies of this type. We must construct all possible divergent triangle diagrams formed by joining a zeroth-order e.m. tensor with two vertices and the latter with each other, by means of three suitable propagators; the e.m. tensor carries an incoming momentum and the other two outgoing lines represent particles created by the same field (two fermions, two scalars, etc.). Then we must regularize the associated integral and extract its finite part. It is easy to see that the triangle diagrams with two outgoing fermions always vanish\footnote{This is true at any loop order, therefore is an exact result. It is due to the fact that all vertices involve either two fermions or none. Therefore in order to form a triangle diagram with one bosonic and two fermionic external legs we need either an even number of fermion vertices and an odd number of fermion propagators or viceversa. In both cases the number of $\gamma$ matrices involved in the diagram is odd, thus the gamma matrix trace vanishes.}. Therefore we cannot have an even trace anomaly of the above type whose density $T[g]$ is the Lagrangian density of the fermion kinetic term. Analogous considerations hold for anomalies with density $T[g]$ corresponding to the Lagrangian density of ghosts, although the vanishing mechanism is different: it is easy to see that there are potentially divergent triangle diagrams with two outgoing { $\overline c^a$ and $c^b$ legs, but the corresponding amplitude vanishes when summed with the crossed amplitude due to anticommutativity and the lack of an appropriate tensor antisymmetric in $a\leftrightarrow b$.} The same holds for diffeomorphism ghosts. However, this is not the case for scalars.

{
Consider, for instance, the triangle diagram very similar to the one that gives rise to the familiar ABJ anomaly, except that the two outgoing legs represent real scalars, say $\Phi$.
It is formed by joining the zeroth-order fermion e.m. tensor $T^{(f)}_{\mu\nu}$ to two fermion-fermion-scalar vertices $V^{(s)}_{ffs}$, and these to each other, by means of three fermion propagators. The only difference with the ABJ calculation is the absence of a $\gamma_5$ matrix and  the trace of four $\gamma$ matrices instead of six. The integral is divergent, it must be regularized and the trace anomaly calculated according to \eqref{Duff}. The result is
{
\be
{\cal A}^{(s)}_\omega = \frac 1{48 \pi^2}\int d^4x\,\omega\, \left(8 \,\Phi\, \square  \Phi-\partial_\mu \Phi\, \partial^\mu \Phi\right)\label{scalartraceanom}
\ee}
This is {the triangle diagram} result. It does not satisfy the WZ consistency conditions. The all-order expression is either\footnote{In the perturbative context of this analysis, the expressions of trace anomalies we need involve only the orders of approximation relevant for the perturbative order we are considering. However the identification of all-order expressions are both a simple way to check the WZ consistency conditions and a practical way to find all the orders of approximation.}
\be
{\Delta}^{1s)}_\omega \sim \int d^4x\sqrt{g}\, \omega\, g^{\mu\nu} \sfD_\mu \Phi \, \sfD_\nu \Phi\label{scalartraceanomfull1}
\ee
or
\be
{\Delta}^{(2s)}_\omega \sim \int d^4x\, \sqrt{g}\,\omega \left(\partial_\mu \Phi\, \partial^\mu \Phi +\frac 1{6} R\, \Phi^2\right)\label{scalartraceanomfull2}
\ee
{ or
\be
{\Delta}^{(3s)}_\omega \sim \int d^4x\, \sqrt{g}\,\omega \left( \Phi\, \square\Phi -\frac 1{6} R\, \Phi^2\right)\label{scalartraceanomfull3}
\ee}
They all satisfy the WZ consistency conditions. What combination of the three is the appropriate one in any specific case can come only from the explicit calculation. {In the case of trace anomaly due to the scalar $\Phi$, interacting with a Dirac fermion, the final result is
\be
{\cal A}^{(s)}_\omega = \frac 1{48 \pi^2}\int d^4x\,\sqrt{g}\,\omega\, \left(8 \,\Phi\, \square  \Phi-\partial_\mu \Phi\, \partial^\mu \Phi- 9 R\, \Phi^2\right)\label{scalartraceanomtrue}
\ee
The difference between this last expression and equations (\ref{scalartraceanomfull2},\ref{scalartraceanomfull2}) is due to higher order diagram contributions, see Appendix D.}

In the round brackets of eqs.(\ref{scalartraceanomfull1},\ref{scalartraceanomfull2}) we can add also a term $\Phi^4$, which form a consistent cocycle by itself and can appear at higher orders of the calculation, but for our present discussion the above formulas are sufficient.

Similar expressions hold with the obvious replacements also for  a complex doublet $H$, instead of the real scalar $\Phi$.}
{
As for the dilaton we can have
\be
{\cal A}^{(\varphi)}_\omega \sim \int d^4x\,\sqrt{g} \,\omega \left(\partial_\mu \varphi\, \partial^\mu \varphi+ \frac 13 R \varphi^2\right)\quad {\rm or} \quad \sim  \int d^4x\,\sqrt{g}\, \omega\,\left(\varphi \square \varphi +\frac 16 R \left(\varphi^2-3\varphi\right)\right)\label{dilatontraceanom}
\ee}
Both terms satisfy the WZ consistency conditions. Other consistent expressions are possible.

A similar result is obtained from the triangle diagram constructed by joining $T^{(g)}_{\mu\nu}$, by means of two gluon propagators, to two vertices $V_{ssg}$ and the latter by means of a scalar propagator, the two outgoing legs being scalars. 

There are many other similar examples. The anomalies (\ref{quadweyl},\ref{gausbonnet}) may come not only from the triangle diagrams of $\langle\!\langle T^{(f)}_{\mu\nu}\rangle\!\rangle$ but also from those of $\langle\!\langle T^{(g)}_{\mu\nu}\rangle\!\rangle$ and $\langle\!\langle T^{(s)}_{\mu\nu}\rangle\!\rangle$ both for $\Phi$ and $H$, and $\langle\!\langle T^{(\varphi)}_{\mu\nu}\rangle\!\rangle$, as well as in $\langle\!\langle T^{(c,\bar c)}_{\mu\nu}\rangle\!\rangle$   and $\langle\!\langle T^{(\xi, \overline \xi)}_{\mu\nu}\rangle\!\rangle$, provided we construct the triangle with the right vertices. Each of the corresponding anomalies will have its own coefficient. 

The anomalies of the type \eqref{trueAomega2} can be produced not only in the  triangle diagram of $\langle\!\langle T^{(f)}_{\mu\nu}\rangle\!\rangle$, but also in $\langle\!\langle T^{(g)}_{\mu\nu}\rangle\!\rangle$, $\langle\!\langle T^{(c,\bar c)}_{\mu\nu}\rangle\!\rangle$ and
$\langle\!\langle T^{(s)}_{\mu\nu}\rangle\!\rangle$, both for $\Phi$ and $H$.

Anomalies of the type (\ref{scalartraceanomfull1},\ref{scalartraceanomfull2},\ref{scalartraceanomfull3}) can be produced in triangle diagrams of $\langle\!\langle T^{(f)}_{\mu\nu}\rangle\!\rangle$, $\langle\!\langle T^{(s)}_{\mu\nu}\rangle\!\rangle$ and $\langle\!\langle T^{(g)}_{\mu\nu}\rangle\!\rangle$.
And so on.

It is out of the scope of this article to explicitly calculate these potential anomalies. They are not really challenging but rather lengthy calculations, where for completeness one should consider also other diagrams besides the ones mentioned before.  For instance, in the case of $\langle\!\langle T^{(f)}_{\mu\nu}\rangle\!\rangle$ one should consider also bubble diagrams involving two fermion propagators and a 2-fermion--2-graviton vertex. Although there are no known explicit examples of non-trivial contributions to anomalies of this type, it is still necessary to examine them. 

The examples already shown should, however, be enough to illustrate the problem. We have plenty of trace anomalies that break Weyl invariance. Considering the wide randomness of their rational coefficients a mutual cancelation, like in the case of the odd trace anomalies (mode (B)), looks impossible in any reasonable theory. There is however another method for that, and we would like now to focus on it: the breaking of conformal invariance by even trace anomalies can be repaired by adding to the effective action suitable WZ terms. Let us briefly discuss them.}
\vskip 0.3 cm
{ Refs. \cite{duff1977,christensen1976,christensen1977,capper,christensen1978,christensen-duff1978,CD2,osborn1,deserschwimmer1993,duff1994,Bertlmann,bast2009,duff2020,I}}

\subsection{WZ terms}
\label{ss:WZ}
Any trace anomaly can be written in the form
\be
{\cal A}_\omega[g,f] = \int d^{\sfd}x \,\sqrt{g}\,\omega\, F[g,f] \label{traceanomaly}
 \ee
 where $g=\{g_{\mu\nu}\}$ is the metric, $\omega$ is the Weyl transformation parameter $\delta_\omega g_{\mu\nu}=2\omega\, g_{\mu\nu}$, $f$ denotes any other field and $F$ is a local function of $g$ and $f$. Assuming $\omega$ to be an anticommuting Abelian field, the anomaly must satisfy the consistency condition
 \be
 \delta_\omega {\cal A}_\omega =0\label{traceanomalyWZ}
 \ee
 which expresses simply the fact that two subsequent Weyl transformations made in reverse order yield the same result. This is in fact an integrability condition. It means that, with the help of an auxiliary field $\sigma$, which transform as $\delta_\omega \sigma=-\omega$, we can construct a local functional ${\cal W}_{WZ}[\sigma,g,f]$, such that
 \be
 \delta_\omega  {\cal W}_{WZ}[\sigma,g,f]= - {\cal A}_\omega[g,f] \label{WZtrace}
 \ee
 The construction parallels the analogous one for chiral anomalies, \cite{I}. Introduce the parameter $t$, $0\leq t\leq 1$ and the interpolating metric
 \be
 g_{\mu\nu} (t) = e^{2\sigma t} g_{\mu\nu}\label{interpolatingmetric}
 \ee
 such that 
 \be
 \delta_\omega g_{\mu\nu}(t)= 2(1-t)\,\omega\,g_{\mu\nu}(t),\label{interpolmetric1}
 \ee
 as well as an interpolating field
 \be
 f(t) = e^{-y\, t\, \sigma} f, \quad\quad \delta_\omega f(t) = -y(1-t)\omega f(t)\label{f(t)}
 \ee
 where $y=0$ for a gauge field, $y=\frac {\sfd-2}2$ for a scalar field. 

We have also
 \be
 \frac d{dt} g_{\mu\nu}(t) =2\,\sigma g_{\mu\nu}(t) , \quad\quad  \frac d{dt}f(t) =-y\,\sigma f(t) 
\label{ddtgt}
 \ee 
 
 For the field $\varphi$ we put $f(t)\equiv\varphi(t)= \varphi+\sigma t$, thus
\be
\delta_\omega \varphi(t) =\omega(1-t), \quad\quad \frac d{dt} \varphi(t)= \sigma\label{varphit}
\ee
 
 Then we construct the functional
 \be
 {\cal W}_{WZ} [\sigma,g,f] &=&\frac 18 \int_0^1 dt\, \int d^\sfd x\,\sqrt{g(t)} F[g(t),f(t)]
 g^{-1}(t) \frac d{dt} g(t)\0\\
 &=& \int_0^1 dt\, \int d^\sfd x\,\sqrt{g(t)} F[g(t),f(t)]\,\sigma \label{WWZ}
 \ee 

Now we notice that the $t$ derivatives of the interpolating fields coincides with their Weyl variation 
except for the replacement of $\sigma$ with $(1-t)\,\omega$.

A simple evaluation yields
\be
\delta_\omega {\cal W}_{WZ} [\sigma,g,f] &=& \int_0^1 dt \frac d{dt}\left( \int d^\sfd x\sqrt{g(t)}\, F[g(t),f(t)]\right)(1-t)\,\omega\0\\
&&- \int_0^1 dt \, \int d^\sfd x\sqrt{g(t)}\, F[g(t),f(t)]\omega  = -{\cal A}_\omega[g,f]
\label{WZverifyied}
\ee

The  WZ terms has the same formal expression as the anomaly with $\omega$ replaced by $\sigma$, $g_{\mu\nu}$ replaced by $g_{\mu\nu}(t)$ and $f$ with $f(t)$. For instance the WZ term for the anomaly with density $\sim F_{\mu\nu} F^{\mu\nu}$ takes the extremely simple form
\be
{\cal W}_{WZ}[\sigma, g,V]\sim \int d^4x \sqrt{g}\, \sigma \, F_{\mu\nu}F^{\mu\nu} \label{WZFF}
\ee
This is the case also  for the anomaly \eqref{scalartraceanomfull1} and for the square Weyl tensor anomaly \eqref{quadweyl}.
\vskip0,3cm
{ Refs.\cite{wesszumino1971,bardeen1984,I}}

\subsubsection{Application of WZ terms to $\cal TW$}
\label{sss:WZtoTW}

From the previous subsection we can see all the even trace anomalies present in $\cal TW$ can be canceled by adding to the effective action the corresponding WZ term constructed by replacing $\sigma$ in the above formulas with $\varphi_+$ or $\varphi_-$ according to whether we refer to $\cal TW_+$ or $\cal TW_-$. 
For instance, the WZ term \eqref{WZFF} takes the form
\be
{\cal W}^{(\pm)}_{WZ}[\varphi_\pm, g_\pm,V]\sim \int d^4x \sqrt{g_\pm}\, \varphi_\pm \,\tr\left( F_{\mu\nu}F_{\mu'\nu'}\right) g_\pm^{\mu\nu'} g_\pm ^{\nu\nu'}\label{WZFFpm}
\ee
to cancel the even trace  anomaly induced by the (common to both sides) SU(2) gauge field. These are new interaction terms in the quantum action.

A  remark is in order should one use, instead of $\varphi_\pm$, the field $\chi_\pm$, see eq.\eqref{ScEH'}. The WZ term would be
\be
{\cal W}^{(\pm)}_{WZ}[\chi_\pm, g_\pm,V]\sim \int d^4x \sqrt{g_\pm}\,\log\left(\ell \chi_\pm\right)  \,\tr\left( F_{\mu\nu}F_{\mu'\nu'}\right)g_\pm^{\mu\nu'} g_\pm ^{\nu\nu'} \label{WZFFchi}
\ee
To write it down one must introduce a scale of length $\ell$, which may create problems for renormalization (see a discussion in Appendix E).

\subsection{Conformal symmetry restoration}

As stressed before, our way to deal with conformal symmetry is not the traditional way of treating gauge symmetries. As pointed out several times, in $\cal TW$ there are several other local symmetries, notably the two diffeomorphism symmetries and the SM gauge symmetries.
Our approach to quantization consists in gauge fixing the latter symmetries, while leaving free the conformal symmetry during quantization. This allows us to apply the perturbative approach as outlined above. In turn this requires introducing the FP ghosts for the former symmetries in order to vindicate the existence of relevant metric and gauge boson propagators and recover the corresponding BRST symmetries, an indispensable implement for unitarity and renormalization. It is important to highlight the different standing of local conformal symmetry on one side and diffeomorphism and/or gauge symmetry on the other. The latter are absolutely needed for renormalization, they are the fundamental symmetries that one has to restore in the form of { (mutually compatible)} BRST symmetries at every step of the perturbative renormalization of the relevant theory. The former does not have the same status -- it is not strictly required to fix the gauge (the relevant propagator exists anyhow).  It may however be violated by even trace anomalies without compromising the consistency of the theory. It is nevertheless possible to restore conformal invariance by adding suitable WZ terms. 

The way to proceed with (one loop) renormalization of $\cal TW$ is therefore the following. In $\cal TW$ at one-loop we meet divergent integrals (in diagrams with low enough number of legs)  and one has to regularize them and carefully define their infinite parts. Let us suppose that this has been done without meeting obstacles. The next step consists in absorbing such infinities by means of field and coupling constant renormalizations. The one-loop trace calculations of the various matter e.m. tensors are not affected by this procedure, but, as we have seen, they also give rise to infinities whose finite parts generate trace anomalies. To recover conformal invariance of the one-loop action we have to add the relevant WZ terms. The price is the addition to the quantum action of  new interaction terms that involve the dilaton, such as \eqref{WZFF} and in general \eqref{WZverifyied}.

At this stage we would like to stress an important point.
As has been noted, see for instance \cite{codello2012}, any classical local field theory is implicitly conformal-invariant. To make the invariance explicit it is enough  to use a dilaton field and introduce suitable expressions of it in the local terms of the action which are not already conformal-invariant. This is the procedure we have adopted above to immerse $\cal T$ in the Weyl geometry. To this end we have introduced dilaton fields, represented by the symbols $\varphi, \widehat \varphi, \varphi_+, \varphi_-$. What we are considering now is different, we are checking if Weyl symmetry can survive quantization. Once the effective action has been one-loop renormalized, the Weyl invariance turns out to be broken by several trace anomalies. However, as we have seen,  such trace anomalies can be canceled by suitable WZ terms, which restore in this way Weyl invariance.  { It should  not be forgotten} that this cancelation cannot be carried out for arbitrary local field expressions. It can be done for trace anomalies because the latter satisfy the WZ consistency conditions. Now, introducing WZ terms entails introducing new interaction terms in the action. Considering $\cal TW_+$ and $\cal TW_-$, this operation can be carried out by identifying $-\sigma$ with $\varphi_+$ and $\varphi_-$, respectively, in the WZ terms of the previous section. In this way we end up with two one-loop conformal  invariant theories, ${\cal TW}^{(1)}_+$ and ${\cal TW}^{(1)}_-$. 

The previous one may hardly be considered a grand result -- WZ terms are more than fifthy years old -- although it is remarkable that the overall cost is the use of a unique AE field $\widehat \varphi$, which is already there in the classical theory. However things become more interesting if we can proceed with the two-loop quantization. At two loops there are many more interaction terms and diagrams to be considered, than at one loop. Due to the number of fields and interactions, and so of diagrams, involved, a direct computational approach is simply daunting. What saves our day are the WZ consistency conditions. For we know that the trace of the matter e.m. tensor must satisfy these conditions at every perturbative order. Now given the fields at hand, the quantum numbers and the dimensions, there exist only certain local expressions of the fields that satisfy the WZ conditions. In our specific case they are the same cocycles that appear at one loop, i.e. the Gauss-Bonnet and square Weyl tensor anomalies and those in eqs.(\ref{trueAomega2}, \ref{trueAomeganonAbelian},\ref{scalartraceanomfull1},\ref{scalartraceanomfull2}) and \eqref{dilatontraceanom}. None else; in particular there are no anomalies whose density $T[g]$ is the fermion or ghost part of the action. The only change at two loops can lie in the renormalization of their coefficients. But, altogether, the result will be again conformal-invariant. And this argument can be repeated at higher orders. So we expect the quantum action to be Weyl-invariant. Of course it remains to examine unitarity and for this see next section and Appendix E.

Finally, let us return to eq.\eqref{Riccieqmod}. Its validity guarantees conformal invariance. We have seen that expanding in $h_{\mu\nu}$, the zeroth-order term vanishes on the LHS and implies that also the RHS must vanish. On the other hand quantizing the RHS we have argued that, by using WZ terms, it is possible to preserve Weyl invariance, which implies a vanishing quantum RHS. We expect therefore that also the quantum version of the LHS should be vanishing. Proceeding as on the RHS, the result should be obvious because the starting operator to be quantized is simply zero. This seems to be confirmed by the formula \eqref{Duff}, because the LHS corresponds to $g^{\mu\nu}\langle\!\langle  R_{\mu\nu}\rangle\!\rangle - \langle\!\langle  R\rangle\!\rangle$.  However we interpret this expression it does not seem to give rise to ambiguities such as those that generate known anomalies. { But no doubt a direct check is necessary in terms of explicit calculations.}

\vskip0.5cm
\noindent See refs.\cite{englert75,englert76,fradkin78,fradkin85,codello2012,ferrara18,scholz2018,rachwal2022}, for other examples of conformal symmetry restoration.

\section{Renormalization and unitarity. A lightning review}
\label{s:unitarity}

Like in any theory involving gravity also in $\cal T$ one cannot avoid the problem of renormalizability and unitarity. In this paper we do not intend to tackle these huge problems. We do not have anything original to add to what can be found in the literature. But we would like to present anyhow a quick summary to put  our proposed theory in perspective against the backdrop of the current and past research.

In $\cal T$ we have two basic ingredients, the SM and gravity.  As for the first, an Abelian and non-Abelian gauge theory coupled to matter, the problem of renormalzability and unitarity has been solved long ago in various stages and by various people, \cite{BRS1,BRS2,BRS3,stora2006}. A detailed proof of renormalization for the GWS model of electroweak interactions can be found in \cite{Kraus97}, where one can also find the relevant literature. For gravity instead the problem is still open. As compared with an enormous amount of literature on the classical aspects of gravity, the one concerning its quantum aspects is, perhaps understandably, rather limited and lacking undisputed conclusions. In fact it became clear from the very beginning that gravity in the form of the EH action is not renormalizable due to the presence of a dimensionful coupling constant. However it was realized that this initial obstacle can be circumvented by introducing in the action other terms quadratic in the curvature \cite{stelle1977,julve-tonin1978}. This renders the theory formally renormalizable, but raises another problem: the gravitational propagator contains a massive component with wrong overall sign (corresponding to a negative norm state) that renders it a physical ghost.  

Such physical ghosts are the thorn in the side of any gravitational theory as long as one cannot exclude that they do not contribute to the internal lines of a Feynman diagram, thus possibly compromising unitarity. We cannot illustrate here all the proposals presented to overcome this  difficulty. We choose to illustrate two of them, which are particularly fitting to our case. The first, \cite{sibold21,sibold23}, studies the renormalization of a gravitational model where  two more square curvature terms have been added to the EH action, each with its own coupling constant. The authors choose as variable the metric fluctuation $h_{\mu\nu}$, fix the DeDonder gauge, introduce the FP ghosts and determine the BRST symmetry of the model. Then they write down the corresponding ST (Slavnov-Taylor) identities and apply the BPHZL (Bogoliubov-Parasiuk-Hepp-Zimmermann-Lowenstein) renormalization scheme. They analyse the double pole in the propagator of $h_{\mu\nu}$, which splits into a massless simple pole and a massive unphysical one, as usual. This is the situation at the tree level. However the authors' claim is that in higher loops the massive component of $h_{\mu\nu}$ does not develop poles, therefore it does not correspond to particles. In this way the S-matrix unitarity turns out to be violated only at the tree level. One of the authors, \cite{sibold24}, has pushed the analysis further by coupling the previous model to a scalar field with quartic interaction, arriving at the same conclusion: the negative norm state shows up only at the tree level. 

The second proposal is due to Oda and collaborators, \cite{oda22a,oda22b,oda22c,oda23,oda24,oda24b}. In \cite{oda22b} the author focuses on a theory defined by the classical action \eqref{scalaraction} in $\sfd=4$, which, as we noted, is Weyl-invariant and is called Weyl-invariant scalar-tensor gravity. They fix both the diffeomorphsm and the Weyl gauge and work out the corresponding BRST symmetries. Unlike the previous example the quantization is carried out in the canonical way. The author determines the equal time (anti)commutation relations among all the fields and their conjugates. On this basis, applying a formalism originally developed by Nakanishi \cite{nakanishi78,nakanishi-ojima}, he proves the existence of a global symmetry the `choral' symmetry, which is spontaneously broken at the quantum level. Its Nambu-Goldstone bosons are the graviton and the dilaton, which are consequently massless.  The author is also able to analyse the physical S-matrix using the method introduced by Kugo and Ojima, \cite{kugo-ojima} and prove that it is unitary. 

 The authors of ref.\cite{oda22c} consider a somewhat related conformal-invariant model (the third proposal). They deal with the Weyl symmetry as a gauge symmetry and introduce the corresponding gauge field (which above we called $C_\mu$). Simultaneously they add to the action \eqref{scalaraction} the kinetic term for $C_\mu$, and three more terms, each with its own independent coupling: a Weyl-covariant kinetic term, a quartic coupling for $\phi$ and a term $\phi^2 \widetilde R^2$, where $\widetilde R$ is the same as in \eqref{Rscalarepsilon} with $\epsilon=0$. They fix both gauges in a mutually compatible way and develop a parallel analysis to \cite{oda22b}. They show that the S-matrix in this model is unitary and its physical spectrum consists of the two components of a massless graviton and the three components of a massive Proca field, which comes from $C_\mu$ having absorbed $\phi$ via the Higgs mechanism.  

The other side of these positive results is that { they do not seem to be unitary}. Once we declare that they are Weyl-invariant, renormalization demands to introduce in the action all terms with the right dimensions that are Weyl-invariant, so also the term  \eqref{ScW}. It is known that this term is likely to carry in the spectrum a massive ghost. This is precisely what is analyzed in ref.\cite{oda23}, where the term \eqref{ScW} is added to the action\eqref{scalaraction}. In order to apply the canonical quantization the presence of four derivatives requires the introduction of an auxiliary (St${\rm \ddot u}$ckelberg) vector field, which entails an additional   (St${\rm \ddot u}$ckelberg) symmetry. There are therefore three symmetries and it turns out to be impossible to find mutually compatible (i.e. mutually anticommuting) BRST transformations for all three. Therefore the authors choose two BRST operators, correspinding to diffeomorphisms and to a mixture of the other two. Then they proceed to the canonical quantization, to extract the equal time commutators and to analyse the asymptotic states. The particle spectrum is made of a massless graviton and a massive ghost with negative norm, together with  zero norm states which form a quartet. Since all these states satisfy the Kugo-Ojima criterion for physicality, this means that unitarity is explicitly violated.

{ 
Let us return now to the $\cal TW$ theory. We can see that several of its features are contained in the previous examples. Therefore we can try to apply some of the previous results to it, well conscious, however, that an {\it ad hoc} careful analysis is still missing. Nevertheless we believe it sensible to expect that this theory is, at least formally, renormalizable, but with the presence in its particle spectrum of negative norm states, unless we exclude by hand a term like \eqref{ScW}, which renders the theory non-unitary (see also the discussion in Appendix E). Squeezed between the almost unreasonable perfection of the SM, which is renormalizble and unitary, and the promised land of superstring theories that lure us with the ultimate goal of UV finiteness, we can only recognize that $\cal TW$ is far away from both, it looks a simple effective field theory, defective from the point of view of UV completeness and unitarity.

To be more precise, looking at the second option above one would be induced to give up Weyl invariance in favor of renormalizability and unitarity. In this perspective Weyl invariance would be an accidental symmetry that manifest itself in approximate form only at the classical level. If instead we consider the first and third options, we would be led to conclude that Weyl symmetry imposes too strong conditions for them to be compatible with unitarity in $\cal TW$. { From this point of view, therefore, there are no happy news. On the other hand the contrary would be quite surprising. It is a rather common belief, and there are arguments in support, that a consistent quantum field theory including quantum gravity cannot be constructed with a finite number of fields \cite{Kundu,Maldacena,Steinacker}.

On the other hand the right attitude is perhaps to consider these two alternatives not as excluding each other, but in physical sequence: they would refer to different ranges of energy, the first to lower and the second to higher energy, the lack of unitarity of the second denoting a UV incompleteness, that is signaling missing degrees of freedom in the spectrum of $\cal TW$. In this regard a reference to \cite{codello2012} is not out of context. In that paper the authors argue that in the space of field configurations encircling a conformal field theory there exists a trajectory of the renormalization group which is entirely made of conformal theories from the UV to the IR. At first sight this seems to contradict our previous conclusions. But this may not be the case. The theories along a renormalization group trajectory do not necessarily have always the same form (same fields, same couplings, ...), but may undergo transitions to different forms with different elementary degrees of freedom.

Leaving this issue unprejudiced,} we believe it makes sense to explore the physical implications of $\cal TW$. After all it is constructed on the basis of coherence between the SM and gravity, and it is perhaps not unreasonable to hope that the incompleteness that affects it can be disregarded at least in a limited region of energy. Although very conjectural, the following does not seem to us totally unreasonable. }

\vskip0.5 cm

See also \cite{brandt24,karananas24,leclair24}

\section{Reduction and gauge fixing}
\label{weylgaugefixing}

The action \eqref{actionchir-symmconf} is a rather general (even though not the most general) chirally symmetric and Weyl-invariant action for $\cal TW$. In the sequel we somewhat simplify it in order to limit  ourselves to a contained, though hopefully still meaningful, discussion. The first simplification is to set $H_+=H_-=H$. That is, not only the $SU(2)$ gauge bosons are common to the left and right sectors, but also the Higgs field. All the formulas introduced above for $H_\pm$ are still valid with the simple substitution of symbols. Next we remark that \eqref{actionchir-symmconf} brings in new fields, beside the ones of gravity and SM, i.e. $\widehat C_\mu$ and $\widehat \varphi$. As we have already remarked, there is no compelling reason to introduce $\widehat C_\mu$, thus, as before,  in order to somewhat simplify the discussion, we will drop it, i.e. we set $\widehat C_\mu =0, \widehat C_{\mu\nu}=0$ and $\alpha=1, \epsilon=1$.  The field $\widehat \varphi$ is sufficient to implement AE Weyl invariance. 
The degrees of freedom contained in it are the same as the dofs in $\widehat \omega$. There is no mismatch, unlike with gauge fixing in the SM, for instance. Consequently fixing the gauge $\widehat \varphi=0$ does not require any FP ghosts. 
{
Choosing the gauge $\widehat \varphi=0$ produces a mass term for $H=H_\pm$ and $\Phi_{\pm}$ given by
\be
M^2_h=M_{h\pm}^2=M^2_\pm + \frac{{\mathfrak c}_\pm \zeta_{h\pm}}{\kappa_\pm} \quad\quad {\rm and} \quad\quad M_{\Phi\pm}^2= m^2_\pm+\frac{{\mathfrak c}_\pm \zeta_{\pm}}{\kappa_\pm},\label{M2Mi2}
\ee}
We will perform this operation in two steps. First we consider a partial breaking of the AE Weyl symmetry, by setting
\be
\varphi_+=0 \label{rightgaugefixing}
\ee
but leaving $\varphi_-$ free. This will leave ${\cal TW}_L$ unchanged and symmetric under the Weyl transformations parametrized by $\omega_-$, while the Weyl symmetry of ${\cal TW}_R$ is broken. This operation produces masses only for $H$ and $\Phi_+$,{
\be
 M_{h}^2= M^2_\pm+\frac{{\mathfrak c}_\pm \zeta_{h\pm}}{\kappa_\pm} \quad\quad {\rm and} \quad\quad M_{\Phi+}^2=m_+^2+\frac{{\mathfrak c}_+ \zeta_{+}}{\kappa_+},\label{M2Mi2+}
\ee}
respectively.

From now on the two theories ${\cal TW}_L$ and ${\cal TW}_R$, split. After the partial Weyl gauge fixing they are not anymore AE conformal-invariant, so that we call them ${\cal TW}_L$ and ${\cal T}_R$, respectively. What is more important, they have two different evolutions. 

\subsection{The fate of the two halves}

\subsubsection{The right-handed ${\cal T}_R$ model and its breaking}

We postulate an early (i.e. high energy) breakdown of the chiral symmetry in the right theory, according to the the pattern described in subsection \ref{ss:singlethiggsmechanism} with the scalar field identified with $\Phi_{+}$\footnote{Actually, in order to determine the minimum of the action we should analyze the global potential involving all the scalar fields. But since the $\Phi_+, \Phi_-$ and $H$ do not have mixed interaction terms and their minima are supposed to differ by several  order of magnitude.}, we think we can analyse them separately.

Setting $\Phi_+={v_++\rho_+}$, where $ \rho_+$ is a real field and $v_+= \sqrt{\frac {2M_+^2}{\lambda_+}}$, a real number, denotes the minimum of the potential, we have
\be
D_\mu \Phi_+ D_\mu \Phi_+&=& \partial_\mu  (v_++\rho_+)\partial^\mu (v_++\rho_+)\label{DmuHDmuH}
\ee
The quadratic terms in $\frac 12 D_\mu \Phi D_\mu \Phi- \sf V$ ($\sf V$ is the potential) are
\be 
\frac 12 \partial_\mu \rho_+ \partial^\mu\rho_+-\frac 12 \lambda v_+^2 \rho_+^2\label{quadraticaction}
\ee
The result is a massive scalar $\rho_+$.

{
Finally, the Yukawa couplings introduced in subsection \ref{s:yukawacoupling}, see eq.\eqref{yukawaR}, $\sqrt{g_+}\,\frac{y_+}2\,\overline \psi_R'  \Phi_+ (\psi'_R)^c +h.c.$, when $\langle \Phi_+\rangle =v_+$, produces a mass term proportional to $v_+$, see eq.\eqref{majod}.
Here $\psi'_R$  is a concise notation for the multiplet \eqref{Rspectrum}. As we have explained in subsection \ref{s:yukawacoupling}, see eq.\eqref{majod}, $\psi'_R$ are fermion that satisfy the Majorana equation, see \eqref{freeMajoranaeq}. }Since all these masses, as well as the mass of $\rho_+$, are proportional to $v_+$, they can be made as large as we wish. This allows us to suppose that the energy at which this Higgs mechanism takes place can be very high, between the Planck and the Grand Unification scales. We can call it the {\it primitive spontaneous chiral {(symmetry breaking scale}}. We recall that the just illustrated breaking does not affect the other gauge symmetries. { As for the symmetry under the AE diffeomorphisms $\widehat \xi^\mu$, it is broken by the Yukawa terms and by the gauge fixing, but a symmetry remains with respect to the diffeomorphisms parametrized by $\xi_-^\mu$ on the left and $\xi^\mu_+$ on the right. The $SU(2)$ gauge fields $W_\mu$ (valued in the Lie algebra ${\mathfrak su(2)}$) are in common to both theories and, up to this stage, untouched by the gauge fixing (the $SU(2)$ gauge symmetry survives the previous breaking). The actions $S^{(+)}_{aeg}$ and $S^{(-)}_{aeg}$ are certainly invariant under the diffeomorphisms parametrized by $\xi^\mu_+$ and $\xi^\mu_-$, respectively, {\it as long as we consider them separate theories}; but, since $W_\mu$ belong to both, only the invariance under the ordinary diffeomorphisms $\xi^\mu$ survives.} 

The AE Weyl symmetry is also limited to the transformations parametrized by $\omega$, it is broken by the gauge fixing in ${\cal TW}_R$, but a Weyl symmetry under the $\omega$ transformations survives in the ${\cal TW}_L$, whose fields remain massless.    

At this stage all the gauge bosons are massless. But there is room for a breakdown of the electroweak symmetry $SU(2)\times U(1)_L$, that generate masses for the three $SU(2)$ gauge bosons and for $H$. We recall that the masses of the gauge bosons and the Higgs field are of course the same  on both sides $M^2_W=\sfg_+^2 v_+^2 =\sfg_-^2 v_-^2$ { ($\sfg_\pm$ are the $SU(2)$ gauge couplings in the two sides)} and $M^2 _{h} =M_\pm^2+\frac{{\mathfrak c}_\pm \zeta_{h\pm}}{\kappa_\pm}$. {In this process both $SU(3)$ and the $U(1)_R$  symmetries remain unbroken.  
}

Let us summarize the situation. Before the chiral symmetry breaking we have a theory, $\cal TW$, split into two halves, with a different particle spectrum and each with its own gravity (mediated by $g_{+\mu\nu}$ and $g_{-\mu\nu}$, respectively) but having in common the $SU(2)$ vector bosons { and $H$ scalars. They communicate through the exchange of these particles: the relative terms  in the action that produce this interaction are the fermion-fermion-boson vertices on the two sides and the cubic and quartic self-interaction terms for the vector bosons, as well as the potential interaction terms involving the $H$ scalars. }Other direct interactions between the two halves are not permitted by the ubiquitous presence of the projectors $P_\pm$.  Of course there remains the possibility of indirect interactions: for instance the metric $g_{+\mu\nu}$ can interact with an $SU(2)$ vector boson, the latter with another $SU(2)$ vector boson, and this with $g_{-\mu\nu}$. These and similar interactions are likely to be extremely weak, except at extremely high energies, {if grand unification of forces is allowed in this theory.}

If this is so then we may suppose that the right and left-handed theories evolve almost independently, because the link between them is represented by the exchange of weak interacting $SU(2)$ bosons and scalars. After the spontaneous symmetry breaking the right-handed fermions  become very massive, and the interaction with the left-handed model even less important. { Needless to say, an analogy with what we know about SM, does not make sense under  these different boundary conditions, and  an {\it ab initio} analysis is necessary for the right-handed  model. Perhaps, however, some general features can be guessed; for instance  since the right-handed quarks and leptons are very heavy any nucleation is unlikely, leaving us with a gas of these massive particles. We could say that while ${\cal T}_R$ becomes almost `fossil' theory, ${\cal TW}_L$ is in full evolution, except for the exchange of weak interacting $SU(2)$ bosons, which can be treated as a perturbation. This `misconceived'} ${\cal T}_R$  theory has however many properties (heavy particles, weak interaction with luminous matter, coupled to its own gravity) that can help describe the dark matter. 

Refs.: \cite{chkareuli24,malekpour24,silva24}

\subsection{The left-handed ${\cal TW}_L$ model and the SM}

After the spontaneous chiral symmetry  breaking the left-side theory  ${\cal TW}_L$ almost decouples from the right- side one, and evolves almost autonomously. Its particle spectrum is that of the MSM plus a right-handed neutrino and a massive $\Phi_-$ scalar with a quartic potential, coupled to a gravity described by the $g_{-\mu\nu}$ metric. We recall that it is described by the action 
\be
S^{(c-)}_{ch-sym}= S^{(-)}_f  + S^{(-)}_g +  S^{(-)}_{aeg} +  ( S_{YdL}+h.c.)+   S^{(c-)}_{EH+aes+aed+C} \label{actionchir-symmconfL}
\ee
It is invariant under diffeomorphisms spanned by $\xi^\mu_-$ and under $SU(3)_L\times SU(2)\times U(1)_L$ gauge transformations. It is also invariant under Weyl transformations spanned by $\omega_-$. { As long as we can treat the interaction with the right-handed half as a disturbance, it would seem that we cannot repeat straightaway what we said for the entire $\cal TW$ model, for ${\cal TW}_L$ in isolation has an overall non-trivial odd-parity trace anomaly. However remember that this anomaly is a topological fact, it is related to the family's index theorem. From the theory and recent experiments of entangled systems we learn that topological features may survive separation. It is therefore likely that the cancelation of these anomalies in $\cal TW$ is not suppressed by the splitting into $\cal TW_+$ and $\cal TW_-$ due to the chiral breaking. We take it as a plausible hypothesis.

At this stage we can fix the gauge $\varphi_-=0$, which exhibits the SM coupled to EH gravity carried by the metric $g_{-\mu\nu}$.  In the course of the evolution it is to be expected that at a certain stage the gravity quanta become irrelevant and the gravitational field starts behaving entirely classically. From that stage on the theory is a renormalizable one on a classical gravitational background (odd trace anomalies have become irrelevant). After a while (in the cosmological evolution) the scale of electroweak breaking is reached, and the $SU(2)\times U(1)_L$ symmetry breaks down to the $U(1)$ of electromagnetism.  A story we already know, apart from the fate of the real $\Phi_-$ scalar field, still to be explored. Needless to say, it is a natural candidate for the inflaton field.  To allow such an interpretation, however, we have to arrange the parameters in such a way that $M^2_{\Phi_-}$ is negative, see eq.\eqref{M2Mi2}, and in combination with $\lambda_-$ satisfies the conditions of slow-roll inflation.

What has been said so far in this section is a conjectural, very qualitative, picture of the evolution of the $\cal TW$ theory. This paper intended to highlight the role of anomalies in the  bottom-up construction of a theory that enlarges the SM to include gravity. Now, starting from the above results and conjectures,  explicit calculations and quantitative checks are necessary to confirm or refute them. An entirely new chapter opens up. { But, perhaps, a comment is necessary from the start concerning the presence in $\cal TW$ of two metrics.  We have already remarked that our entire construction works even if we consider only a unique metric $g_{+\mu\nu}= g_{-\mu\nu}$. But we have maintained throughout the more general setting of two distinct metrics. In this more general case it would seem that we have two ways of measuring distances and times. In a sense this is true, but not for us. We, as observers, see only one metric, the other is seen only by a right-handed observer, if it can exist. The interaction between the two metrics is extremely feeble, although one cannot exclude that it may lead to observable consequences.

\section{Conclusions}

In the construction of the SM chiral gauge anomalies have played a prominent role. In a theory constructed with Weyl fermions  the absence of consistent chiral anomalies was one the basic principles, and looking at how it has been realized one cannot but feel admiration for its creators. But once we immerse this jewel in a background of gravity it seems that the fortune that accompanied them in their endeavor has abandoned us. New anomalies appear, and, precisely, what we prove in this paper is that it is impossible to get rid of them with the spectrum of fields offered by the SM. We have to be more specific. The anomalies we are talking about, the dangerous ones, which we have named type O (O stands for obstruction), are not only the traditional consistent chiral gauge anomalies, which vanish in the MSM, but also by those generated by the diffeomorphisms and the mixed ones, which fortunately also vanish in the MSM. But among the type O anomalies there is another kind of them, which has apparently been disregarded in the literature. They are the chiral trace anomalies which appear in general in models of Weyl fermions coupled to a metric or a gauge potential. Like all the type O anomalies they signal the fundamental drawback of these models, the corresponding Weyl propagators are missing. This is due to a topological obstruction  (a non-perturbative fact) guaranteed by the family's index theorem. 

Once the appropriate anomaly hygiene is applied to the SM and its extensions, minimally coupled to a background gravitational field, we found out that none of them is free of these anomalies. In fact the more `resistant' ones are the chiral trace anomalies induced by the $SU(2)$ gauge fields of the SM. These are the results of part I of our paper. In the second part we have proposed a scheme to overcome these difficulties. We have noticed first that a left-right symmetric model is fit to tame this problem. If it is not in absolute the only way, it is certainly the simplest and more elegant. Therefore we proposed a L-R model which is free of all type O anomalies -- as an example --, but in fact it can be enlarged to an entire family of similar models, this is the reason why we prefer to call it a scheme. This model has a left-right symmetry, which we named primitive chiral symmetry. It almost splits into two models, one with left-handed fermions, the other with right-handed ones, each of them almost a copy of an enlarged MSM. The almost refers to the fact that the $SU(2)$ gauge potentials are in common to both. Each of them is coupled to a different gravity theory, characterized by its own metric. At this stage we came across a dilemma: to be or not to be conformal. 

As it is well-known by now, it is rather simple to make a local field theory Weyl-invariant by simply adding a new field, a dilaton, to the spectrum. Then the problem is twofold: is this enlargement required by physics? is it compatible with quantization? We do not have definite answers to either question. Tentatively, we are oriented to think that it would be quite natural (and even experimentally supported by the Bjorken scaling limit) that at very high energies, such as those of the very early universe, the physics be insensitive to a dimensionful scale. This is what an approach \`a la Wilson would suggest, if we are allowed to be inspired by the analogy with low energy statistical model: the high energy limit of the universe regressing toward the beginning should be a UV fixed point, most likely a conformal fixed point. On the other hand it should be in accord with quantization. Here we have found a real difficulty: while it is possible to figure out conformal models, similar to ours, that are renormalizable, there are some indications that unitarity is not respected, suggesting a likely incompleteness of the spectrum of fields.
 
Notwithstanding these difficulties we believe that our scheme, dictated on the basis of mathematical consistency, should be able to describe at least a limited stretch of the evolution of the universe. We have envisaged an early breaking of the primitive chiral symmetry, which should have split the history of the two L and R models, the latter describing the physics of a gas of very massive quarks and leptons,  the former evolving towards a model of ordinary classical gravity coupled to the ordinary SM, while keeping, so to speak, the two in touch via the weak interacting $SU(2)$ gauge fields. 

Several aspects of the above scheme are still too crudely described or incomplete. For instance we do not have an explanation for the mechanism that can triggers the early chiral symmetry breaking, but from this side we are in good company with many phenomena that we can describe but not explain. In this paper we have not considered the consequences of our model involving the cosmological constants as well as other cosmological aspects. { Neither have we analysed the evolution of the right model.} Needless to say many explicit checks and calculations are needed. At the same time though, this scheme is not only mathematically consistent (it is anomaly-free) but it { may describe} the `strange'  chiral asymmetry of the SM by exhibiting it as the existence of the other face, the dark matter world, and leads naturally at low enough energy to the SM coupled to classical gravity equipped with inflation.  

Refs.: \cite{bambi2021,Baumgart2022,bianchi24,coriano24b,mukhanov2005,Mukhanov2016,Weinberg2008}
\vskip 1cm

{\bf Note added.} We have used the axially-extended or hypercomplex formalism in order to formulate the $\cal T$ model in a very general form, where both vector and axial potentials, metric and pseudometric as well as scalars and pseudoscalars appear. The latter are needed only in the hatted formulation of the action. Once we split $\cal T$ into ${\cal T}_L\cup {\cal T}_R$ we can assume even parity metrics, potentials and scalars everywhere. Therefore when the model ${\cal T}_L$ is required to contain the SM, we missed in the text to specify that the metrics and the gauge potentials as well as the scalars, carrying $\pm$ subscripts, are to be understood as parity even fields. 

 A further important supplementary remark is: cancelation of left-right gauge induced $SU(2)$ trace anomalies requires that the $SU(2)$ left and right couplings be equal: $\sfg_+=\sfg_-$. 

\vskip 1cm
{\bf Acknowledgements.} We would like to thank Predrag Dominis Prester and Maro Cvitan for their past collaboration and recent discussions. 

}

 \section{Appendices}

\subsection*{A. Gamma matrices in flat spacetime}

In absence of gravity we refer to theories on a flat Minkowski spacetime with 
flat metric $\eta_{\mu\nu}$ with mostly $ (-) $ signature. The gamma matrices satisfy $\{\gamma^\mu,\gamma^\nu\}= 2 \eta^{\mu\nu}$ and
\be
\gamma_{\mu}^\dagger= \gamma_0\gamma_\mu\gamma_0\0  {.}
\ee

The charge conjugation operator $C$ is defined to satisfy
\be
\gamma_{\mu}^T = -C^{-1} \gamma_\mu C, \quad\quad CC^*=-1, \quad\quad
CC^\dagger=1\label{C}.
\ee 
The chiral matrix $\gamma_5=i\gamma^0\gamma^1\gamma^2\gamma^3$ has the
properties
\be
\gamma_5^\dagger=\gamma_5, \quad\quad (\gamma_5)^2=1, \quad\quad C^{-1} \gamma_5
C= \gamma_5^T\0  {.}
\ee
and
\be
\tr (\gamma_5 \gamma_\mu\gamma_\nu\gamma_\lambda\gamma_\rho)= -4 i \varepsilon_{\mu\nu\lambda\rho}\label{tracegamma5}
\ee
The chiral projectors
\be
P_L\equiv P_-= \frac {1-\gamma_5}2, \quad\quad P_R\equiv P_+= \frac {1+\gamma_5}2,\quad\quad {\rm with} \quad\quad P_++P_-=1,\quad\quad P_\pm^2=P_\pm .\label{gammaproj}
\ee
In terms of a Dirac fermion $\psi$ Weyl fermions are defined by $\psi_{L,R} = P_{L,R} \psi$.

The  generators of the Lorentz group in the `anti-hermitean' version are $\Sigma_{\mu\nu} =\frac 14
[\gamma_\mu,\gamma_\nu]$.

Under parity a Dirac spinor transforms as
\be
\EP \psi({\vec x},t) \EP^{-1} = e^{i\beta_p} \gamma_0 \psi(-{\vec x},t)\label{paritypsi}
\ee
Under charge conjugation it transform as
\be
\EC \psi({\vec x},t) \EC^{-1} = e^{i\beta_c} \gamma_0 C\psi^*(\vec x, t) \label{chargeconjpsi}
\ee
and for a complex scalar field we have
\be
\EP  \phi({\vec x},t) \EP^{-1} = e^{i\alpha_p} \phi(-{\vec x},t), \quad\quad \EC \phi({\vec x},t) \EC^{-1} = e^{i\alpha_c}\phi^\dagger(\vec x, t) \label{paritychargephi}
\ee
where $\alpha_p,\alpha_c,\beta_p,\beta_c$ are arbitrary phases, which we usually set equal zero.

A vector field $V_\mu$ transform as
\be
\EP  V_\mu(\vec x,t)  \EP^{-1} =  V^\mu(-{\vec x},t), \quad\quad \EC V_\mu({\vec x},t) \EC^{-1} = -V_\mu(\vec x, t) \label{paritychargeVmu}
\ee
{
and for a pseudovector field $A_\mu$

\be
\EP  A_\mu(\vec x,t)  \EP^{-1} = - A^\mu(-{\vec x},t), \quad\quad \EC A_\mu({\vec x},t) \EC^{-1} = A_\mu(\vec x, t) \label{paritychargeVmu}
\ee

A metric $g_{\mu\nu}$ transforms as
\be
\EP  g_{\mu\nu}(\vec x,t)  \EP^{-1} =  g^{\mu\nu} (-{\vec x},t), \quad\quad \EC g_{\mu\nu}({\vec x},t) \EC^{-1} = g_{\mu\nu}(\vec x, t) \label{paritychargegmunu}
\ee
while 
\be
\EP  f_{\mu\nu}(\vec x,t)  \EP^{-1} =-  f^{\mu\nu} (-{\vec x},t), \quad\quad \EC f_{\mu\nu}({\vec x},t) \EC^{-1} =- f_{\mu\nu}(\vec x, t) \label{paritychargefmunu}
\ee
}

\subsection*{B. Euclidean field theory calculations}

Most classical and quantum field theories, such a the SM, are formulated in a Minkowski metric (both flat and non-flat). However true calculations with Minkowski metrics are often not available and, in order  to extract concrete results, theoretical physicists have been obliged to sell their souls to the Euclidean devil. This is done via Wick rotations. At the end of the calculations the results are turned Minkowski via inverse Wick rotations.

A Wick rotation means: $x^0\to \tilde x^0= i x^0, k^0\to \tilde k^0=i k^0$ and $\gamma^0 \to \tilde \gamma^0= i\gamma^0$, while $x^i, k_i, \gamma^i$  remain unchanged.  A tilde on top of an object represents its Wick-rotated correspondent.  We have in particular
\be
\{ \tilde \gamma^\mu, \tilde \gamma^\nu\}=2 \tilde \eta^{\mu\nu}\equiv - 2 \delta^{\mu\nu}\label{Euclgammagamma}
\ee
and
\be
\tilde \gamma_\mu^\dagger = -\tilde \gamma_\mu \label{tildegammadagger}
\ee
The chiral matrix $\gamma_5$ changes sign under a Wick rotation, $\tilde\gamma_5=-\gamma_5$.

In this article Wick rotations are used for perturbative and non-perturbative calculations (via the SDW method) in the intermediate steps. A particular caution has to be employed with Weyl fermions. It is not possible to write down a full Euclidean action for Weyl fermions using the Osterwalder-Schrader prescription, \cite{Osterwalder1972}. The attempt to Wick rotate a Lagrangian like $i \overline \psi\gamma^\mu \partial_\mu\psi$, where $\psi$ is Dirac, ends up in a doubling of fermionic degrees of freedom: $\overline \psi$ can only be a spinor independent of $\psi$. More elaborated approaches where a Wick rotation of a Weyl spinor $\psi$ is accompanied by a rotation of  the same cannot comply with the hermiticity, see \cite{I}. In sum it is a nonsense to make a Euclidean calculation for Weyl fermions starting from an action formulated in terms of Euclidean Weyl fermions proper, because this would change the nature of the problem. The best we can do is use the Wick rotation to make sense of the Feynman and path integrals wherever they are met in the course of the calculations. In perturbative calculations we set up the Feynman diagram calculations in Minkowski terms and switch to Euclidean when evaluating the Feynman integrals. In the SDW method we Wick-rotate the square Dirac operator to transform it into a self-adjoint operator and make sense of the $s$-integrals. In both cases we use Wick rotations only in such junctures, without any attempt to identify the Euclidean field theories underlying them, which is impossible.

Finally, in the third approach we use, which is based on the family's index theorem, we start at the very beginning from the Euclidean formulation of the Dirac operator, because  a proof of that theorem in a Lorentzian metric is not available. Anyhow, this does not meet the just mentioned difficulties, because  what we analyse is the invertibility of the Dirac operator in isolation, without reference to a particular action it is inserted in. We assume that the results of the theorem are true also for the Minkowski version.

\subsection*{C. Gauge fixing and ghosts}

The gauge fixing and ghost action for a non-Abelian YM theory coupled to a metric, in the Lorenz gauge, is
\be
S^{c,\bar c}_{g.f.} = \int d^4 x\, \sqrt{g}\, \left(b^a \widetilde\nabla_\mu V^a_\nu g^{\mu\nu} +\frac \alpha 2 b^a b^a - g^{\mu\nu} \bar c^a\widetilde\nabla_\mu (D_\nu c)^a \right)\label{Scbarcgf}
\ee
summed over $a$, where $\widetilde\nabla$ denotes the metric covariant derivative with Christoffel symbols $\widetilde \Gamma$, \eqref{Christ3} with $\alpha=1$, and $D_\mu c = \partial_\mu c + [V_\mu, c]$. This action is invariant under diffeomorphisms and also under Weyl transformations, provided we assign Weyl weight -2 to $b^a$ and $\overline c^a$, and 0 to $c^a$. Eq.\eqref{Scbarcgf} is of course invariant under the gauge BRST transformations \eqref{deltall} and $\delta_c \overline c^a= b^a, \delta_c b^a=0$.

A full gauge fixing and ghost action { for diffeomorphisms in} the DeDonder gauge is
\be
S^{(\xi,\bar \xi)}_{g.f.}= \int d^4 x\, \sqrt{g}\,\left( b_\lambda\widetilde \Gamma^\lambda_{\mu\nu} g^{\mu\nu} + \frac {\alpha}2 b_\lambda g^{\lambda \rho} b_\rho\, e^{2\varphi} - \overline \xi_\lambda g^{\mu\nu} \partial_\mu \partial_\nu \xi^\lambda\right) \label{xixibargf}
\ee
which is BRST invariant. For the $g_{\mu\nu}$ and $\xi^\mu$ BRST transformations see (\ref{deltaxigmunu},\ref{deltaximu}), for $\overline \xi_\lambda $: $\delta_\xi \overline\xi_\lambda= - b_\lambda +\xi\!\cdot\! \partial \overline \xi_\lambda+ \partial _\lambda \xi^\rho \overline \xi_\rho$, while $b_\lambda$ transforms as an ordinary covariant vector. Eq.\eqref{xixibargf} is also Weyl-invariant provided we assign Weyl weight -2 to $b_\lambda$ and $\overline\xi_\lambda$ and weight 0 to $\xi^\mu$.

{
\subsection*{D. Consistency of scalar cocycles}

Let us consider, for instance, the cocycle \eqref{scalartraceanomfull2}. To understand its origin from the second term in  \eqref{scalartraceanom}, which comes from an explicit triangle diagram calculation, and its relation with other diagrams, we must abandon the identification of $\Phi$ as the fluctuating field and write instead $\Phi=\Phi_0+\phi$, where $\Phi_0$ is a classical constant background and $\phi$ is the fluctuating field. We split the cocycle   \eqref{scalartraceanomfull2} in a series expansion based on the number $i$ of fluctuating (infinitesimal) fields:
\be
\Delta_\omega = \sum_{i=0}^\infty \Delta_\omega^{(i)}\label{seriescocycles}
\ee
where
\be 
&&\Delta_\omega^{(0)}=0, \quad\quad \Delta_\omega^{(1)}= \frac 13 \int d^4x\, \omega\left(\partial_\mu\partial_\nu h^{\mu\nu} -\square h_\lambda^\lambda\right) \Phi_0^2,\0\\
&& \Delta_\omega^{(2)}= \int d^4x\, \omega\left(\eta^{\mu\nu} \partial_\mu \phi \partial_\nu \phi+
\frac 23 \partial_\mu\partial_\nu h^{\mu\nu} \Phi_0 \phi -\frac 23 \square h_\lambda ^\lambda \Phi_0 \phi\right),  \quad {\rm etc.}\label{Deltai}
\ee
In parallel we must change the Weyl transformation in accordance with this series expansion
\be
&&\delta_\omega^{(0)}h_{\mu\nu} =  \omega\eta_{\mu\nu},  \quad \delta_\omega^{(1)}h_{\mu\nu}=2\omega h_{\mu\nu} ,\quad \delta_\omega^{(2)}h_{\mu\nu}=0, ...\label{seriesdeltah}\\
&&\delta_\omega^{(0)}\phi=- \omega \Phi_0, \quad \delta_\omega^{(1)}\phi=-\omega \phi, \quad \delta_\omega^{(2)}\phi=0,\quad...\label{seriesdeltaphi}
\ee
and expand the WZ consistency condition $\delta_\omega \Delta_\omega=0$. The first two non-trivial terms are 
\be
&&\delta^{(0)}_\omega \Delta^{(1)}_\omega + \delta^{(1)}_\omega \Delta_\omega^{(0)}=0\0\\
&&\delta^{(1)}_\omega \Delta^{(1)}_\omega + \delta^{(0)}_\omega \Delta_\omega^{(2)}=0 \label{descentWZ}
\ee
It is not hard to prove that they hold true. The series goes on with
\be 
\delta^{(1)}_\omega \Delta^{(2)}_\omega + \delta^{(0)}_\omega \Delta_\omega^{(3)}=0, \quad \ldots \label{descentWZ}
 \ee
but { in this case} we must introduce new approximation terms for $\Delta_\omega$, which we dispense with here. It is interesting to notice that, apart from the first terms of $ \Delta_\omega^{(2)}$, all the other terms do not come from the triangle diagram calculation mentioned before eq.\eqref{scalartraceanom}. They are in fact higher order terms which can be obtained from the fourth point diagrams where the fermion e.m. tensor is attached to three vertices ($V_{ffh}, V_{ff\phi}, V_{ff\Phi_0}$), by means of four fermion propagators.  { This is another example of a trace calculation which is undecidable at the lowest perturbative order}.
}

{
\subsection*{E. Dilaton and dimensions}
\label{ss:dimensions}

The field $\varphi$ (the dilaton), introduced in section \ref{s:weylgeometry}, is not the only possible choice. We could have set $\varphi = \ell \phi$, where $\ell$ is a constant with the dimension of a length and $\phi$ is a dimension one field. In this case $\phi$ becomes an ordinary real scalar field with a canonical kinetic term (after normalization). The problem with this choice is its identification with the field $\sigma$ in the WZ terms: $\sigma=\ell \phi$. It transforms the WZ term into a five dimension operator expression multiplied by $\ell$. In the framework of renormalization theory such a term { renders the theory} formally non-renormalizable.

A similar  problem arises also with the identification $\sigma= \log(\ell \chi)$, as already pointed out in subsection \ref{sss:WZtoTW}. The choice $\sigma=\varphi$ seems the most logical, but it is itself not exempt of problems in subsequent developments. Once we have introduced a dimensionless field $\varphi$ in a theory, upon renormalization we must allow for the appearance of all the terms, with the right dimensions and symmetries, involving also $\varphi$. In our case we can construct for instance the expression
\be
Q = \square \varphi -\partial_\mu \varphi \partial^\mu \varphi +\frac 16 R\label{exprQ}
\ee
Under a Weyl transformation it transforms as $Q \rightarrow e^{-2\omega} Q$. Therefore
\be
\int d^4 x\, \sqrt{g}\, Q^2 \label{intQ2}
\ee
is Weyl-invariant, has the right dimensions and thus it can arise in the renormalization process. Unfortunately it contains a quartic kinetic term, which added to the quadratic kinetic term, splits the propagator into a massless plus a massive one, the latter representing either a tachyon or a ghost, thus possibly affecting unitarity.}

\end{document}